\documentclass{article}

\usepackage{arxiv}

\usepackage[utf8]{inputenc} 
\usepackage[T1]{fontenc}    
\usepackage{hyperref}       
\usepackage{url}            
\usepackage{booktabs}       
\usepackage{amsfonts}       
\usepackage{nicefrac}       
\usepackage{microtype}      
\usepackage{lipsum}		
\usepackage{graphicx}
\usepackage{natbib}
\usepackage{doi}

\usepackage{hyperref}
\usepackage{subfig}
\usepackage{microtype}
\usepackage{amsmath}
\usepackage{amssymb}
\usepackage{tikz}
\usetikzlibrary{positioning}
\usepackage{pgfplots,pgfplotstable}
\pgfplotsset{compat=1.3} 

\newcommand{\mC}{\mathbf{C}}

\newcommand{\mQ}{\mathbf{Q}}
\newcommand{\xx}{\mathbf{x}}

\newcommand{\vb}{\mathbf{b}}
\newcommand{\vc}{\mathbf{c}}
\newcommand{\ve}{\mathbf{e}}
\newcommand{\vf}{\mathbf{f}}
\newcommand{\vg}{\mathbf{g}}

\newcommand{\vp}{\mathbf{p}}
\newcommand{\vt}{\mathbf{t}}
\newcommand{\vu}{\mathbf{u}}
\newcommand{\vx}{\mathbf{x}}
\newcommand{\vv}{\mathbf{v}}
\newcommand{\cB}{\mathcal{B}}
\newcommand{\cC}{\mathcal{C}}
\newcommand{\cD}{\mathcal{D}}
\newcommand{\cE}{\mathcal{E}}
\newcommand{\cF}{\mathcal{F}}
\newcommand{\cG}{\mathcal{G}}

\newcommand{\cP}{\mathcal{P}}
\newcommand{\cU}{\mathcal{U}}
\newcommand{\cV}{\mathcal{V}}
\newcommand{\Transp}{\mathrm{T}}
\newcommand{\vNull}{\mathbf{0}}

\definecolor{lineOne}{HTML}{1B9E77}
\definecolor{lineTwo}{HTML}{D95F02}
\definecolor{lineThree}{HTML}{7570B3}
\definecolor{lineFour}{HTML}{E7298A}

\title{Unified Smooth Vector Graphics: Modeling Gradient Meshes and Curve-based Approaches Jointly as Poisson Problem}

\author{ 
Xingze Tian\\
	Department of Computer Science\\
	Friedrich-Alexander-Universit{\"a}t Erlangen-N{\"u}rnberg\\
	Erlangen, Germany \\
	\texttt{xingze.tian@fau.de} \\
	\And
	\href{https://orcid.org/0000-0002-3020-0930}{\includegraphics[scale=0.06]{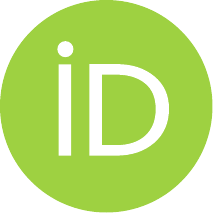}\hspace{1mm}Tobias G{\"u}nther} \\
	Department of Computer Science\\
	Friedrich-Alexander-Universit{\"a}t Erlangen-N{\"u}rnberg\\
	Erlangen, Germany \\
	\texttt{tobias.guenther@fau.de} \\
}

\renewcommand{\shorttitle}{Unified Smooth Vector Graphics}

\hypersetup{
pdftitle={Unified Smooth Vector Graphics: Modeling Gradient Meshes and Curve-based Approaches Jointly as Poisson Problem},
pdfsubject={cs.GR},
pdfauthor={Xingze Tian, Tobias G{\"u}nther},
pdfkeywords={Smooth vector graphics, diffusion curve, gradient mesh, Poisson problem},
}

\begin{document}
\maketitle

\newcommand{\resultrow}[7]{%
\begin{tikzpicture}%
    \node[anchor=south west,inner sep=0] (image) at (0,0) {\includegraphics[width=0.19\linewidth]{#2}};%
    \begin{scope}[x={(image.south east)},y={(image.north west)}]%
        \node[inner sep=0pt] (crop) at (0.1843,0.8157)
    {\includegraphics[width=.07\textwidth]{#1}};
    \end{scope}%
\end{tikzpicture}\hfill\hfill%
\begin{tikzpicture}%
    \node[anchor=south west,inner sep=0] (image) at (0,0) {\includegraphics[width=0.19\linewidth]{#4}};%
    \begin{scope}[x={(image.south east)},y={(image.north west)}]%
        \node[inner sep=0pt] (crop) at (0.1843,0.8157)
    {\includegraphics[width=.07\textwidth]{#3}};
    \end{scope}%
\end{tikzpicture}\hfill\hfill%
\includegraphics[width=0.19\linewidth]{#5}\hfill%
\includegraphics[width=0.19\linewidth]{#6}\hfill%
\includegraphics[width=0.19\linewidth]{#7}%
}%

\begin{abstract}
Research on smooth vector graphics is separated into two independent research threads: one on interpolation-based gradient meshes and the other on diffusion-based curve formulations. 
With this paper, we propose a mathematical formulation that unifies gradient meshes and curve-based approaches as solution to a Poisson problem. 
To combine these two well-known representations, we first generate a non-overlapping intermediate patch representation that specifies for each patch a target Laplacian and boundary conditions. 
Unifying the treatment of boundary conditions adds further artistic degrees of freedoms to the existing formulations, such as Neumann conditions on diffusion curves.
To synthesize a raster image for a given output resolution, we then rasterize boundary conditions and Laplacians for the respective patches and compute the final image as solution to a Poisson problem. 
We evaluate the method on various test scenes containing gradient meshes and curve-based primitives.
Since our mathematical formulation works with established smooth vector graphics primitives on the front-end, it is compatible with existing content creation pipelines and with established editing tools.
Rather than continuing two separate research paths, we hope that a unification of the formulations will lead to new rasterization and vectorization tools in the future that utilize the strengths of both approaches.
\end{abstract}


\begin{figure*}[t]
  \centering%
  \resultrow
  {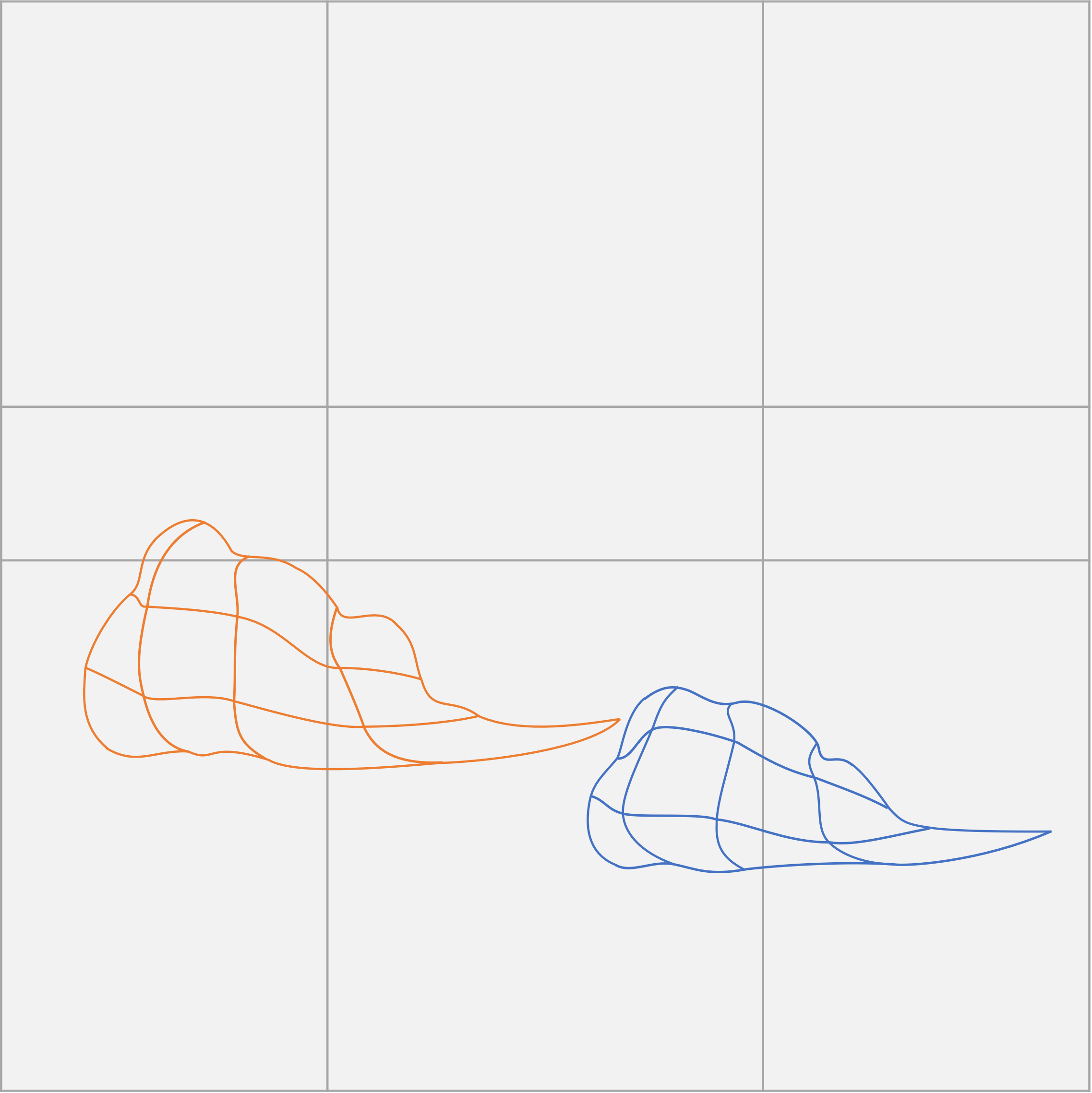} 
  {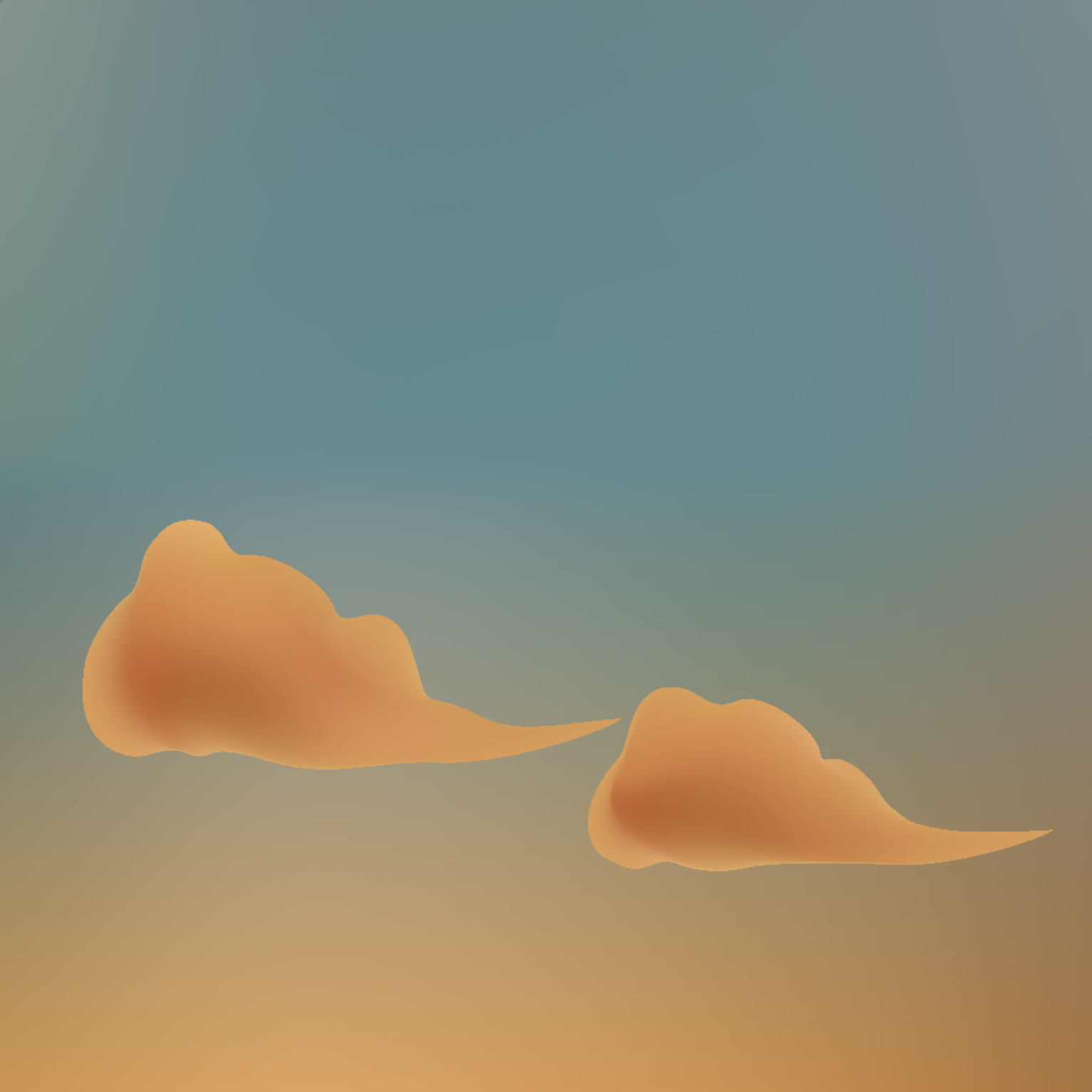} 
  {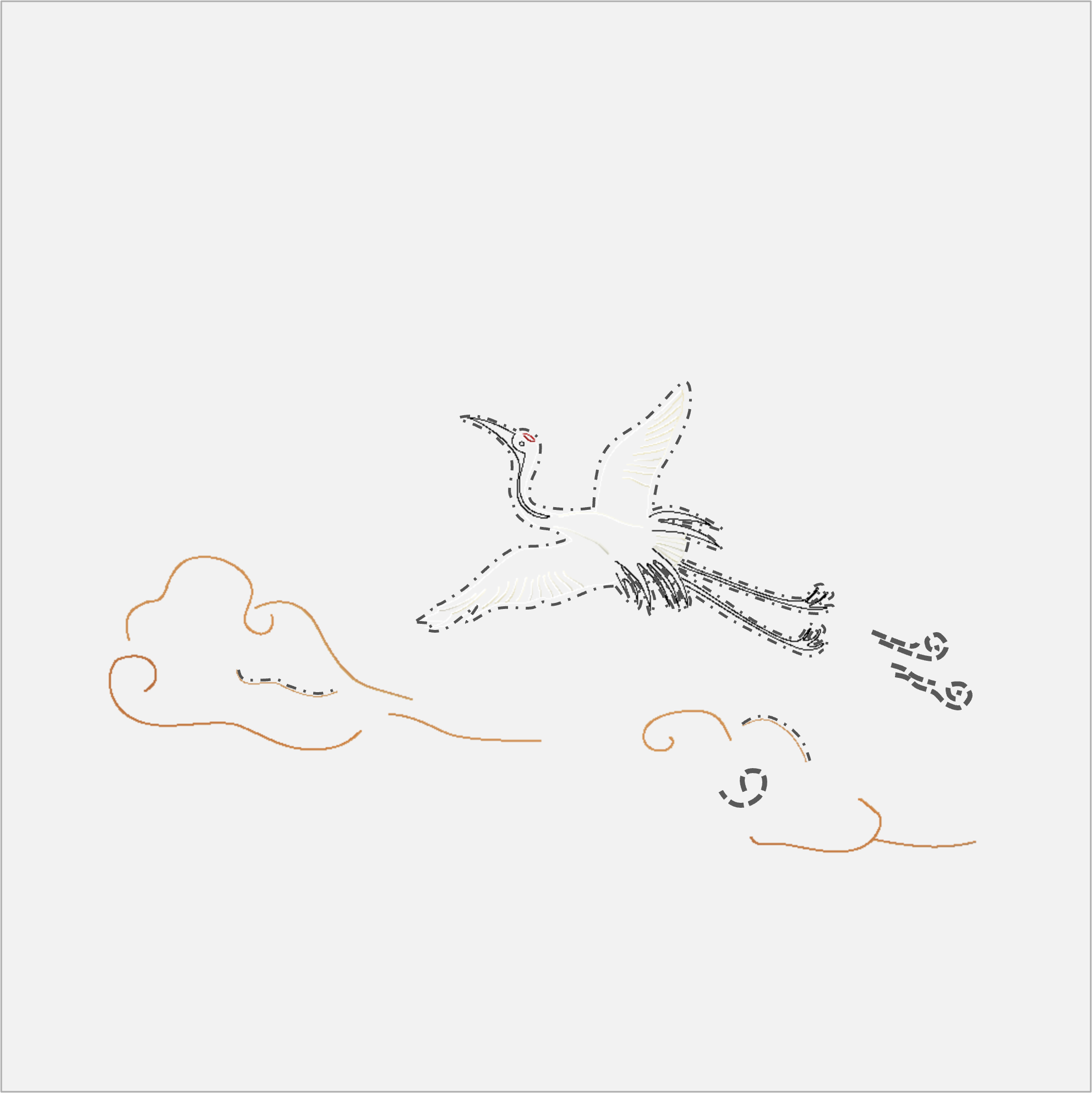} 
  {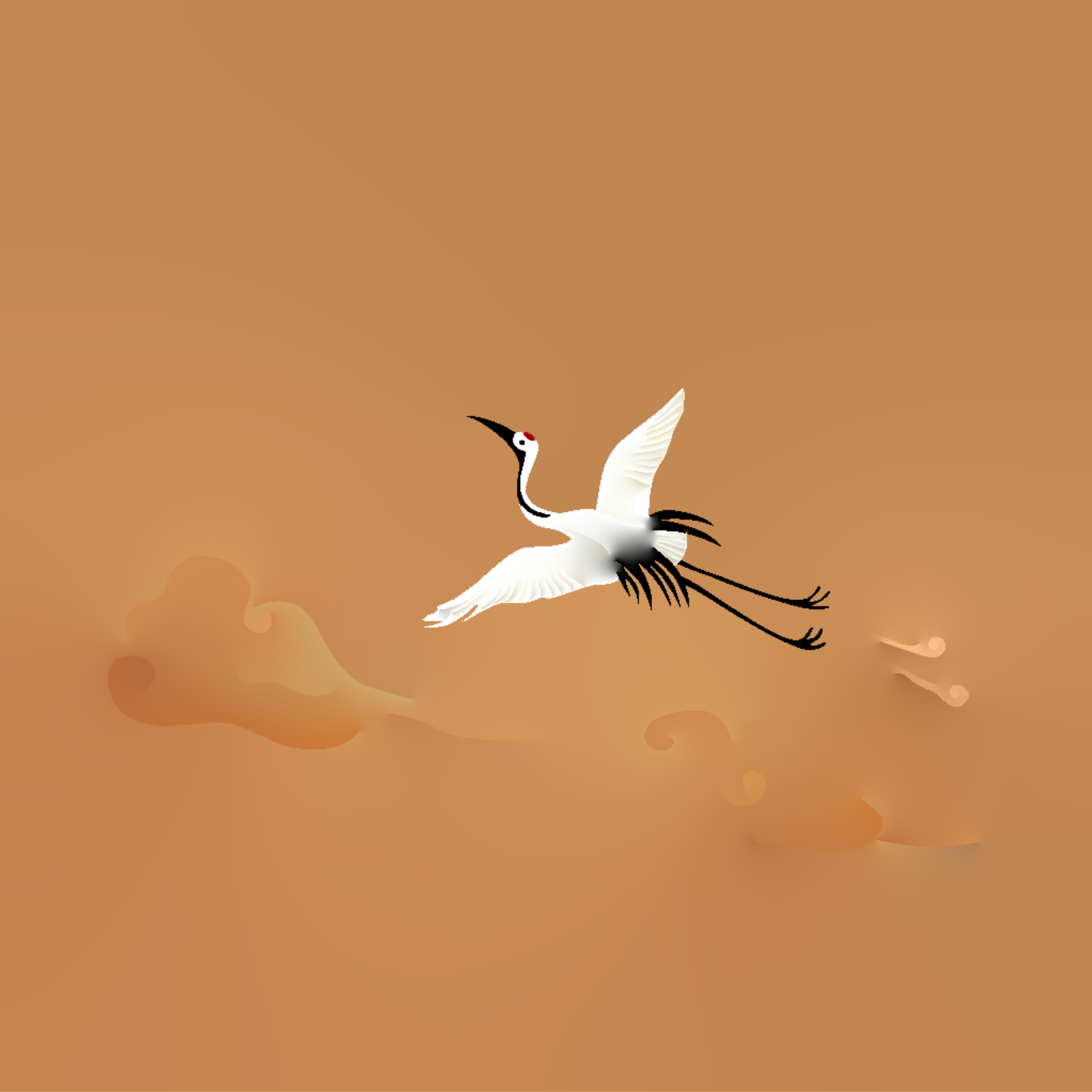} 
  {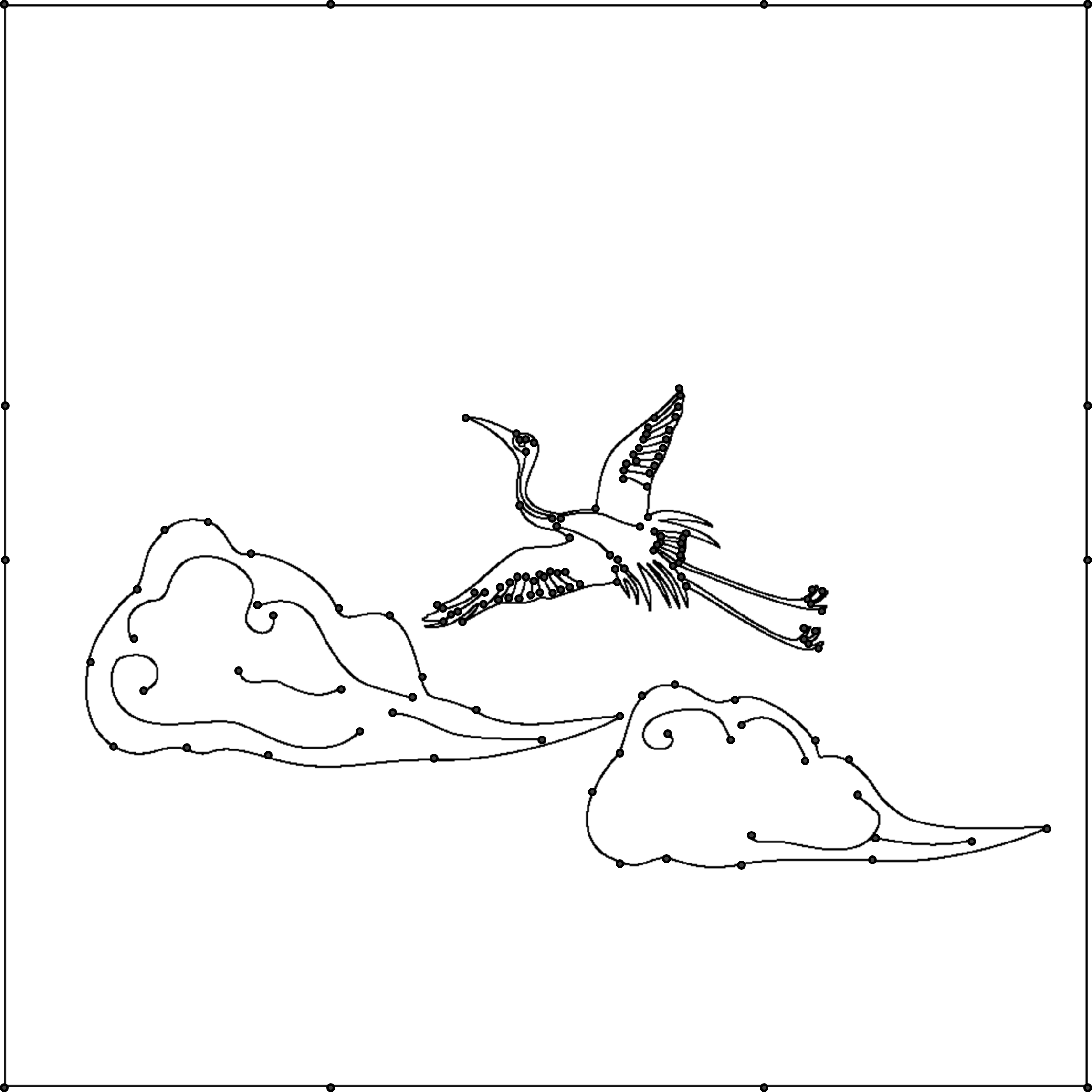} 
  {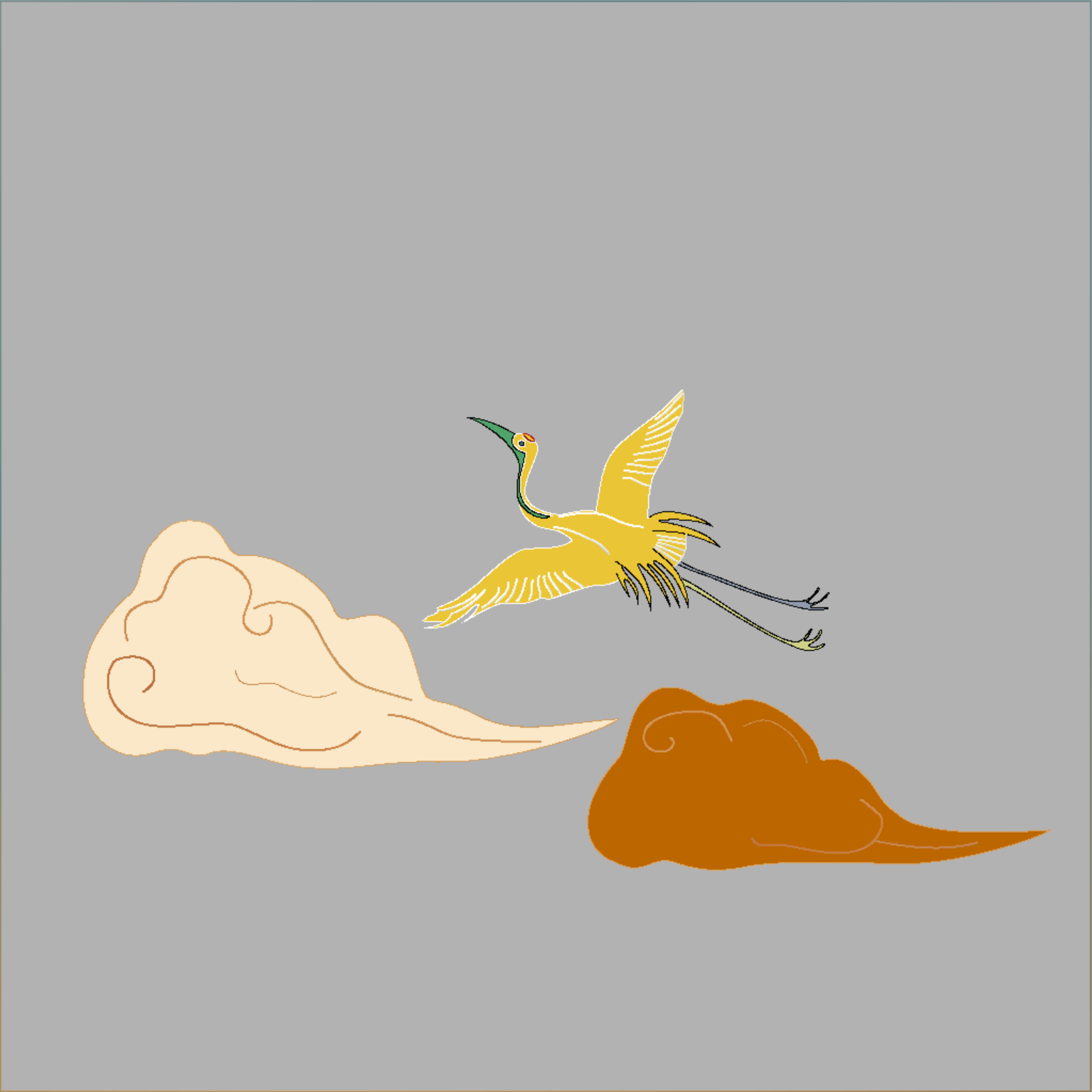} 
  {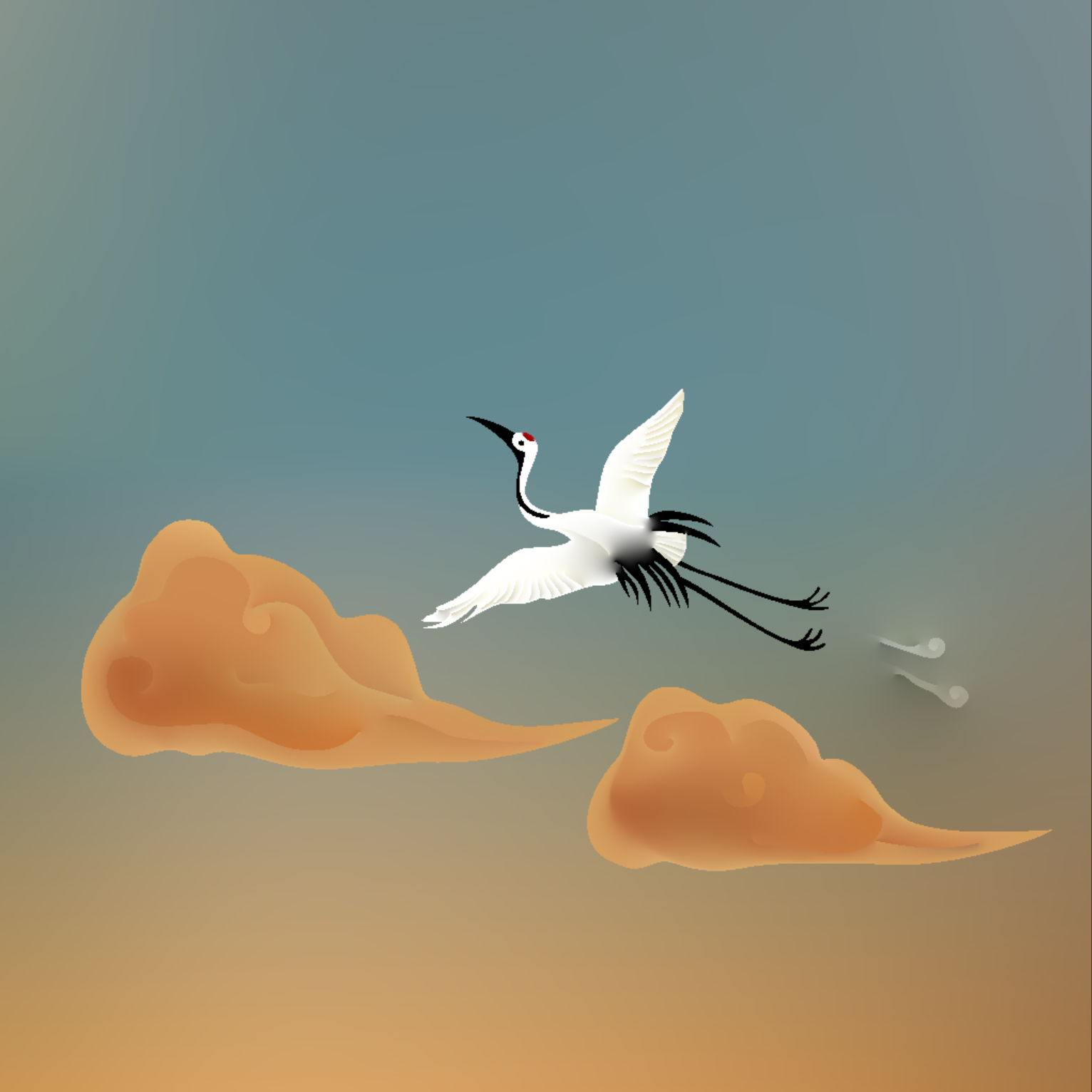} 
  \begin{minipage}{0.19\linewidth}%
      \centering%
      \scriptsize%
      (a) input meshes
  \end{minipage}\hfill\hfill%
  \begin{minipage}{0.19\linewidth}%
      \centering%
      \scriptsize%
      (b) input curves
  \end{minipage}\hfill\hfill%
  \begin{minipage}{0.19\linewidth}%
      \centering%
      \scriptsize%
      (c) undirected edge graph
  \end{minipage}\hfill%
  \begin{minipage}{0.19\linewidth}%
      \centering%
      \scriptsize%
      (d) unified patch representation
  \end{minipage}\hfill%
  \begin{minipage}{0.19\linewidth}%
      \centering%
      \scriptsize%
      (e) final synthesized image
  \end{minipage}%
  \\
  \caption{In this paper, we present an algorithm that unifies the image synthesis of smooth vector graphics containing, gradient meshes, diffusion curves, and Poisson curves in the same scene. Given a set of gradient meshes (a) and diffusion and Poisson curves (b), we first resolve the geometric intersections by forming an undirected edge graph (c), from which a unified patch representation is constructed that divides the domain into separate regions with well-defined Dirichlet and/or Neumann boundary conditions (d). Lastly, the image can be computed with an off-the-shelf Poisson solver (e). 
  }%
  \label{fig:teaser}
\end{figure*}

\section{Introduction}

\begin{figure}%
    \centering%
    \captionsetup{justification=centering}%
    \begin{minipage}{0.24\linewidth}%
    \centering%
    \includegraphics[width=\linewidth]{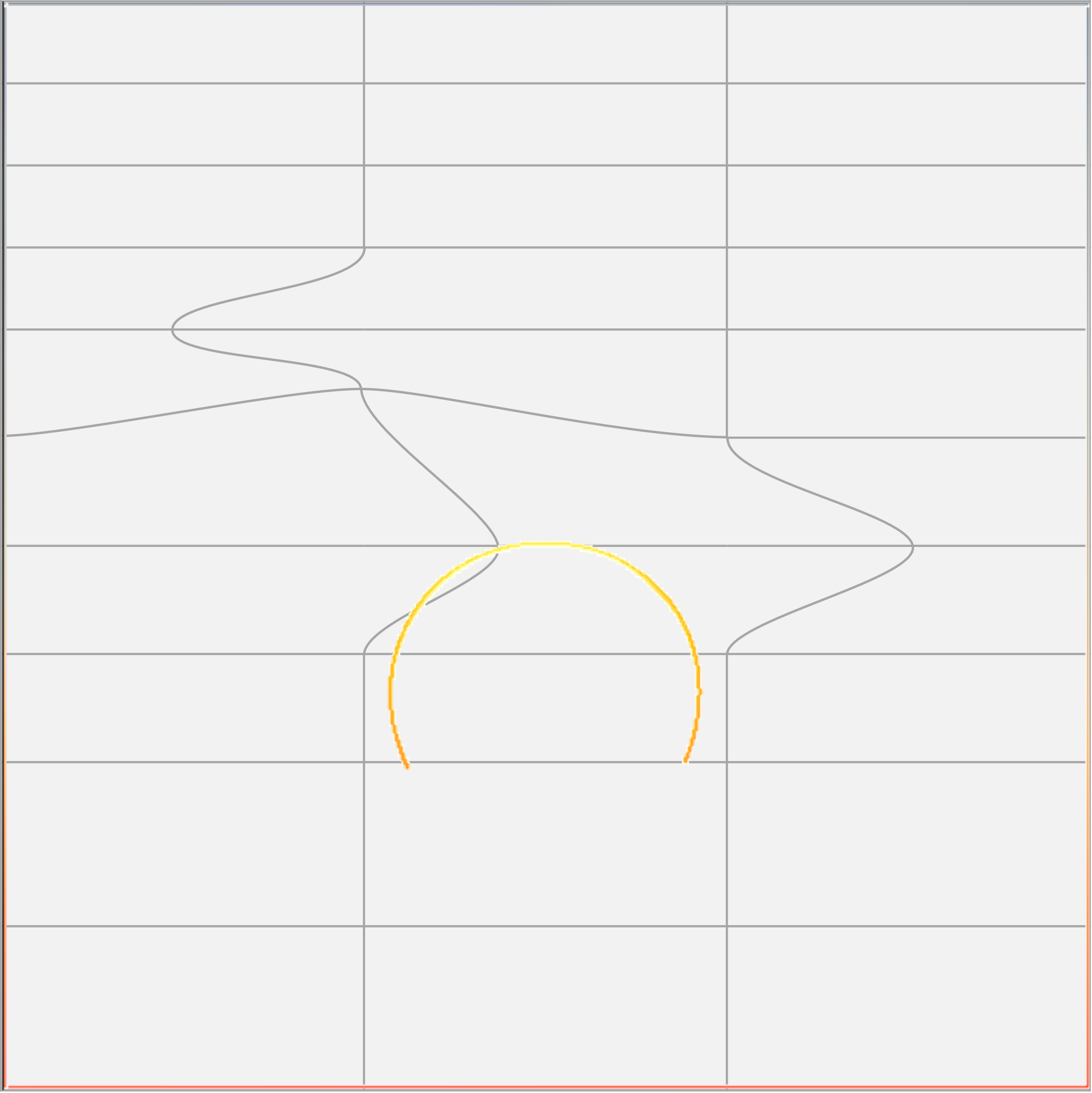}\\%
    \scriptsize%
    (a) Input primitives%
    \end{minipage}\hfill%
    \begin{minipage}{0.24\linewidth}%
    \centering%
    \includegraphics[width=\linewidth]{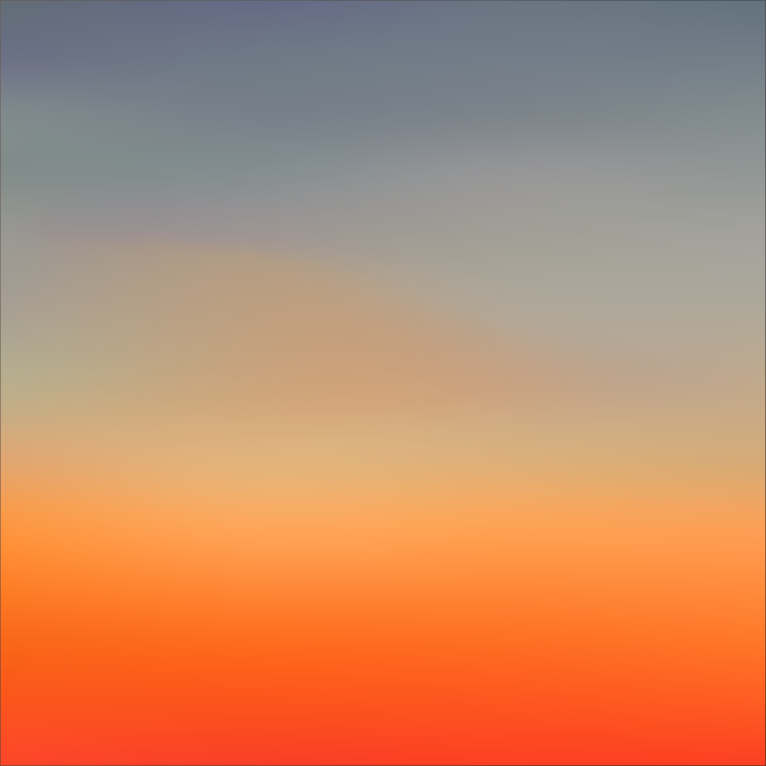}\\%
    \scriptsize%
    (b) Mesh-based only%
    \end{minipage}\hfill%
    \begin{minipage}{0.24\linewidth}%
    \centering%
    \includegraphics[width=\linewidth]{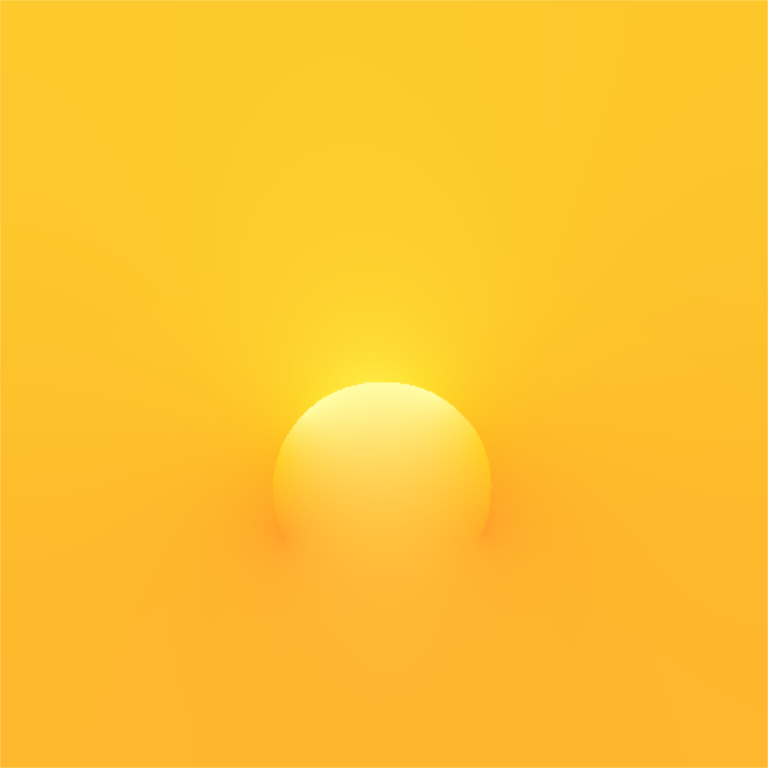}\\%
    \scriptsize%
    (c) Curve-based only%
    \end{minipage}\hfill%
     \begin{minipage}{0.24\linewidth}%
    \centering%
    \includegraphics[width=\linewidth]{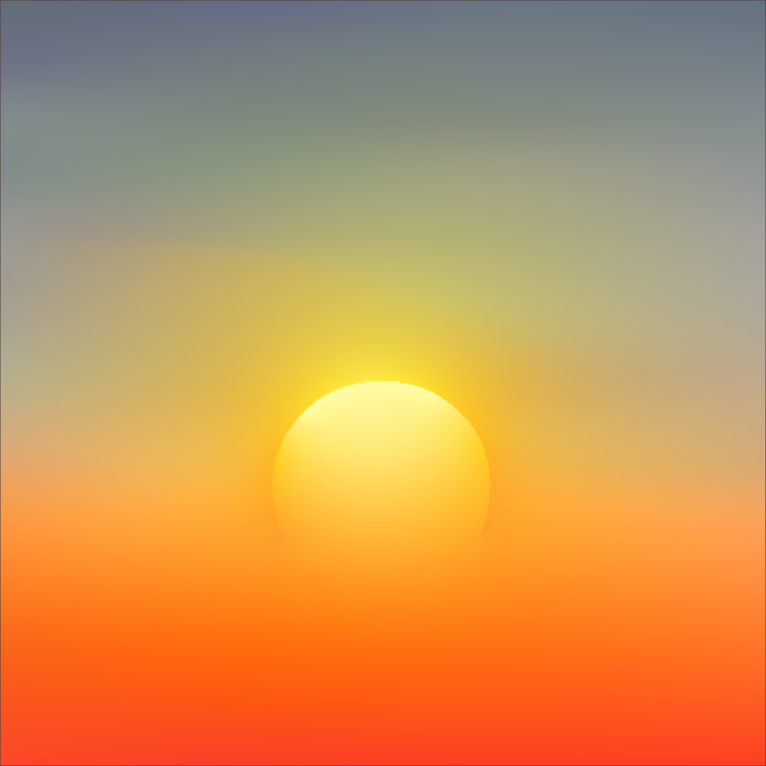}\\%
    \scriptsize%
    (4) Patches (ours)%
    \end{minipage}\hfill%
    \caption{This figure shows the benefits of combining gradient meshes with diffusion curves. Gradient meshes are useful for controlling large multi-hue color gradients (the sky), while curve-based methods can add details (the Sun).
    From left to right, we show the input primitives (gradient meshes and diffusion curves) (a), the result of only rasterizing the gradient meshes (b), the result of only diffusing diffusion curves (c), and lastly our combined representation that contains both (d). Note that the sun diffuses into the sky, mimicking scattering.}
    \label{fig:motivation-combination}
\end{figure}

The field of smooth vector graphics describes image content based on geometric primitives, such as curves or meshes. 
This is in contrast to raster graphics, which store the color of each pixel explicitly. 
Smooth vector graphics are a popular tool for the modeling of scale-independent content, such as icons, schematic illustrations, or web content~\citep{Hsiao23:Img2Logo}, since images can be synthesized on any output resolution. 
Basic vector graphics primitives are standardized in the SVG format, which includes the most basic shapes and color gradients. 
In research, two conceptually different approaches have been developed to extend the artistic control in smoothly-shaded regions. 
On the one hand, there are gradient meshes~\citep{sun2007image,lai2009automatic}, which interpolate colors from tensor product surfaces efficiently. 
And on the other hand, there are curve-based approaches such as diffusion curves~\citep{orzan2008diffusion,jeschke2016generalized} and Poisson curves~\citep{hou2018poisson} that model the final image as solution to a diffusion problem from user-defined boundary conditions, e.g., curve colors. 
Both approaches remained incompatible, and research on editing, rasterization, and vectorization, has continued separately on the two research threads~\citep{Tian23}. 
With this paper, we propose a mathematical formulation that allows combining gradient meshes and curve-based approaches consistently for the first time. 
Fig.~\ref{fig:teaser} gives an example, where gradient meshes are combined with diffusion curves and Poisson curves.
To maximize the compatibility with existing methods we take a set of gradient meshes, diffusion curves, and Poisson curves as input, which are well-known primitives that the user interacts with on the front-end. 
To unify the modeling, we convert those representations first into an intermediate patch representation that specifies for each patch a target Laplacian function and boundary conditions.
This intermediate representation is not visible on the user interface and is only used internally to assemble the Poisson problems.
We propose to model the pixel color within gradient meshes as solution to a partial differential equation (PDE) in order to match their mathematical treatment with that of curve-based methods. 
In the conversion process to our patch representation, geometric intersections of gradient meshes and curves get resolved automatically to yield a non-overlapping domain decomposition. 
In our new formulation, additional types of boundary conditions are added to the existing primitives, which extends the artistic freedom.
For example, Neumann conditions can be specified on diffusion curves, as in \cite{Bang23:Multipole}.
To synthesize a raster image for a given output resolution, we rasterize boundary conditions and Laplacians for the respective patches and compute the final image as solution to a Poisson problem. 
We evaluate the method on various test scenes containing gradient meshes and curve-based primitives. 
In summary, our contributions are:
\begin{itemize}
    \item We propose a unified mathematical formulation of gradient meshes and curve-based methods as a Poisson problem.
    \item We unify the treatment of boundary conditions, which adds (optional) degrees of freedom for artistic control.
    \item We introduce an automatic conversion algorithm that divides the input primitives into non-overlapping patches.
    \item We describe a rasterization algorithm for the novel patch formulation that synthesizes images for a given resolution.
    \item We provide an open source implementation that reads the scenes, constructs the patches, and renders them.
    \item We compare the unified scenes with vectorizations using gradient meshes only, and using diffusion curves only.
\end{itemize}
For the first time, our approach enables the unified mathematical treatment of gradient meshes and curve-based methods, which have been developed independently for over a decade. 
Since we take existing gradient mesh and curve formulations as input, our approach achieves high compatibility with existing content creation pipelines. 
Further, the final algorithm leads to a standard Poisson problem, for which highly optimized numerical solvers readily exist. 
The benefits of combining gradient meshes and curve-based formulations in the same scene are shown in Fig.~\ref{fig:motivation-combination}.
While curve-based approaches are useful to create intricate details, it is easier to create larger color gradients with mesh-based approaches~\citep{Tian23}.
By combining both methods, we are able to generate expressive results that are difficult to obtain with either representation on its own.

In the following section, we first introduce into the necessary background on curve modeling before covering related work on gradient meshes and curve-based formulations. 
Afterwards, we introduce our unified mathematical description, which is followed by our introduction of the automatic conversion algorithm from standard primitives into our patch formulation. 
Then, we describe the image synthesis process, user interactions, and lastly we evaluate the approach on numerous scenes, containing both gradient meshes and curve representations.

\section{Background}
In this section, we formally introduce the notation used to describe mesh-based and curve-based vector graphics primitives.
For a detailed coverage, we refer to a recent survey~\citep{Tian23}.

\subsection{B\'ezier Curves}
In this paper, we denote B\'ezier curves of degree $n$ as parametric curves $\gamma(t) : [0,1] \rightarrow \mathbb{R}^m$ in $m$-dimensional space:
\begin{align}
    \gamma(t) = \sum_{i=0}^n B_i^n(t) \cdot \vb_i \;\; \textrm{with} \;\; B_i^n(t) = \binom{n}{i} t^i (1-t)^{n-i},
    \label{eq:bezier-curve}
\end{align}
where $B_i^n(t)$ are the Bernstein basis functions, and $\vb_i\in\mathbb{R}^m$ are the B\'ezier control points.
The curve tangent is denoted by $\dot\gamma(t) = \frac{\mathrm{d}\gamma(t)}{\mathrm{d}t}$.
A cubic B\'ezier curve ($n=3$) in a two-dimensional plane ($m=2$) has four control points $\vb_0$, $\vb_1$, $\vb_2$, and $\vb_3$, and has by convention a right-oriented curve normal $\gamma_n(t) : [0,1] \rightarrow \mathbb{R}^2$:
\begin{align}
    \gamma_n(t) = \begin{pmatrix}
    0 & 1 \\
    -1 & 0
    \end{pmatrix}
    \frac{\dot\gamma(t)}{\|\dot\gamma(t)\|}.
    \label{eq:bezier-normal}
\end{align}
Thus, given the control points $\vb_0$, $\vb_1$, $\vb_2$, $\vb_3$, the curve point $\gamma(t)$, tangent $\dot\gamma(t)$, and normal $\gamma_n(t)$ can be evaluated for each $t$.
Multiple B\'ezier curves can be concatenated with suitable continuity conditions to assemble a B\'ezier-spline curve~\citep{farin2002curves}.

\subsection{Mesh-based Methods}
\label{sec:related-mesh-based}
Mesh-based vector graphics primitives specify color gradients inside connected regions by using interpolation.
Depending on the mesh topology, they are categorized into triangular, rectangular, and irregular meshes.
Our work focuses on rectangular meshes.

\paragraph{Rectangular Mesh}
By generalization of the B\'ezier curves in Eq.~\eqref{eq:bezier-curve}, a B\'ezier tensor-product surface $f(u,v) : [0,1]^2 \rightarrow \mathbb{R}^m$ is
\begin{align}
    f(u,v) = \sum_{i=0}^{n_1} \sum_{j=0}^{n_2} B_i^{n_1}(u) \cdot B_j^{n_2}(v) \cdot f_{i,j}.
    \label{eq:parametric-surfaces}
\end{align}
Later, we use the symbol $\partial_u f_{i,j}:=\frac{\partial f(u,v)}{\partial u}_{|u=u_i, v=v_j}$ to abbreviate the evaluation of a derivative at a discrete location.
Each tensor-product surface is bound by four B\'ezier curves $f(u,0)$, $f(u,1)$, $f(0,v)$ and $f(1,v)$ with $u,v\in[0,1]$.

\begin{figure}[t]%
    \centering
    \captionsetup[subfloat]{labelfont=scriptsize,textfont=scriptsize}
    \subfloat[Gradient Mesh \label{fig:existing-primitives-gradient-mesh}]{\includegraphics[height=0.2\linewidth]{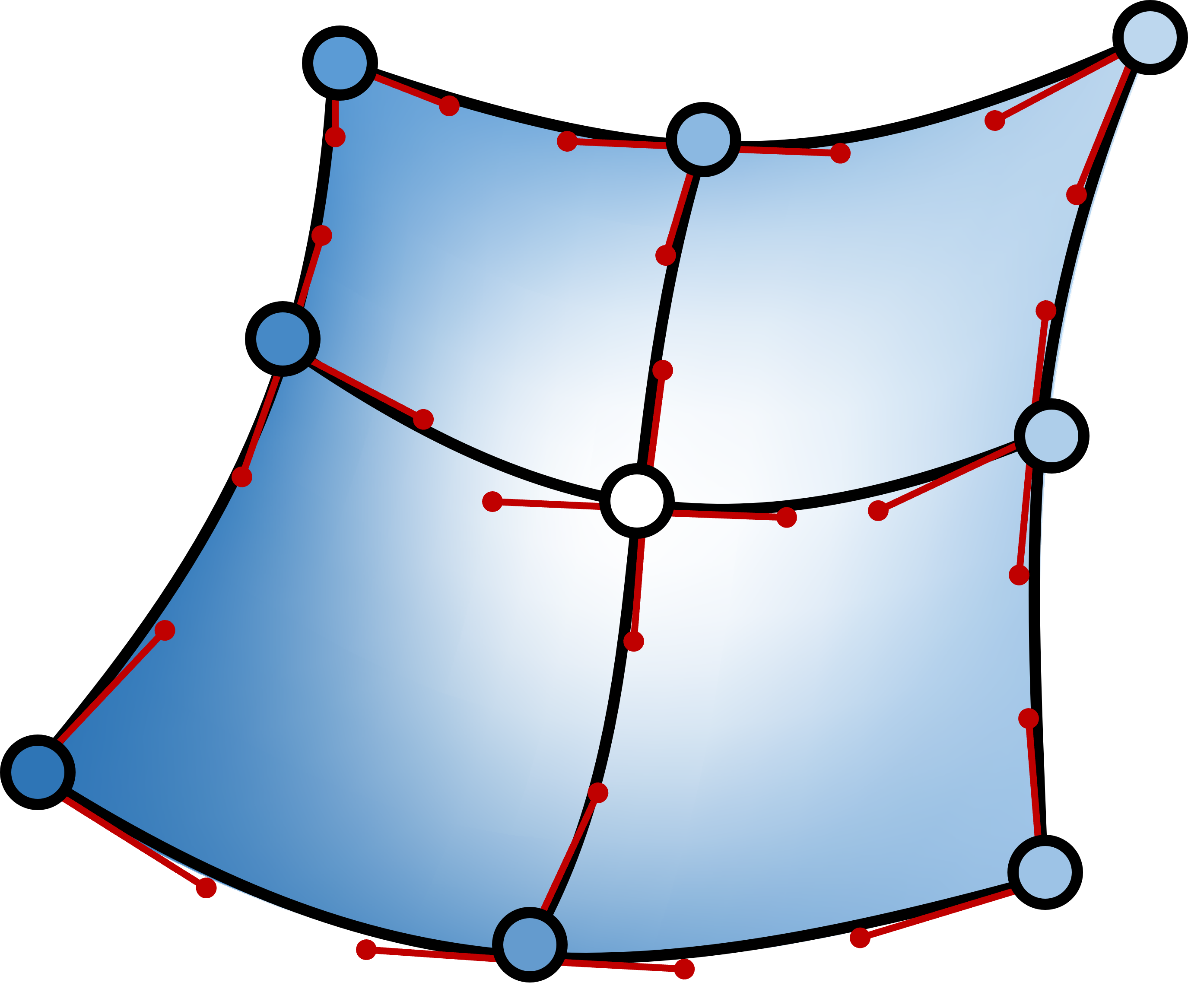}}\quad%
    \subfloat[Diffusion Curve \label{fig:existing-primitives-diffusion-curve}]{\quad\includegraphics[height=0.2\linewidth]{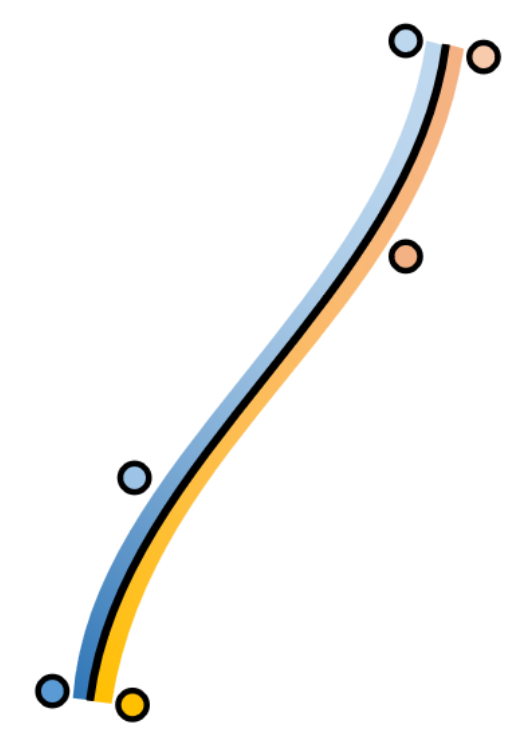}\quad}\quad%
    \subfloat[Poisson Curve \label{fig:existing-primitives-poisson-curve}]{\qquad\includegraphics[height=0.2\linewidth]{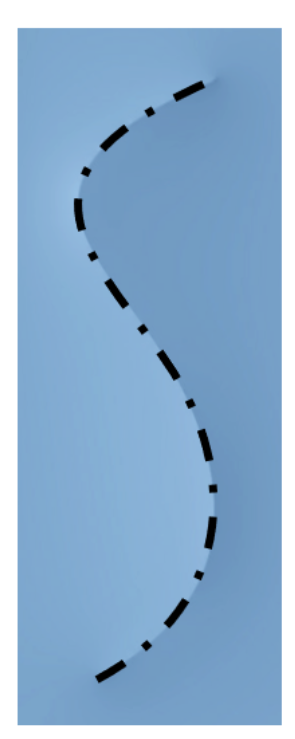}\qquad}%
    \caption{Overview of the three mesh and curve primitives that are used in our framework. Gradient meshes (a) are assembled from Ferguson patches, which contain smooth color gradients and are controlled by colors at the corners and tangent handles. Diffusion curves (b) define a color on the left and right side, typically by piecewise linear interpolation among color stops. Poisson curves (c) allow for the modeling of cusps and highlights by locally adjusting the Laplacian of the underlying color field.}
    \label{fig:existing-primitives}
\end{figure}

\cite{price2006object} used bi-cubic surfaces, which have a degree of $n_1=n_2=3$. 
The direct editing of its 16 control points $f_{0,0}, \dots, f_{3,3}$ is cumbersome.
Fortunately, a bi-cubic tensor-product surface can be expressed as a Ferguson patch, cf. \cite{sun2007image}:
\begin{align}
    f(u,v) = \vt(u)^\Transp \cdot \mC^\Transp \cdot \mQ \cdot \mC \cdot \vt(v)
    \label{eq:gradient-mesh}
\end{align}
with the monomial basis $\vt(t) = (1,t,t^2,t^3)^\Transp$ and the matrices:
\begin{align}
    \begingroup
\setlength\arraycolsep{1.5pt}
    \mC \hspace{-0.2em} = \hspace{-0.2em} \begin{pmatrix}
    1 & 0 & -3 & 2 \\
    0 & 0 & 3 & -2 \\
    0 & 1 & -2 & 1 \\
    0 & 0 & -1 &1
    \end{pmatrix}\hspace{-0.2em},
    \mQ \hspace{-0.2em} = \hspace{-0.2em} \begin{pmatrix}
    f_{0,0} & f_{0,3} & \partial_v f_{0,0} & \partial_v f_{0,3} \\
    f_{3,0} & f_{3,3} & \partial_v f_{3,0} & \partial_v f_{3,3} \\
    \partial_u f_{0,0} & \partial_u f_{0,3} & \partial_u\partial_v f_{0,0} & \partial_u\partial_v f_{0,3} \\
    \partial_u f_{3,0} & \partial_u f_{3,3} & \partial_u\partial_v f_{3,0} & \partial_u\partial_v f_{3,3}
    \end{pmatrix}
    \endgroup
    \nonumber
\end{align}
To simplify the modeling, the mixed partial derivatives are commonly set to zero, i.e., $\partial_u\partial_v f_{i,j}=0$, and thus, a function $f(u,v)$ can be controlled by the user by specifying the values $f_{i,j}$ and the tangent handles $\partial_u f_{i,j}$ and $\partial_v f_{i,j}$ at the four corners $i,j\in\{0,3\}$, see Fig.~\ref{fig:existing-primitives-gradient-mesh}.
A \emph{gradient mesh} consists of multiple Ferguson patches that are stitched together with suitable continuity condition.
The functions $f(u,v)$ that are modeled by the gradient mesh are the position $\vx(u,v)$ and the color $\vc(u,v)$.
Thus, the degrees of freedom are $\vx_{i,j}$, $\vc_{i,j}$ and the tangent handles $\partial_u \vx_{i,j}$, $\partial_u \vc_{i,j}$ and $\partial_v \vx_{i,j}$, $\partial_v \vc_{i,j}$ at the four corners $(i,j)\in\{(0,0), (0,3), (3,0), (3,3)\}$.
To model holes in surfaces, \cite{lai2009automatic} proposed to split gradient meshes along discontinuities.
\cite{barendrecht2018locally} enabled local refinement of cubic rectangular patches.
\cite{Baksteen21} fitted mesh colors~\citep{Yuksel10:MeshColors} for colorization of gradient meshes.
In line with many other papers~\citep{xiao2012example,xiao2015optimization,wan2018scribble,wei2019field}, we likewise base our description of gradient meshes on \cite{sun2007image}, see Eq.~\eqref{eq:gradient-mesh}.

\paragraph{Triangular and Irregular Meshes}
As an alternative to rectangular meshes, triangle-based patches have been explored, where the patch boundaries are straight~\citep{Lawonn19:ImageTri} or curved edges~\citep{xia2009patch,xiao2022image,Wang24:CurveImageTri}.
Control over the continuity across adjacent patches is desirable~\citep{liao2012subdivision,zhou2014representing}.
To this end, quadratic triangle configuration B-splines (TCBs)~\citep{cao2019finite,schmitt2019bivariate} have been used by \cite{zhu2022} to model color variations while staying $C^1$ continuous.
An example for irregular patches are the polygonal patches of 
\cite{swaminarayan2006rapid}, who used constant colors in the interior.
With the introduction of B\'ezigons~\citep{yang2015effective} closed regions have been bounded by B\'ezier curves.
For color interpolation in the interior, several options are available~\citep{hettinga2019colour,Hettinga21:MeshColorVectorization}, including cubic mean value coordinates~\citep{li2013cubic}.
Since the methods above use differentiable interpolants, they are compatible with our approach.

\subsection{Curve-based Methods}
With curve-based methods, color gradients are the result of a diffusion process from colors given at the domain boundaries.

\paragraph{Diffusion Curves} 
Originally, diffusion curves~\citep{orzan2008diffusion} are B\'ezier curves $\gamma(t)$ with a piecewise linear color on the left and right side, i.e., $\vc_l(t)$ and $\vc_r(t)$.
The colors serve as Dirichlet boundary conditions in a diffusion process that produces smooth color gradients, see Fig.~\ref{fig:existing-primitives-diffusion-curve}.
The reformulation of \cite{jeschke2009gpu} treats diffusion curves as domain boundaries $\partial\Omega$ of a smoothly shaded region $\Omega$. 
The color field of the shaded region $\vc(\xx)$ is as smooth as possible and meets the boundary color $\vg(\xx)$, as specified by the colors on the left and right hand side of the diffusion curve.
\begin{align}
\vc(\xx) = 
    \begin{cases}
	    \Delta \vc(\xx) = 0, & \xx\in\Omega\backslash \partial \Omega \\
        \hphantom{\Delta} \vc(\xx) = \vg(\xx), & \xx\in\partial\Omega
	\end{cases}
    \label{eq:diffusion-curves}
\end{align}
Follow-up research on rendering~\citep{pang2011fast,sun2014fast,Bang23:Multipole}, vectorization~\citep{zhao2017inverse,lu2019depth} and editing~\citep{jeschke2011estimating,lu2020shape} utilized this formulation.
To provide more control over the color gradients, \cite{bezerra2010diffusion} controlled the strength and direction of the diffusion.
Alternatively, \cite{sun2012diffusion} introduced textures in diffusion curves.
Several renderers have been proposed by now that are able to rasterize diffusion curves, including ray tracing approaches~\citep{bowers2011ray,prevost2015vectorial}, Monte Carlo methods~\citep{Sawhney20:WoS,Sawhney23:WoSt,Sugimoto23:WoB,Li23:NeuralCaches,Miller23:Caching,Sugimoto24:PWoS,Miller24:Robin,Czekanski24:walkingspherestalkingneighbors}, and boundary element methods~\citep{ilbery2013biharmonic,Bang23:Multipole,Chen24Lightning}.
Lately, Monte Carlo methods have been phrased in a differentiable manner, which is useful for inverse problems such as vectorizations~\citep{Yilmazer24,Yu24:DiffWoS,Miller24:DiffWoS}.

\paragraph{Poisson Curves} 
By introducing \emph{Poisson vector graphics}, \cite{hou2018poisson} edited the gradient in the domain by introducing a desired (piecewise-constant) Laplacian $\vf(\xx)$:
\begin{align}
\vc(\xx) = 
    \begin{cases}
	    \Delta \vc(\xx) = \vf(\xx), & \xx\in\Omega\backslash \partial \Omega \\
        \hphantom{\Delta} \vc(\xx) = \vg(\xx), & \xx\in\partial\Omega
	\end{cases}
    \label{eq:poisson-curves}
\end{align}
Thus, rather than looking for the smoothest possible color gradient (via $\vf(\xx)=0$), bumps and discontinuities can be introduced in the color gradient, which are controlled locally by the user by drawing curve and region primitives, see Fig.~\ref{fig:existing-primitives-poisson-curve}.
Non-constant Laplacians in Poisson regions have been described by raster images, and were utilized for color transfer between images~\citep{Fu19:PVGColorTransfer} and for image vectorization~\citep{Fu24:PVGFaceVectorization}.
Alternatively, \cite{finch2011freeform} proposed to constrain high-order derivatives, seeking solutions that are harmonic in their Laplacian domain, i.e., $\Delta^2 \vc(\xx) = \mathbf{0}$.
With this formulation, extrapolation may create new extrema at unconstrained positions~\citep{finch2011freeform,boye2012vectorial}, which requires manual adjustment of control points~\citep{finch2011freeform} or a non-linear optimization~\citep{jacobson2012smooth}. Further, the diffused color range is tied to the scene geometry, which may lead to oversaturation~\citep{jeschke2016generalized} when scaling objects.
In our work, we concentrate on the Poisson formulation in Eq.~\eqref{eq:poisson-curves}, which includes the diffusion curves in Eq.~\eqref{eq:diffusion-curves} as special case.
Poisson vector graphics~\citep{hou2018poisson} have led to a new line of work on the control over color gradients within smooth vector graphics.
In this paper, we seek to utilize the well-established gradient meshes~\citep{sun2007image}, which so far have not been incorporated into this framework.

\section{Method Overview}
\begin{figure*}[t]%
    \centering%
    \captionsetup[subfloat]{labelfont=scriptsize,textfont=scriptsize}
    \subfloat[Input Primitives]{\includegraphics[width=0.24\linewidth]{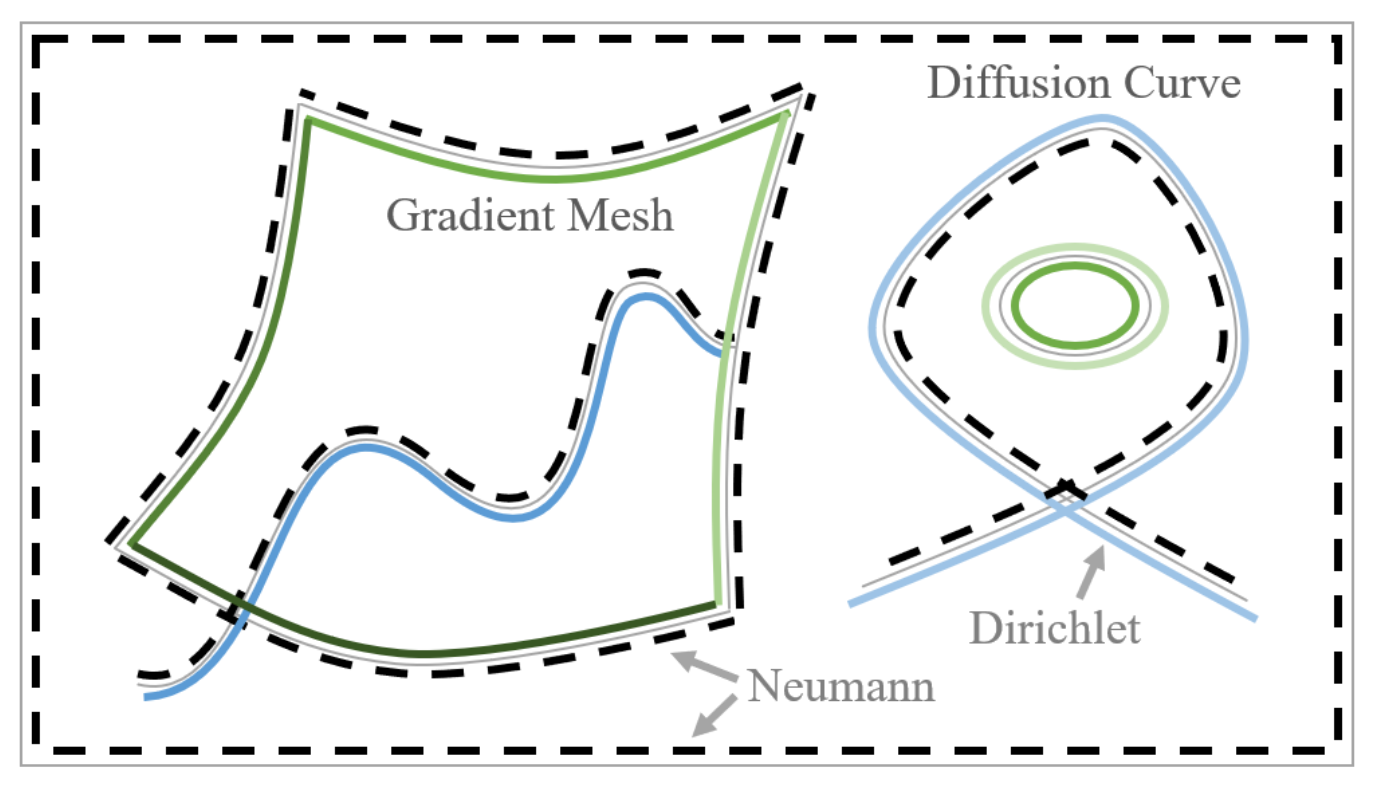}}\hfill%
    \subfloat[Edge Graph]{\includegraphics[width=0.24\linewidth]{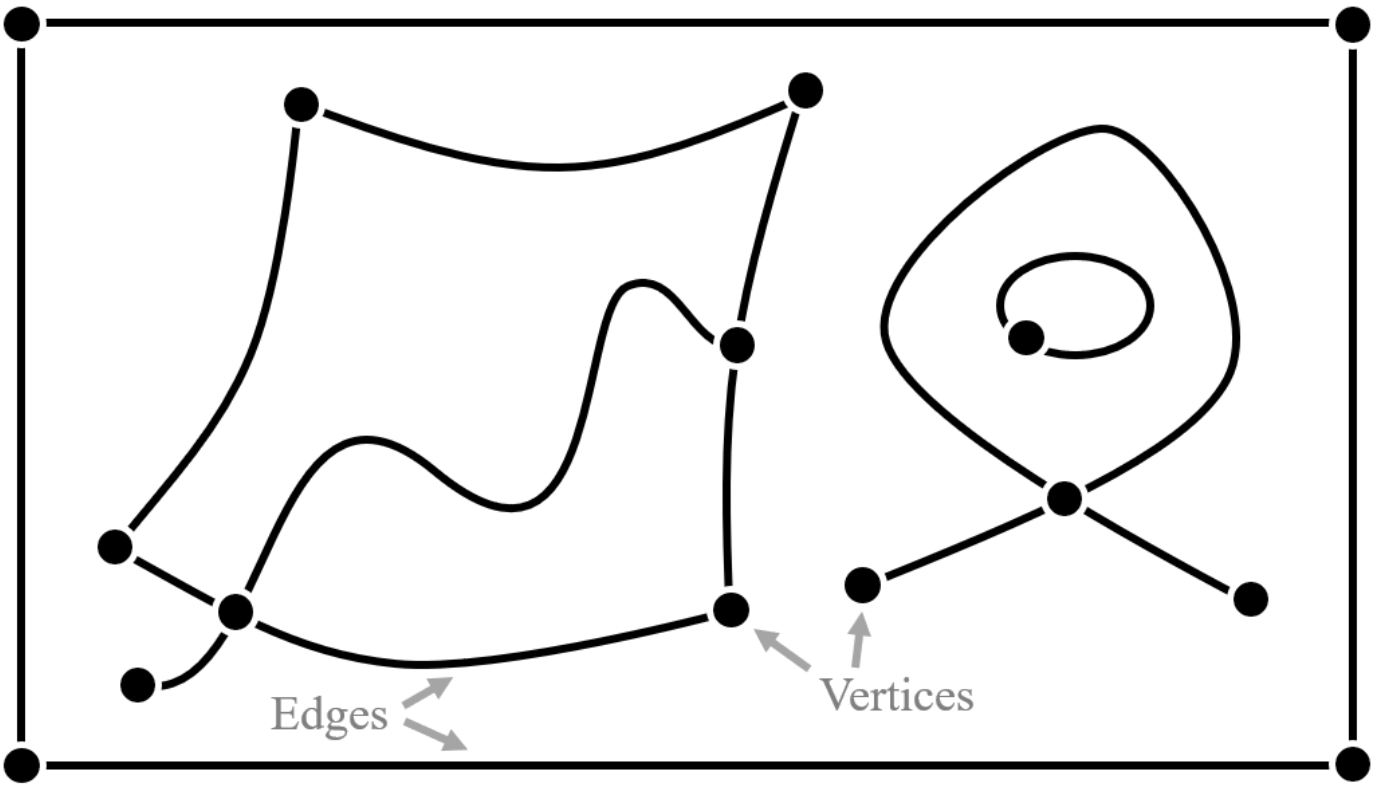}}\hfill%
    \subfloat[Unified Patches]{\includegraphics[width=0.24\linewidth]{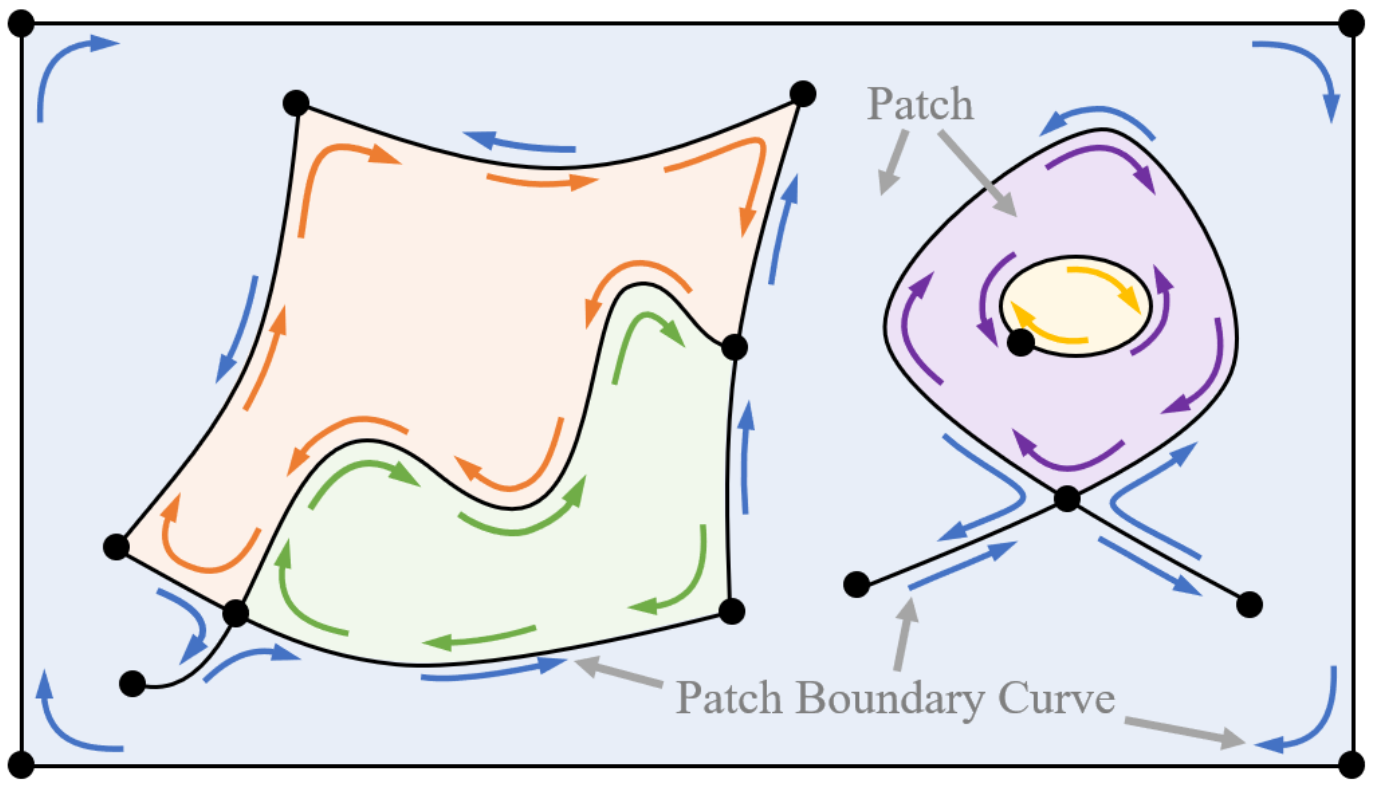}}\hfill%
    \subfloat[Image Synthesis]{\includegraphics[width=0.24\linewidth]{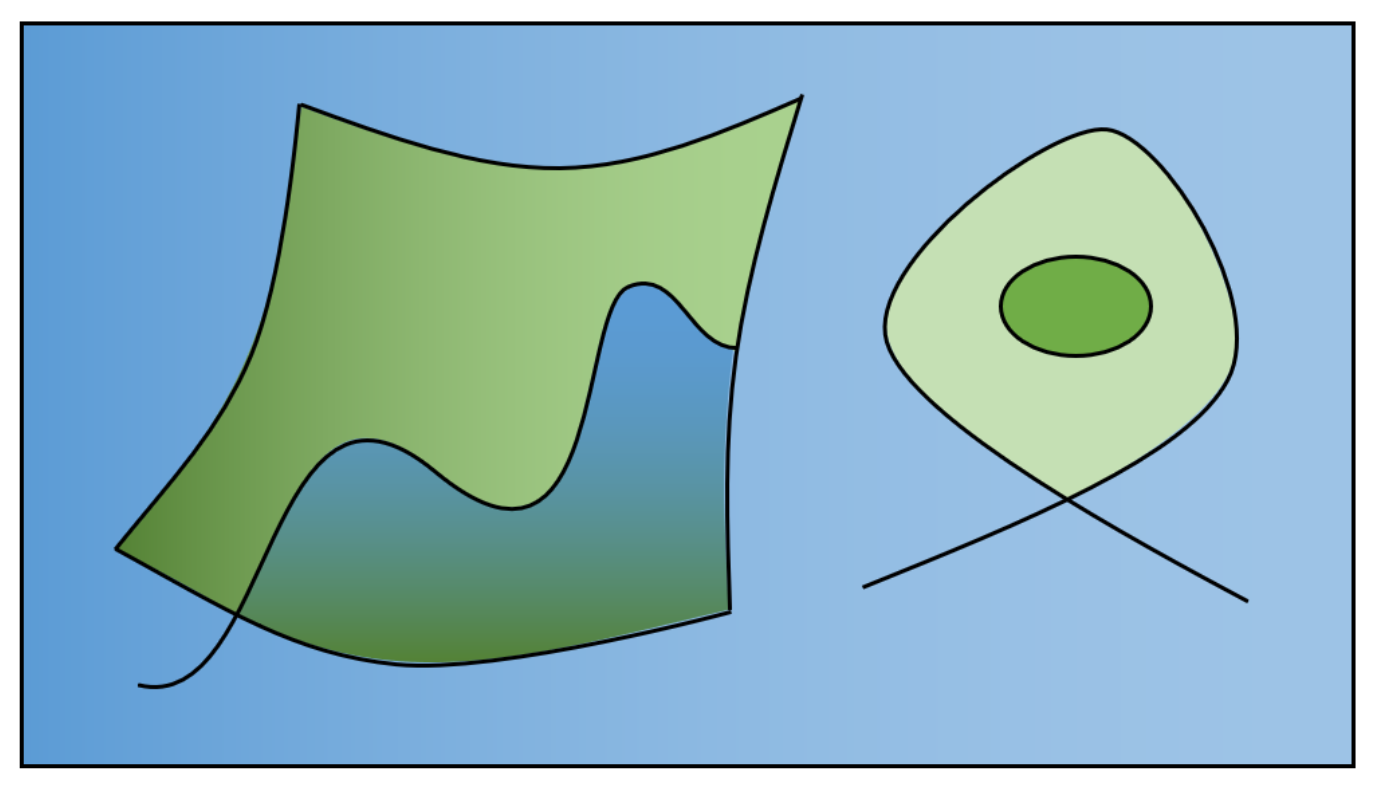}}%
    \caption{Schematic overview of our generalized smooth vector graphics pipeline. (a) Input to our method is a collection of (extended) gradient meshes, (extended) diffusion curves, and Poisson curves. (b) The input primitives are converted and inserted into an edge graph that resolves the intersection of input primitives. (c) From the edge graph a unified patch data structure is formed that defines patches as closed regions with boundary conditions according to the underlying input primitives. (d) The unified patch data structure specifies the ingredients for a Poisson problem that is solved to synthesize the final image.}
    \label{fig:overview}
\end{figure*}

Our goal is to synthesize raster images for vector graphics scenes that contain mesh-based and curve-based primitives at the same time.
Usually, mesh-based and curve-based methods are rasterized in fundamentally different ways.
While mesh-based approaches use \emph{interpolation}, curve-based methods solve a \emph{Poisson problem}.
To model both approaches consistently, we describe the mesh-based approaches as solution to a Poisson problem that is identical to interpolation in the absence of any curve-based primitives.
A user of our system may decide to draw curve-based primitives that intersect each other or intersect with mesh-based primitives.
To formulate the Poisson problem, the scene is decomposed into non-overlapping regions for which boundary conditions are defined and for which a Laplacian is determined that controls the color gradient in the interior.
We formulate the rasterization of smooth vector graphics as a pipeline consisting of multiple steps, see Fig.~\ref{fig:overview}.
\begin{enumerate}
    \item \emph{Input Primitives.} Input to our method are standard smooth vector graphics primitives, such as gradient meshes, diffusion curves, and Poisson curves, which are created and edited by the user. We extend gradient meshes and diffusion curves by allowing the user to specify Dirichlet conditions and homogeneous Neumann conditions on either side.
    \item \emph{Edge Graph.} Both the extended gradient meshes and the extended diffusion curves are inserted into an undirected edge graph data structure~\citep{Gangnet89:PlanarMaps}, which records and resolves all intersections of boundary curves.
    \item \emph{Unified Patches.} By traversing the edge graph, we define closed, non-overlapping regions that have consistent boundary conditions. The interior of each patch is equipped with a Laplacian function that is either homogeneous, or sampled from a gradient mesh and/or multiple Poisson curves.
    \item \emph{Image Synthesis.} Given the patches with their boundary conditions, we solve the Poisson problem for each patch  with an off-the-shelf Poisson solver.
\end{enumerate}
A C++ reference implementation of our method can be found at \citep{Tian25:USVG-code}.
In the following, the four steps are explained in more detail.

\section{Input Primitives}
Input to our approach is a collection of extended gradient meshes $\cG$, a collection of extended diffusion curves $\cD$ and a collection of Poisson curves $\cP$.
Those three types of input primitives are drawn by the user and can be edited and controlled by manipulation of control points, tangent handles, and color stops, identical to the editing processes of conventional diffusion curves, gradient meshes, and Poisson curves~\citep{Tian23}.
In the following, we explain how the primitives are defined and how they are extended in our framework.

\begin{figure}[b]%
    \centering%
    \includegraphics[width=0.8\linewidth]{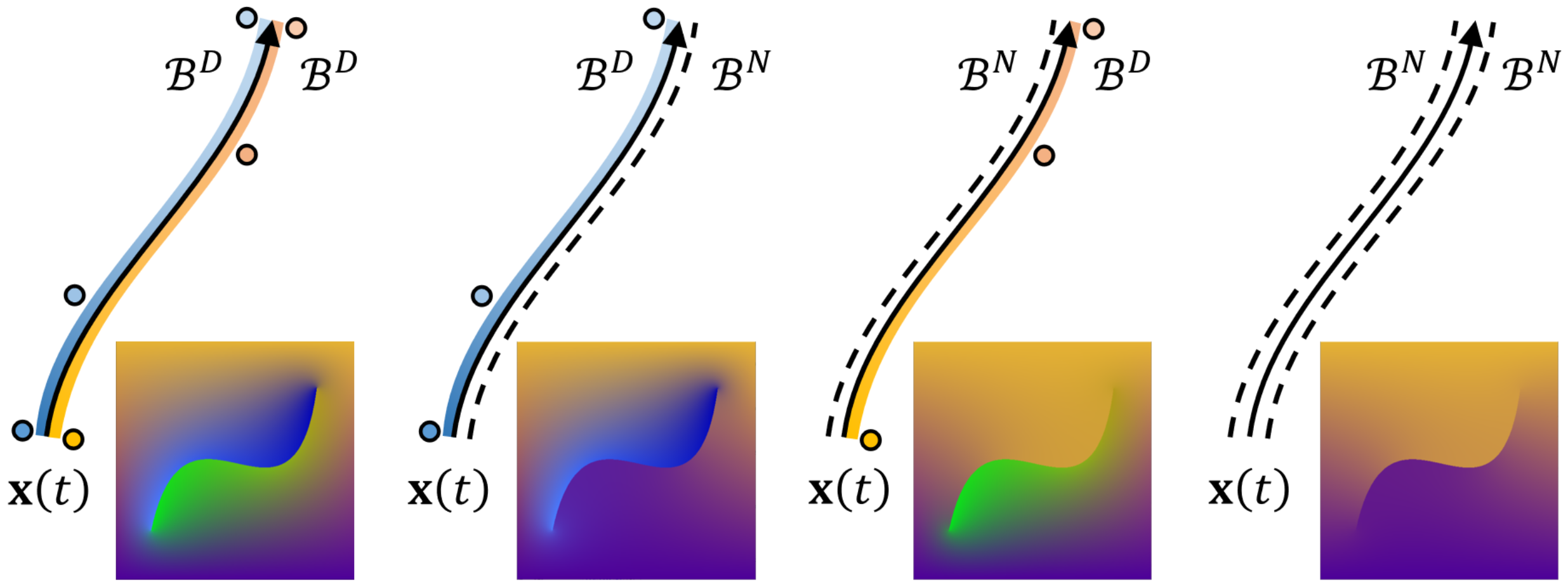}%
    \caption{Input boundary curves consist of a spatial curve $\vx(t)$ and boundary conditions on the left and right side, which may either be a Dirichlet boundary condition $\cB^D$ (shown as colored line with control points) or a homogeneous Neumann boundary condition $\cB^N$ (shown as a dashed line).}%
    \label{fig:input-boundary-curve}%
\end{figure}

\subsection{Definition}
First, we introduce an \emph{input boundary curve}, which is a triplet $(\vx(t), \cB_l, \cB_r)$, as shown in Fig.~\ref{fig:input-boundary-curve}. It consists of:
\begin{itemize}
    \item the spatial curve $\vx(t)$,
    \item a left boundary condition $\cB_l$,
    \item and a right boundary condition $\cB_r$.
\end{itemize}
The spatial location of the input boundary curve is given by a cubic B\'ezier spline $\vx(t) : \mathbb{R} \rightarrow \cU$, where $\cU \subset \mathbb{R}^2$ denotes the image domain.
The curve parameter range $t\in [t_0, t_1]$ defines the \emph{orientation} of the curve.
We require $t_0 < t_1$, where the curve begins at $t_0$ and ends at $t_1$. This gives rise to the notion of a \emph{left} side and a \emph{right} side of the curve.
Each input boundary curve will later define a boundary condition in the field $\vc(\vx):\cU \rightarrow \cC$, where $\cC \subset \mathbb{R}^3$ is the color space.
The boundary condition is either a Dirichlet boundary condition $\cB^D$ with color $\vc(t)$ 
or a homogeneous Neumann boundary condition $\cB^N$ with boundary normal $\vx_n(t)$, cf. Eq.~\eqref{eq:bezier-normal}:
\begin{align}
\cB^D &= \begin{cases}
    \vc(\vx)_{|\vx=\vx(t)} = \vc(t) & \,\qquad \textrm{Dirichlet with $\vc(t)$}
\end{cases}
\label{eq:boundary-condition-dirichlet}
\\
\cB^N &= \begin{cases}
    \frac{\partial \vc(\vx)}{\partial \vx}_{|\vx = \vx(t)} \cdot \vx_n(t) = 0 & \textrm{homogen. Neumann}
\end{cases}
\label{eq:boundary-condition-neumann}
\end{align}
For notational convenience, we name the \emph{left} boundary condition $\cB_l \in \{\cB^D, \cB^N\}$ and the \emph{right} boundary condition $\cB_r\in \{\cB^D, \cB^N\}$. 
Adding inhomogeneous Neumann conditions would be a straightforward extension that would, however, add further complexity to the user interface.

\begin{figure}[t]%
    \centering%
    \includegraphics[width=0.33\linewidth]{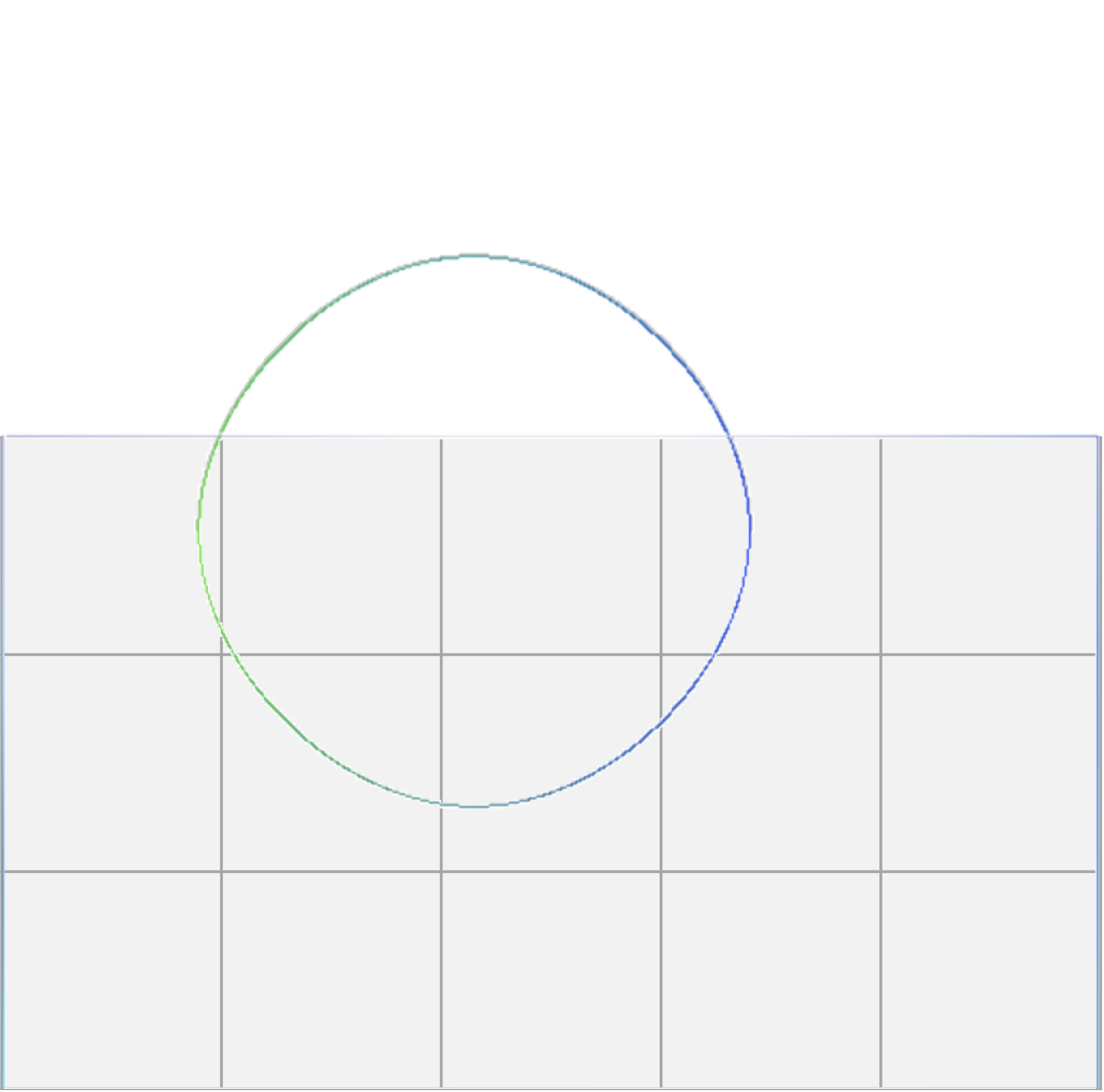}\hfill%
    \includegraphics[width=0.33\linewidth]{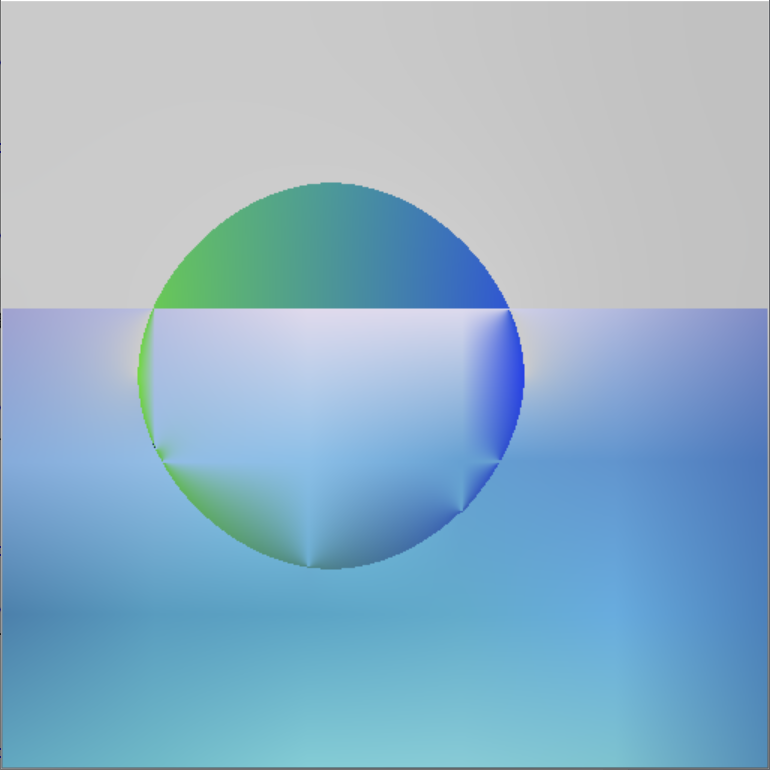}\hfill%
    \includegraphics[width=0.33\linewidth]{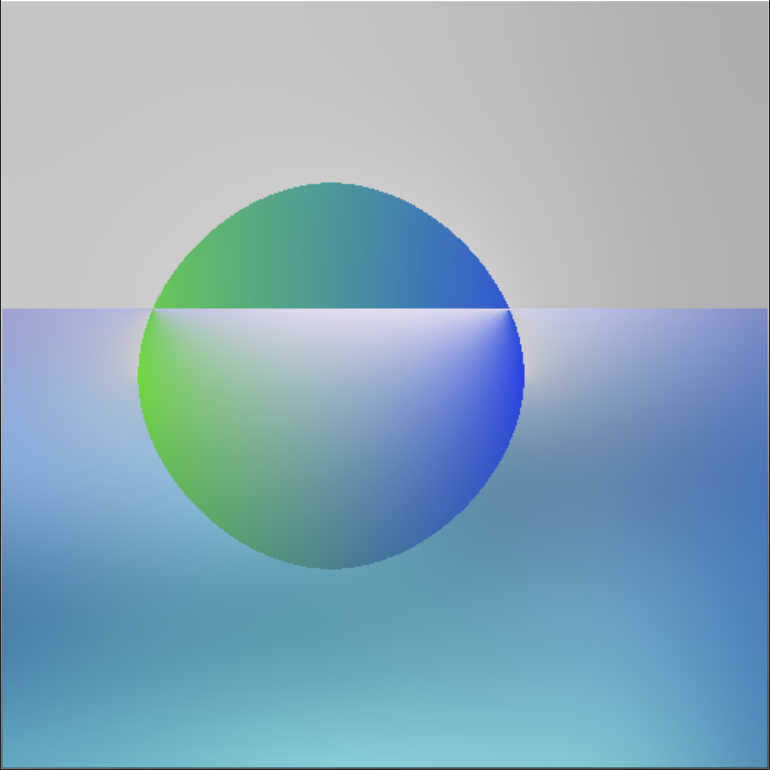}\\%
    \begin{minipage}{0.33\linewidth}%
    \centering
    \scriptsize
    input primitives
    \end{minipage}\hfill%
    \begin{minipage}{0.33\linewidth}%
    \centering
    \scriptsize
    w/ inner boundaries
    \end{minipage}\hfill%
    \begin{minipage}{0.33\linewidth}%
    \centering
    \scriptsize
    w/o inner boundaries
    \end{minipage}%
    \caption{When constructing input boundary curves, the inner edges of gradient meshes are not added, since they would create diffusion barriers. The inner edges are not needed, since the color gradients are fully determined by the gradient mesh Laplacian, which is considered in the Poisson problem.}
    \label{fig:gradient-mesh-lines}
\end{figure}

\paragraph{Extended Gradient Mesh}
A gradient mesh that consists of a single Ferguson patch as in Eq.~\eqref{eq:gradient-mesh}, with position $\vx(u,v)$ and color $\vc(u,v)$, gives rise to four clock-wise oriented boundary curves with positions $\vx(t) \in \{ \vx(1,1-t), \vx(1-t,0), \vx(0,t), \vx(t,1) \}$.
Using the clock-wise orientation of the boundary curves, we set the \emph{right} boundary conditions (i.e., the inside of the gradient mesh) to a Dirichlet condition $\cB_r=\cB^D$ by using the corresponding colors $\vc(t) \in \{ \vc(1,1-t), \vc(1-t,0), \vc(0,t), \vc(t,1) \}$.
Our \emph{extension} compared to a regular gradient mesh is that we add the definition of a boundary condition on the \emph{outside} of the gradient mesh, as well.
The \emph{left} boundary condition (i.e., the outside of the gradient mesh) is by default set to a homogeneous Neumann condition $\cB_l=\cB^N$.
The user may change this to a Dirichlet condition if desired.
For a gradient mesh that consists of multiple connected Ferguson patches, input boundary curves are only generated for the outermost Ferguson patch boundaries, which coincide with the boundary of the gradient mesh.
Adding the interior Ferguson patch boundaries is neither necessary nor useful, since the smooth color gradient in the interior of the gradient mesh will be fully defined by the Laplacian of its Ferguson patches.
Fig.~\ref{fig:gradient-mesh-lines} demonstrates this at the example of a closed diffusion curve (circle), which intersects a gradient mesh (blue).
If interior patch boundaries were added, it would introduce unwanted diffusion barriers that are counter-intuitive.

\paragraph{Extended Diffusion Curve}
A diffusion curve $\gamma(t) : \mathbb{R} \rightarrow \cU$ with left color $\vc_l(t) : \mathbb{R} \rightarrow \cC$ and right color $\vc_r(t): \mathbb{R} \rightarrow \cC$  translates to an input boundary curve by setting $\vx(t) = \gamma(t)$ and by applying Dirichlet conditions on the left and right with $\cB_l=\cB^D$ and $\cB_r=\cB^D$.
Same as \cite{Bang23:Multipole}, we \emph{extend} diffusion curves by allowing to replace the Dirichlet conditions with homogeneous Neumann conditions.

\paragraph{Poisson Curve}

\begin{figure}[t]%
    \centering%
    \begin{minipage}{0.16\linewidth}%
    \includegraphics[width=\linewidth]{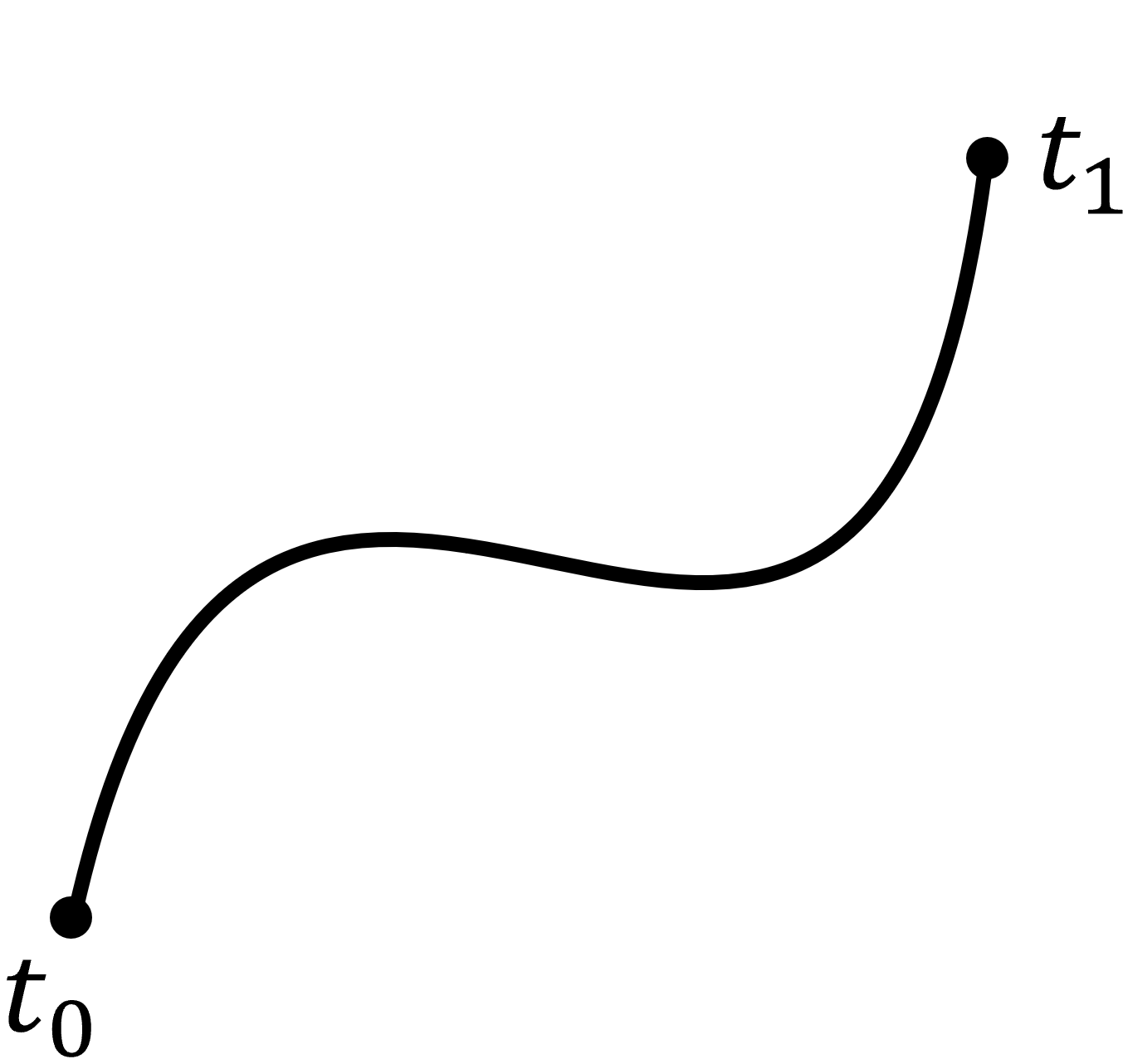}
    \end{minipage}\hfill%
    \begin{minipage}{0.16\linewidth}%
    \includegraphics[width=\linewidth]{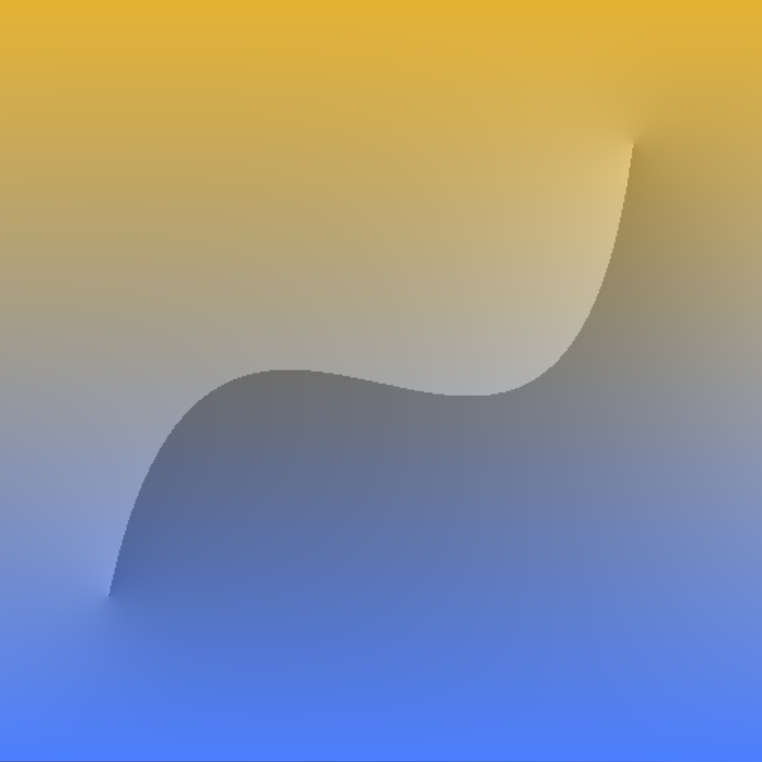}
    \end{minipage}\hfill%
    \begin{minipage}{0.16\linewidth}%
    \includegraphics[width=\linewidth]{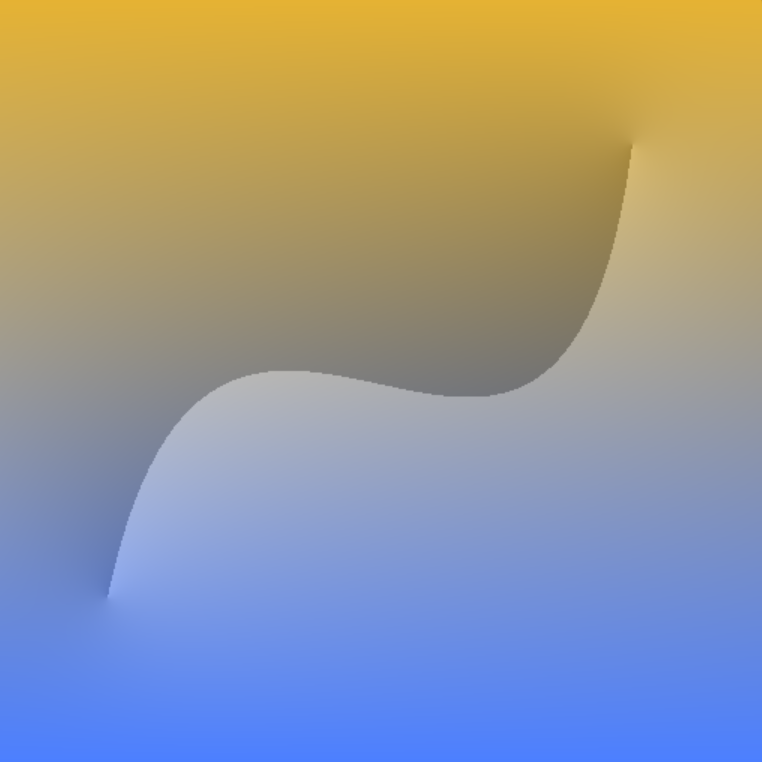}
    \end{minipage}\hfill%
    \begin{minipage}{0.16\linewidth}%
    \includegraphics[width=\linewidth]{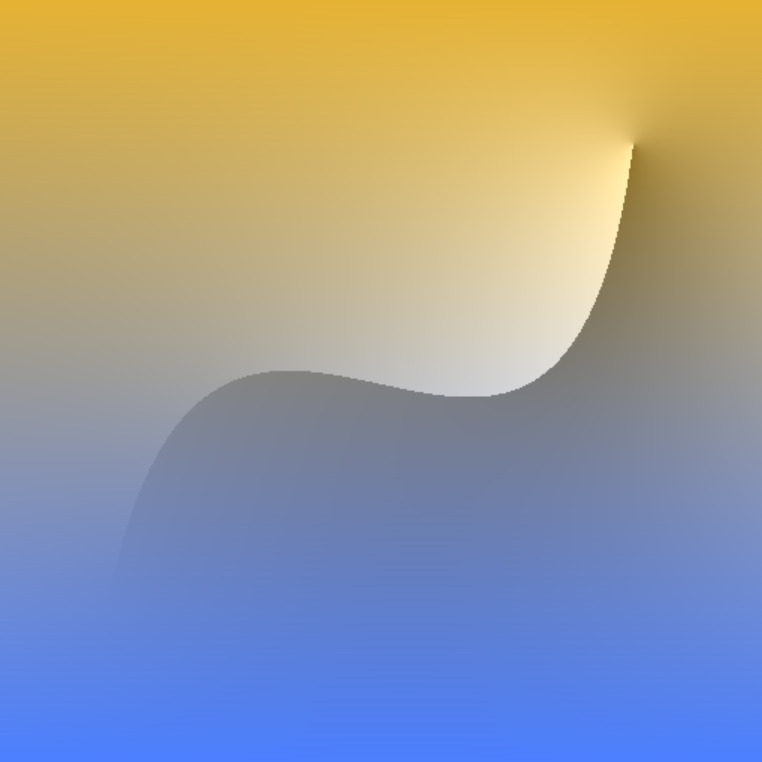}
    \end{minipage}\hfill%
    \begin{minipage}{0.16\linewidth}%
    \includegraphics[width=\linewidth]{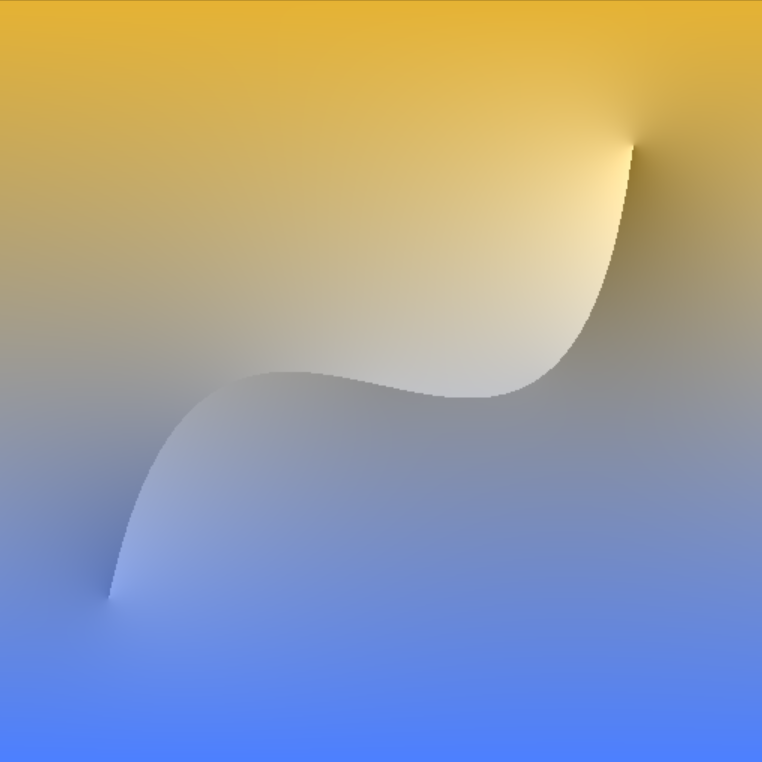}
    \end{minipage}\hfill%
    \begin{minipage}{0.16\linewidth}%
    \includegraphics[width=\linewidth]{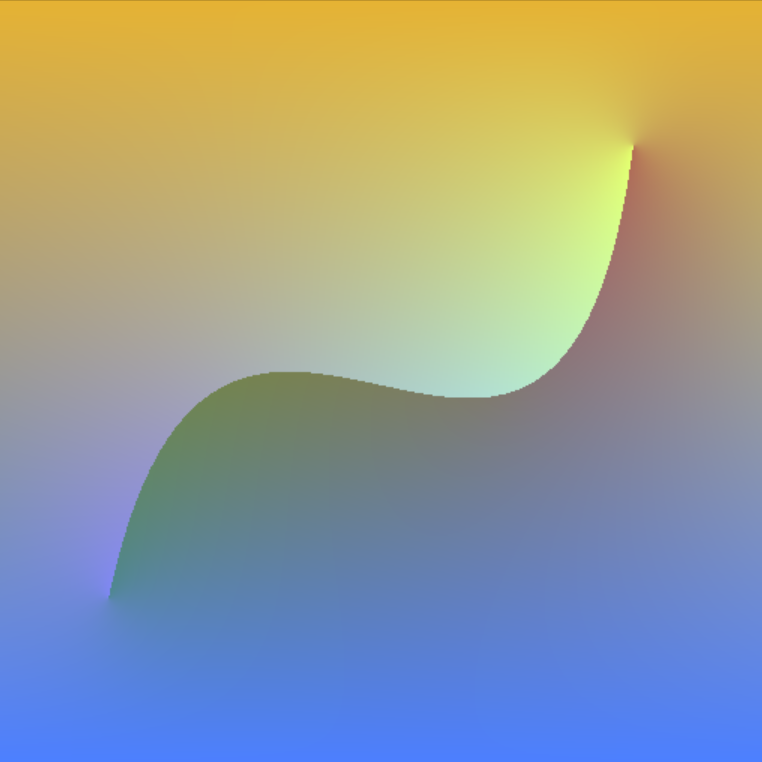}
    \end{minipage}\\%
    \begin{minipage}{0.16\linewidth}%
    $~$
    \end{minipage}\hfill%
    \begin{minipage}{0.16\linewidth}%
    \includegraphics[width=\linewidth]{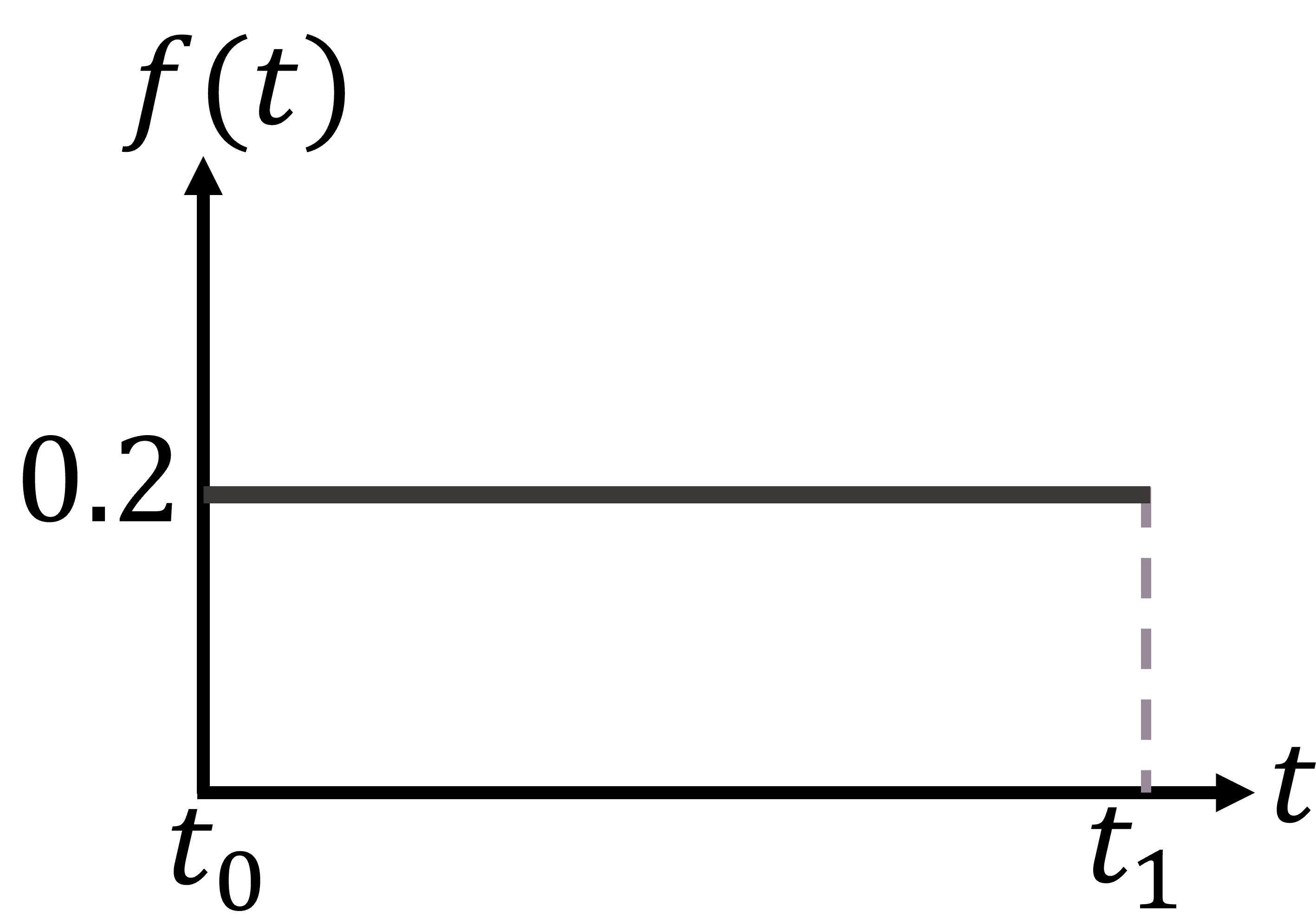}
    \end{minipage}\hfill%
    \begin{minipage}{0.16\linewidth}%
    \includegraphics[width=\linewidth]{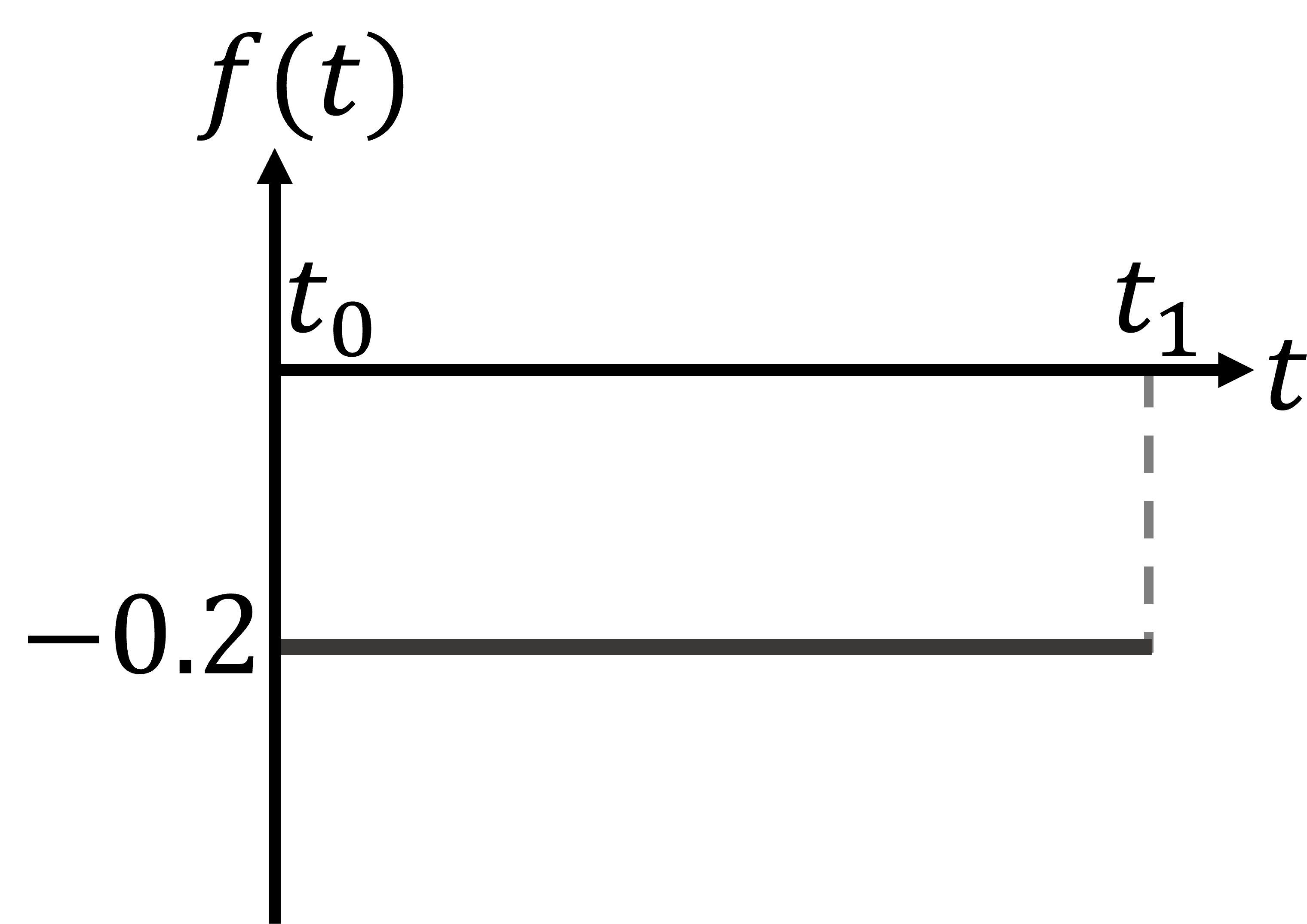}
    \end{minipage}\hfill%
    \begin{minipage}{0.16\linewidth}%
    \includegraphics[width=\linewidth]{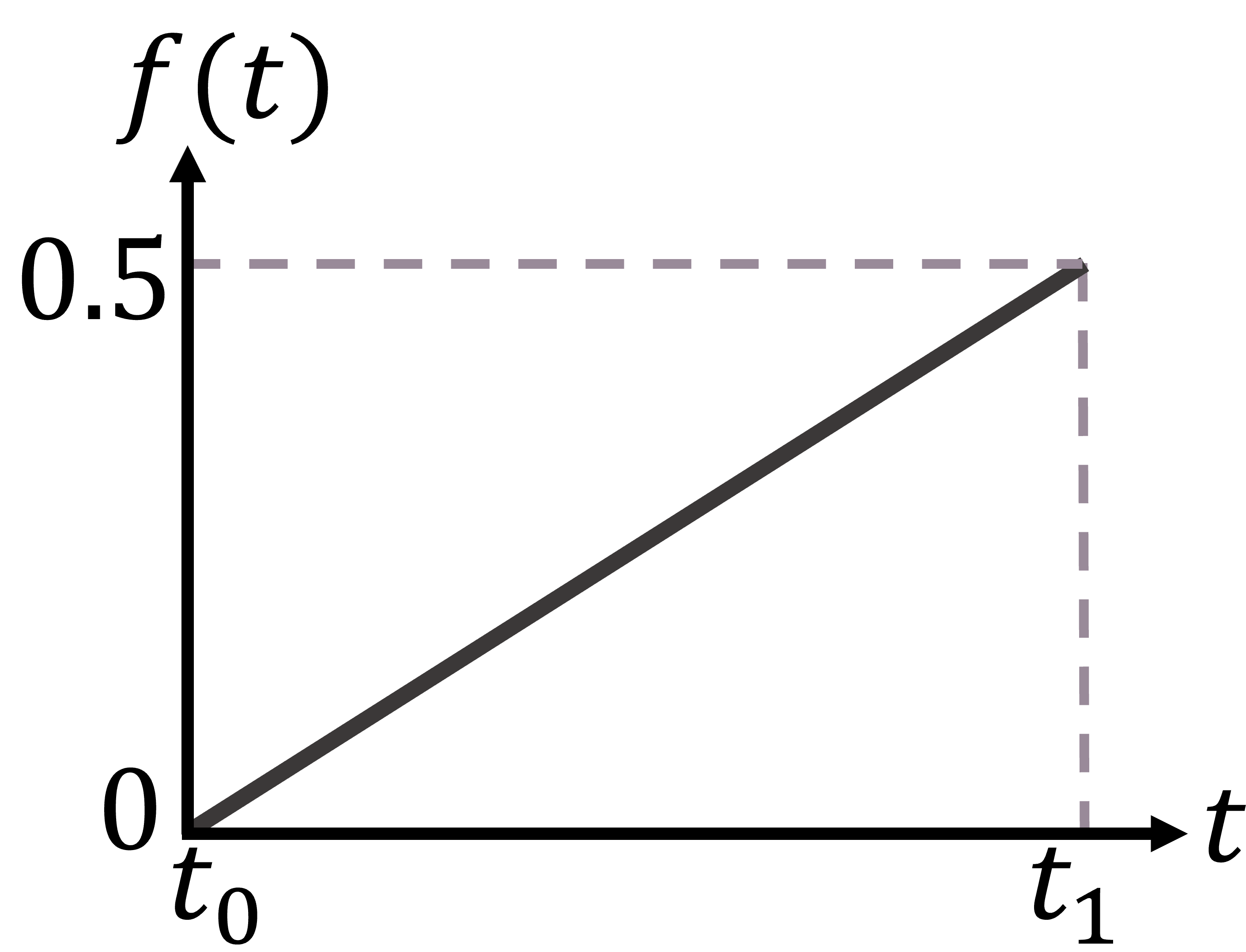}
    \end{minipage}\hfill%
    \begin{minipage}{0.16\linewidth}%
    \includegraphics[width=\linewidth]{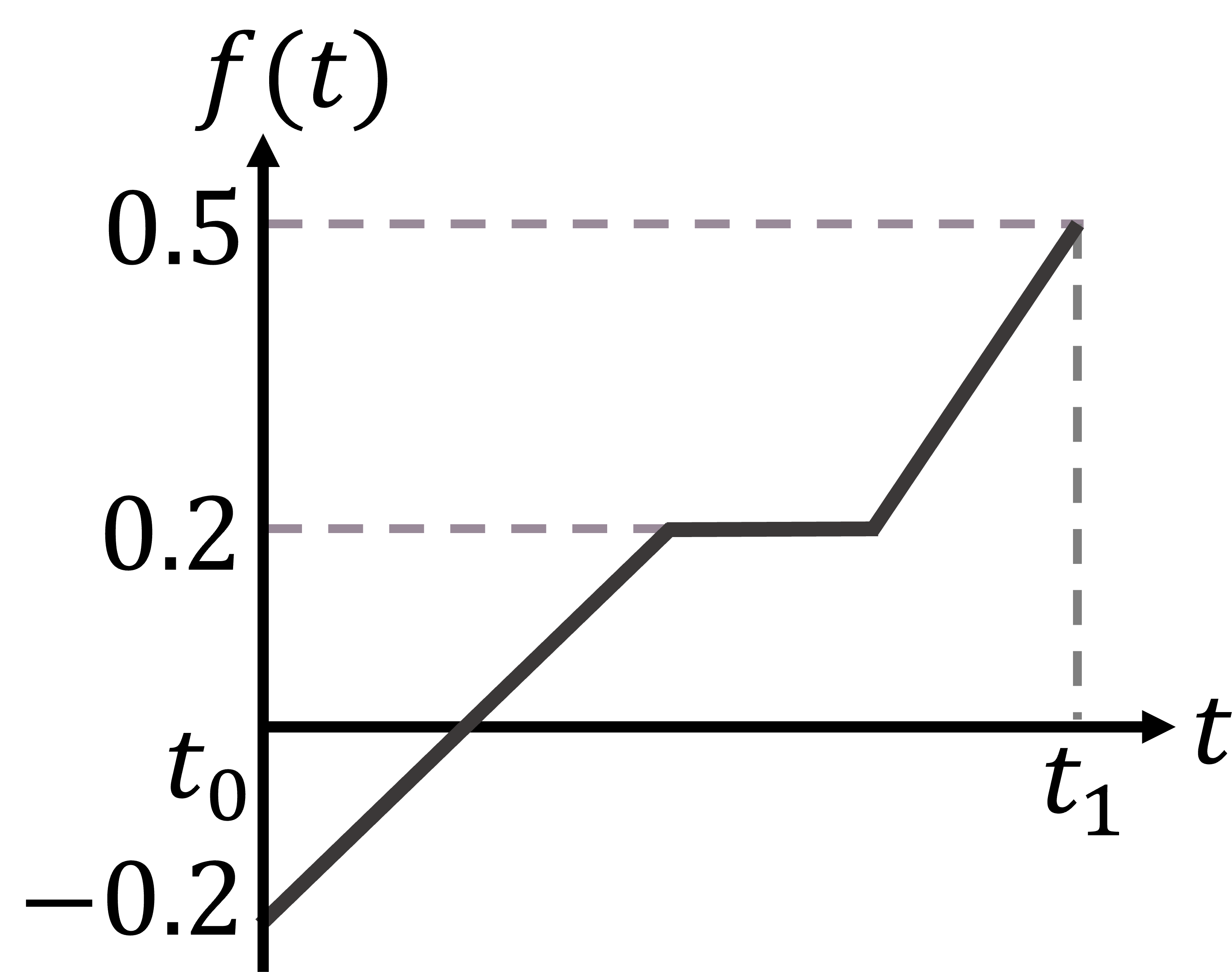}
    \end{minipage}\hfill%
    \begin{minipage}{0.16\linewidth}%
    \includegraphics[width=\linewidth]{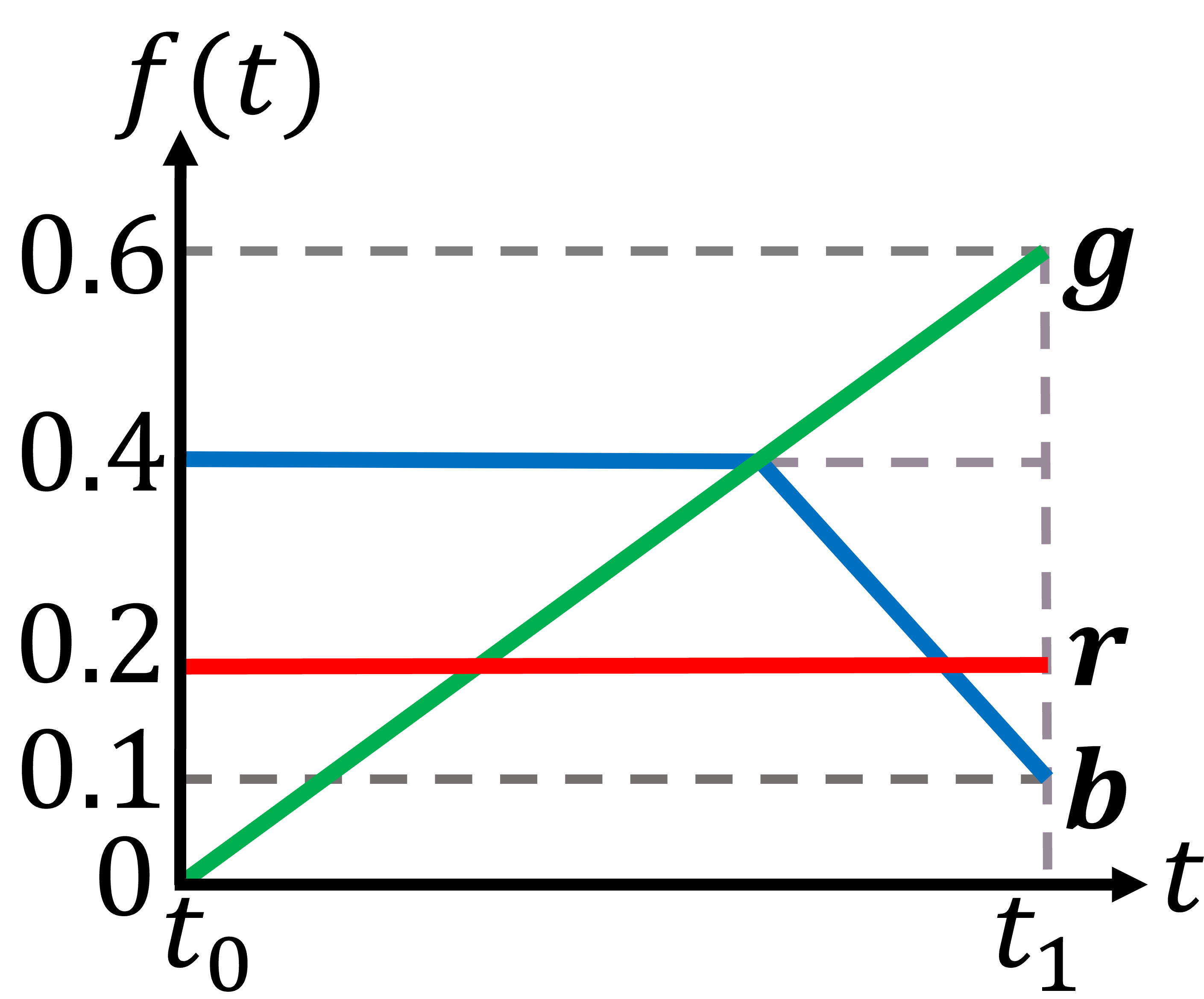}
    \end{minipage}\\%
    \begin{minipage}{0.16\linewidth}%
    $~$
    \end{minipage}\hfill%
    \begin{minipage}{0.16\linewidth}%
    \centering
    \scriptsize
    $\vf_l=0.2$
    \end{minipage}\hfill%
    \begin{minipage}{0.16\linewidth}%
    \centering
    \scriptsize
    $\vf_l=-0.2$
    \end{minipage}\hfill%
    \begin{minipage}{0.16\linewidth}%
    \centering
    \scriptsize
    linear
    \end{minipage}\hfill%
    \begin{minipage}{0.16\linewidth}%
    \centering
    \scriptsize
    piecewise
    \end{minipage}\hfill%
    \begin{minipage}{0.16\linewidth}%
    \centering
    \scriptsize
    rgb
    \end{minipage}%
    \caption{Overview of supported target Laplacians along Poisson curves, including constant values, linear slopes, piecewise linear functions, as well as the specification of a target Laplacian per color channel.}
    \label{fig:poisson-falloffs}
\end{figure}

Analogous to diffusion curves, we define a Poisson curve as a spatial curve $\vp(t): \mathbb{R} \rightarrow \cU$ with a target Laplacian function on the left $\vf_l(t): \mathbb{R} \rightarrow \mathbb{R}^3$ and on the right $\vf_r(t): \mathbb{R} \rightarrow \mathbb{R}^3$.
Upon rasterization, the sides cover the pixels $\cF_l \subseteq \cU$ and $\cF_r \subseteq \cU$, respectively, with $\vf_l(\vx) : \cF_l \rightarrow \mathbb{R}^3$ and $\vf_r(\vx): \cF_r \rightarrow \mathbb{R}^3$ being the target Laplacian at spatial coordinates $\vx \in \cD$.
To obtain shading consistency, \cite{hou2018poisson} required the functions $\vf_l(t)$ and $\vf_r(t)$ to sum to zero along the curve, i.e., $\forall \vx: \vf_l(t)=-\vf_r(t)$.
For notational convenience, we collectively refer to a sampling from the left or right side of the target Laplacian as $\vf_{\vp(t)}(\vx) : \cD \rightarrow \mathbb{R}^3$:
\begin{align}
    \vf_{\vp(t)}(\vx) = \begin{cases}
        \vf_l(\vx) \cdot \gamma_l(\vx) & \vx \in \cF_l \\
        \vf_r(\vx) \cdot \gamma_r(\vx) & \vx \in \cF_r \\
        \vNull & \textrm{else}
    \end{cases}
\end{align}
where $\gamma_l(\vx)$ is the number of adjacent pixels that belong to $\cF_r$ and $\gamma_r(\vx)$ is the number of adjacent pixels in $\cF_l$, which are used for normalization, cf. \cite{hou2018poisson}.
An overview of the supported target Laplacians $\vf_l(t)$ is shown in Fig.~\ref{fig:poisson-falloffs}.

\section{Edge Graph}
\label{sec:edge-graph}
Diffusion curves and gradient meshes are placed and edited by the user.
In order to formulate the image synthesis as a diffusion process with well-defined boundary conditions, it is necessary to subdivide the image into closed, non-overlapping regions.
When one or multiple possibly-closed or nested diffusion curves intersect with a gradient mesh, it is necessary to consistently determine a Laplacian for the interior of each separate region.
In the following, we utilize a planar map data structure~\citep{Gangnet89:PlanarMaps} that we use to resolve intersections.
The subsequent section will use this data structure to define connected regions and to determine a Laplacian for each region.

\subsection{Definition}
We introduce the undirected edge graph $(\cV, \cE)$, which consists of vertices $\vv\in\cV$ and edges $\ve\in\cE$.
The set of vertices $\cV$ contains all end points of the input boundary curves, as well as their intersections.
The set of edges $\cE$ expresses how the vertices are connected via the input boundary curves.
Thereby, each edge $\ve\in\cE$ stores a reference to the underlying input boundary curve that it was created for, such that it can access the orientation of the input boundary curve and its boundary conditions on the left side and right side.

\subsection{Construction}
\begin{figure}[t]%
    \centering%
    \includegraphics[width=0.8\linewidth]{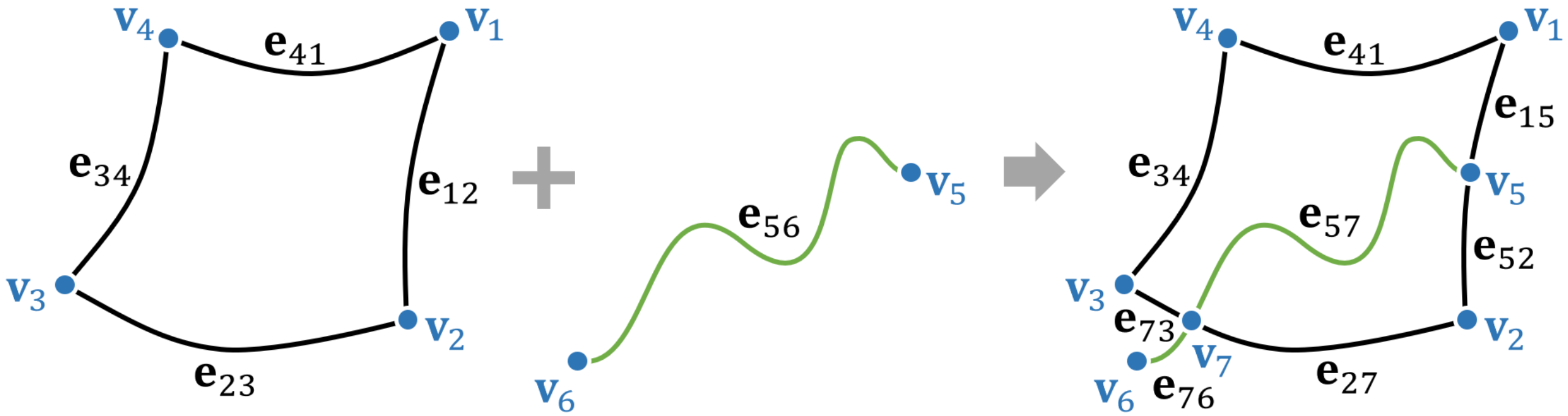}%
    \caption{%
    Given a graph with vertices $\cV=\{\vv_1, \vv_2, \vv_3, \vv_4\}$ and edges $\cE=\{\ve_{12}, \ve_{23}, \ve_{34}, \ve_{41}\}$.
    Inserting a new edge $\ve_{56}$ with end points $\vv_5$, $\vv_6$ leads to a new node $\vv_7$ at the edge intersection and a splitting of edges.
    }%
    \label{fig:edge-graph-construction}%
\end{figure}

\begin{figure}[t]%
    \centering%
    \includegraphics[width=0.8\linewidth]{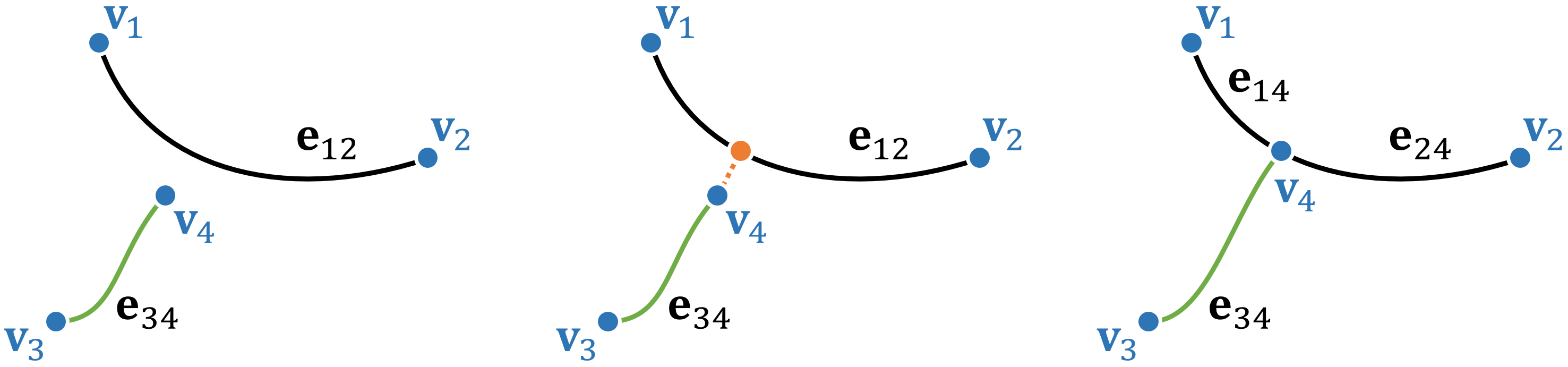}\\%
    \begin{minipage}{0.26\linewidth}%
        \centering \small input configuration
    \end{minipage}\quad%
    \begin{minipage}{0.26\linewidth}%
        \centering \small find closest point
    \end{minipage}\quad%
    \begin{minipage}{0.26\linewidth}%
        \centering \small snap $\vv_4$ onto curve
    \end{minipage}%
    \caption{\small When the endpoint of the next edge ($\ve_{34}$, green) is close to an already existing edge ($\ve_{12}$, black), then we snap the endpoint of the new edge onto the existing edge.}%
    \label{fig:scene-clean-up-schematic-1}%
\end{figure}

\begin{figure}[!t]%
    \centering%
    \includegraphics[width=0.25\linewidth]{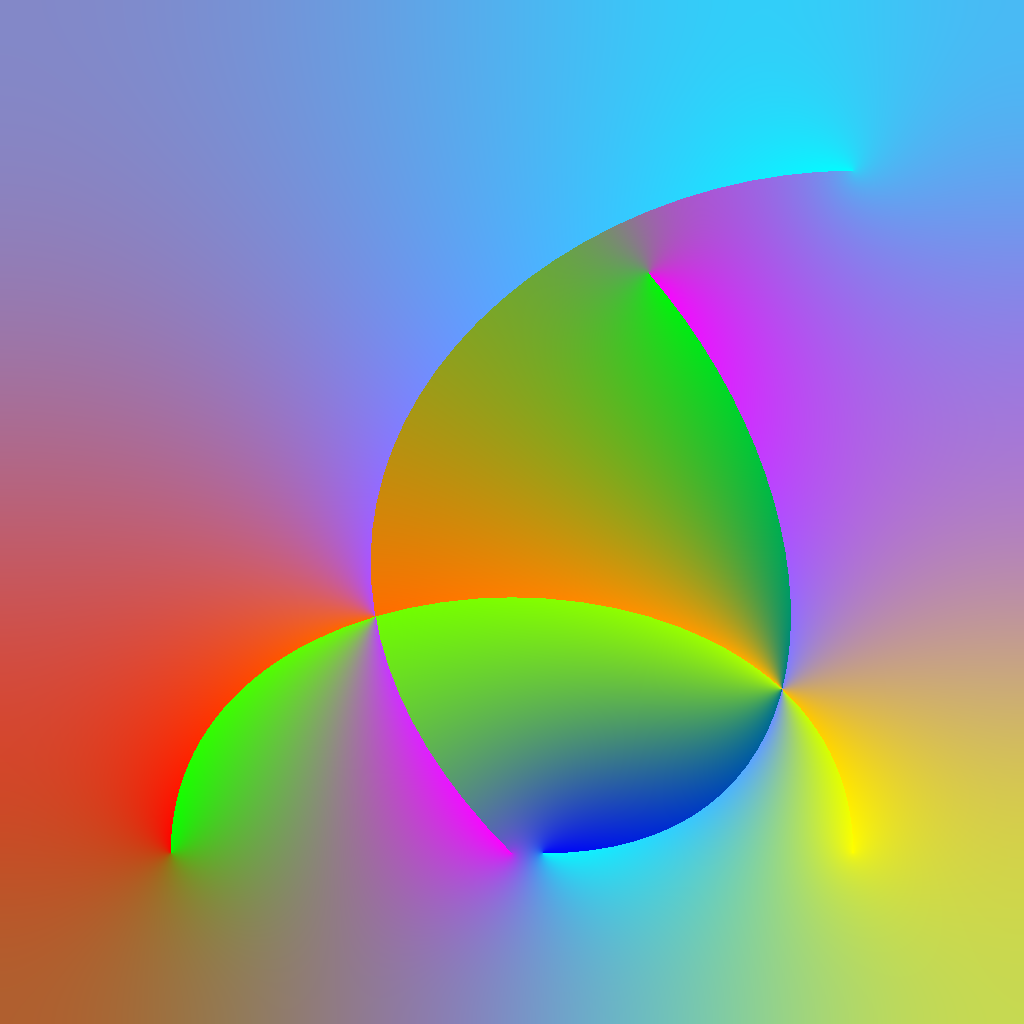}~%
    \includegraphics[width=0.25\linewidth]{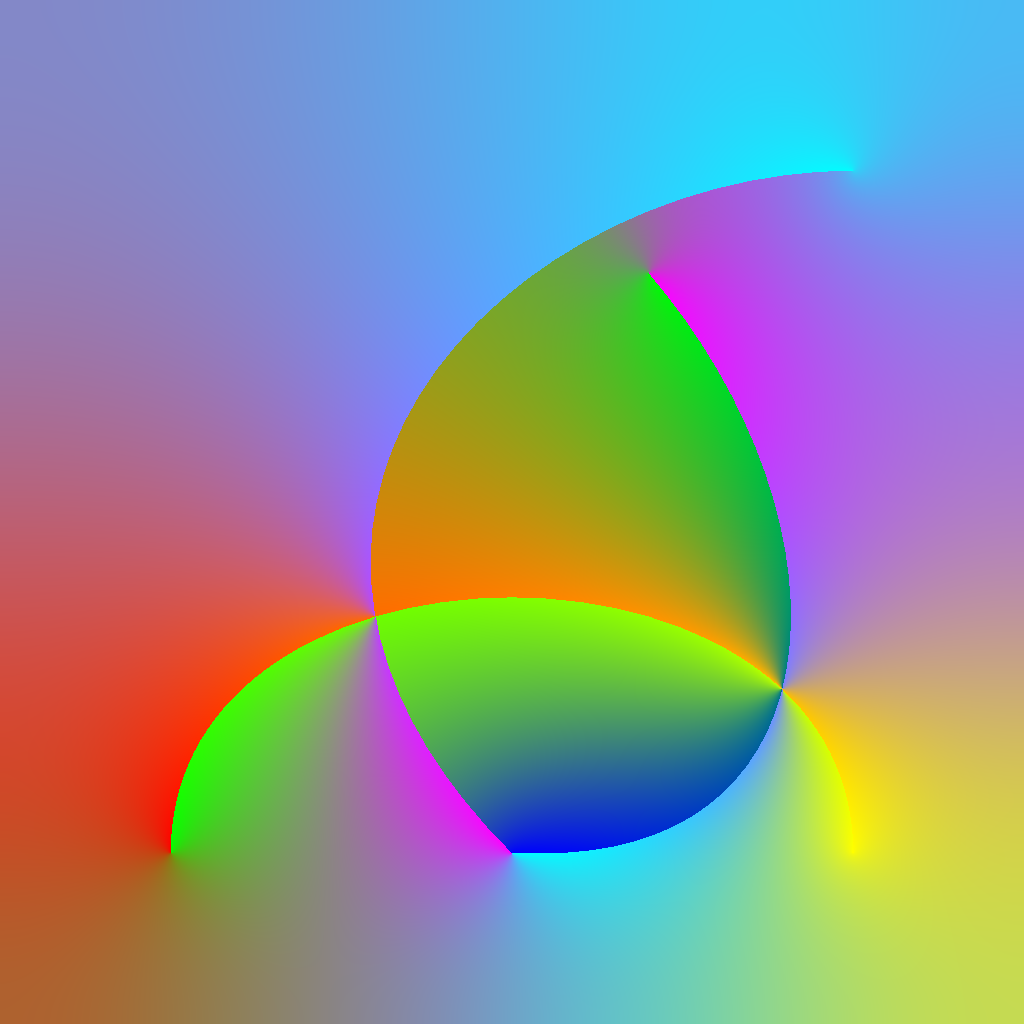}~%
    \includegraphics[width=0.25\linewidth]{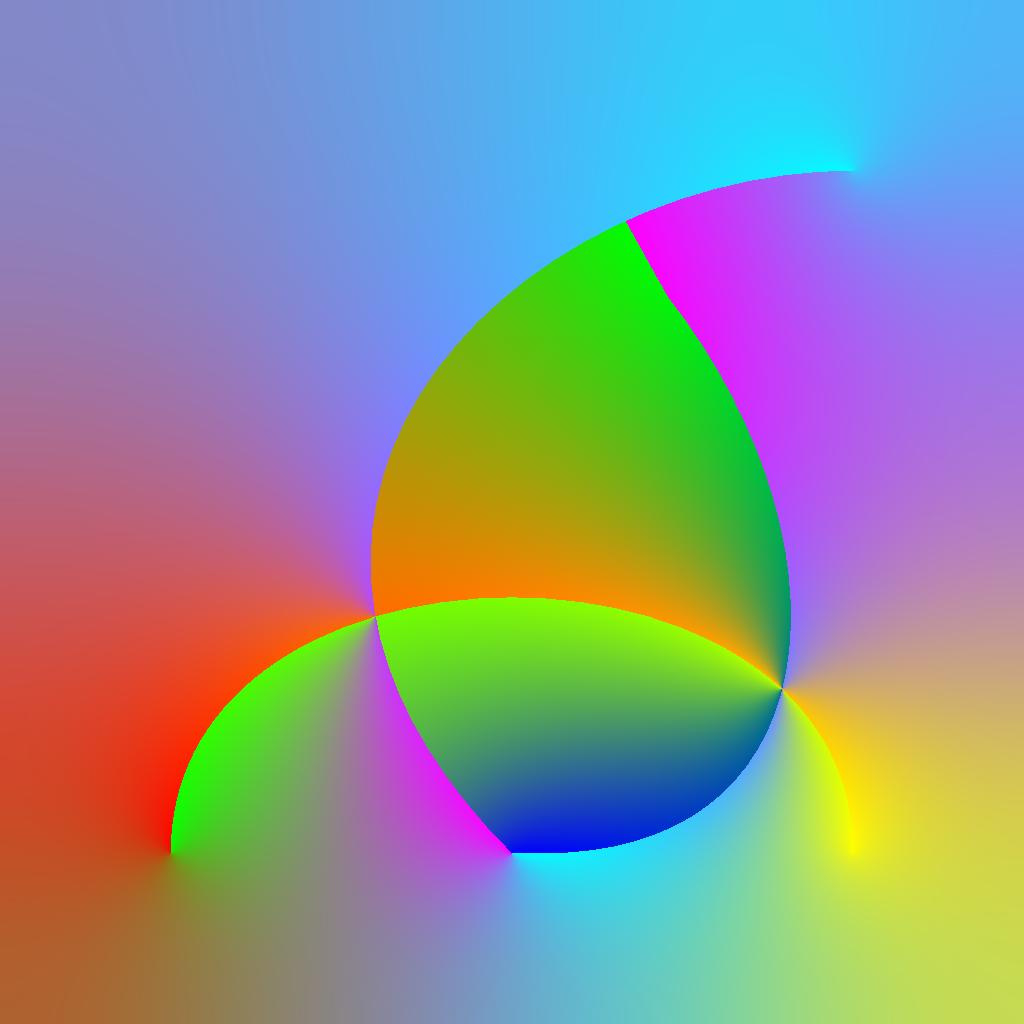}\\%
    \begin{minipage}{0.25\linewidth}%
        \centering \small $\tau=0.0$%
    \end{minipage}~%
    \begin{minipage}{0.25\linewidth}%
        \centering \small $\tau=0.1$%
    \end{minipage}~%
    \begin{minipage}{0.25\linewidth}%
        \centering \small $\tau=0.2$%
    \end{minipage}%
    \caption{\small Here, the snapping threshold $\tau$ is varied. On the left, the result without clean-up is shown. In the middle, the two endpoints at the bottom merged. On the right, the gap at the top was closed. This simple test scene is defined over the domain $[-0.5, 2.5]^2$.}%
    \label{fig:scene-clean-up-comparison}%
\end{figure}

An example of the edge graph construction is given in Fig.~\ref{fig:edge-graph-construction}.
We begin with an empty graph, i.e., $\cV = \varnothing$, $\cE = \varnothing$.
For each input boundary curve $(\vx(t), \cB_l, \cB_r) \in \cG \cup \cD$ from all extended gradient meshes $\cG$ and extended diffusion curves $\cD$, we take the following steps:
\begin{enumerate}
    \item Allocate two vertices $\vv_i = \vx(t_0)$, $\vv_j=\vx(t_1)$ and connect them by a curved edge $\ve_{ij}$ using the curve, i.e., $\ve_{ij} = \vx(t)$.
    \item If there is any other existing point $\vv_k \in \cV$ with $\|\vv_i - \vv_k\|^2<\tau$ or $\|\vv_j - \vv_k\|^2<\tau$, then merge nearby points.
    \item If there is any other edge $\ve_{kl} \in \cE$ with $\ve_{ij}\cap\ve_{kl} \neq \varnothing$, then find the intersection and split the connected edges.
\end{enumerate}
An inserted input boundary curve may have multiple intersections with other curves or with itself. 
The intersections are resolved by inserting vertices until no further intersections can be found.
With this, the graph is brought back to a consistent state and the next input boundary curve can be processed.
The parameter $\tau$ allows for closing gaps that would otherwise cause unwanted color leaking. 
Fig.~\ref{fig:scene-clean-up-schematic-1} illustrates the snapping of an edge endpoint onto an existing curve.
For this, the closest point on the existing edge is determined, which is either one of the end points or a point on the edge.
In case of the latter, the existing edge is subdivided, and the vertex of the new edge is snapped onto the curve.
Fig.~\ref{fig:scene-clean-up-comparison} shows renderings for different thresholds $\tau$ for a simple test scene.

\subsection{Implementation Detail}
\label{sec:edge-graph-implementation-detail}
\begin{figure}[t]%
    \centering%
    \includegraphics[width=0.8\linewidth]{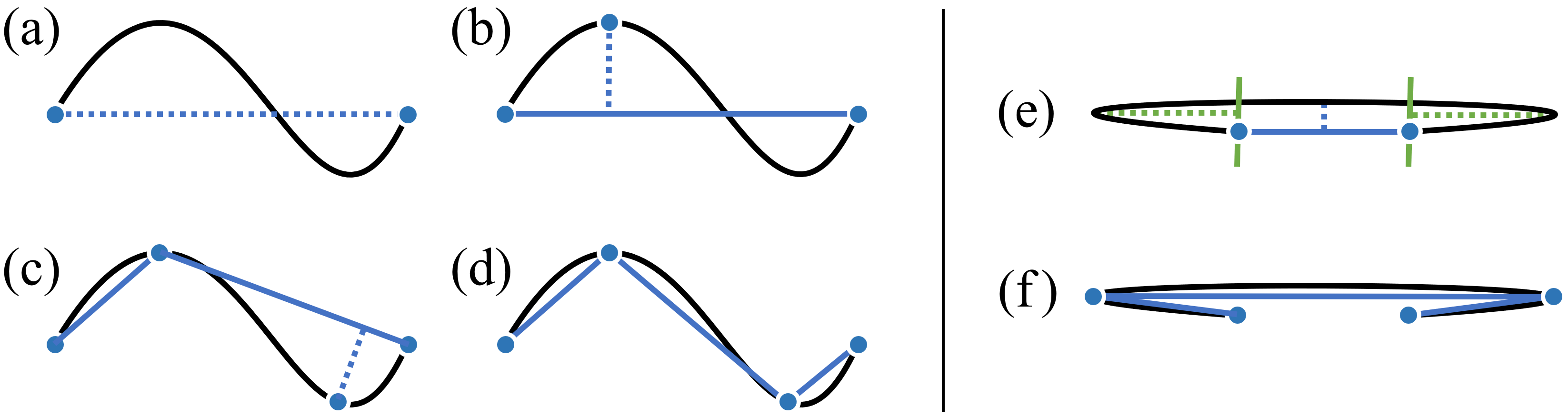}%
    \caption{Illustration of the curve discretization method. In (a)-(d), the recursive subdivision is shown, while (e)-(f) illustrates at an example why measuring vertical distances alone is not sufficient.}%
    \label{fig:top-down}%
\end{figure}

\begin{figure}[t]%
    \centering%
    \includegraphics[width=0.25\linewidth]{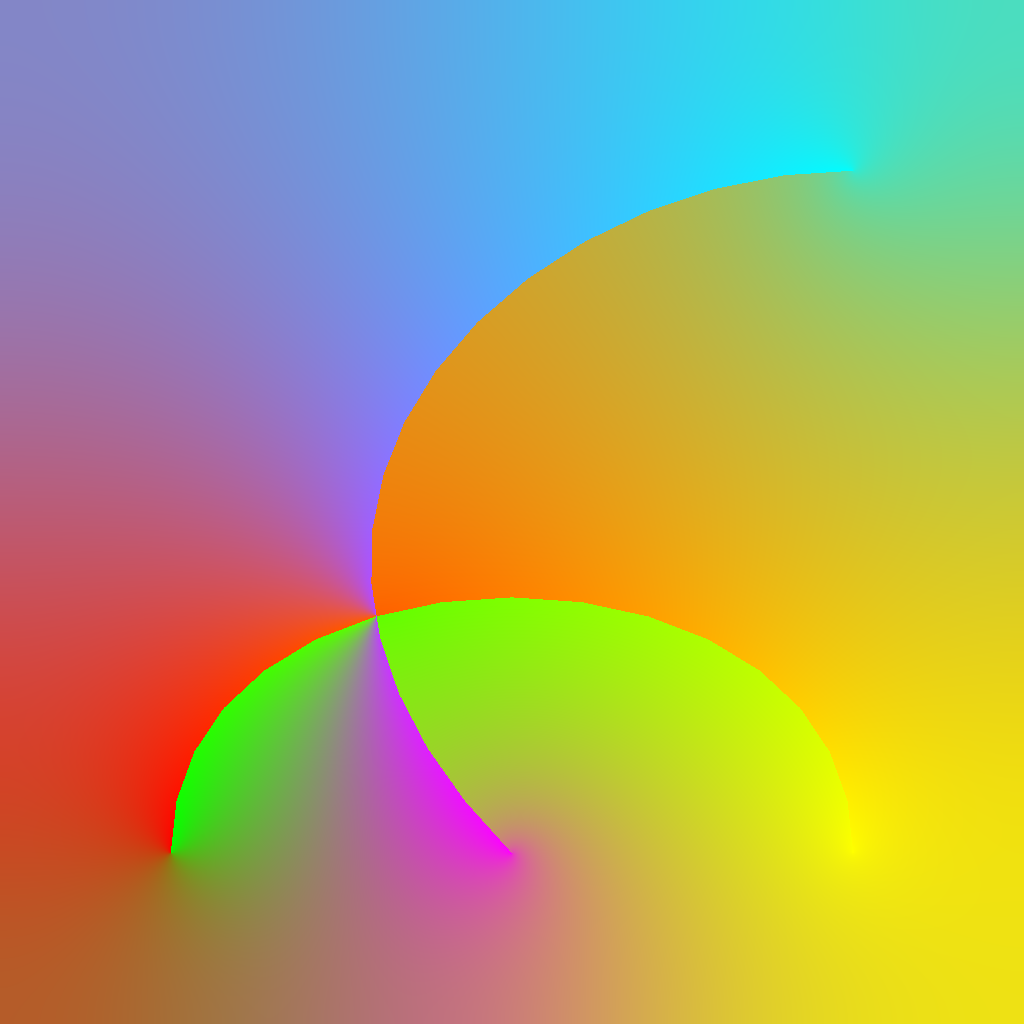}~%
    \includegraphics[width=0.25\linewidth]{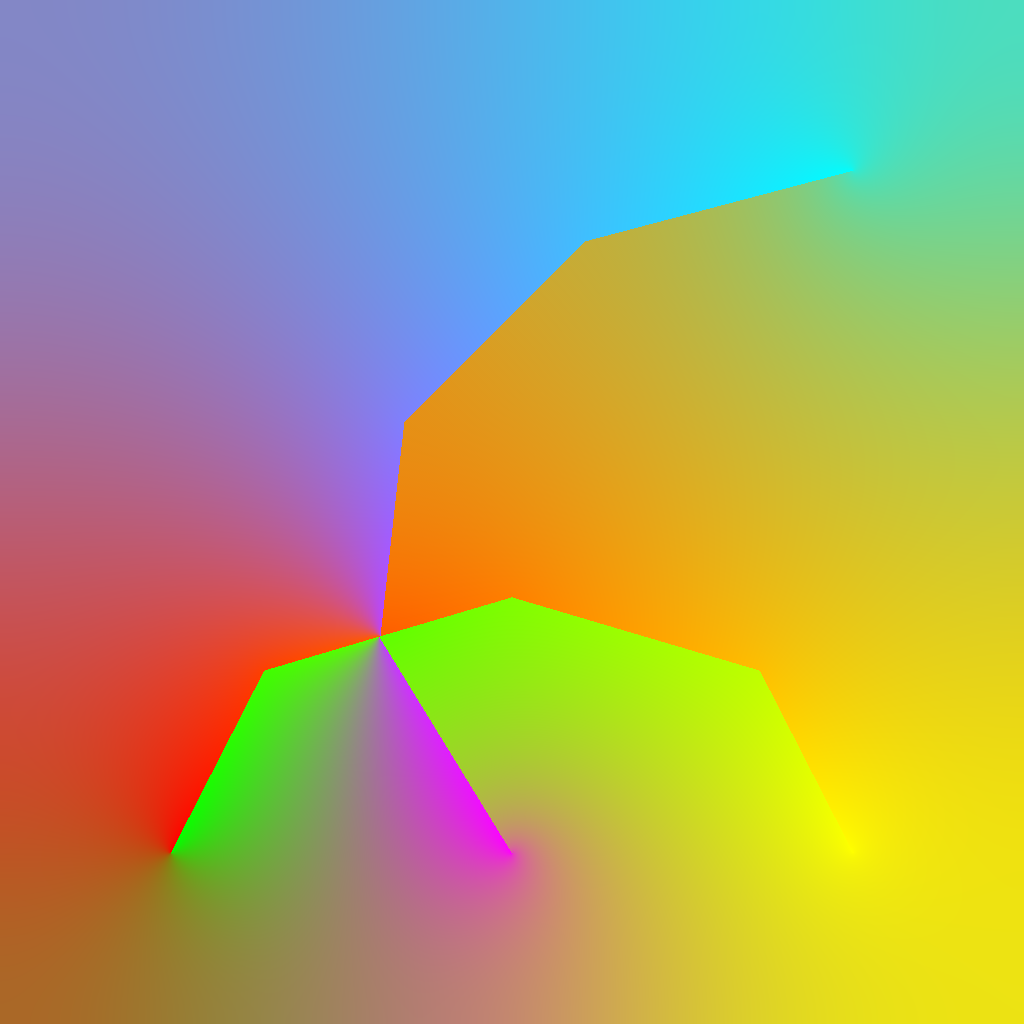}~%
    \includegraphics[width=0.25\linewidth]{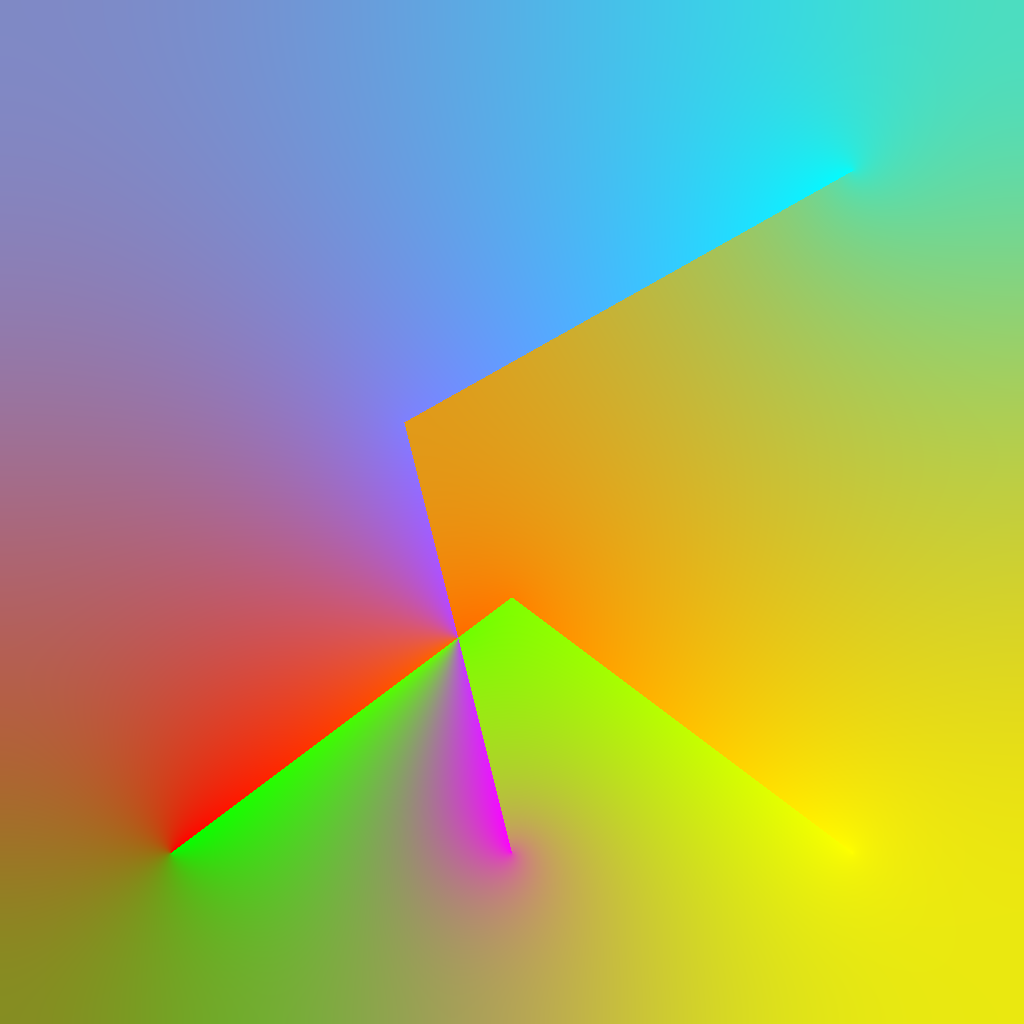}\\%
    \begin{minipage}{0.25\linewidth}%
        \centering \small $\epsilon=0.01$%
    \end{minipage}~%
    \begin{minipage}{0.25\linewidth}%
        \centering \small $\epsilon=0.1$%
    \end{minipage}~%
    \begin{minipage}{0.25\linewidth}%
        \centering \small $\epsilon=0.5$%
    \end{minipage}%
    \caption{Here, the results for different curve discretizations are shown, which are obtained when choosing different error thresholds $\epsilon$. This simple test scene is defined over the domain $[-0.5, 2.5]^2$.}%
    \label{fig:discretizations}%
\end{figure}

During the graph construction process, input boundary curve intersections need to be calculated. 
To accelerate the intersection tests, we internally discretize the input boundary curves into piece-wise linear polylines in a top-down  construction that is based on the Ramer-Douglas-Peucker algorithm~\citep{Ramer72,DouglasPeucker73}. 
However, instead of starting from a polyline, we start directly with the cubic B\'ezier curve. In Fig.~\ref{fig:top-down}(a), we define a line that connects the first and last vertex of the curve. In (b), we determine the curve point, which has the largest vertical distance from the line. This is done recursively using B\'ezier clipping~\citep{Sederberg90:BezierClipping}. In (c), we insert a new vertex at that point and recursively proceed on the segments left and right. In (d), we see the result after termination, which happens when the distance of the farthest point falls below an error threshold $\epsilon$. In (e), we can see that it is not sufficient to only measure the vertical distance to the line (blue). Instead, we also compute the (signed) horizontal distance from the start and end point (green). For subdivision, the maximum of the three distances is computed. 
In (g), the result is shown, which successfully approximates the curve.
Examples of different discretizations are shown in Fig.~\ref{fig:discretizations} for a simple test scene.
Finding curve intersections then reduces to a line-line intersection test. 
When a new curve is added by the user or when the control points of an existing curve are moved, the curve is re-discretized.

\section{Unified Patches}
\label{sec:unified-patches}
Using the edge graph, we subdivide the image domain into non-overlapping patches with well-defined boundary conditions and a Laplacian in the interior.
The image synthesis can then be performed for each patch independently.
In the following, we explain the patch generation process.

\subsection{Definition}
Formally, we denote the image domain as $\cU \subset \mathbb{R}^2$.
Our goal is to split the image domain into connected regions $\Omega_i \in \mathit{\Omega}$:
\begin{align}
    \bigcup_{\Omega \in \mathit{\Omega}} \Omega = \cU
\end{align}
with the requirement that for every $\Omega_i, \Omega_j \in \mathit{\Omega}$ with $\Omega_i \neq \Omega_j$, the intersection of the regions $\Omega_i \cap \Omega_j \subset \cE$ is either empty, a 0-simplex (point), or a 1-simplex (line) in the set of graph edges $\cE$. In other words, the boundaries of regions are sections of the input boundary curves. 
For each region $\Omega$, we denote the desired color field as $c(\xx): \Omega \rightarrow \cC$ for $\xx = (x,y) \in \Omega$, and its boundary as $\partial \Omega \subset \cE$, which might comprise just one closed loop or multiple closed loops if there are holes inside a region. 
We refer to the boundary curves of a patch as \emph{patch boundary curves}, which are clockwise-oriented curves with a boundary condition $\cB$ on only the \emph{right} side.
Further, each patch $\Omega$ is equipped with a target Laplacian function $\vf(\vx) : \Omega \rightarrow \mathbb{R}^3$ that controls the smooth color gradient inside the patch.

\begin{figure}[t]%
    \centering
    \includegraphics[width=0.5\linewidth]{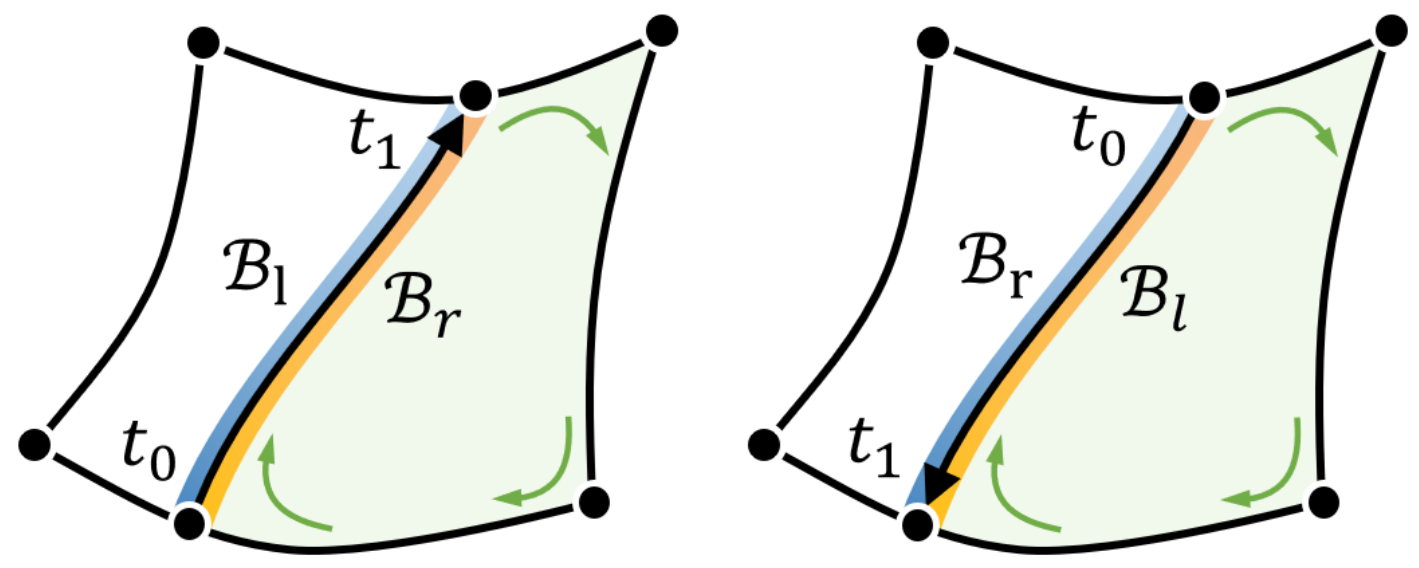}%
    \caption{Each edge is associated with an underlying input boundary curve, which has an orientation (based on the parameter range $[t_0, t_1]$) and a left and right boundary condition ($\cB_l$, $\cB_r$). When traversing the edge during construction of a patch (here green), the left or right boundary condition is chosen based on the orientation of the underlying input boundary curve.}
    \label{fig:patch-construct-orientation}
\end{figure}

\subsection{Construction}
\label{sec:unified-patch-construction}
In the following, we elaborate on how patch boundary curves and patches are created from the edge graph, and how the patch Laplacian function $\vf(\vx)$ is defined for each patch.

\paragraph{Patch Boundary Curves}
Each edge in the edge graph corresponds to an \emph{input boundary curve}.
Further, every edge is adjacent to two patches (except for image domain boundaries, where one side would be outside of the image).
Thus, when constructing patches each edge will be visited up to two times when constructing \emph{patch boundary curves} (once for the left side and once for the right side).
The \emph{patch boundary condition} $\cB$ of a patch boundary curve is chosen from the underlying input boundary curve based on its orientation, as illustrated in Fig.~\ref{fig:patch-construct-orientation}, i.e., we choose from the left and right boundary conditions $\cB_l$, $\cB_r$ of the input boundary curve.
Note that the two adjacent patches can be the \emph{same} patch, for example, when a diffusion curve is nested inside a patch and when it is not intersecting with the outer patch boundary.

\begin{figure}[t]%
    \centering
    \includegraphics[width=0.35\linewidth]{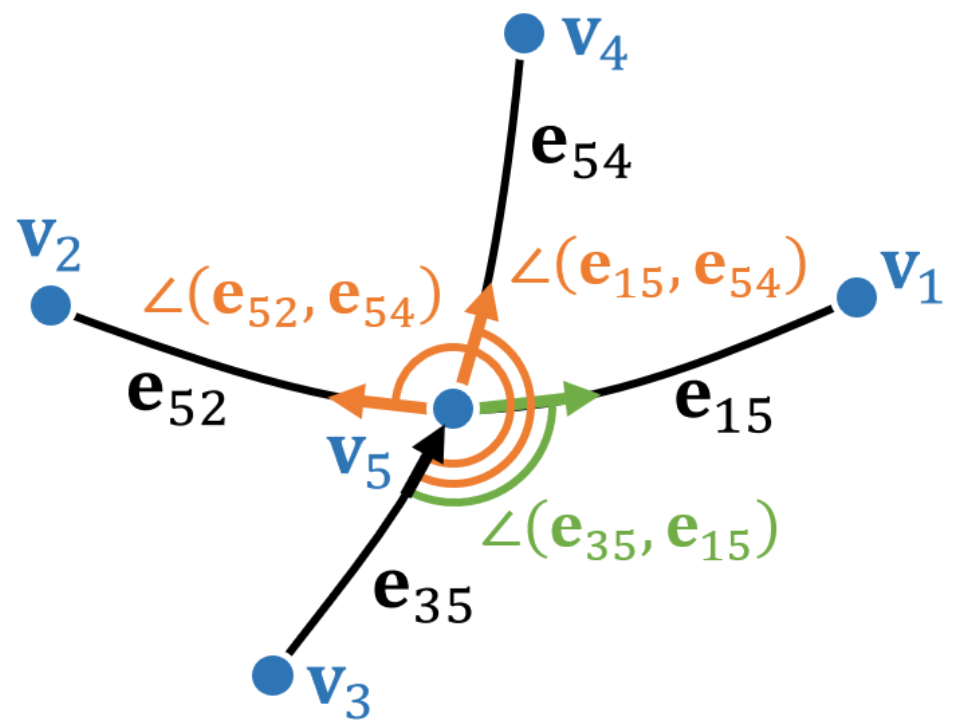}~~~
    \includegraphics[width=0.25\linewidth]{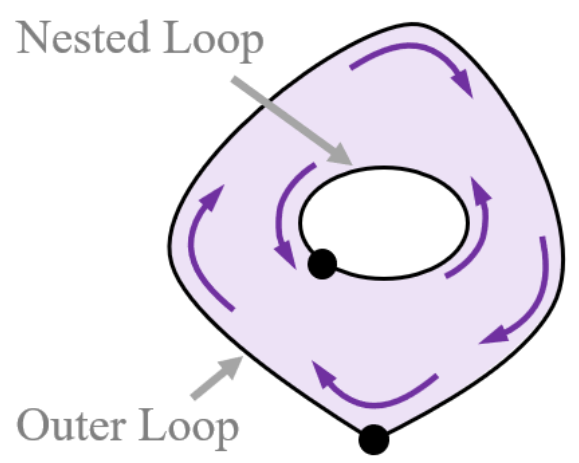}%
    \caption{Left: When traversing from edge $\ve_{35}$ to vertex $\vv_5$, there are three directions, in which the traversal could continue ($\ve_{15}$, $\ve_{54}$, $\ve_{52}$). Following edges, we always turn \emph{right}, i.e., we select the edge with smallest angle (green). Right: When a loop is completely nested inside another loop, the patch boundaries created from the inner loop are added to the patch.}
    \label{fig:patch-construct-angle-nested}
\end{figure}

\begin{figure}[t]%
    \centering%
    \includegraphics[width=0.4\linewidth]{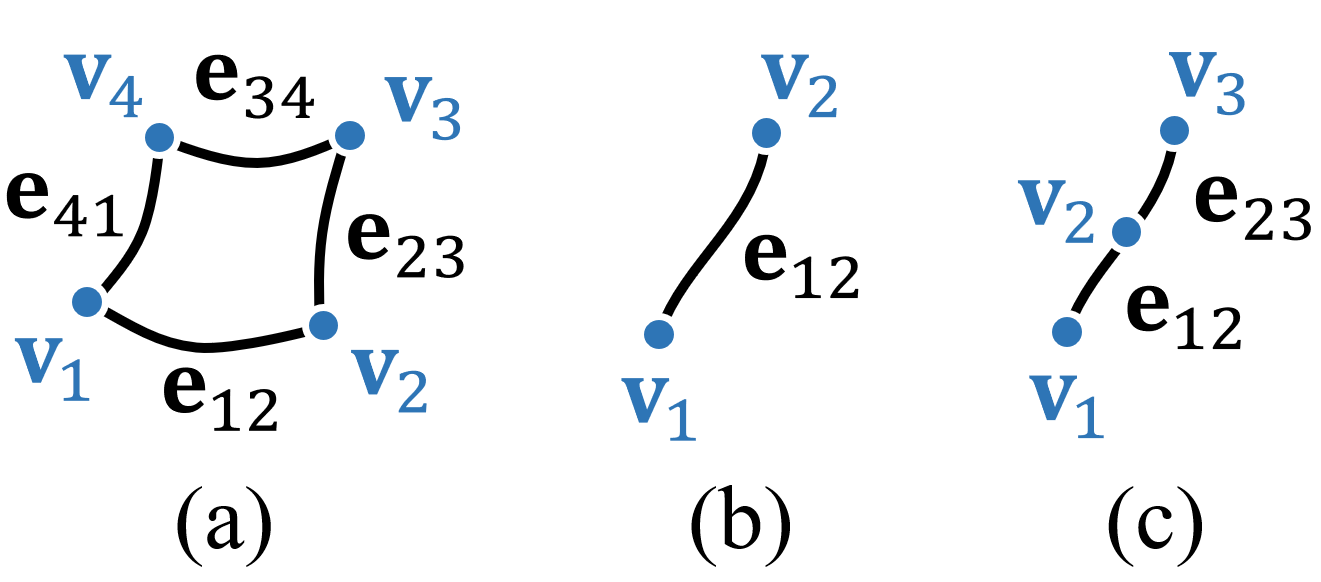}~~~~%
    \includegraphics[width=0.4\linewidth]{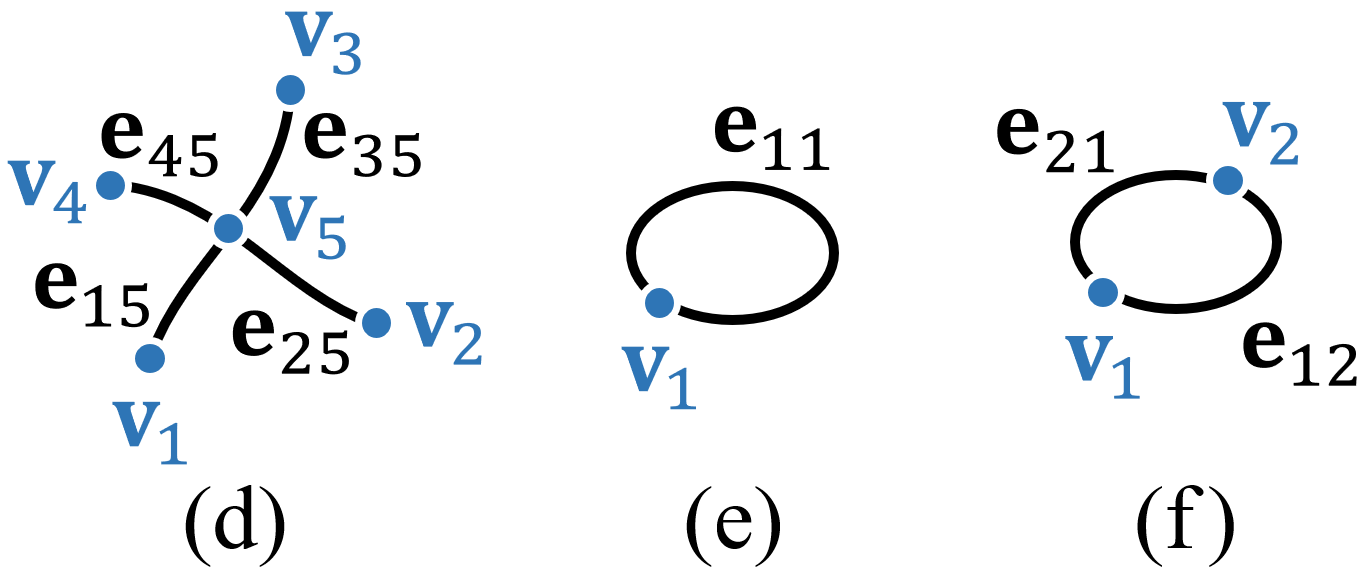}%
    \caption{Exemplary edge graphs to demonstrate the traversal order. Starting from an arbitrary edge in an arbitrary direction, the traversal continues always taking right turns until the loop is closed.}%
    \label{fig:edge-graph-traversal}%
\end{figure}

\paragraph{Patches}
The valence $|\vv_i|$ of vertex $\vv_i\in\cV$ in the graph, i.e., the number of edges connected to it, tells how often the vertex can be visited when traversing the graph in the search for closed patches, namely $2\cdot|\vv_i|$ (with the exception of image boundary vertices for which it is $2\cdot|\vv_i|-2$).
Starting from the first vertex and its first edge in the graph, we traverse the edges from vertex to vertex tracing out a sequence of patch boundary curves that encloses a patch.
A closed patch is formed by always turning \emph{right} when visiting a junction, i.e., a vertex.
This is illustrated in Fig.~\ref{fig:patch-construct-angle-nested} (left).
For the required angle measurements, the curve tangents $\dot\vx(t)$ are evaluated at the end vertex of the edge, i.e., at $t_0$ or $t_1$, respectively.
The edge traversal terminates when a loop was found, i.e., when the traversal returns to the start vertex.
By recording how often a vertex was visited, we repeat the process until all patches are inserted and no vertex is expecting another visit.
Fig.~\ref{fig:edge-graph-traversal} shows a number of exemplary edge graphs.
In the following, we list the corresponding traversal sequences.

(a) Traverse outer edge of a closed domain:
$\vv_1$
$\rightarrow$ $\ve_{12}$
$\rightarrow$ $\vv_2$
$\rightarrow$ $\ve_{23}$
$\rightarrow$ $\vv_3$
$\rightarrow$ $\ve_{34}$
$\rightarrow$ $\vv_4$
$\rightarrow$ $\ve_{41}$
$\rightarrow$ $\vv_1$

(b) Traverse single edge:
$\vv_1$
$\rightarrow$ $\ve_{12}$
$\rightarrow$ $\vv_2$
$\rightarrow$ $\ve_{12}$
$\rightarrow$ $\vv_1$

(c) Traverse sequence of edges:
$\vv_1$
$\rightarrow$ $\ve_{12}$
$\rightarrow$ $\vv_2$
$\rightarrow$ $\ve_{23}$
$\rightarrow$ $\vv_3$
$\rightarrow$ $\ve_{23}$
$\rightarrow$ $\vv_2$
$\rightarrow$ $\ve_{12}$
$\rightarrow$ $\vv_1$

(d) Traverse crossing of edges:
$\vv_1$
$\rightarrow$ $\ve_{15}$
$\rightarrow$ $\vv_5$
$\rightarrow$ $\ve_{25}$
$\rightarrow$ $\vv_2$
$\rightarrow$ $\ve_{25}$
$\rightarrow$ $\vv_5$
$\rightarrow$ $\ve_{35}$
$\rightarrow$ $\vv_3$
$\rightarrow$ $\ve_{35}$
$\rightarrow$ $\vv_5$
$\rightarrow$ $\ve_{45}$
$\rightarrow$ $\vv_4$
$\rightarrow$ $\ve_{45}$
$\rightarrow$ $\vv_5$
$\rightarrow$ $\ve_{15}$
$\rightarrow$ $\vv_1$

(e) Traverse closed loop with one vertex:
$\vv_1$
$\rightarrow$ $\ve_{11}$
$\rightarrow$ $\vv_1$

(f) Traverse closed loop with multiple vertices:
$\vv_1$
$\rightarrow$ $\ve_{12}$
$\rightarrow$ $\vv_2$
$\rightarrow$ $\ve_{21}$
$\rightarrow$ $\vv_1$

\paragraph{Nested patches}
A nested edge loop, as shown in Fig.~\ref{fig:patch-construct-angle-nested} (right), has an interior patch boundary curve sequence, which bounds the patch that is contained by the edge loop, and an exterior patch boundary curve sequence, which denotes a patch boundary of the parent patch that contains the nested edge loop.
Due to the convention of always turning \emph{right} during the edge graph traversal, the two sequences are distinguished by their turning number.
For all interior curve sequences, the boundary curve tangent performs a clockwise rotation (negative turning number), while the tangent of the exterior curve sequence performs a counter-clockwise rotation (positive turning number) when traversing the edge loop.
For an arclength parameterized closed edge sequence $\ve(s) : [s_0,s_1] \rightarrow \partial\Omega$, the integer-valued turning number is calculated by integrating the signed curvature $k(s)$ along the closed patch boundary curve:
\begin{align}
\textrm{turn}(\ve(s)) = \frac{1}{2\pi}\int_{s_0}^{s_1} k(s) \, \mathrm{d}s,
\;\; \textrm{with} \;\;
k(s) = \frac{\mathrm{det}(\dot\ve(s), \ddot\ve(s))}{\|\dot\ve(s)\|^3}.
\label{eq:turning-number}
\end{align}

\begin{figure}[t]%
    \centering%
    \includegraphics[width=0.25\linewidth]{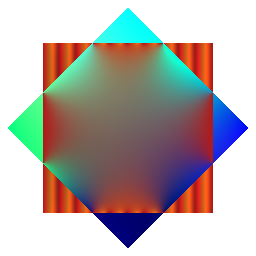}\quad%
    \includegraphics[width=0.25\linewidth]{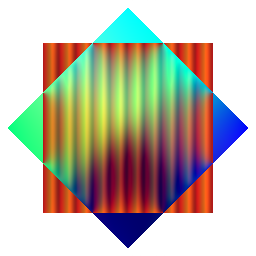}\quad%
    \includegraphics[width=0.25\linewidth]{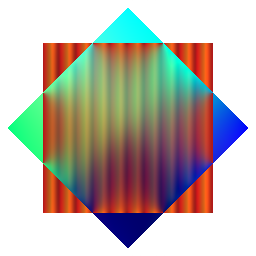}\\%
    \begin{minipage}{0.25\linewidth}%
        \centering \small Zero%
    \end{minipage}\quad%
    \begin{minipage}{0.25\linewidth}%
        \centering \small Sum%
    \end{minipage}\quad%
    \begin{minipage}{0.25\linewidth}%
        \centering \small Average%
    \end{minipage}\hfill%
    \includegraphics[width=0.25\linewidth]{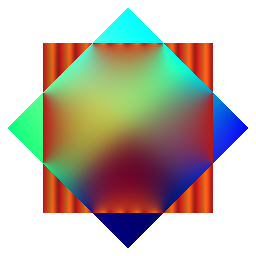}\quad%
    \includegraphics[width=0.25\linewidth]{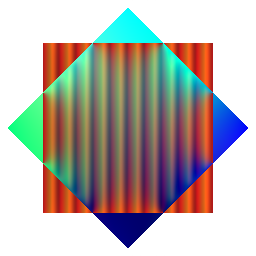}\\%
    \begin{minipage}{0.25\linewidth}%
        \centering \small First a)%
    \end{minipage}\quad%
    \begin{minipage}{0.25\linewidth}%
        \centering \small First b)%
    \end{minipage}%
    \caption{Different options for handling overlapping gradient meshes. 'Zero' sets the Laplacian to zero, which gives the smoothest possible interpolation of the boundary conditions. 'Sum' sums up the Laplacians, which can lead to high values when many gradient meshes overlap. 'Average' computes the average of the Laplacians, which scales better to larger numbers of gradient meshes. 'First' only takes one of the Laplacians, for which we show both options: choosing the smooth square in a) and the striped square in b).}%
    \label{fig:overlapping-gradient-meshes}%
\end{figure}

\paragraph{Patch Laplacian}
The patch Laplacian $\vf(\vx)$ defines the color gradient inside a patch, which is controlled by evaluating and adding the Laplacian $\vf_{\vp(t)}(\vx)$ of Poisson curves $\vp(t)\in\cP$ and by sampling the color derivatives of gradient meshes, cf. Section~\ref{sec:related-mesh-based}.
\begin{align}
    \vf(\vx) = 
    \underbrace{\sum_{\vp(t)\in\cP} \vf_{\vp(t)}(\vx)}_{\textrm{Poisson curves}}
    + 
    \underbrace{\sum_{\vc_i(\vu) \in \cG}
     \lambda_i \cdot \Delta_\vx \vc_i(\vu(\vx)) }_{\textrm{Gradient meshes}}.
    \label{eq:patch-Laplacian}
\end{align}%
If a patch boundary curve of a patch is part of a gradient mesh interior, we add the gradient mesh to a list. Each Laplacian $\Delta_\vx \vc(\vu(\vx))$ can be evaluated analytically as explained in Appendix~\ref{sec:gradient-mesh-fill} (including the coordinate transformation $\vu(\vx)$ that takes image coordinates to a $(u,v)$ coordinate in the gradient mesh).
When a patch is formed from one gradient mesh, then the diffused color gradient inside the patch matches exactly the color gradient of the interpolated gradient mesh.
In the event that multiple gradient meshes are overlapping at the pixel coordinate $\vx$, we use a weighted sum of the gradient mesh Laplacians to compute the source term $\vf(\vx)$.
We offer four choices for how the weights $\lambda_i$ are chosen.
In the following, let $n$ be the number of gradient meshes, i.e., $i \in \{1,\dots,n\}$.
\begin{enumerate}
    \item All weights are $\lambda_i=0$, i.e., the Laplacians are ignored.
    \item All weights are $\lambda_i=1$, i.e., all Laplacians are summed.
    \item All weights are $\lambda_i=\frac{1}{n}$, i.e., the Laplacian's are averaged.
    \item Only one Laplacian is chosen with $\lambda_1=1$, and the others are set to zero $\lambda_2=\dots=\lambda_n=0$.
\end{enumerate}
Options 1) -- 3) are order-independent, while option 4) requires the specification of an order. 
We assume that the gradient meshes $\cG$ are given in a particular order by the user, i.e., we choose the gradient mesh that is sitting 'on top' of all others.
For this, the user interface contains a list of all gradient meshes, which can be reordered by the user.
Fig.~\ref{fig:overlapping-gradient-meshes} gives examples.

\subsection{Implementation Detail}
The turning number in Eq.~\eqref{eq:turning-number} is calculated from the polyline discretization, previously described in Section~\ref{sec:edge-graph-implementation-detail}, by summing up the discrete turning angles. 
A loop is nested inside another loop if one of its vertices is contained in the other loop, since loops are by construction intersection free.
To determine containment bounding boxes are used to find candidates and a detailed check is performed by using the winding angle theorem~\citep{Jacobson13,Spainhour24}, which is a numerically robust way of testing containment.  

\section{Image Synthesis}
\label{sec:image-synthesis}
In the following, we discuss how to rasterize a patch $\Omega$ into a raster image that covers the image domain $\cU$.

\subsection{Definition}
The color field $\vc(\xx)$ of a region $\Omega$ is determined by solving a partial differential equation, i.e., a well-known Poisson problem:
\begin{align}
    \Delta \vc(\xx) = \vf(\xx), \quad s.t. \;\; \cB \in \partial\Omega
    \label{eq:poisson-problem}
\end{align}
with the patch Laplacian $\vf(\vx)$ as given by Eq.~\eqref{eq:patch-Laplacian} and subject to the patch boundary conditions ${\cB}$ that arise from the patch boundary curves on $\partial\Omega$, as introduced in Eqs.~\eqref{eq:boundary-condition-dirichlet}--\eqref{eq:boundary-condition-neumann}.
There are many ways how the Poisson problem could be solved, including direct solvers, iterative solvers, or Monte Carlo methods.
We refer to a survey~\citep{Tian23} for an overview of solvers in the smooth vector graphics literature and also point out several recent solvers, which appeared after the survey~\citep{Bang23:Multipole,Sawhney23:WoSt,Sugimoto23:WoB,Miller24:Robin,Chen24Lightning}.
Since the PDE solver is orthogonal to our contributions, we utilized a simple (multi-grid) Jacobi relaxation~\citep{jeschke2009gpu}:
\begin{align}
    \vc_{i,j}^{(n+1)} = \frac{
        \sum_{x\in\{-1,1\}}
        \sum_{y\in\{-1,1\}}
        w_{i+x,j+y} \cdot c_{i+x, j+y}^{(n)}
        - h^2 \vf_{i,j}
    }
    {w_{i+1,j} + w_{i-1,j} + w_{i,j+1} + w_{i,j-1}}
    \label{eq:Jacobi-relax}
\end{align}
which iteratively calculates the pixel color $\vc_{ij}$ for a pixel $(i,j)$ from its neighbors such that it meets the target Laplacian $\vf_{ij}$ and where $h$ is the grid spacing of pixels in the image domain $\cU$.
The weights $w_{ij}$ are used to handle boundary conditions, which is explained in the implementation details below.

\subsection{Implementation Detail}
\begin{figure}[t]%
    \centering%
    \includegraphics[width=0.5\linewidth]{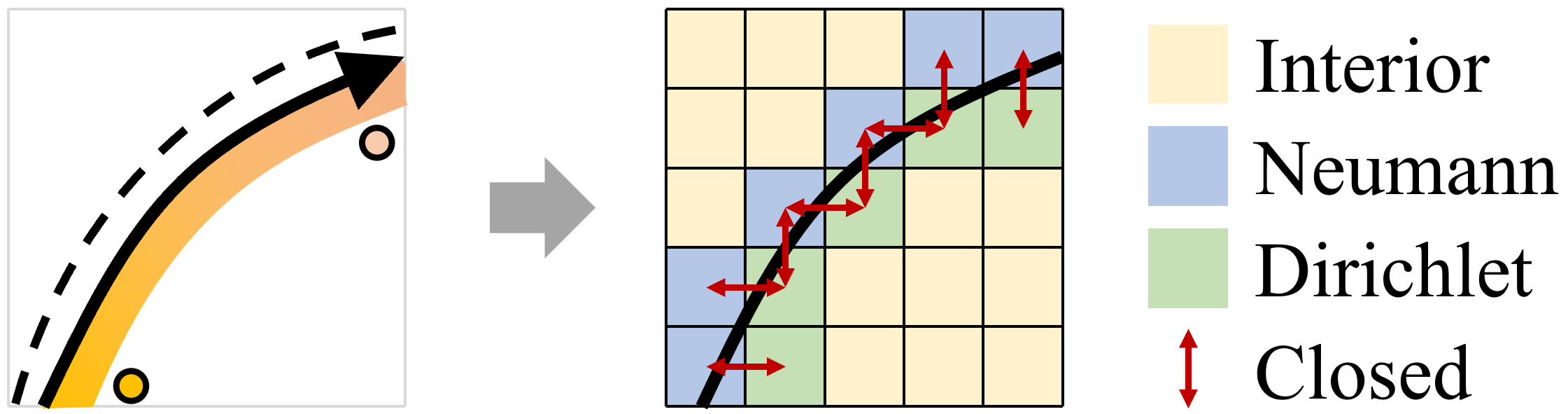}%
    \caption{Left: Given is a diffusion curve with Neumann condition ($\cB_l$) and Dirichlet condition ($\cB_r$). Right: depiction of the pre-computed masks. We determine for each pixel, whether it is in the interior, at a Neumann boundary, or a Dirichlet boundary, which is stored in $k_{i,j}$. Further, we identify if adjacent pixels are separated (horizontally or vertically) by a patch boundary curve, and store this as binary flag in staggered grids $s^h_{i,j}$, $s^v_{i,j}$.}%
    \label{fig:masks}%
\end{figure}

We implemented the Jacobi relaxation on the GPU, which requires a texture containing the target Laplacians $\vf_{i,j}$, a texture containing the type of the pixel $k_{i,j}\in\{\textrm{interior}, \textrm{Dirichlet}, \textrm{Neumann}\}$, a bit mask on a staggered grid that indicates whether horizontally and vertically adjacent pixels are separated by a patch boundary curve $s^h_{i,j}, s^v_{i,j} \in \{\textrm{open}, \textrm{closed}\}$, and a texture that stores the final pixel color $\vc_{i,j}$, see Fig.~\ref{fig:masks}.
If the pixel $(i,j)$ is a:
\begin{enumerate}
    \item \emph{interior pixel} ($k_{i,j} = \textrm{interior}$), then the weights of all adjacent pixels are set to $1$, i.e.,  $w_{i+1,j}=w_{i-1,j}=w_{i,j+1}=w_{i,j-1}=1$.
    \item \emph{boundary pixel with Dirichlet condition} ($k_{i,j} = \textrm{Dirichlet}$), then Eq.~\eqref{eq:Jacobi-relax} is not executed. Instead, the color of the pixel is determined by the boundary condition.
    \item \emph{boundary pixel with Neumann condition} ($k_{i,j} = \textrm{Neumann}$), then the weight of the adjacent pixel $w_{i+1,j}$ is set to $1$ (otherwise $0$) if it is non-separated by a patch boundary curve $s^h_{i,j} = \textrm{open}$ (analogous for the other neighbor pixels). 
\end{enumerate}
We identify if a pixel is a boundary pixel in a pre-process by determining the pixels that belong to the patch using the winding angle theorem~\citep{Jacobson13}, and flag pixel $(i,j)$ as boundary if one of its neighbors $(i+1,j)$, $(i-1,j)$, $(i,j+1)$, $(i,j-1)$ is not in the patch, or if a ray cast between adjacent pixels intersects a patch boundary curve. We determine the closest patch boundary curve and set the pixel type $k_{i,j}$ and the openness $s_n$, $s_e$, $s_s$, $s_w$ accordingly.
If the pixel has a Dirichlet condition, then we set $\vc_{i,j}$ according to the closest hit location on the patch boundary curve. 
If the Dirichlet condition originated from an input boundary curve that belonged to a gradient mesh, then we interpolate the pixel color in the gradient mesh at the location of the pixel center $(i,j)$, rather than taking the color on its boundary curve.
The texture with the target Laplacian $\vf_{i,j}$ is set by evaluating the gradient mesh Laplacian $\Delta_\vx \vc(\vu(\vx))$ (if a gradient mesh was contained, cf. Eq.~\eqref{eq:patch-Laplacian}), and the Laplacians $\vf_{\vp(t)}$ of Poisson curves $\vp(t)$ are additively rasterized on top.

\section{Results}
In the following section, we present our results using scenes that contain gradient meshes, diffusion curves, and Poisson curves.
We made all test scenes available online~\citep{Tian25:USVG-code}.
All images are rendered at a resolution of $1024\times 1024$.

\begin{figure*}[t]%
  \centering%
  \begin{minipage}{0.19\linewidth}%
      \centering%
      \scriptsize%
      (a) input meshes only
  \end{minipage}\hfill\hfill%
  \begin{minipage}{0.19\linewidth}%
      \centering%
      \scriptsize%
      (b) input curves only
  \end{minipage}\hfill\hfill%
  \begin{minipage}{0.19\linewidth}%
      \centering%
      \scriptsize%
      (c) undirected edge graph
  \end{minipage}\hfill%
  \begin{minipage}{0.19\linewidth}%
      \centering%
      \scriptsize%
      (d) unified patch representation
  \end{minipage}\hfill%
  \begin{minipage}{0.19\linewidth}%
      \centering%
      \scriptsize%
      (e) final synthesized image
  \end{minipage}%
  \\
  \resultrow
  {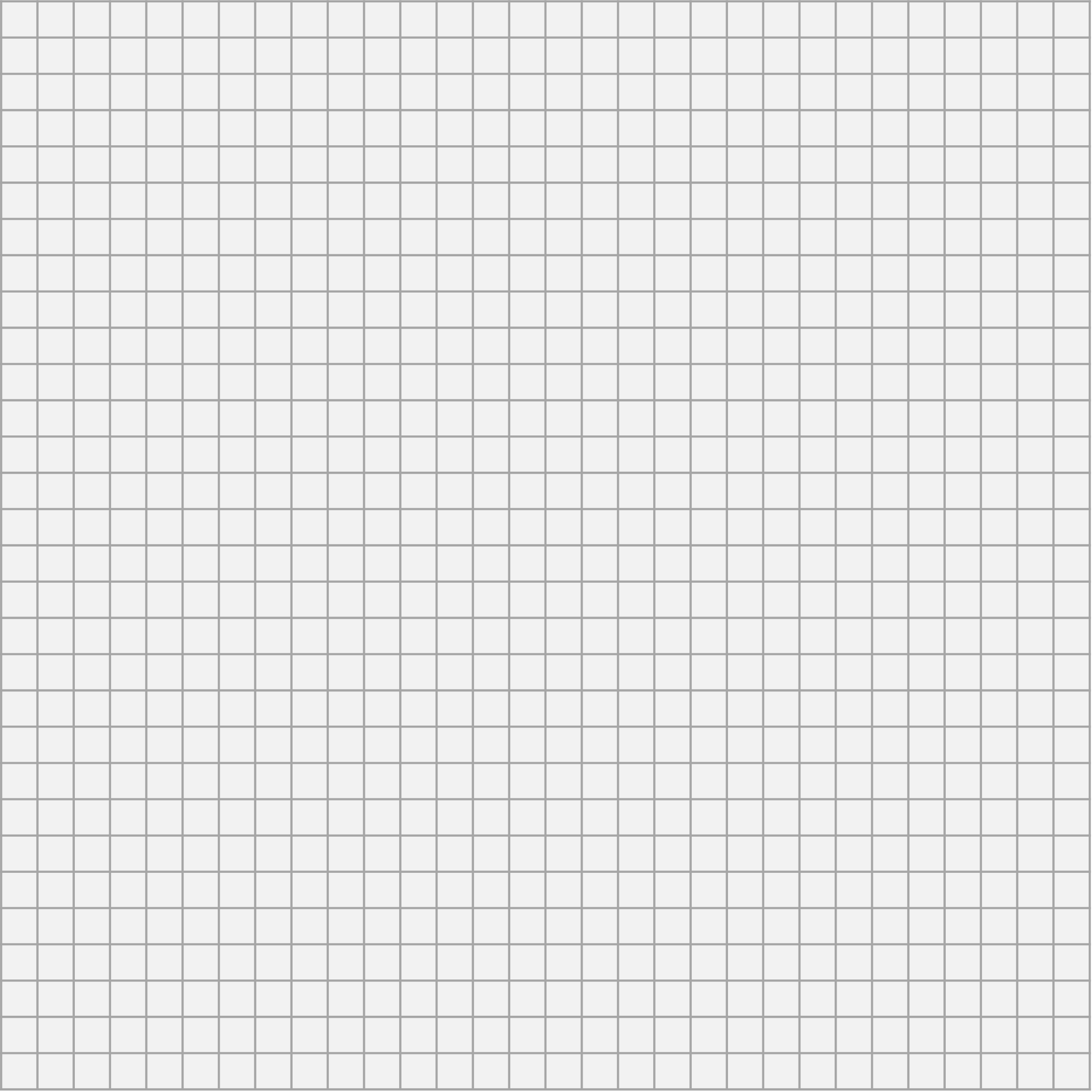} 
  {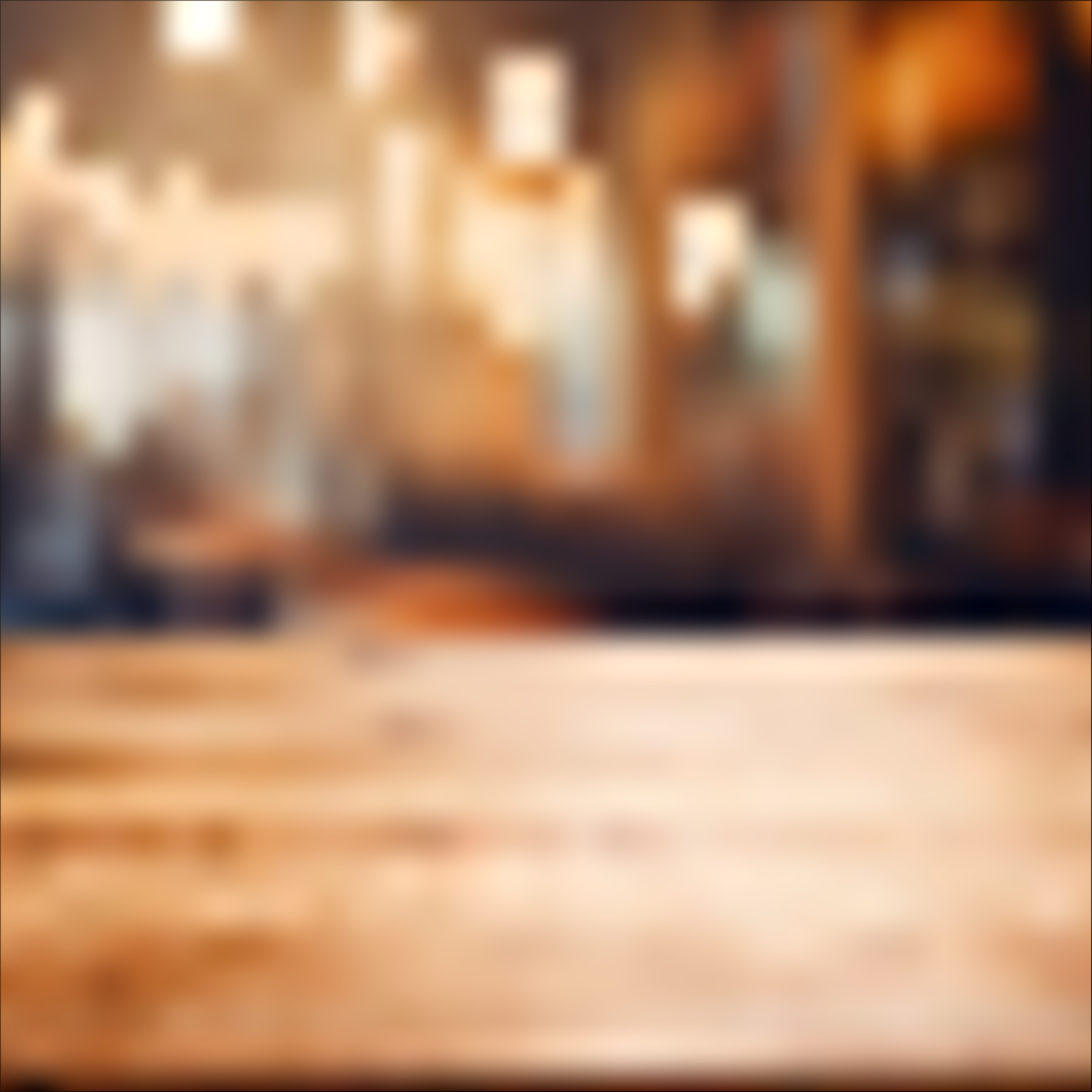} 
  {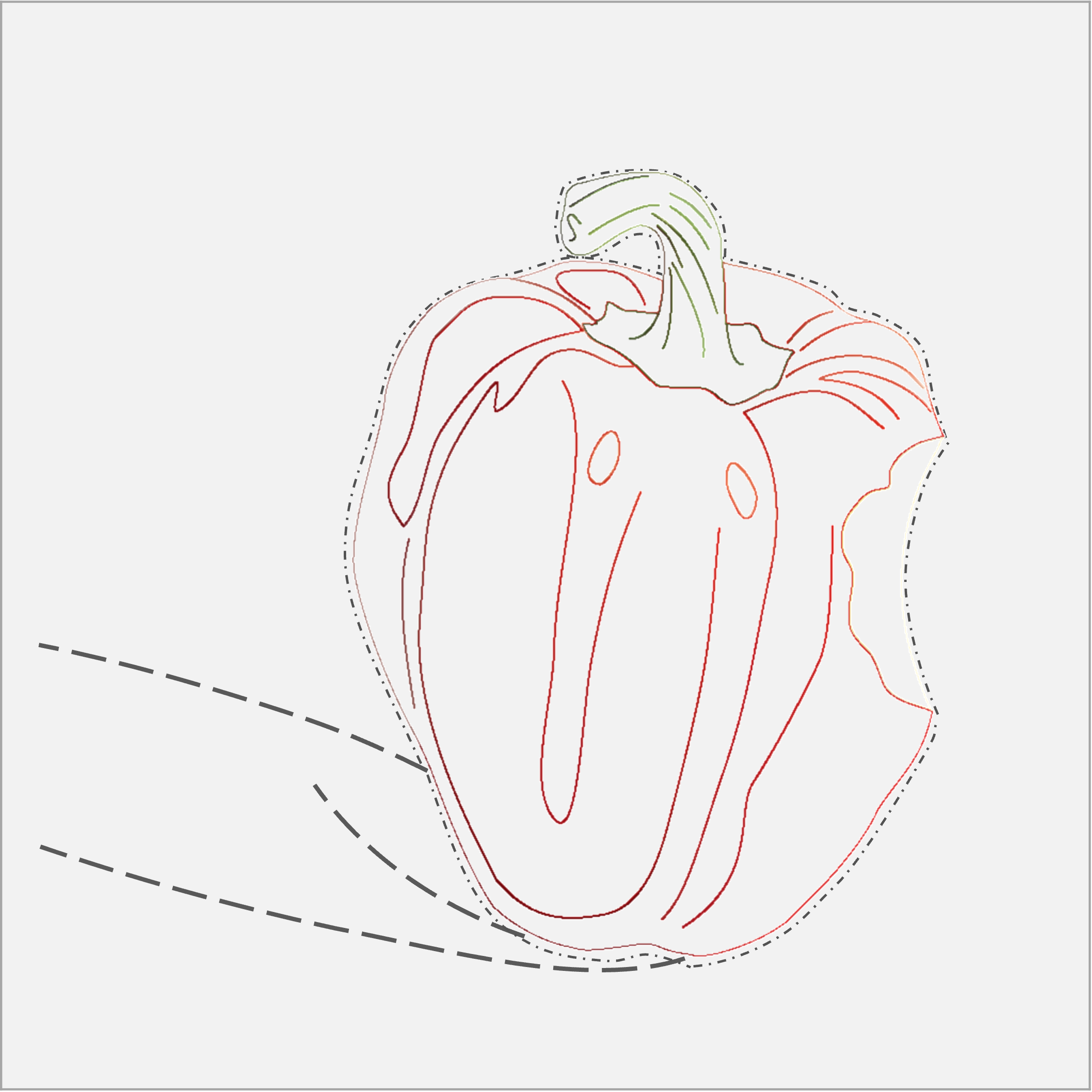} 
  {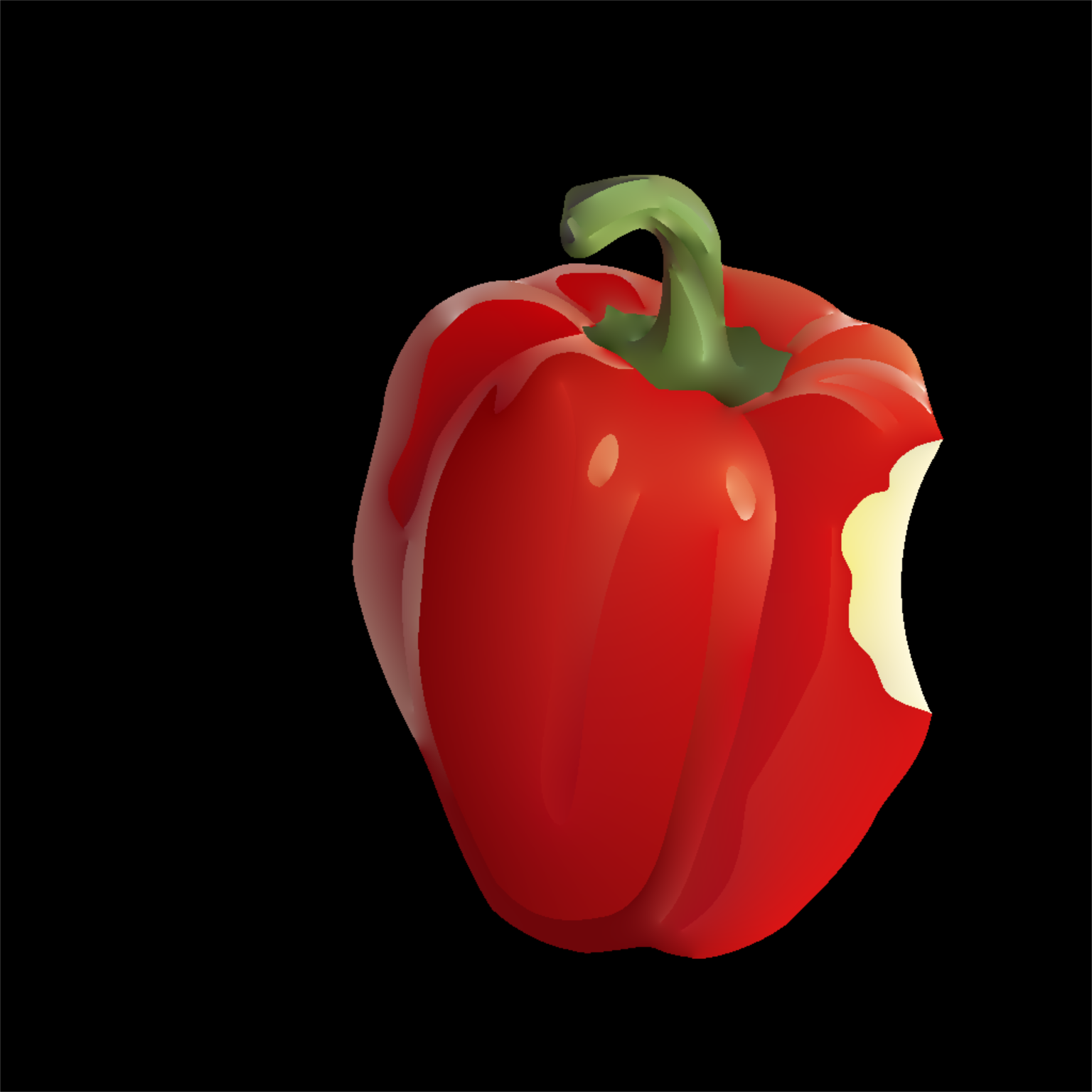} 
  {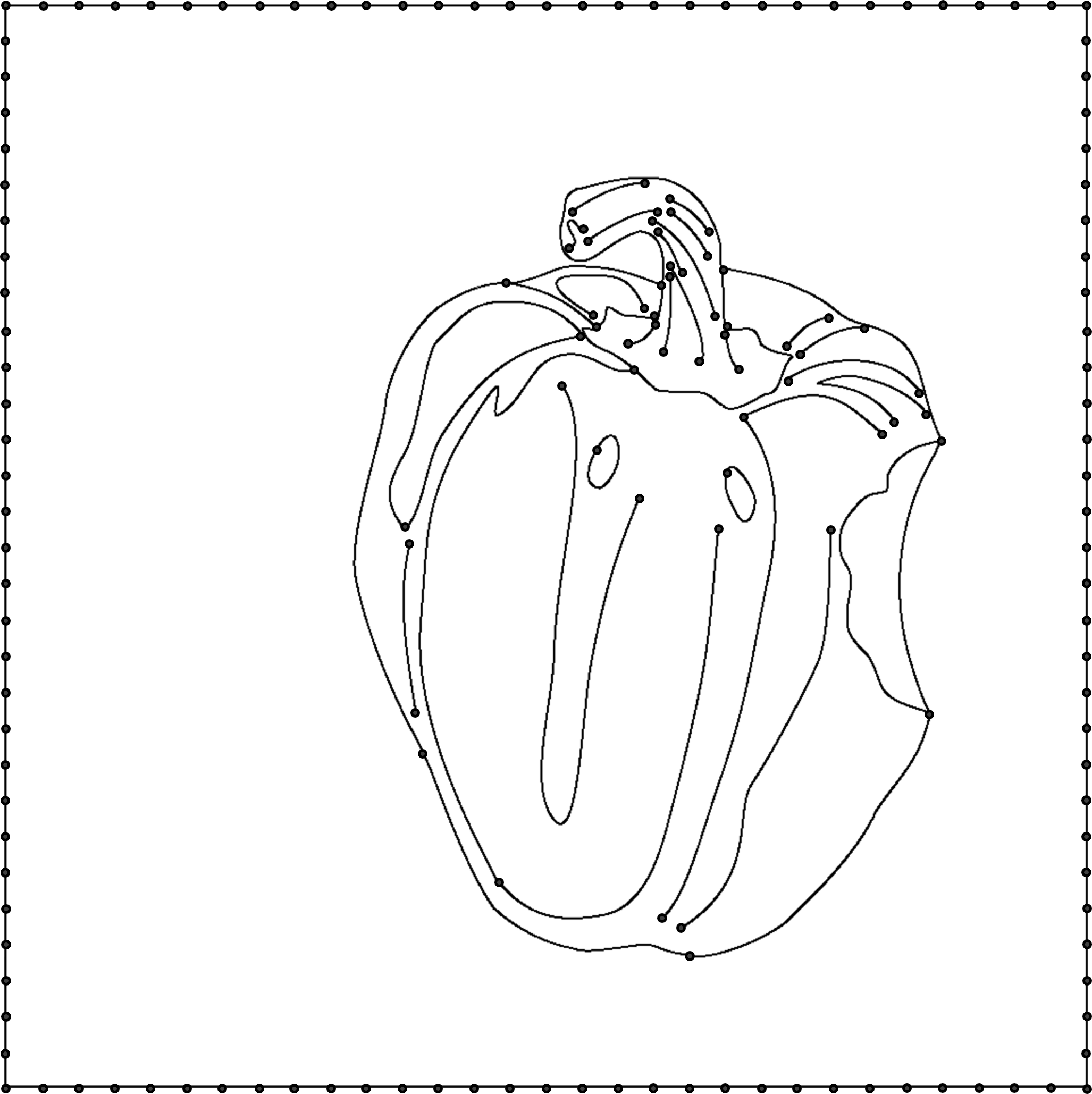} 
  {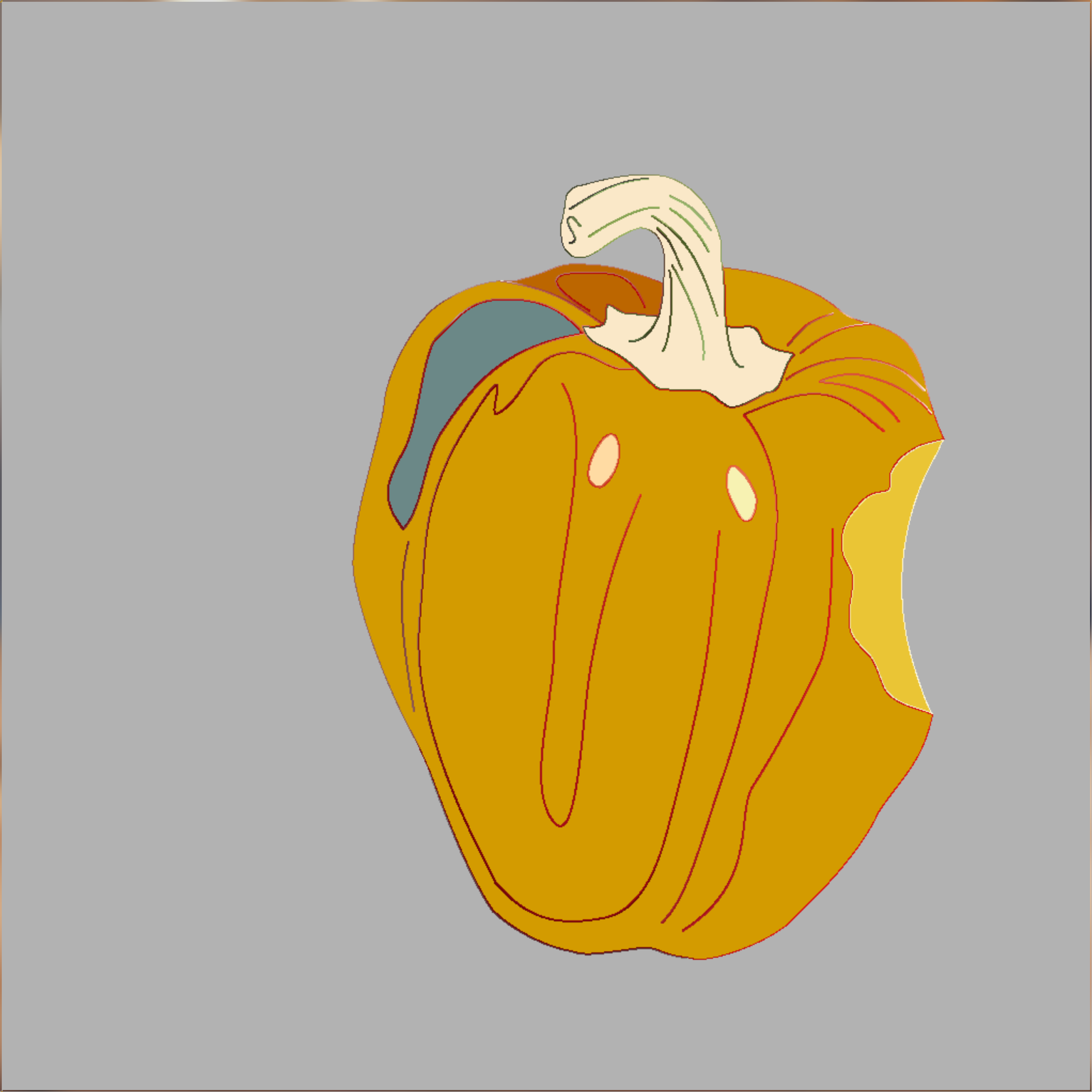} 
  {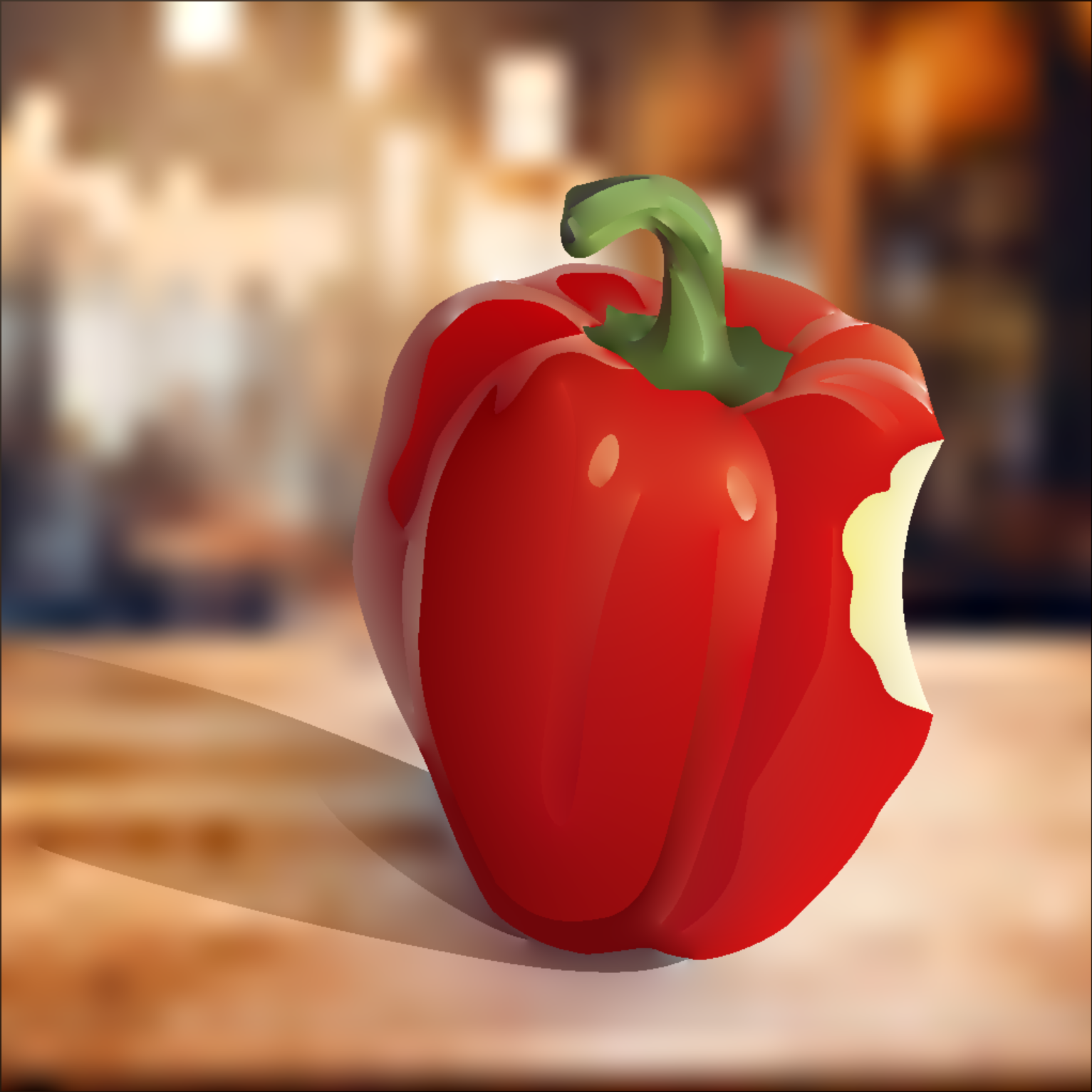} 
  \\%
  \vspace{0.4em}%
  \resultrow
  {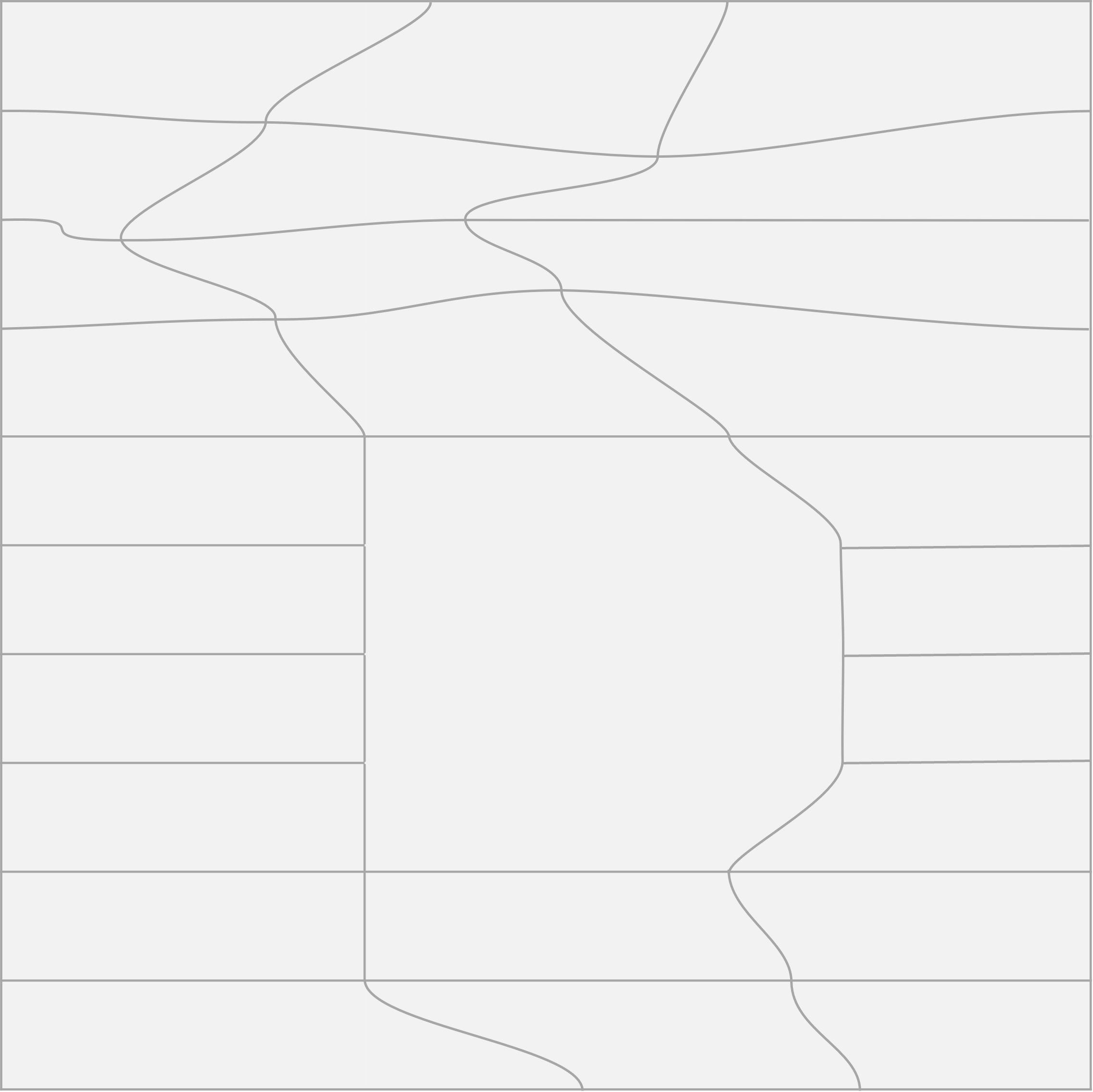} 
  {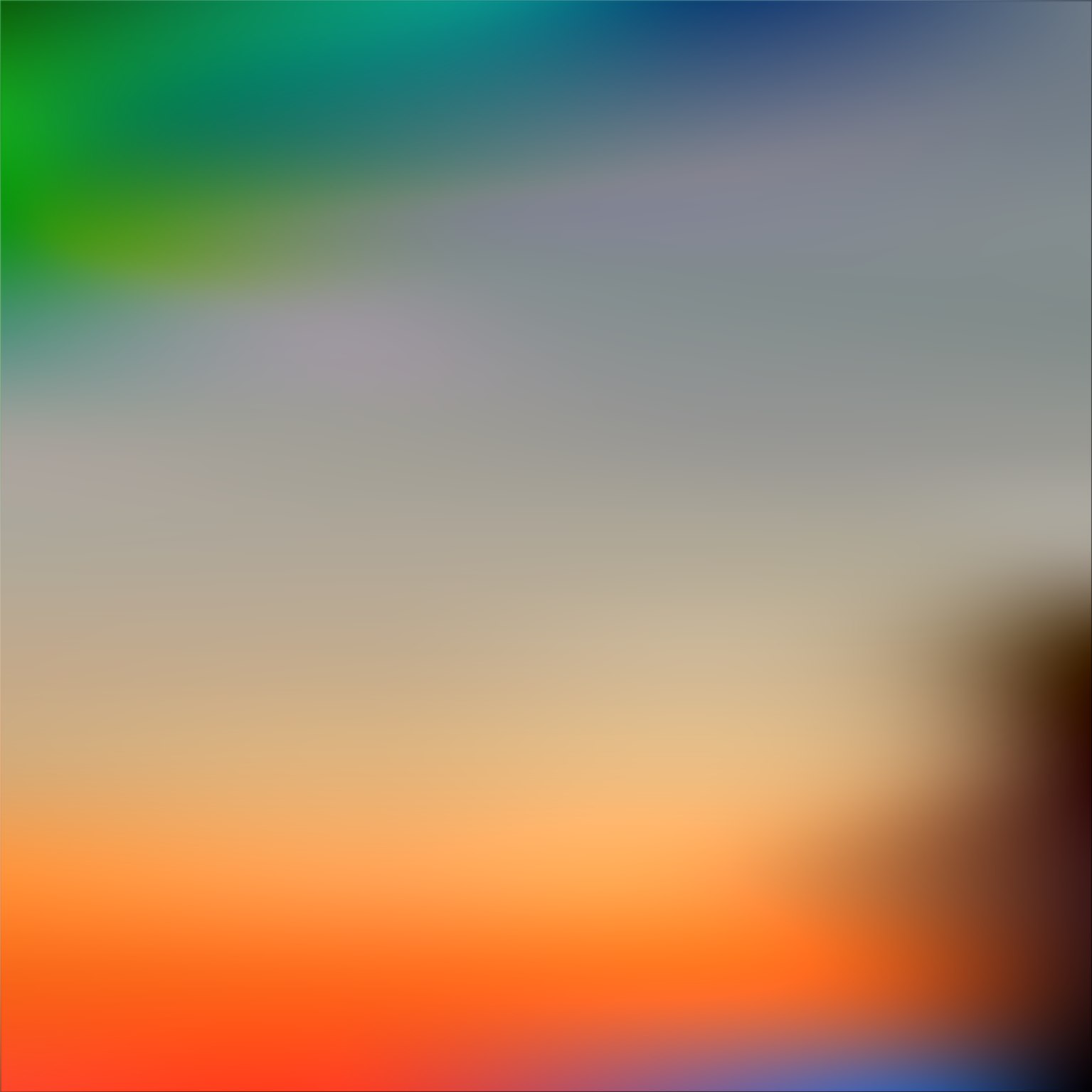} 
  {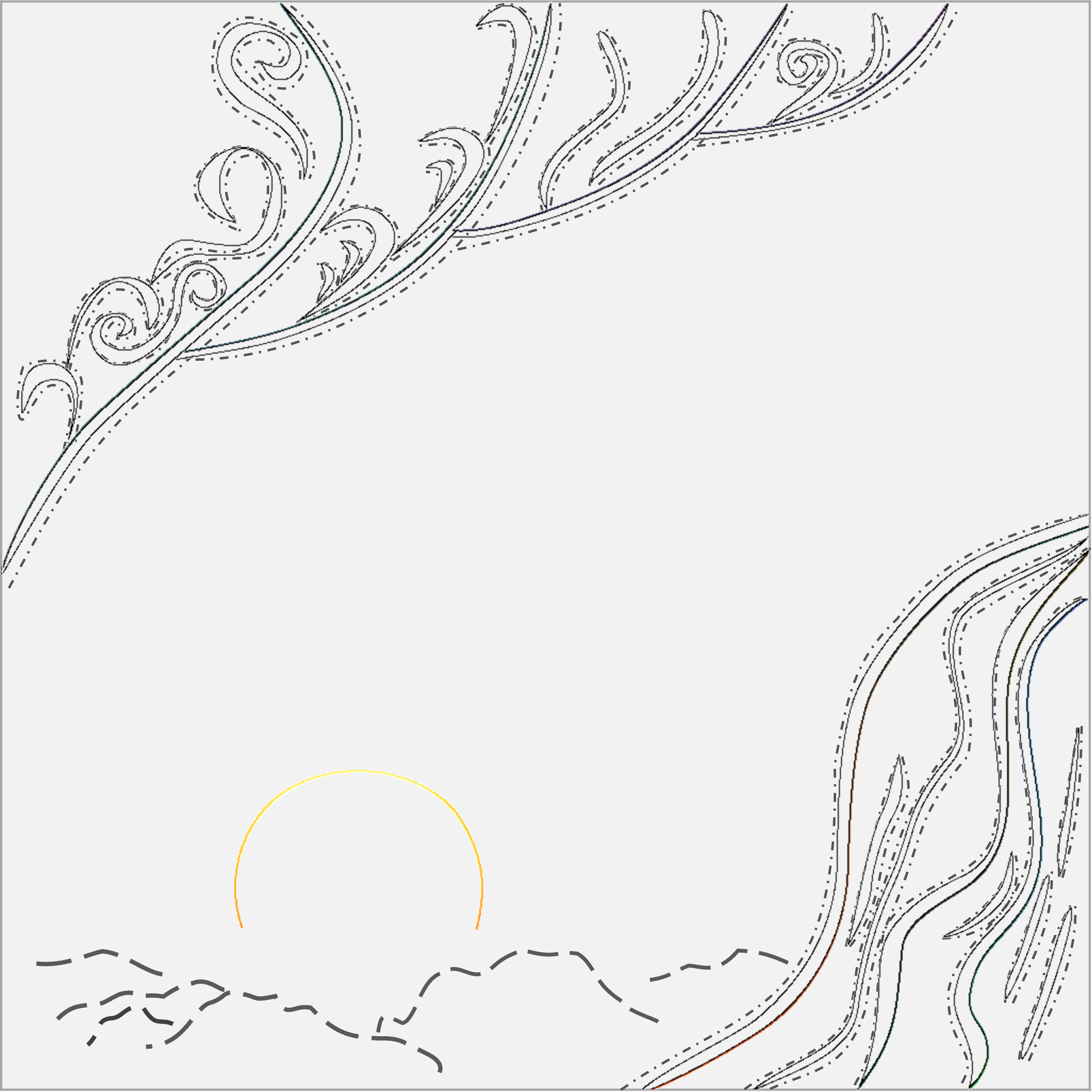} 
  {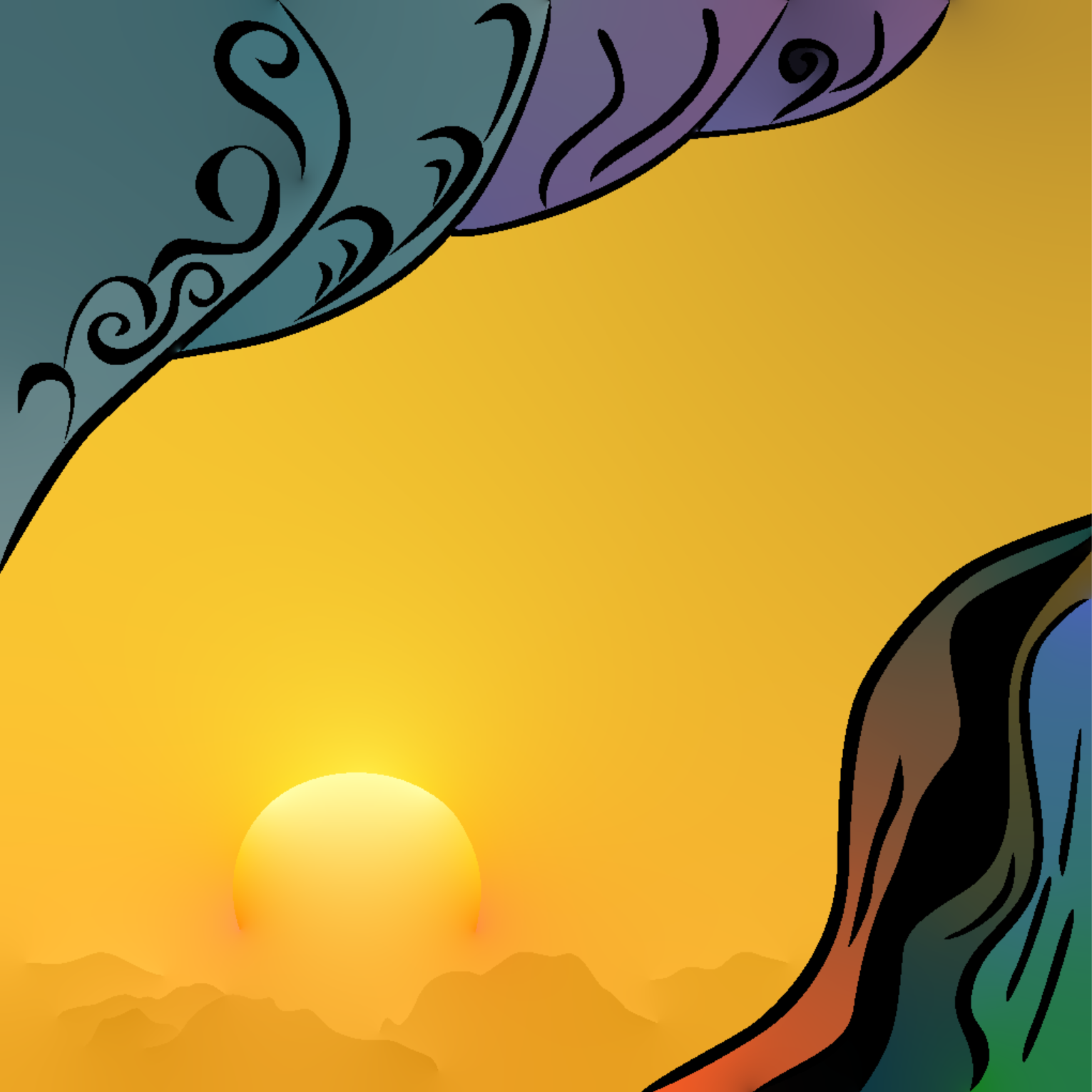} 
  {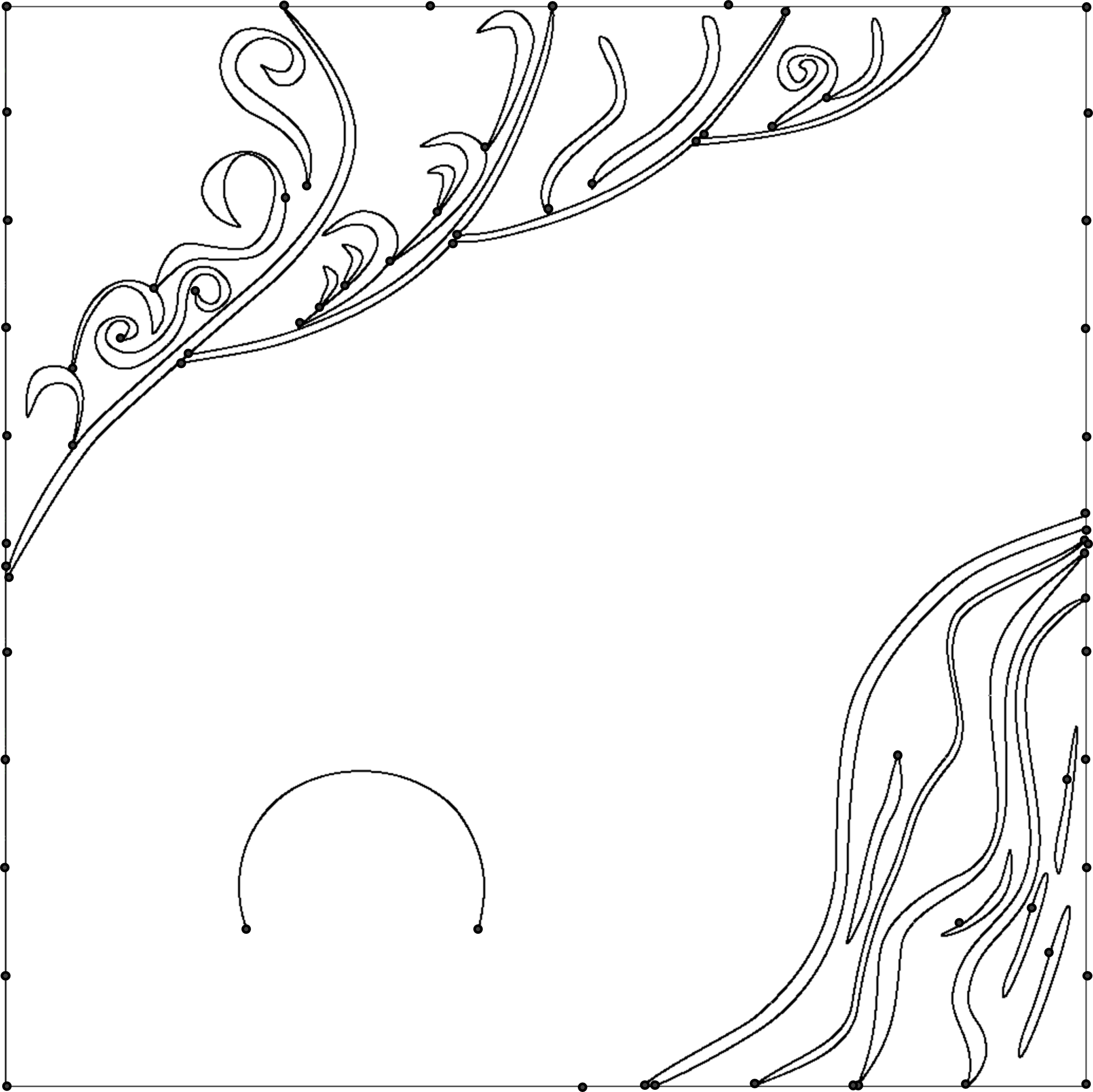} 
  {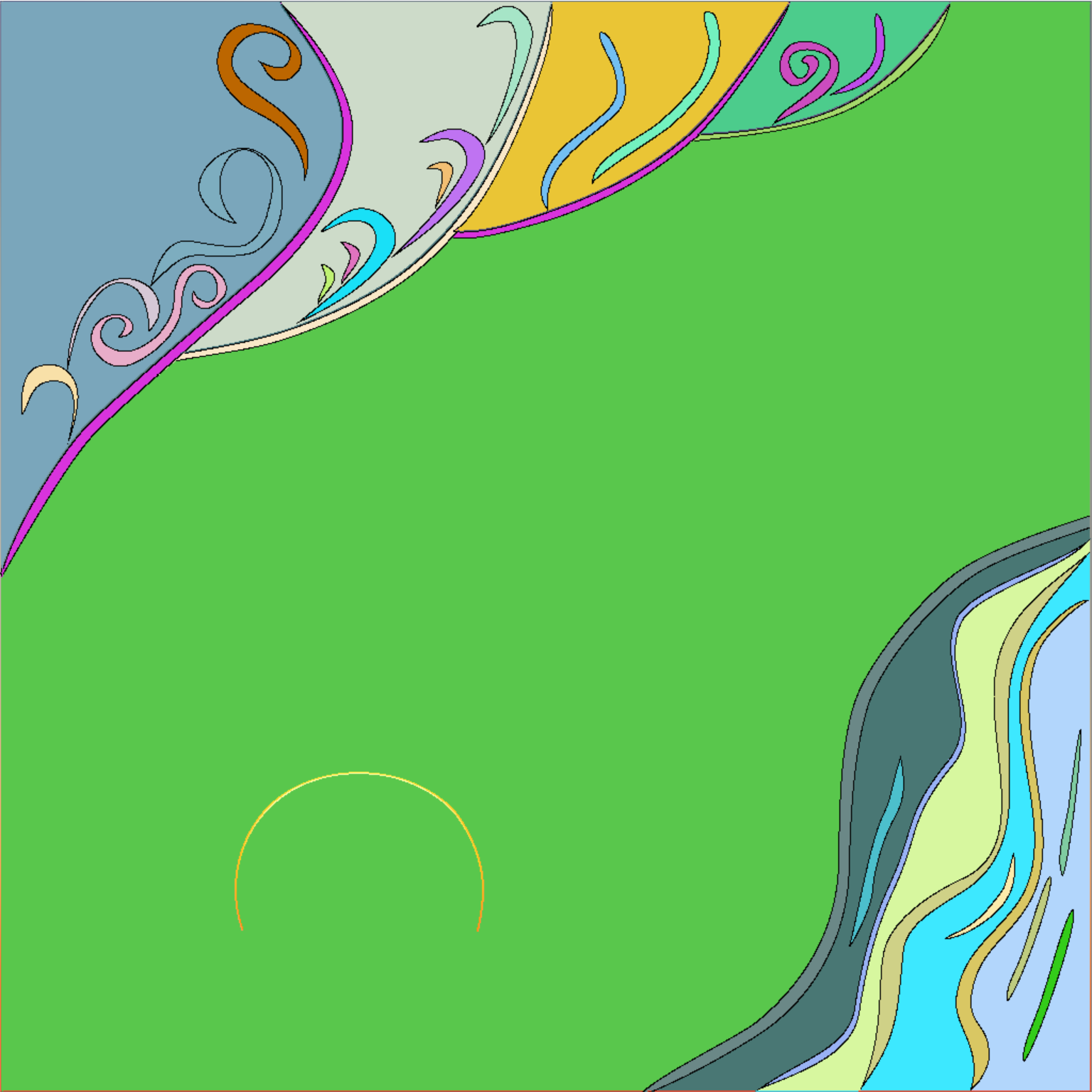} 
  {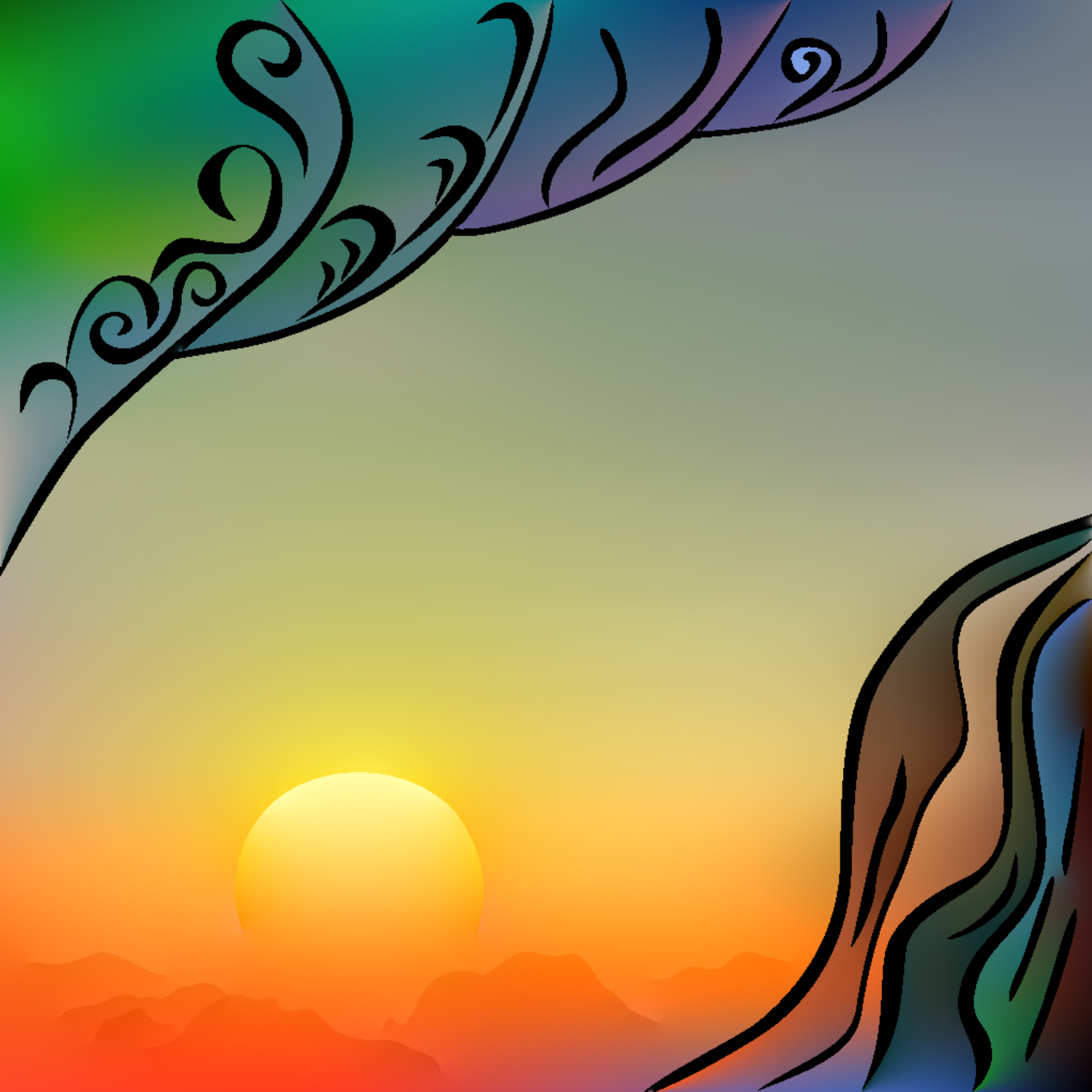} 
  \\%
  \vspace{0.4em}%
  \resultrow
  {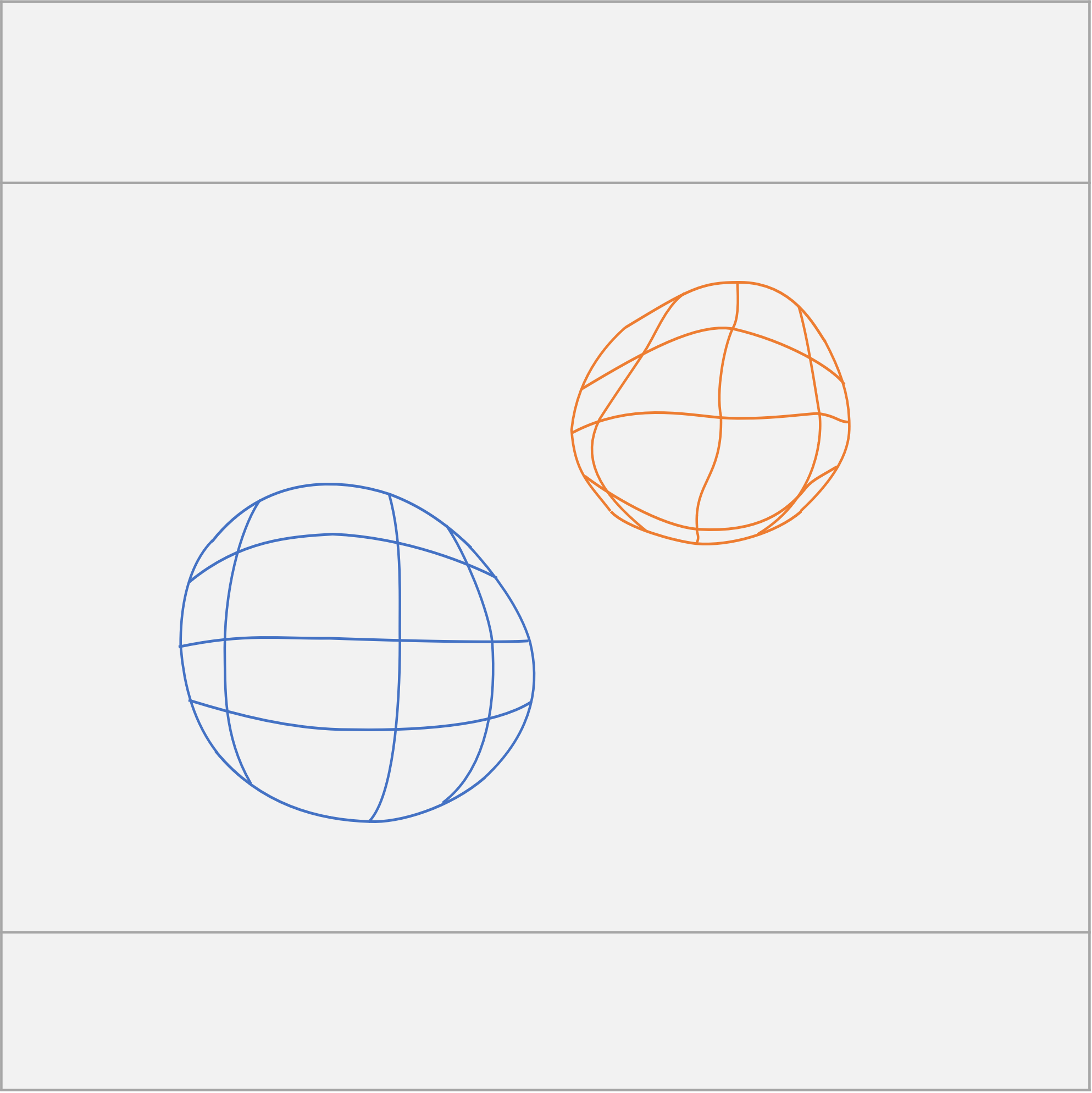} 
  {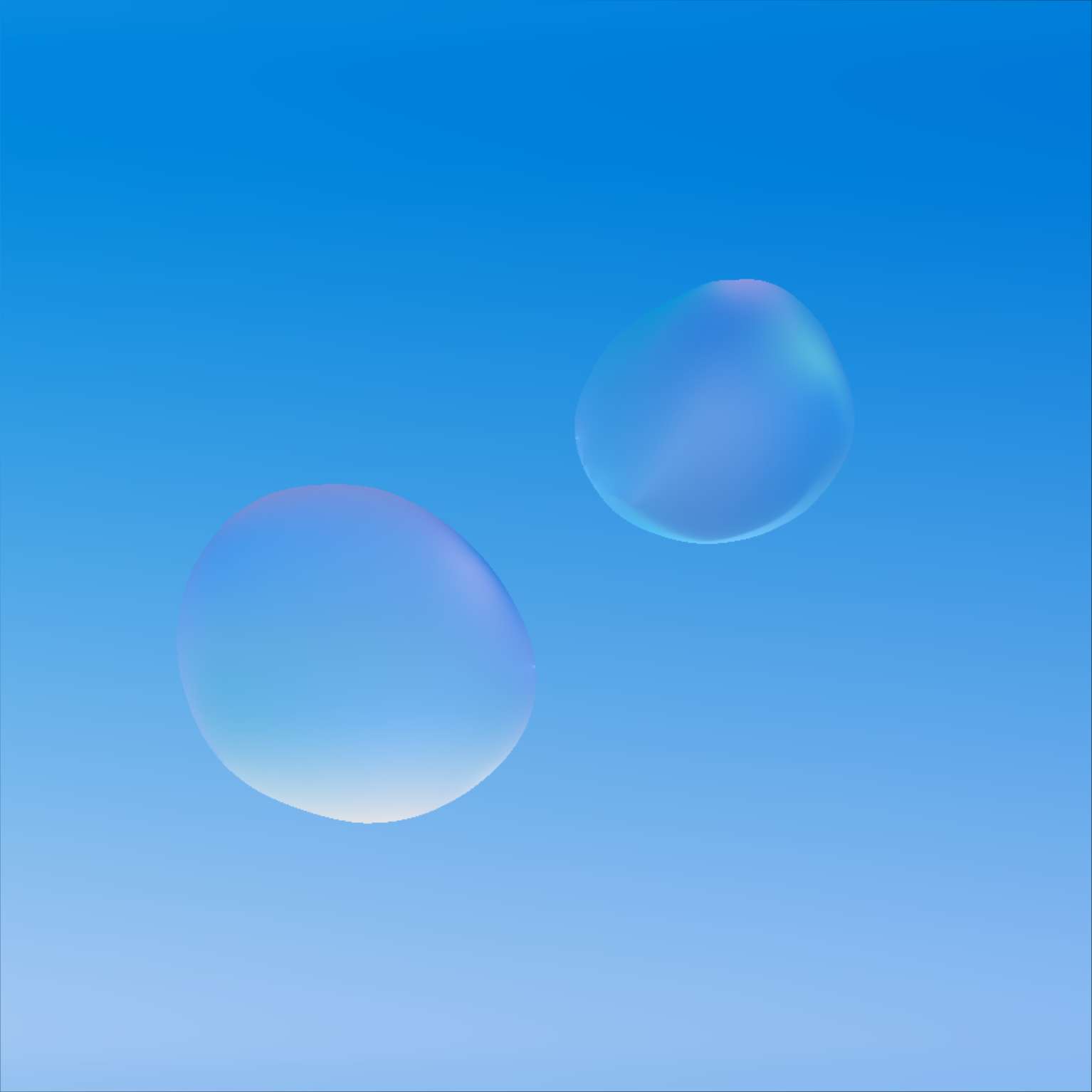} 
  {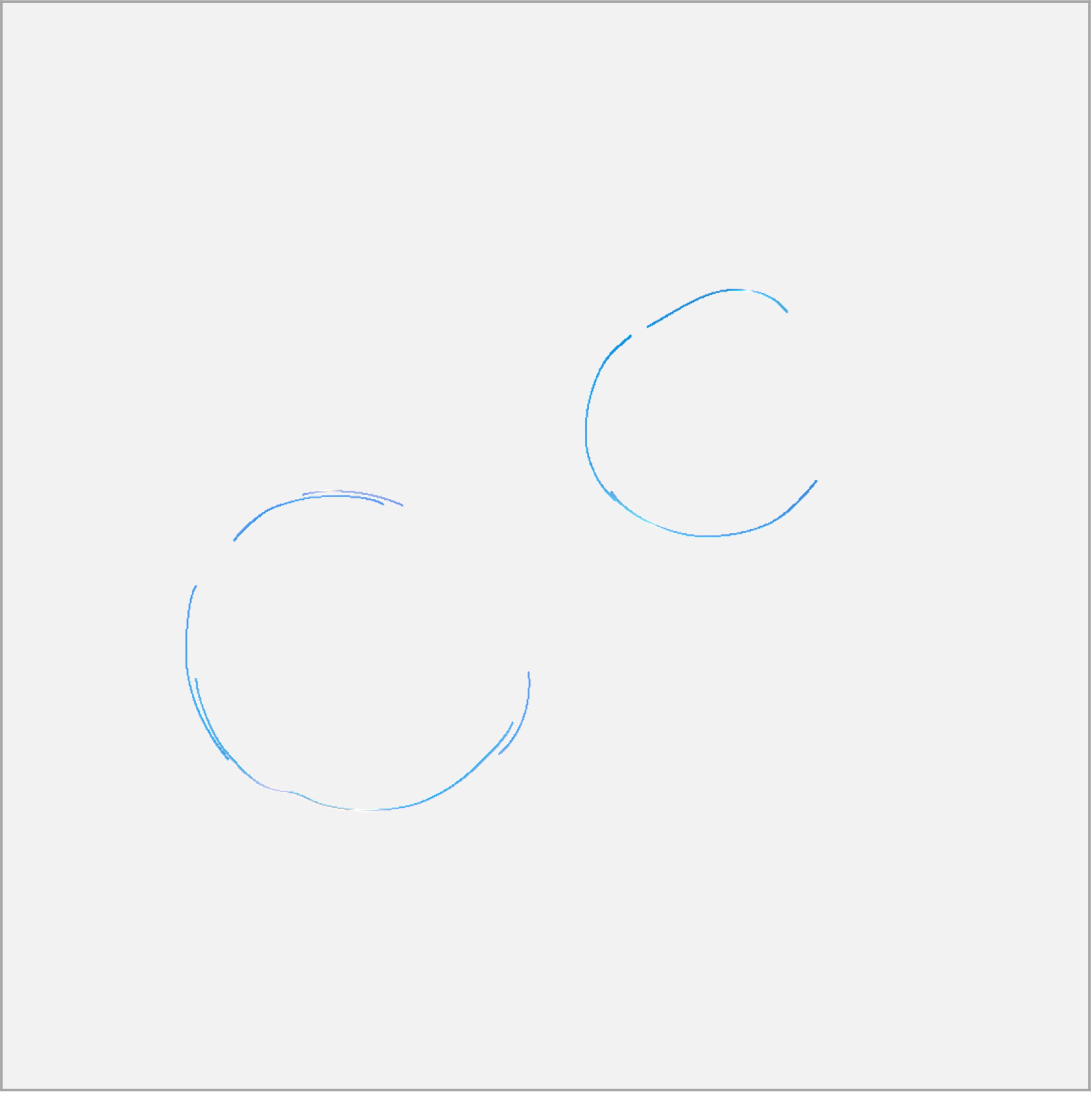} 
  {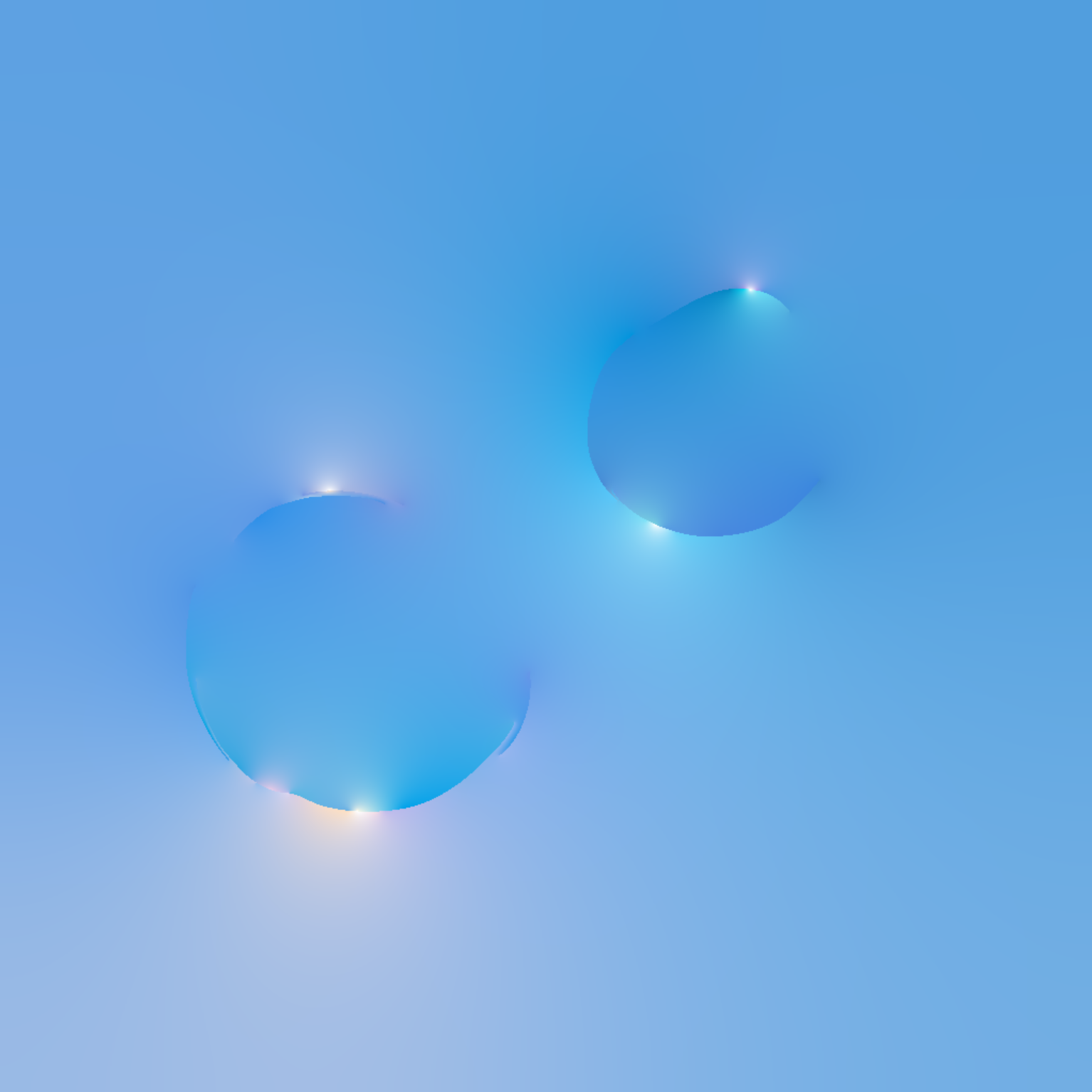} 
  {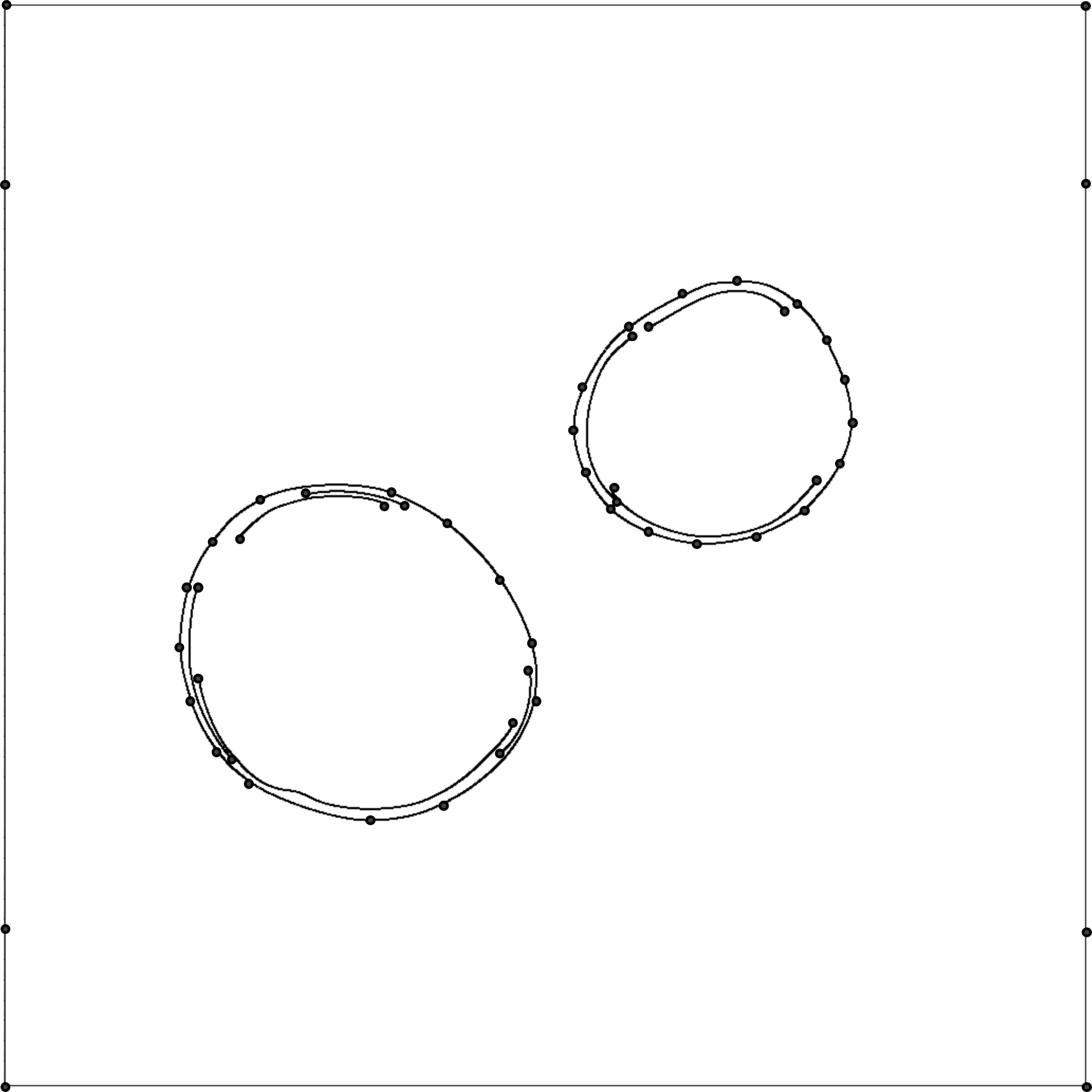} 
  {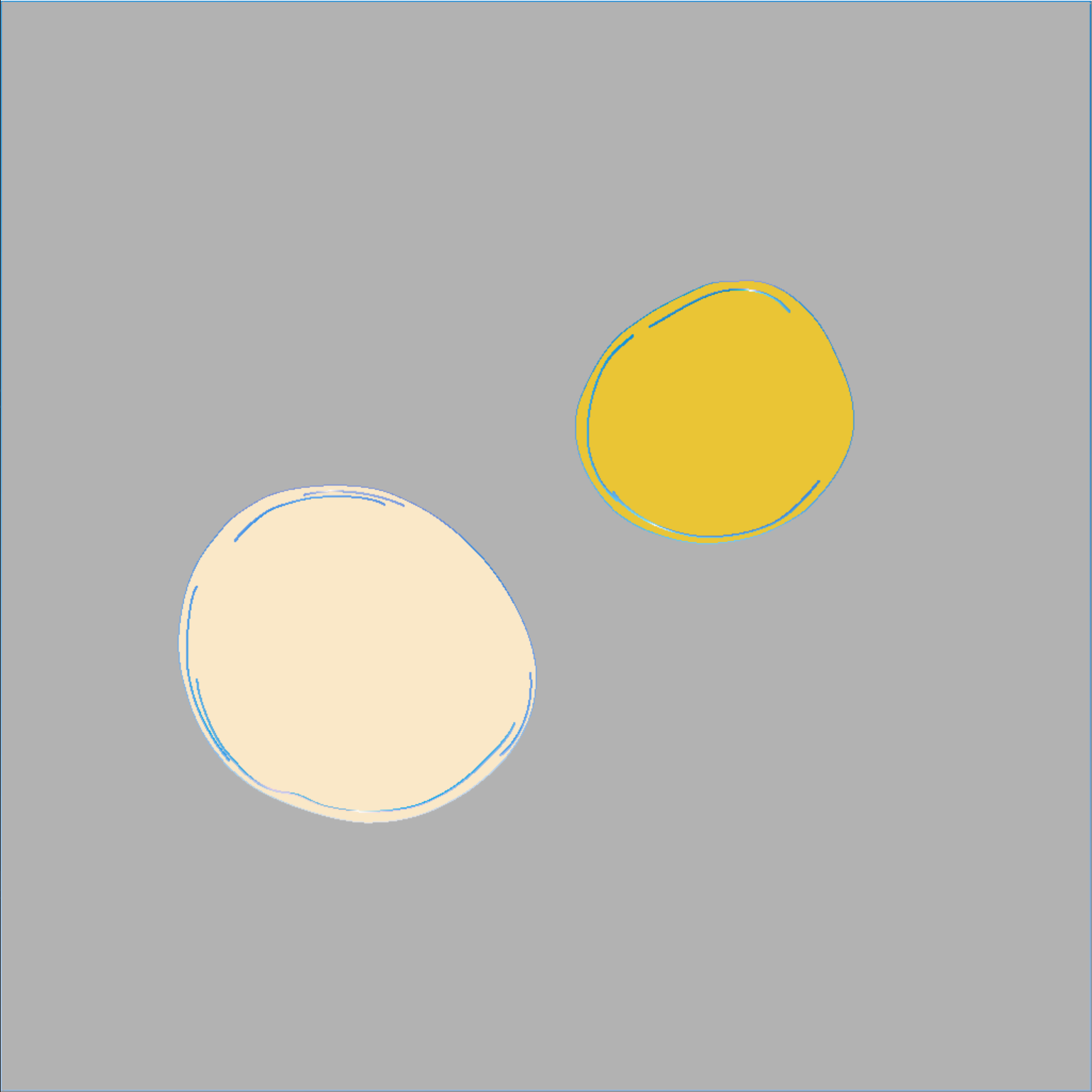} 
  {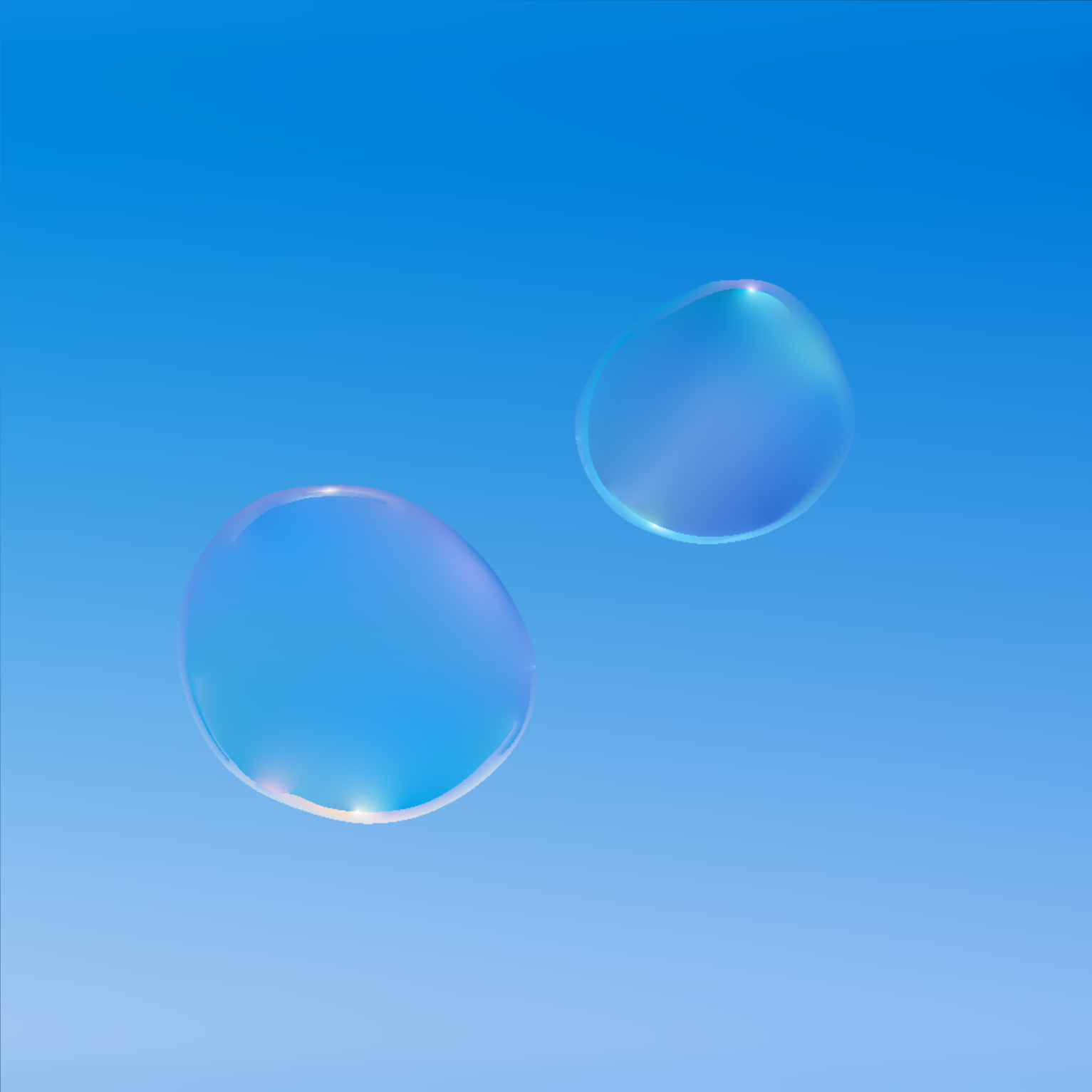} 
  \\%
  \vspace{0.4em}%
  \resultrow
  {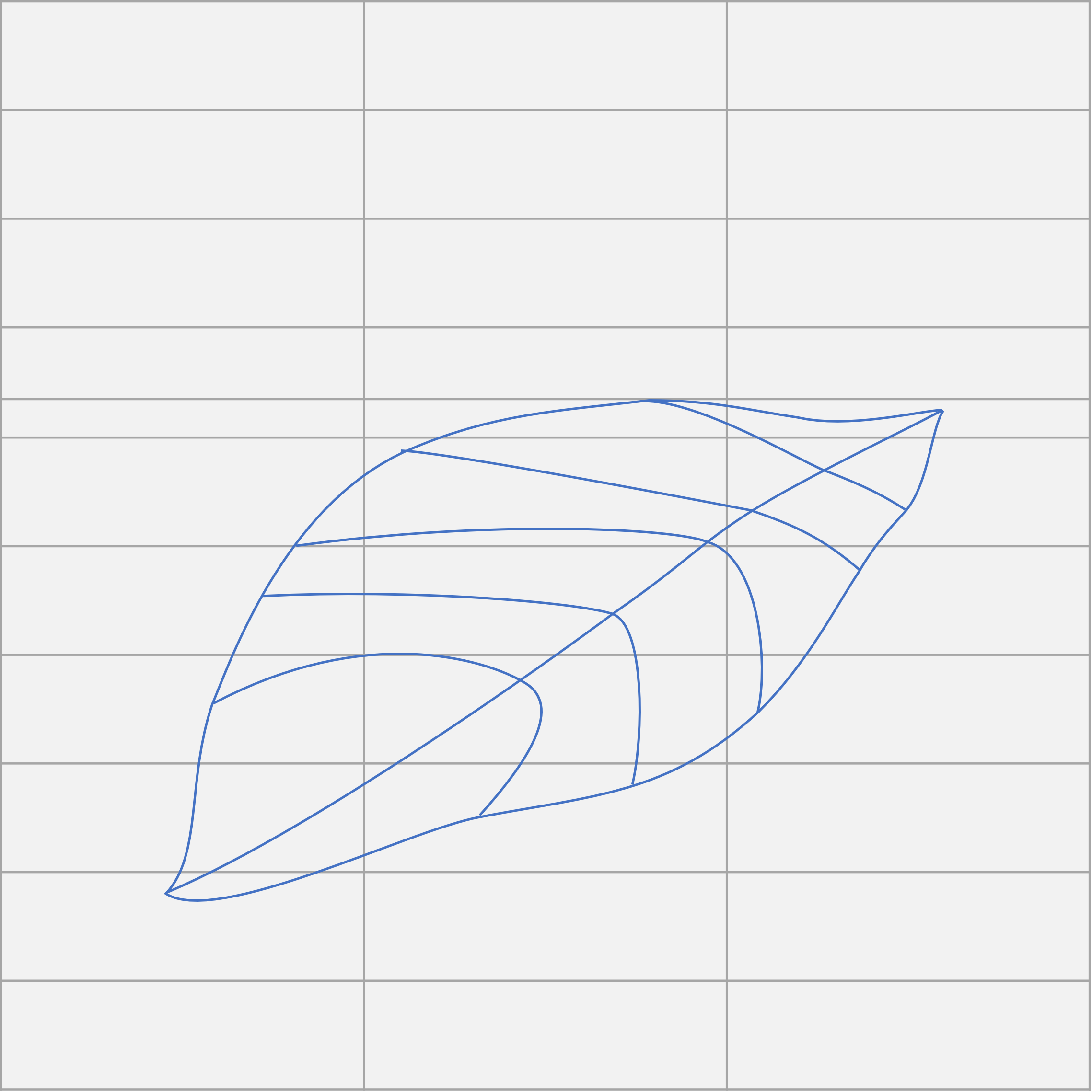} 
  {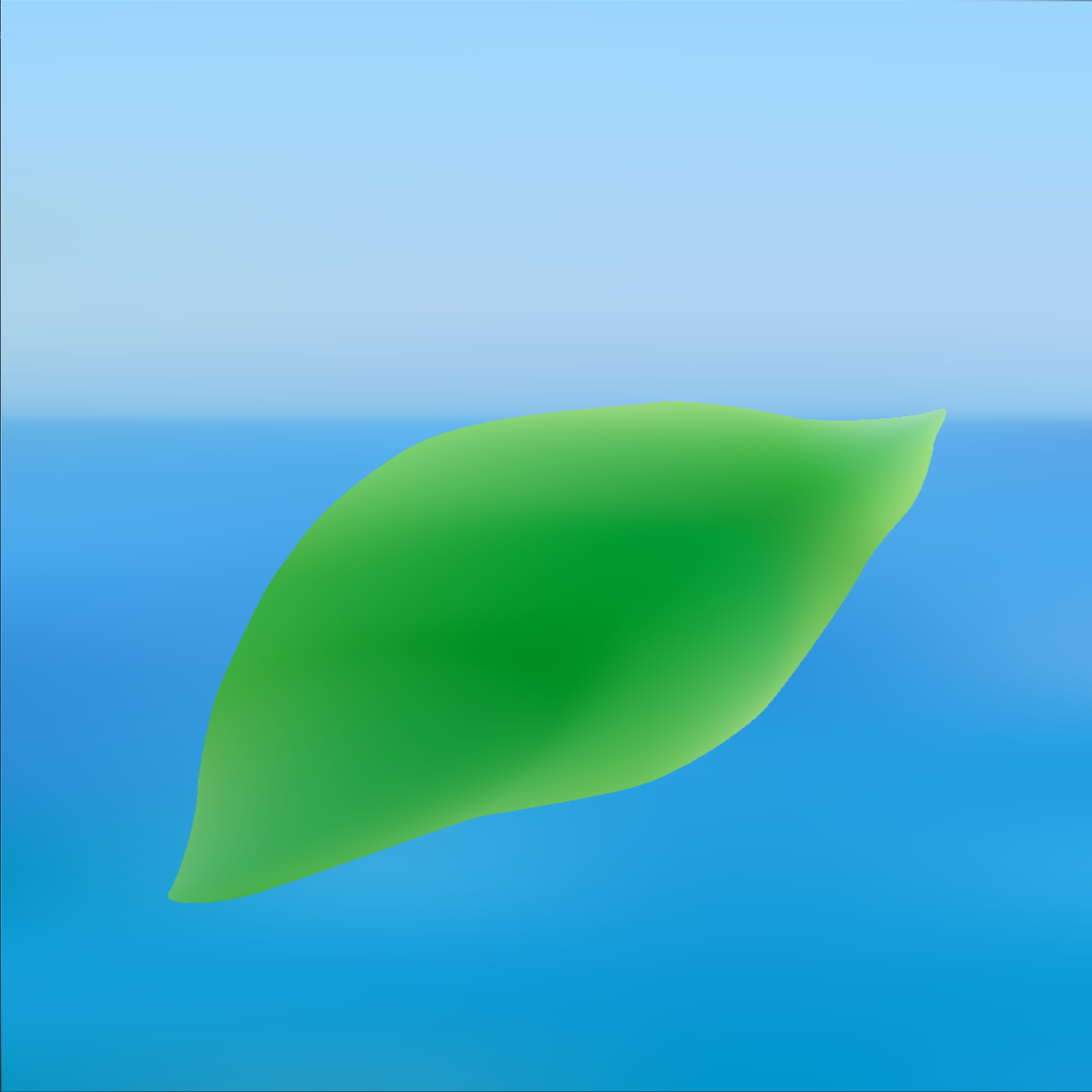} 
  {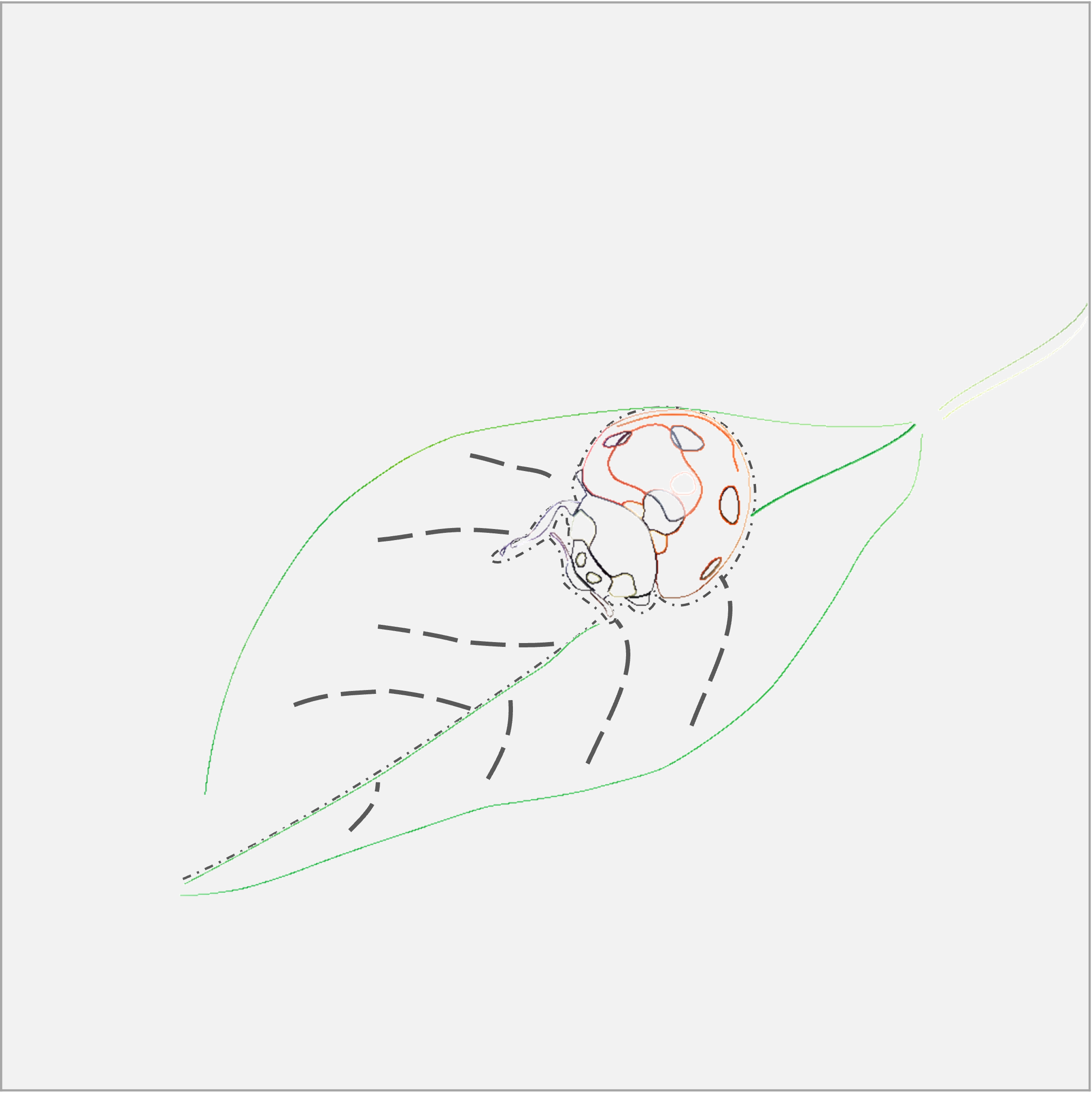} 
  {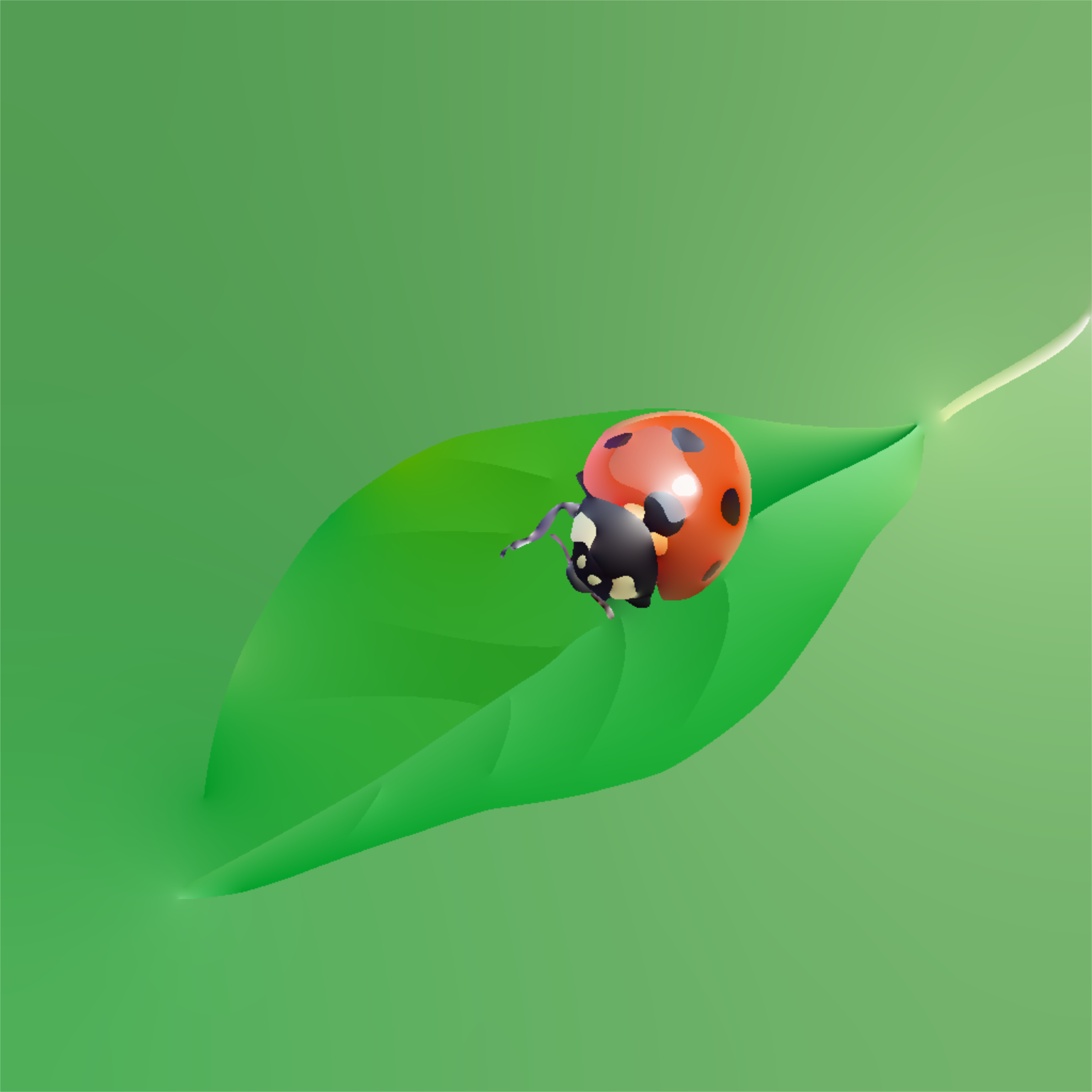} 
  {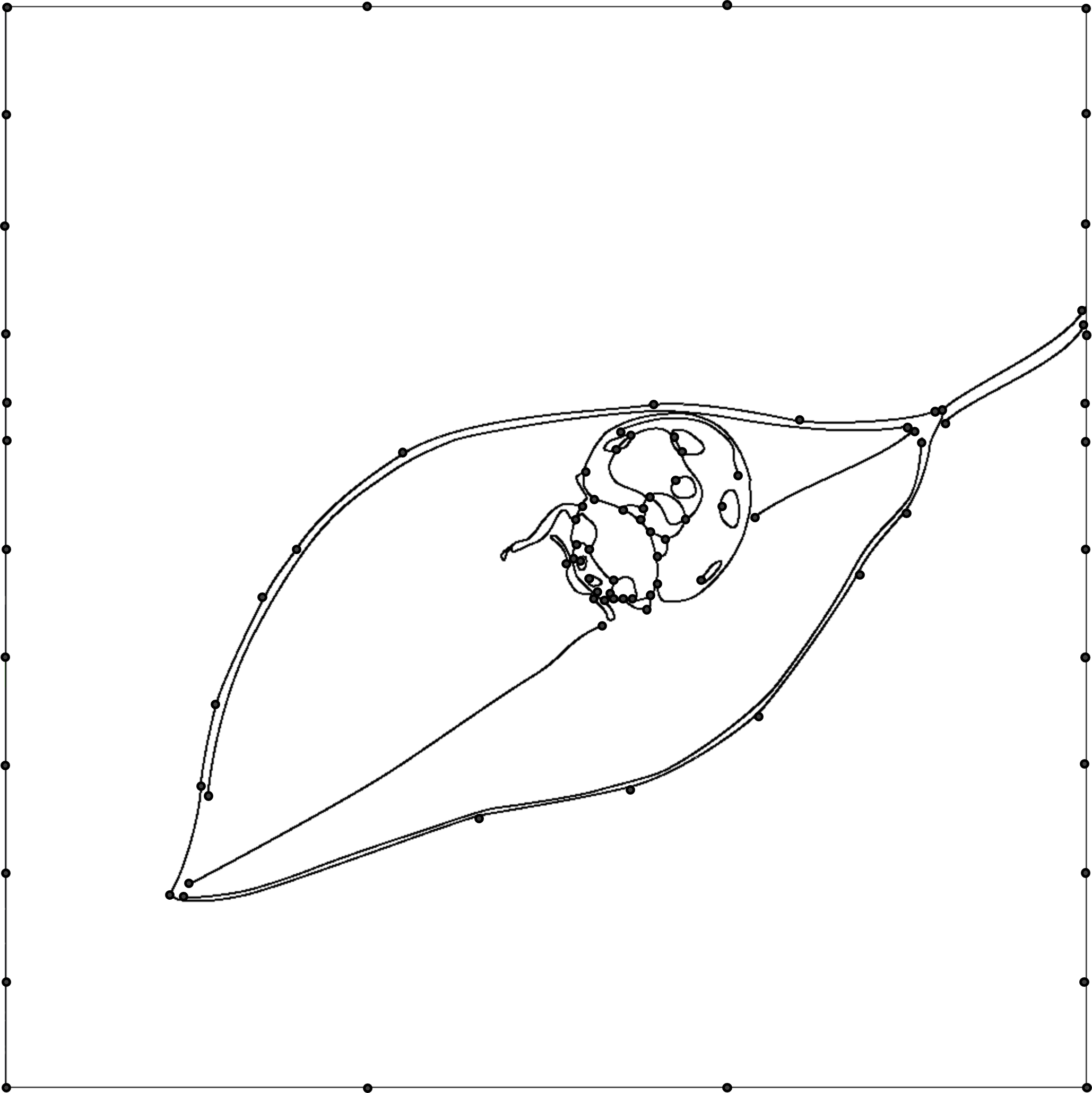} 
  {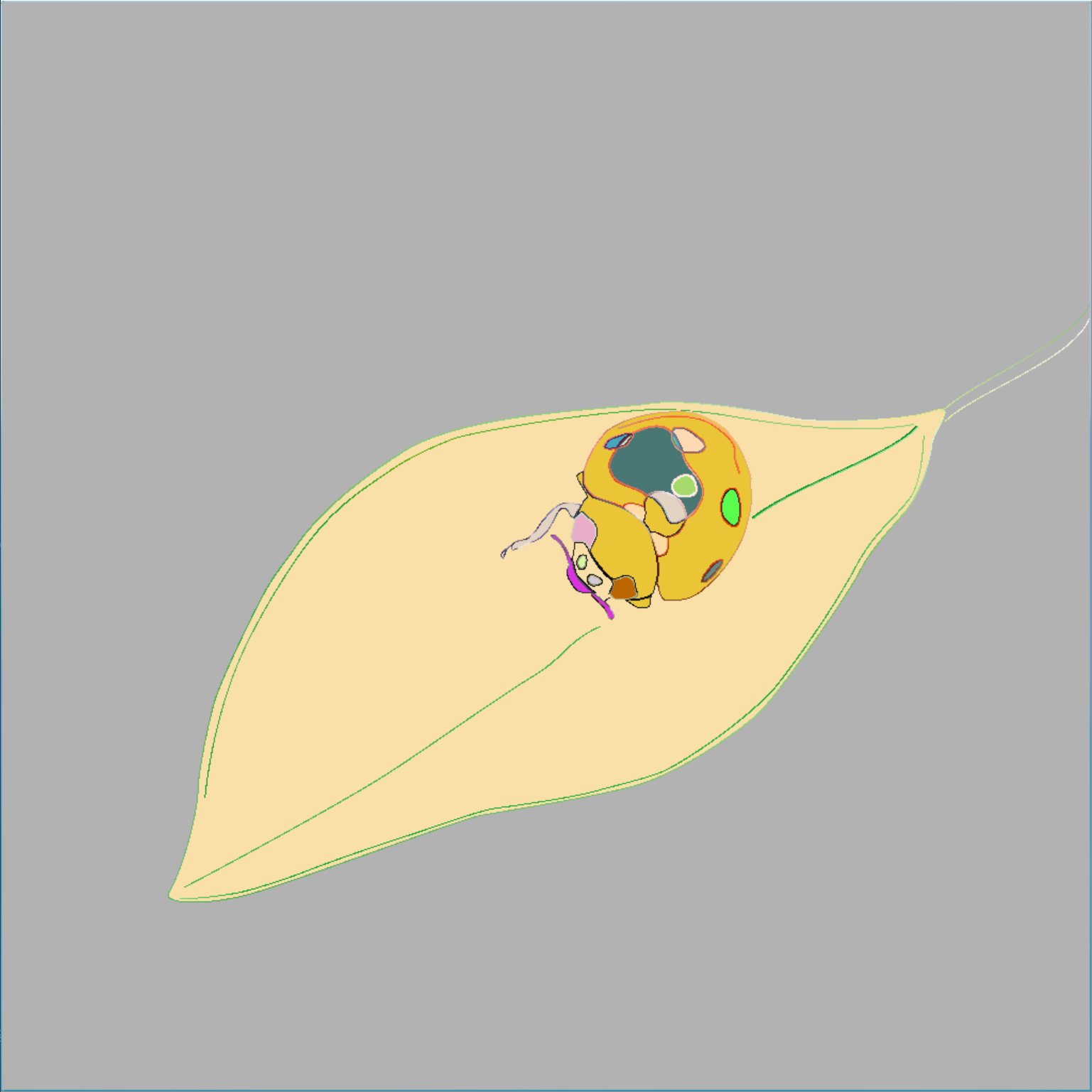} 
  {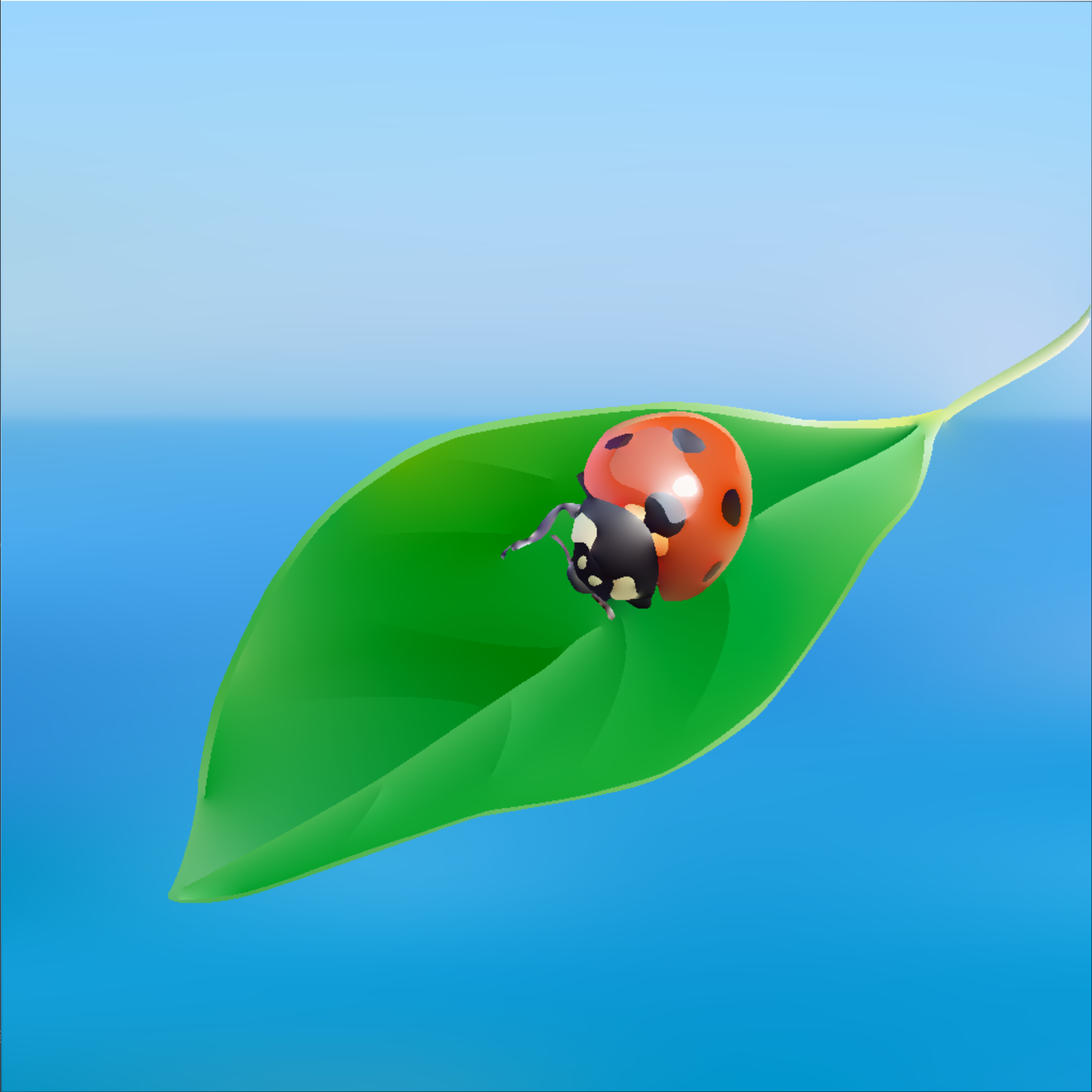} 
  \\%
  \vspace{0.4em}%
  \resultrow
  {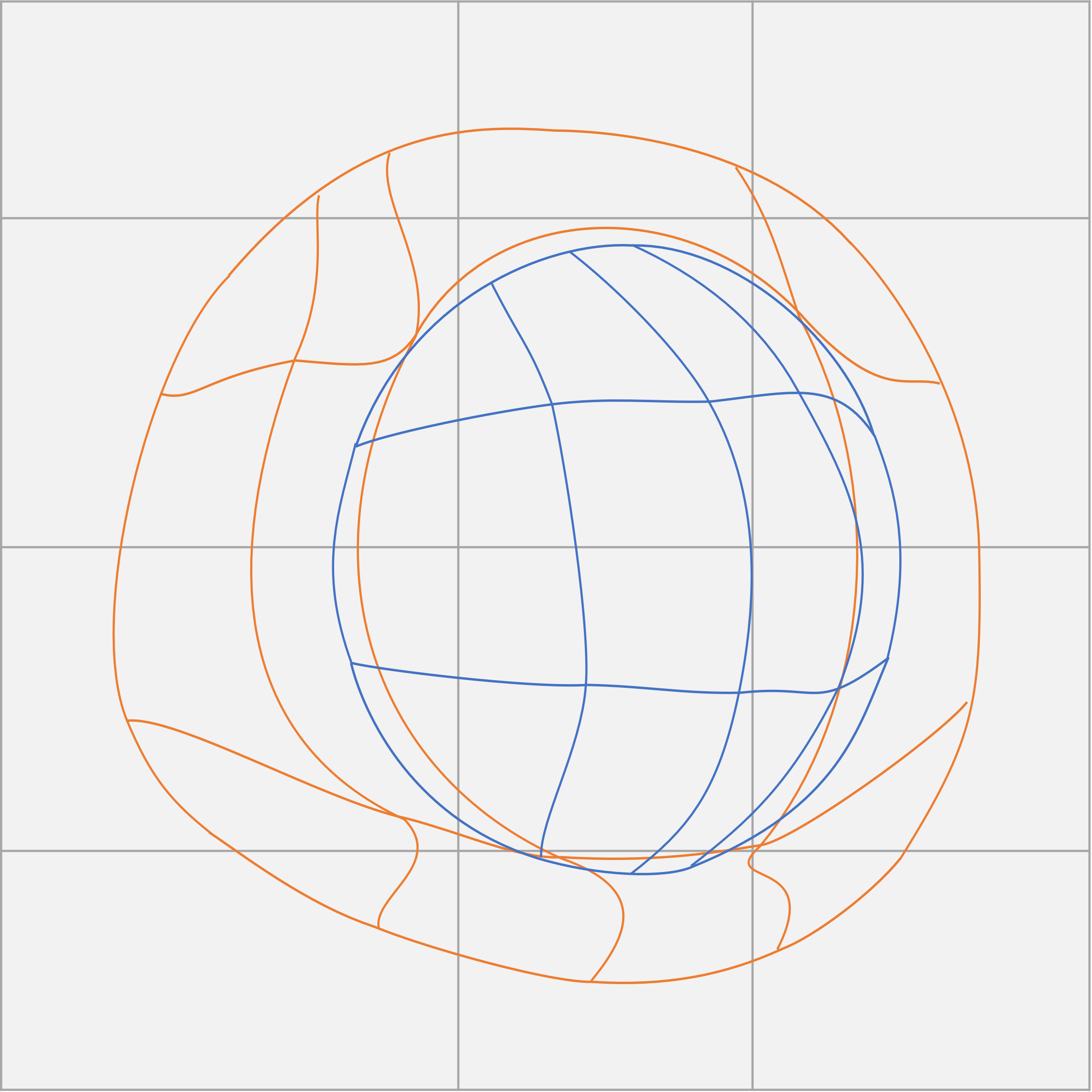} 
  {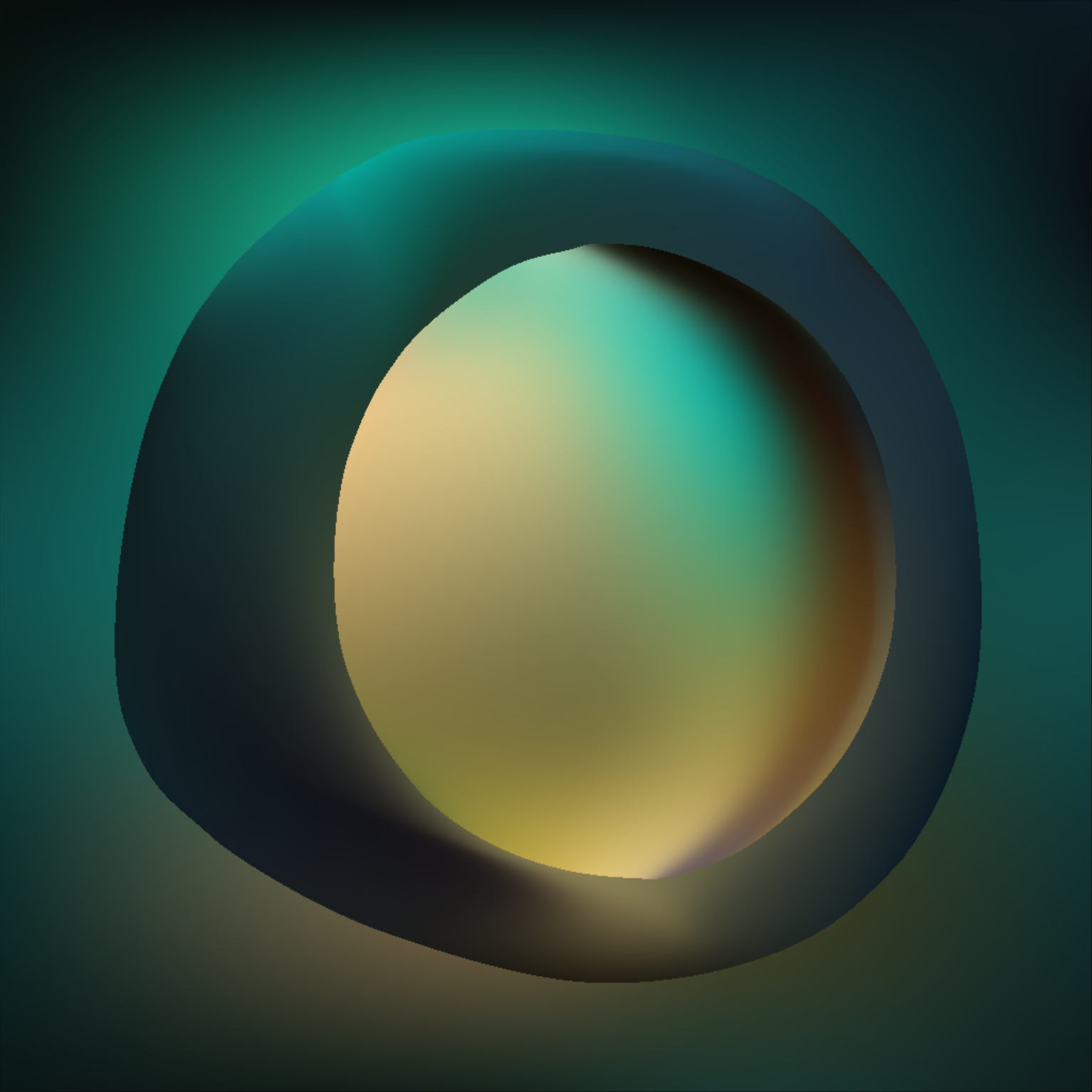} 
  {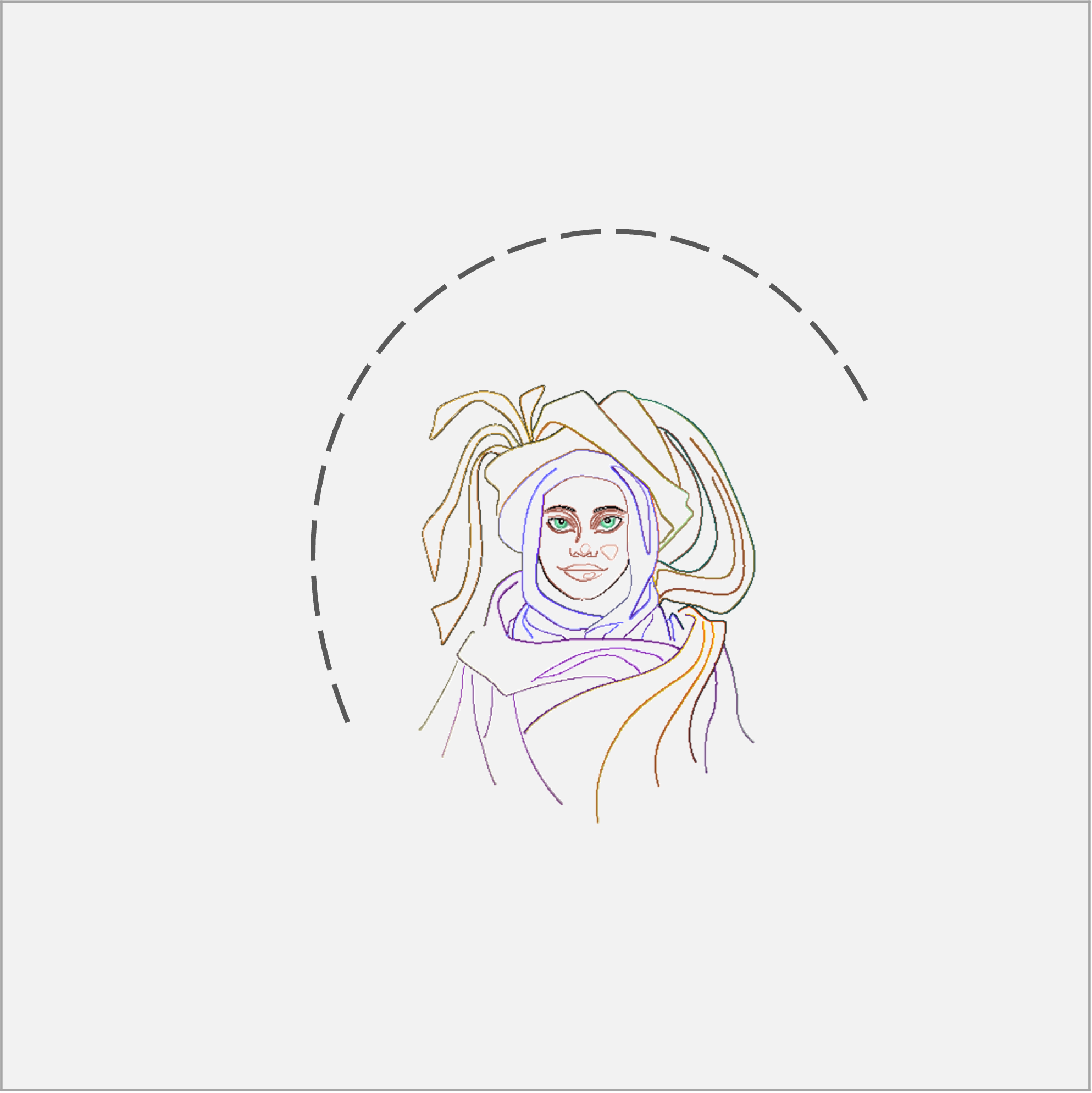} 
  {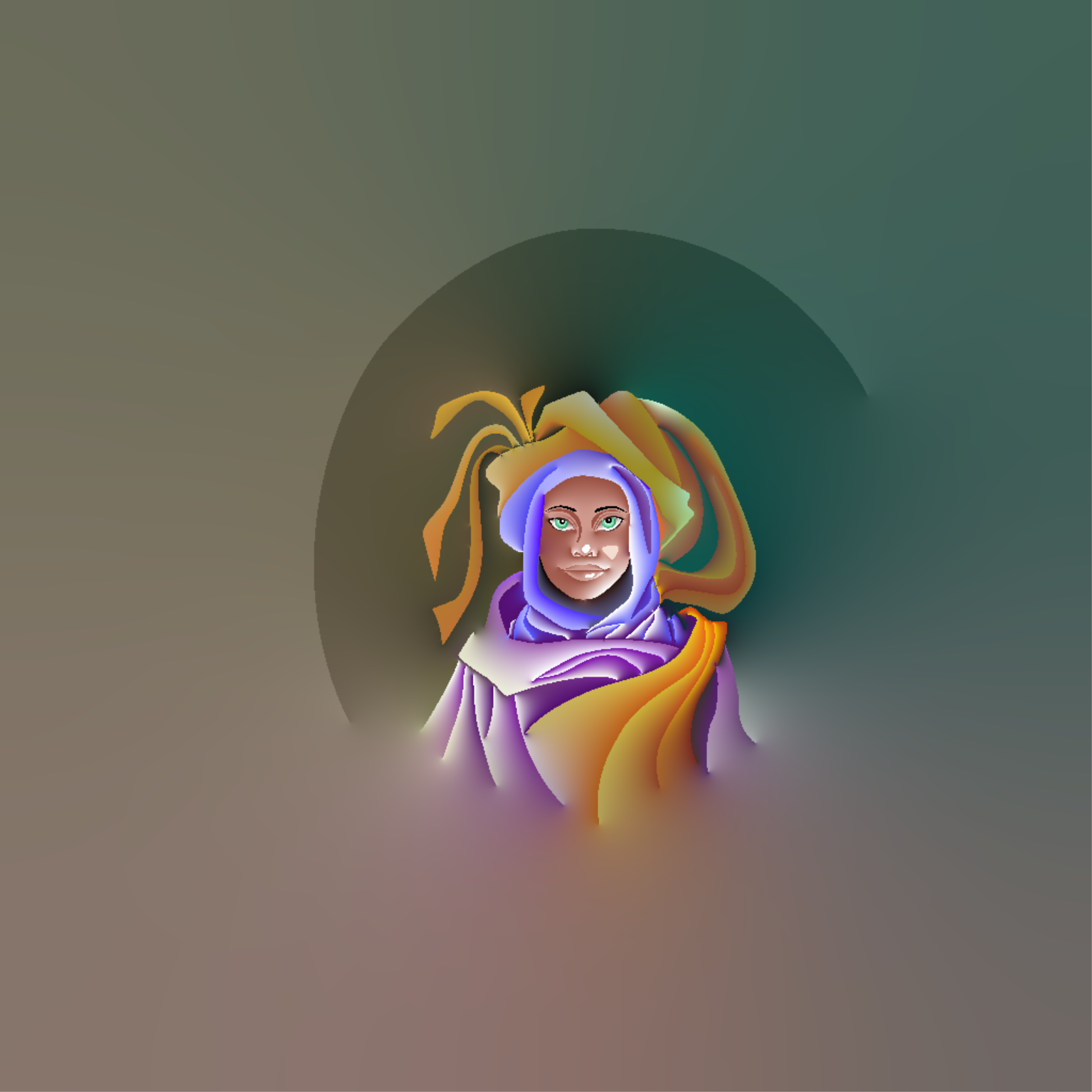} 
  {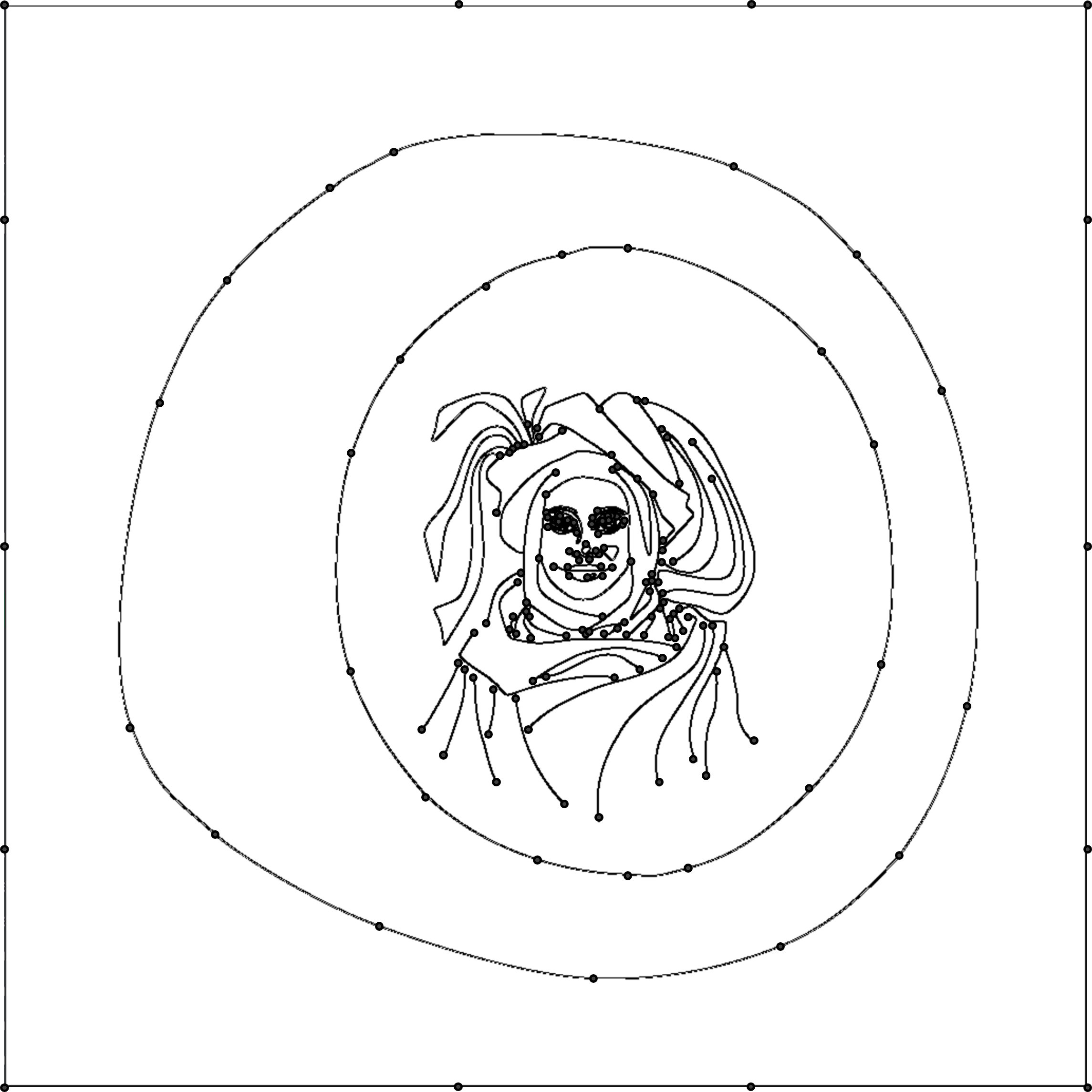} 
  {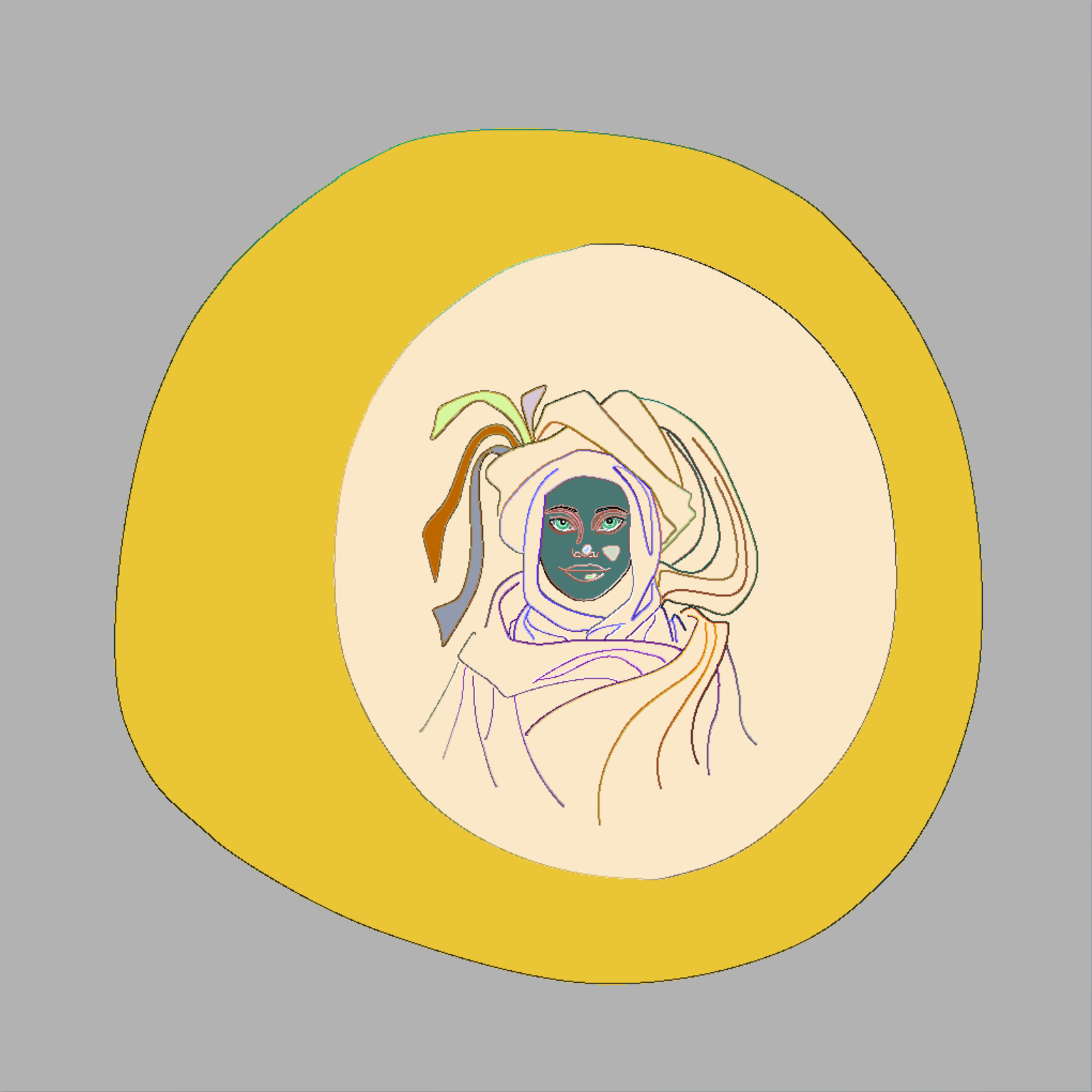} 
  {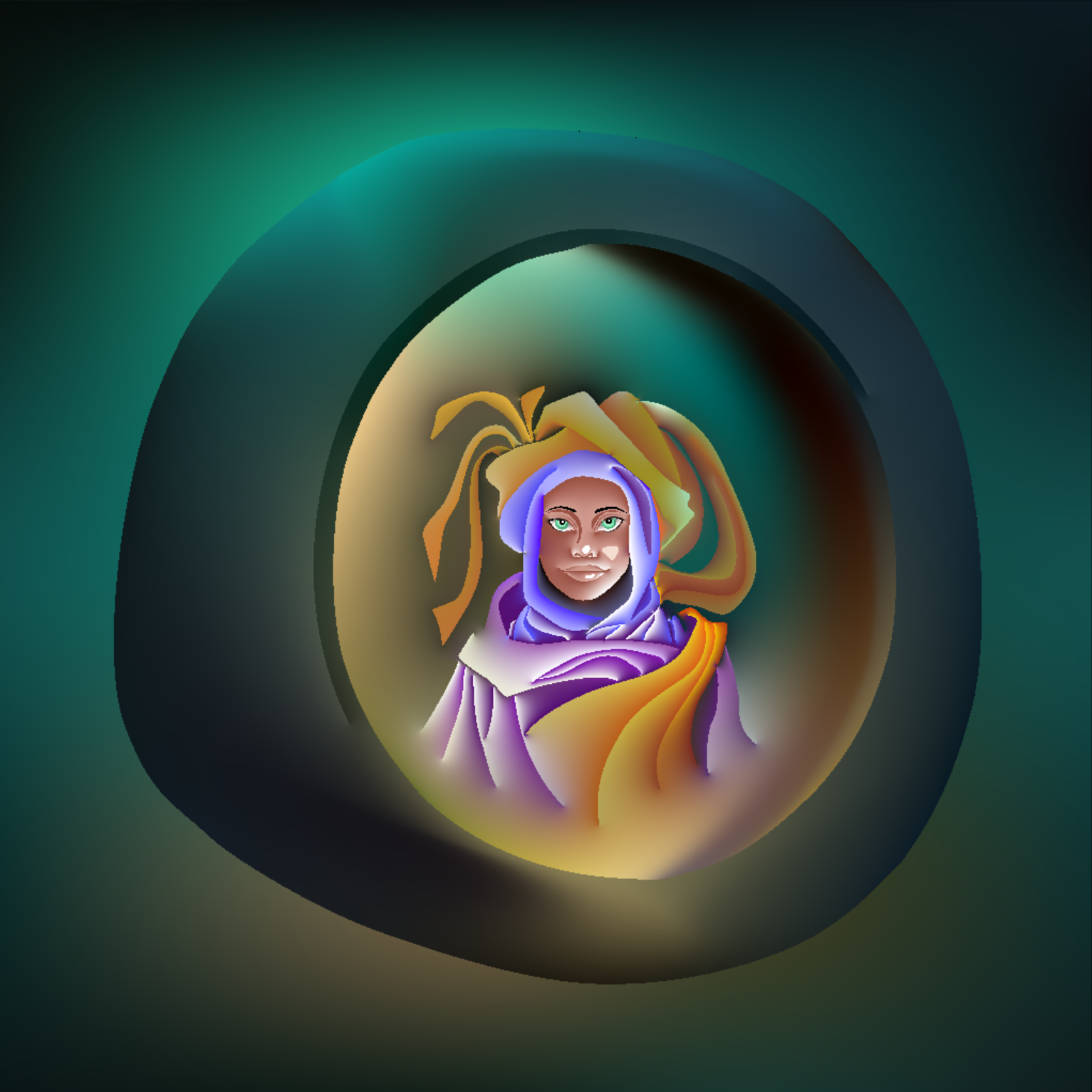} 
  \caption{Qualitative results of the pepper, sunset, bubbles, ladybug, and portal scene. From left to right, we show the input primitives (a) and (b), the undirected edge graphs (c), the unified patch representations (d) that were formed by our algorithm and the final rendering results (e).}%
    \label{fig:qualitative-eval}%
\end{figure*}

\subsection{Qualitative Results}

For all test scenes, we show results of only rasterizing the gradient meshes in Fig.~\ref{fig:qualitative-eval}(a) and only rasterizing the input curves in (b). This way, we can see which parts of the image are modeled with gradient meshes or diffusion/Poisson curves, respectively.
To illustrate the input primitives, we depict the gradient mesh control polygon with solid lines, where different colors are applied for each mesh.
Input boundary curves are shown as double-sided curve with an illustration of the respective boundary condition on the left and right.
Dirichlet conditions directly show the corresponding color, while Neumann conditions are illustrated with a dash-dot pattern.
Poisson curves are visualized with a long dash pattern.
In (c), the undirected edge graph is visualized by solid black curves and the vertices are depicted with black circles.
Note that Poisson curves are not part of the edge graph, since they are rendered additively onto the target Laplacian function.
The unified patch representation in (d) gives all pixels that belong to the same region a common color.
The colors are assigned from a custom color table such that the regions are visually distinguishable.
The color is not related to the image content.
The last column in (e) shows the final result.

\paragraph{Crane}
Inspired from an ancient Chinese painting (the Auspicious Cranes by Emperor Huizong), we modeled a crane with diffusion curves, which is shown in Fig.~\ref{fig:teaser}.
The boundary of the crane is largely modeled with Neumann conditions.
We modeled the clouds with two $3\times 4$ gradient meshes. In addition, we added diffusion curves and Poisson curves on the gradient meshes for the layered and detailed effects of clouds.
The background is formed from a $3\times 3$ gradient mesh that mimics the similar color transition of the inspirational image.

\paragraph{Pepper}
Fig.~\ref{fig:qualitative-eval} (first row) shows an extension of the \emph{poivron} scene from \cite{orzan2008diffusion}.
We replaced the background with a $30\times 30$ gradient mesh that was fitted with Adobe Illustrator to a photograph inside a restaurant.
The gradient mesh's smoothness creates the impression of defocus blur, putting emphasis on the pepper in the foreground.
We took a bite out of the pepper by adding further diffusion curves.
The outer boundary of the pepper is modeled with homogeneous Neumann conditions, such that it can be placed in the scene without causing color diffusion into the background.
The shadow on the table is modeled with Poisson curves.

\paragraph{Sunset}
Fig.~\ref{fig:qualitative-eval} (second row) shows a colorful sunset.
The color gradient in the background contains multiple hues and is modeled with a $10\times 3$ gradient mesh. We also deformed the spatial control points for artistic effects.
Since it is interrupted by other curve primitives in-between, the color gradient would have been difficult to model with diffusion curves only.
The sun is drawn using a diffusion curve, which allows for the modeling of the faint glow  into the background.
On the top and bottom of the image, we placed diffusion curves that contrast the nature background with expressive wavy shapes.

\paragraph{Bubbles}
In Fig.~\ref{fig:qualitative-eval} (third row), we modeled two soap bubbles.
We used a $1\times 3$ gradient mesh to model smooth color transitions in the sky.
The bubbles are created by two $4\times 4$ gradient meshes. The smooth color transitions are difficult to model directly with diffusion curves. 
This example shows that a vivid scene can be represented by a small number of primitives when combining mesh-based and curve-based methods.

\newcommand{\compareentry}[4]{%
\begin{minipage}{0.16\linewidth}%
\includegraphics[width=\linewidth]{#1}%
\end{minipage}
\begin{minipage}{0.08\linewidth}%
\includegraphics[width=\linewidth]{#2}\\%
\begin{tikzpicture}%
    \node[anchor=south west,inner sep=0] (image) at (0,0) {\includegraphics[width=\linewidth]{#3}};%
    \begin{scope}[x={(image.south east)},y={(image.north west)}]%
        \node[overlay, anchor=north west, color=white] at (0.0,1.0) {\tiny RMSE: {#4}};
    \end{scope}%
\end{tikzpicture}%
\end{minipage}
}%

\paragraph{Ladybug}
The test scene in Fig.~\ref{fig:qualitative-eval} (fourth row) is based on the \emph{ladybug} diffusion curve model of \cite{orzan2008diffusion}.
We made minor adjustments to the original curve endpoints to remove edge crossings that would otherwise cause color leakage. In addition, we set the outer boundary of the ladybug to Neumann condition.
We placed the ladybug in a beach scene, which contains an ocean and a blue sky. The background is modeled using a $11\times 3$ gradient mesh.
In addition, we modeled a leaf on which the ladybug sits with a $6\times 2$ gradient mesh, which captures the subtle color changes without smoothing.
The leaf veins are created with Poisson curves, and the outer boundary is modeled with diffusion curves to add scattering.

\paragraph{Portal}
The last test scene in Fig.~\ref{fig:qualitative-eval} (fifth row) shows a portal.
We created the portal with three gradient meshes: a $4\times 3$ background mesh, and two $3\times 4$ gradient meshes for the outer and internal parts of the portal.  
The rendered portal contains smooth color transitions to represent light and shadows. 
Inside the portal, We modified the \emph{zephyr} diffusion curve model of \cite{orzan2008diffusion}. The diffusion curves diffuse color into the portal. The mesh Laplacian provides richer effects and more potential for artistic control compared to curve-only results. 

\begin{figure*}[t]%
  \centering%
  \begin{minipage}{0.16\linewidth}%
      \centering%
      \scriptsize%
      (a) reference image (ours)
  \end{minipage}\hfill%
  \begin{minipage}{0.24\linewidth}%
      \centering%
      \scriptsize%
      (b) gradient meshes (manual)
  \end{minipage}\hfill\hfill%
  \begin{minipage}{0.24\linewidth}%
      \centering%
      \scriptsize%
      (c) diffusion curves (Orzan)\\\cite{orzan2008diffusion}
  \end{minipage}\hfill\hfill%
  \begin{minipage}{0.24\linewidth}%
      \centering%
      \scriptsize%
      (d) diffusion curves (isocontours)\\\cite{zhao2017inverse}
  \end{minipage}%
  \\%
  \begin{minipage}{0.16\linewidth}%
      \includegraphics[width=\linewidth]{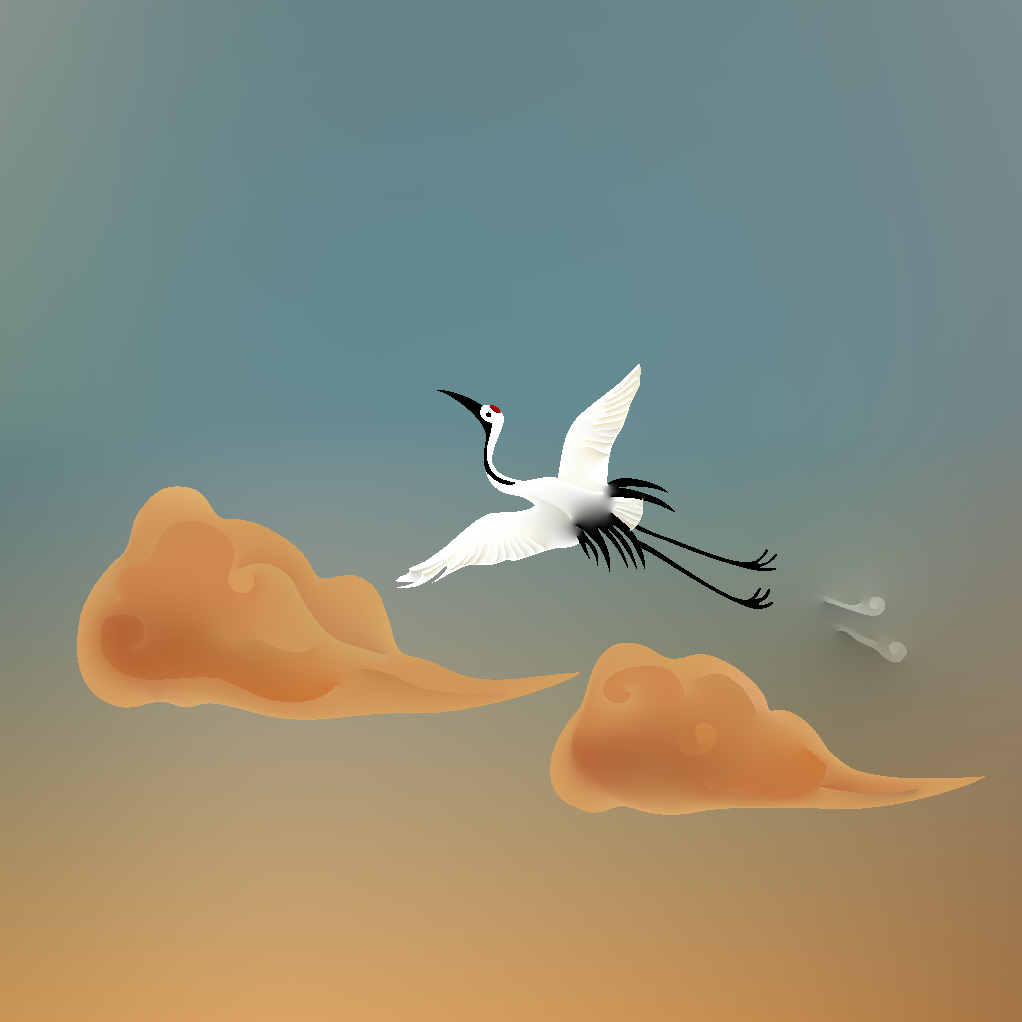}
  \end{minipage}\hfill%
  \compareentry
  {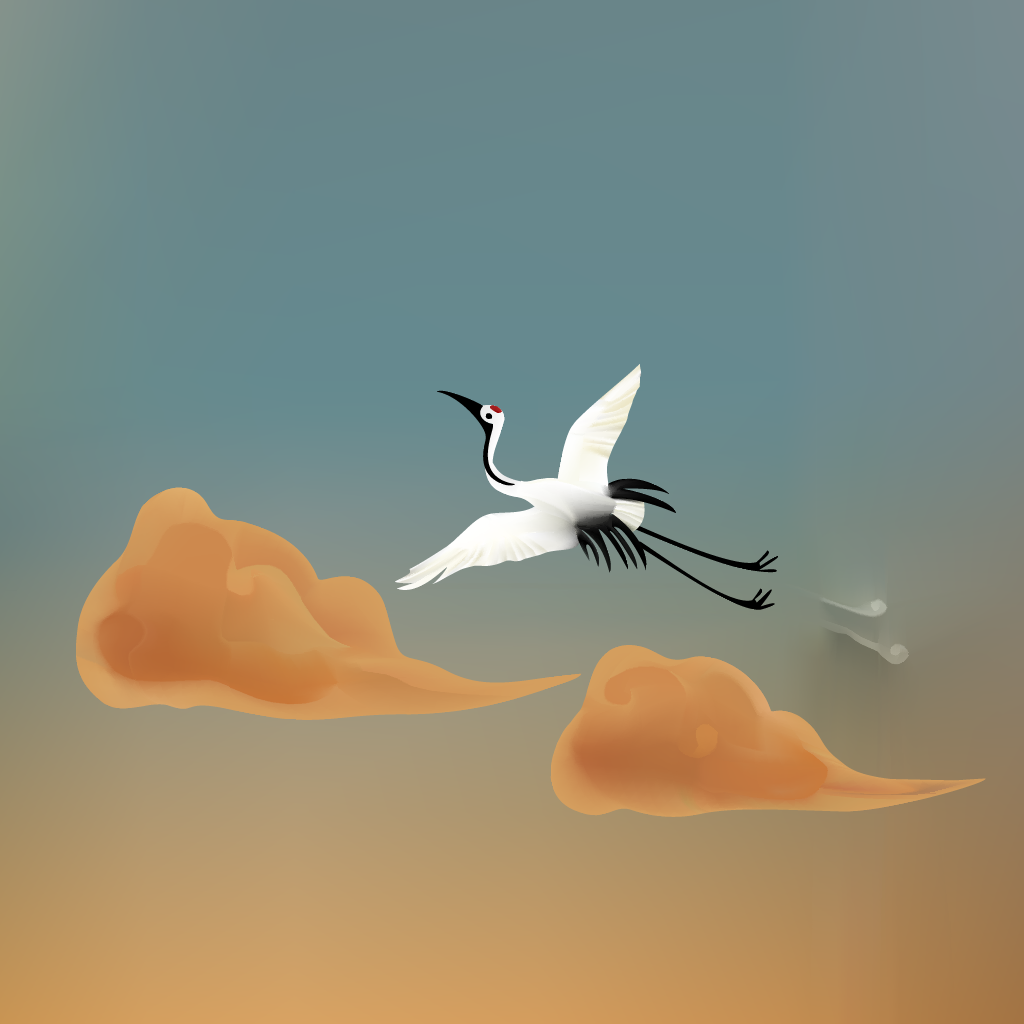}
  {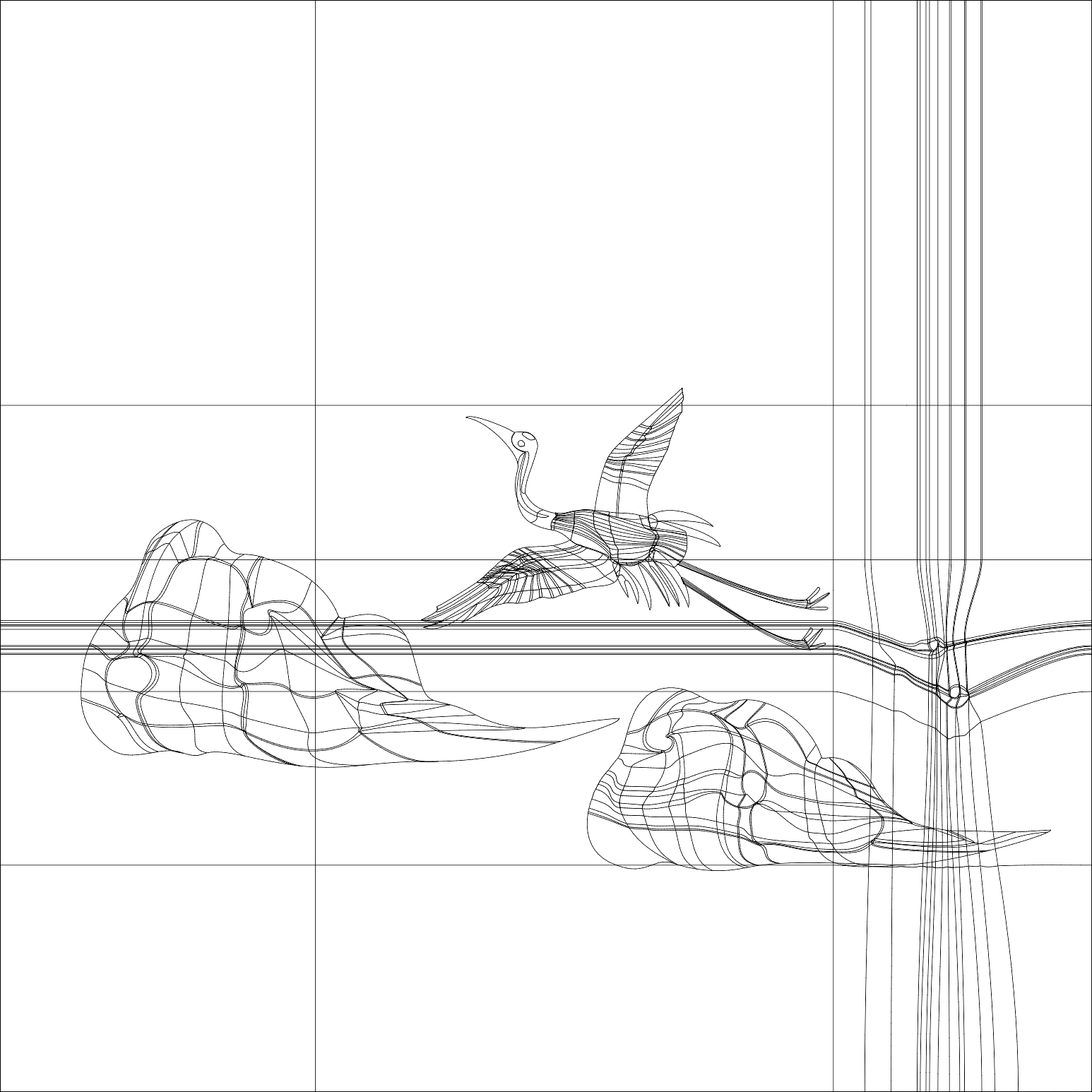}
  {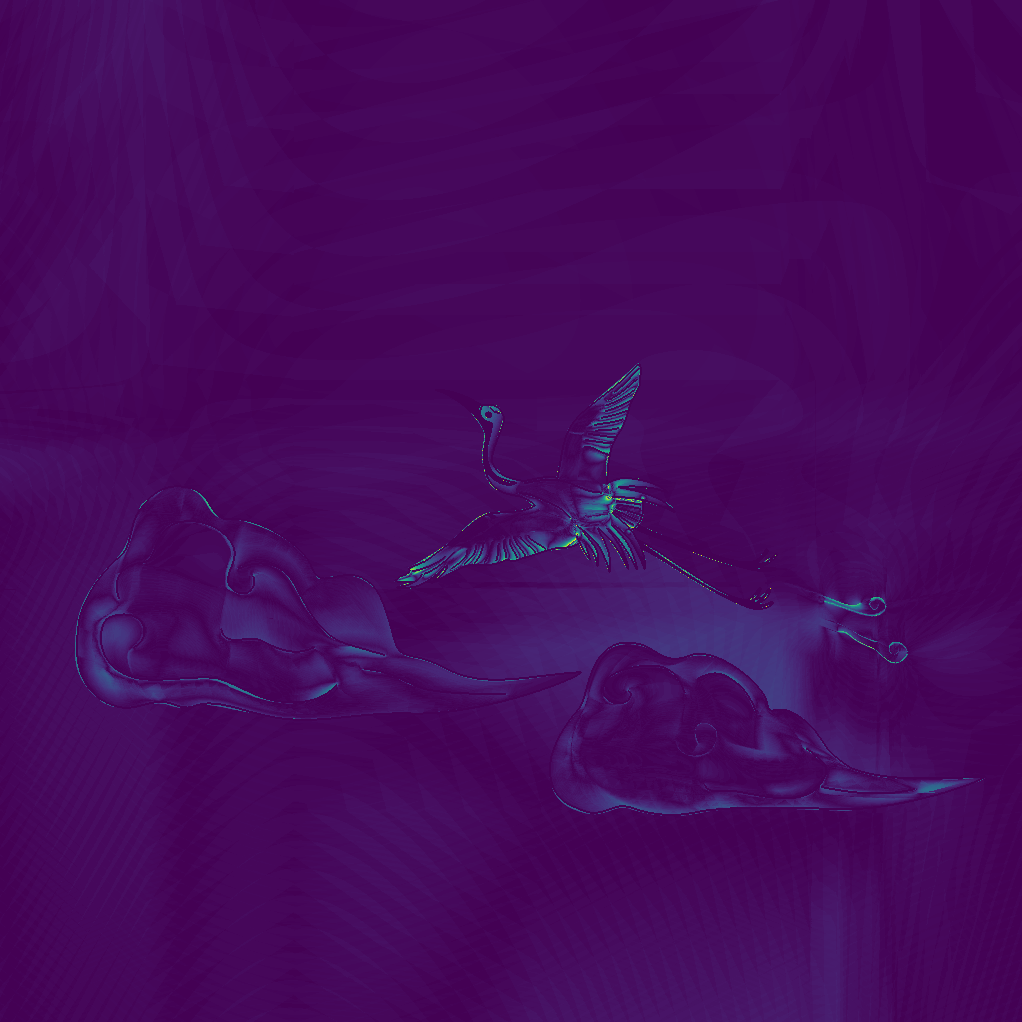}
  {0.013}
  \hfill%
  \compareentry
  {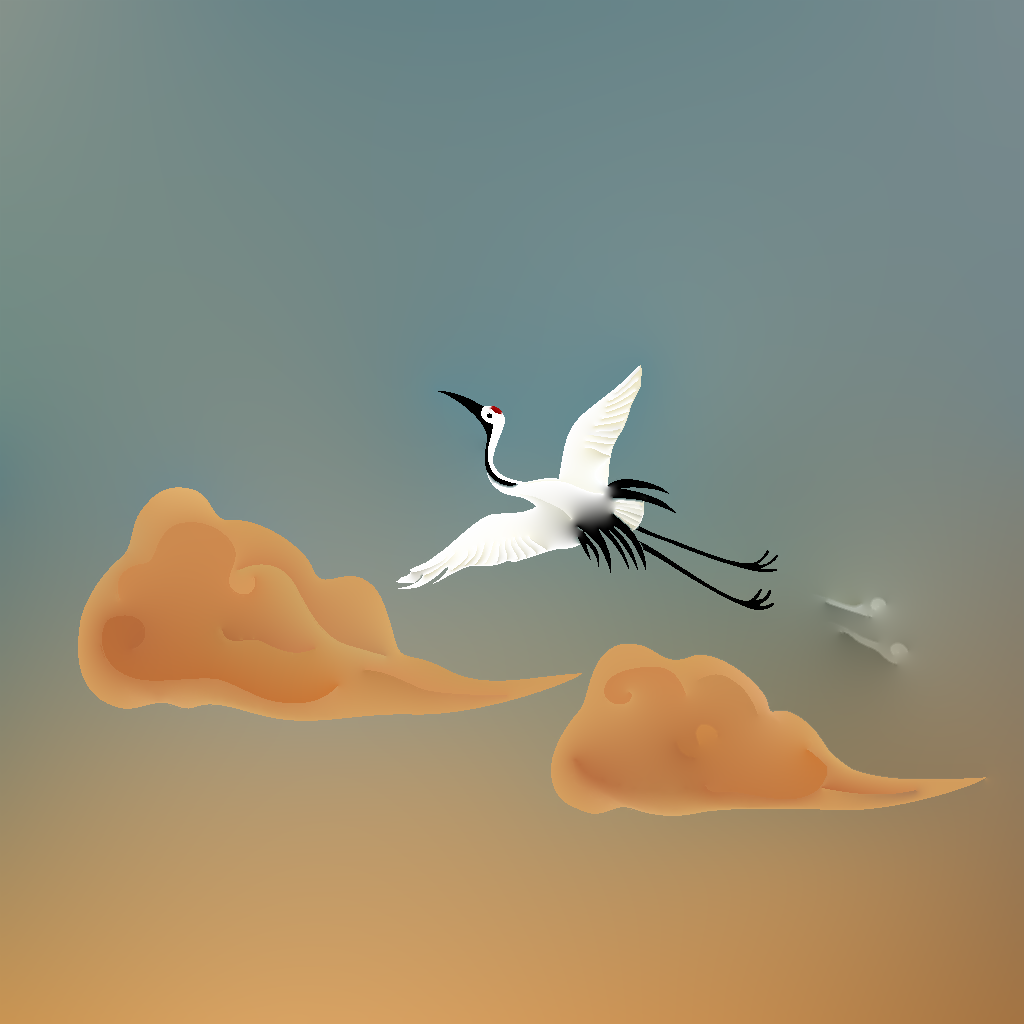}
  {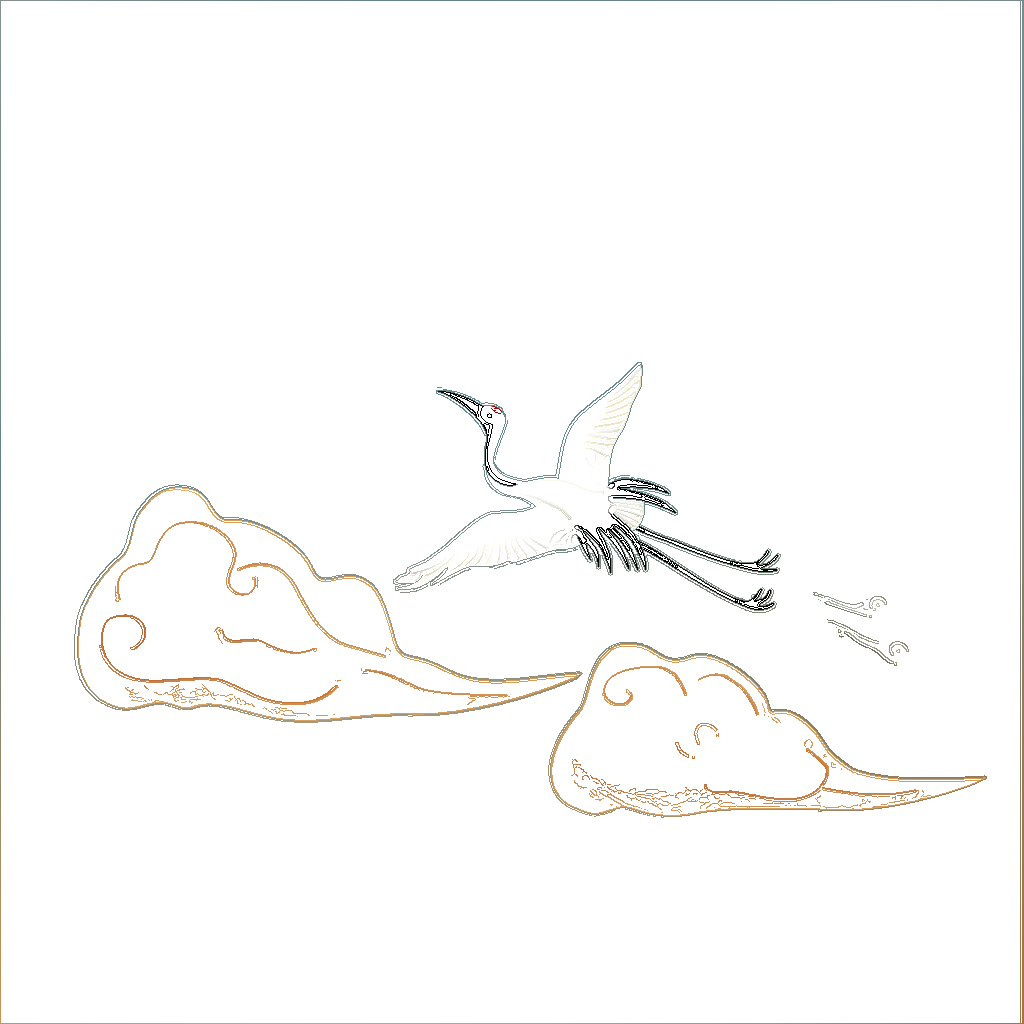}
  {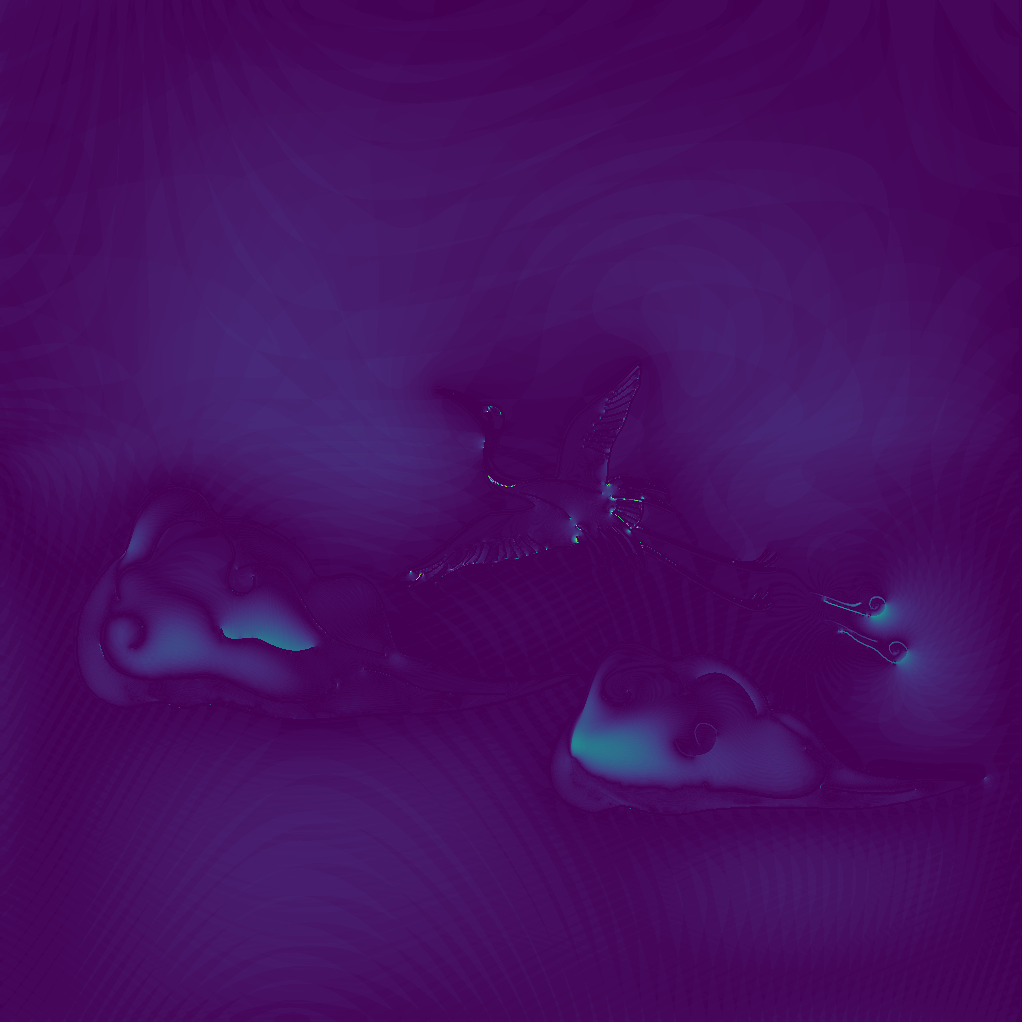}
  {0.018}
  \hfill%
  \compareentry
  {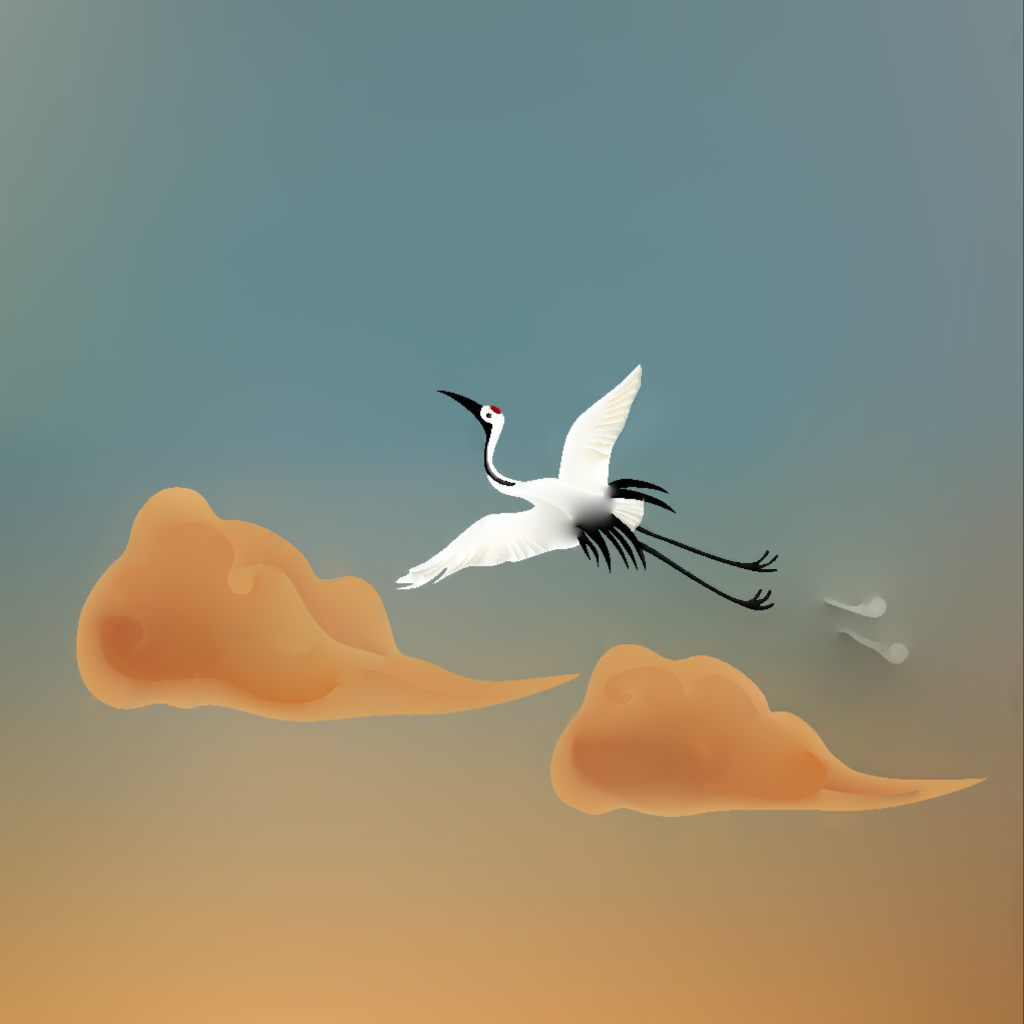}
  {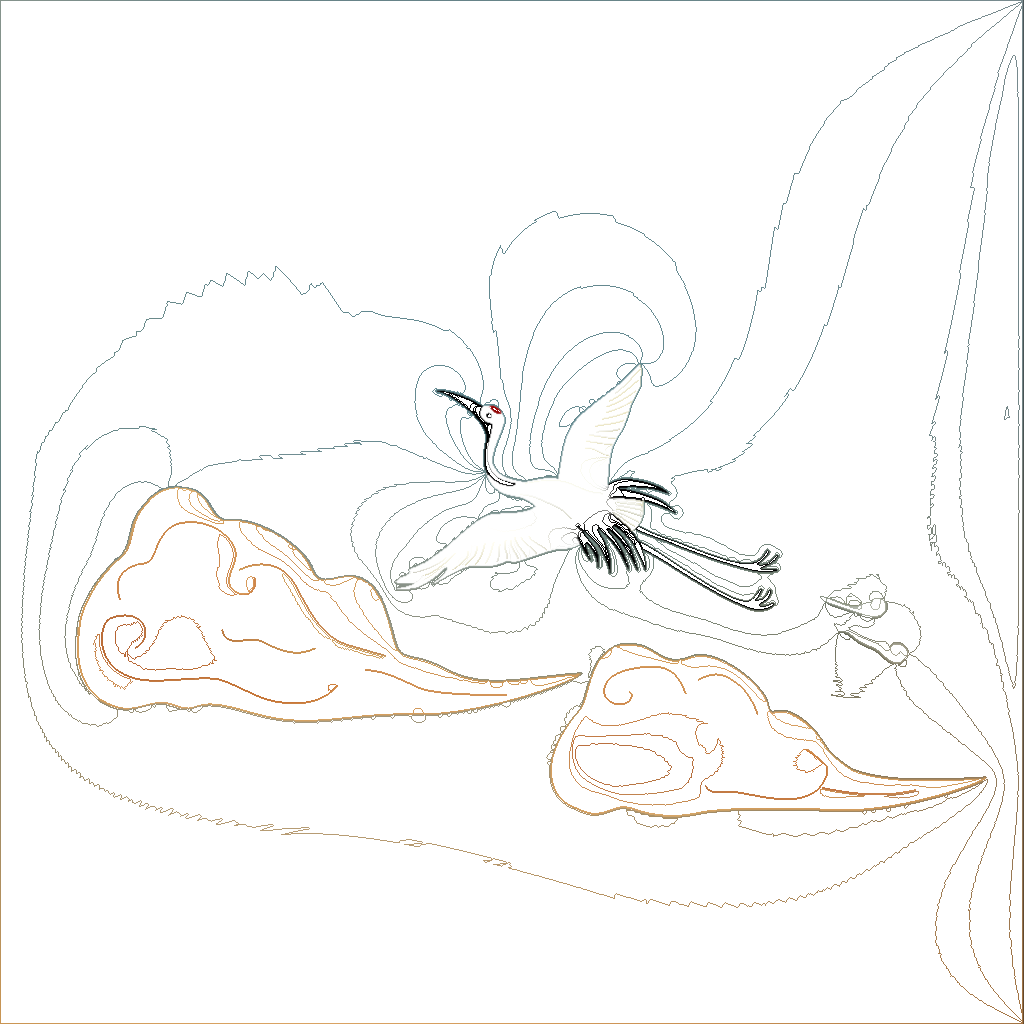}
  {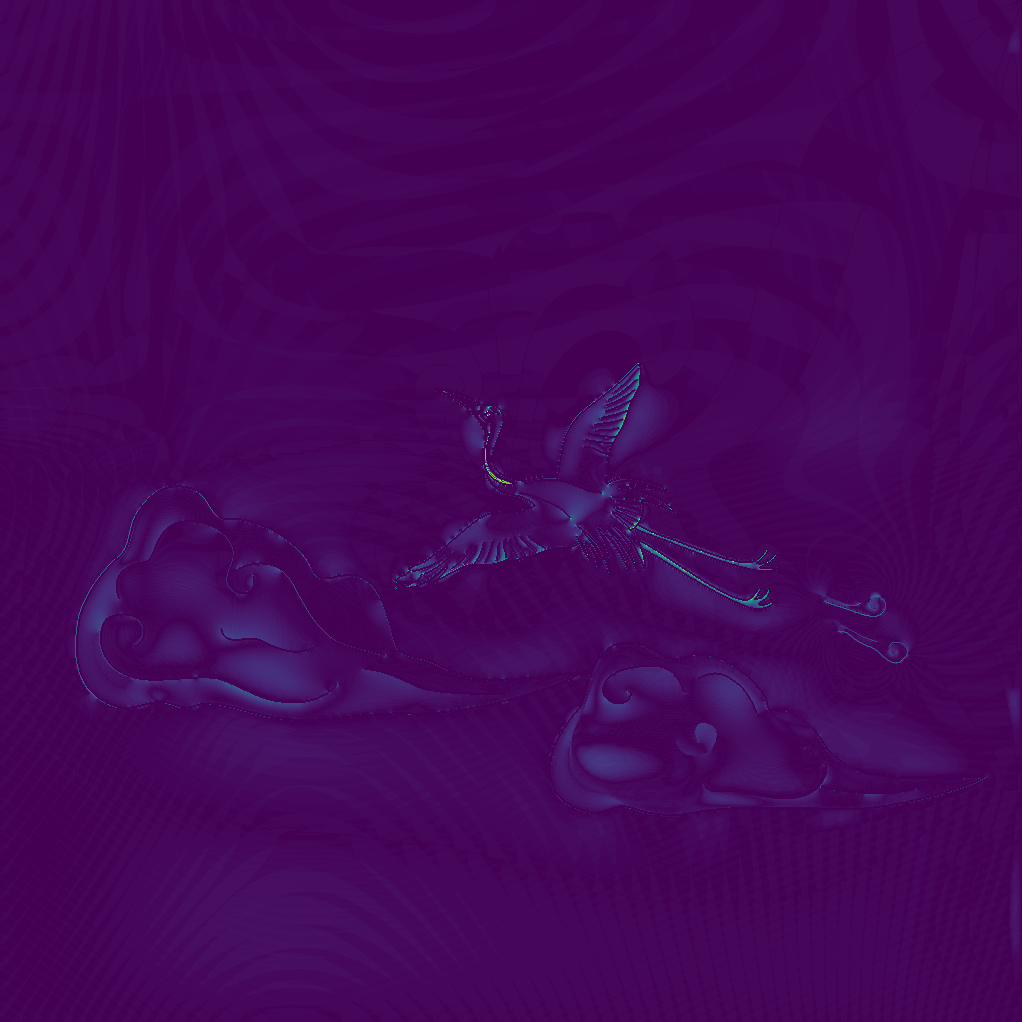}
  {0.008}
  \\%
  \vspace{0.4em}%
  \begin{minipage}{0.16\linewidth}%
      \includegraphics[width=\linewidth]{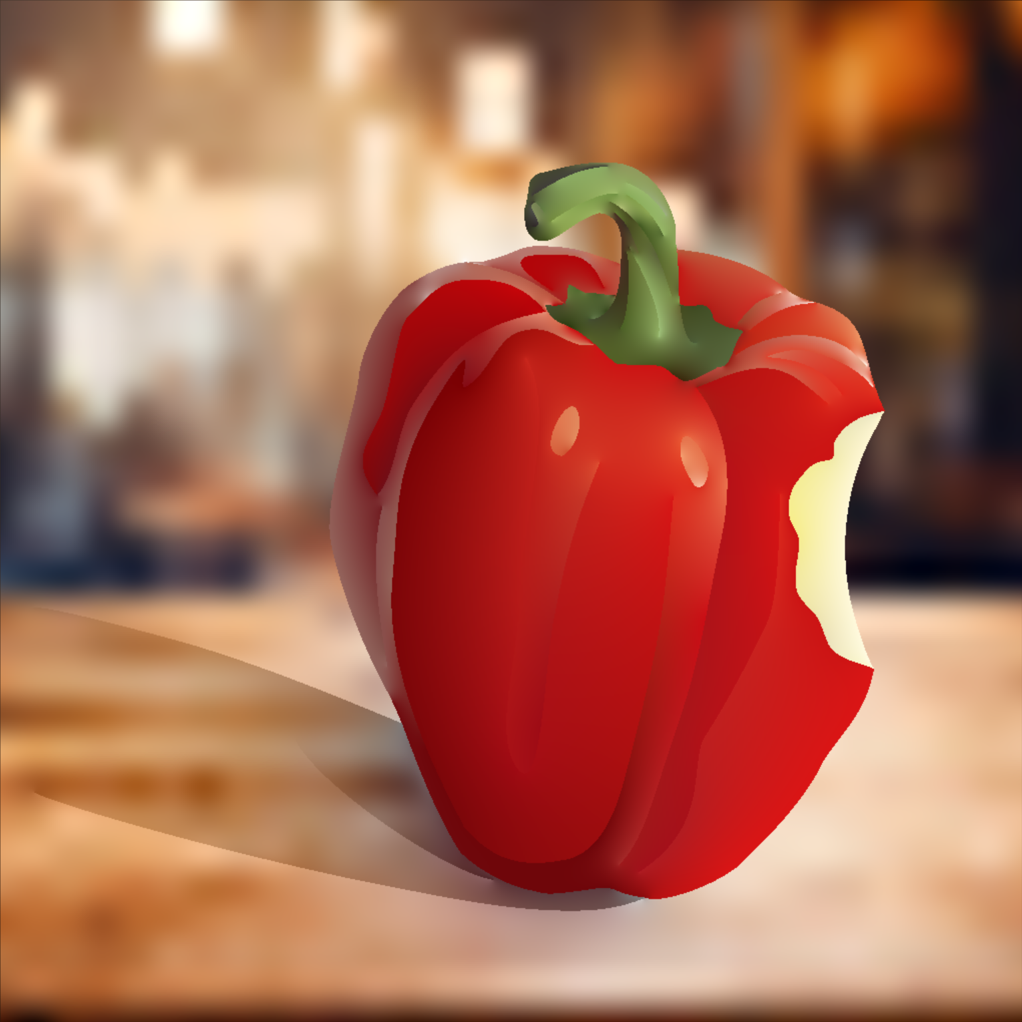}
  \end{minipage}\hfill%
  \compareentry
  {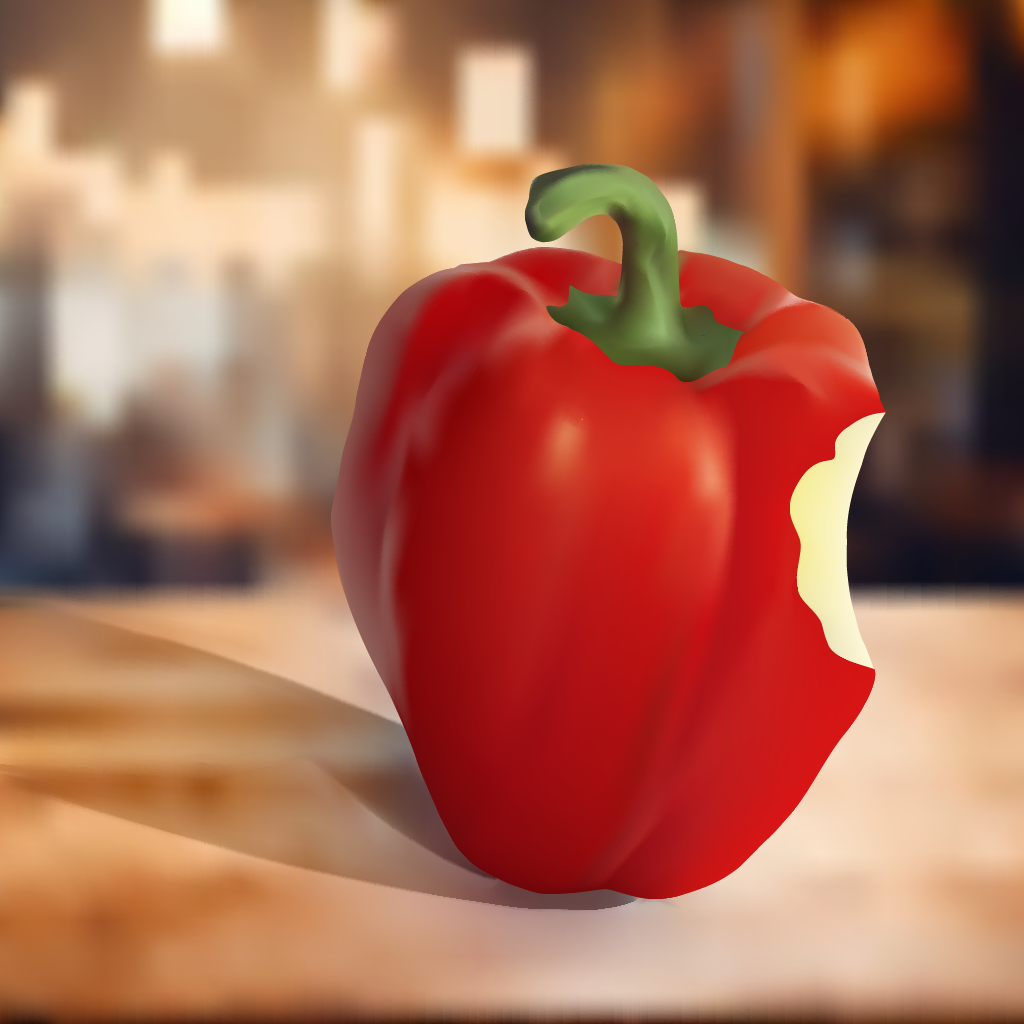}
  {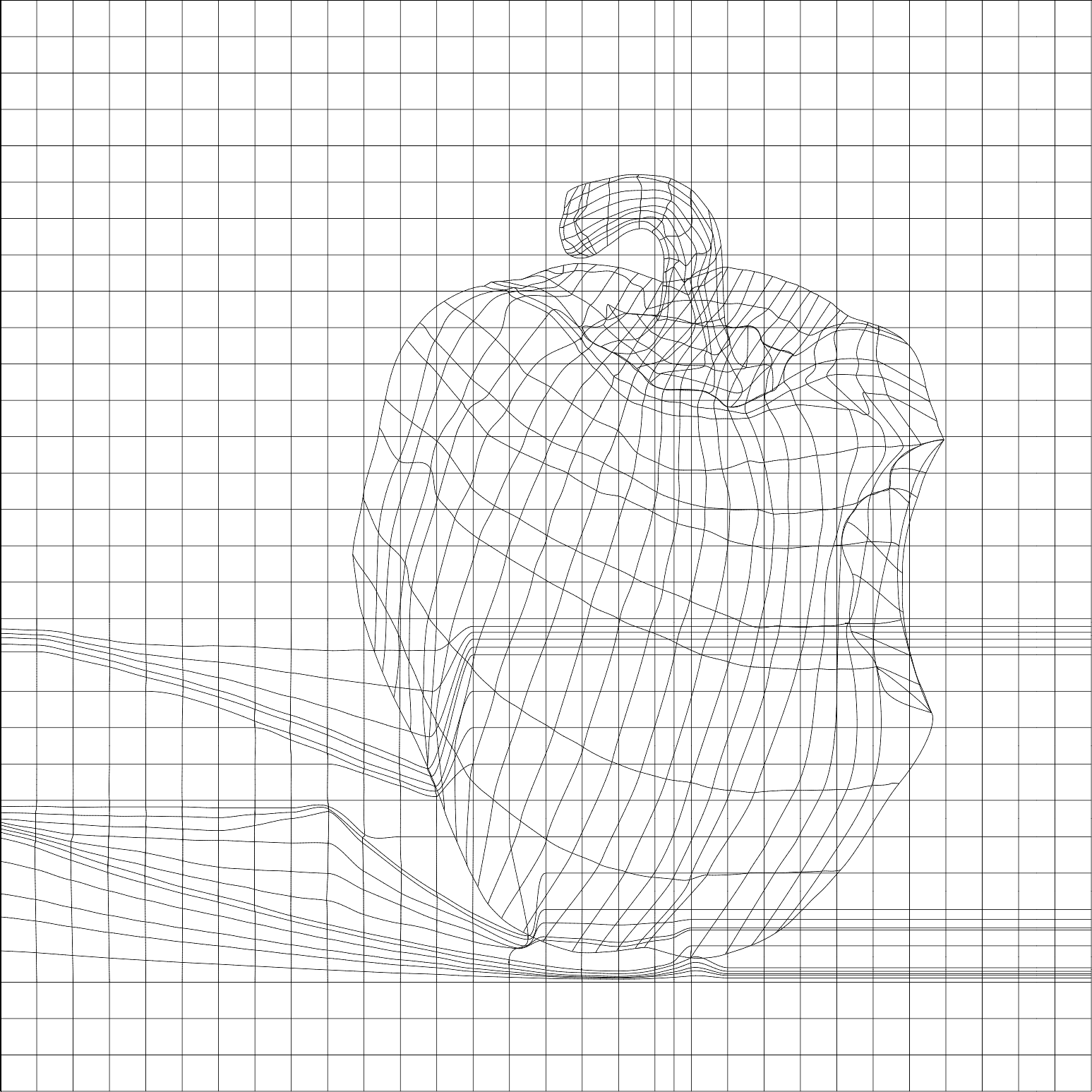}
  {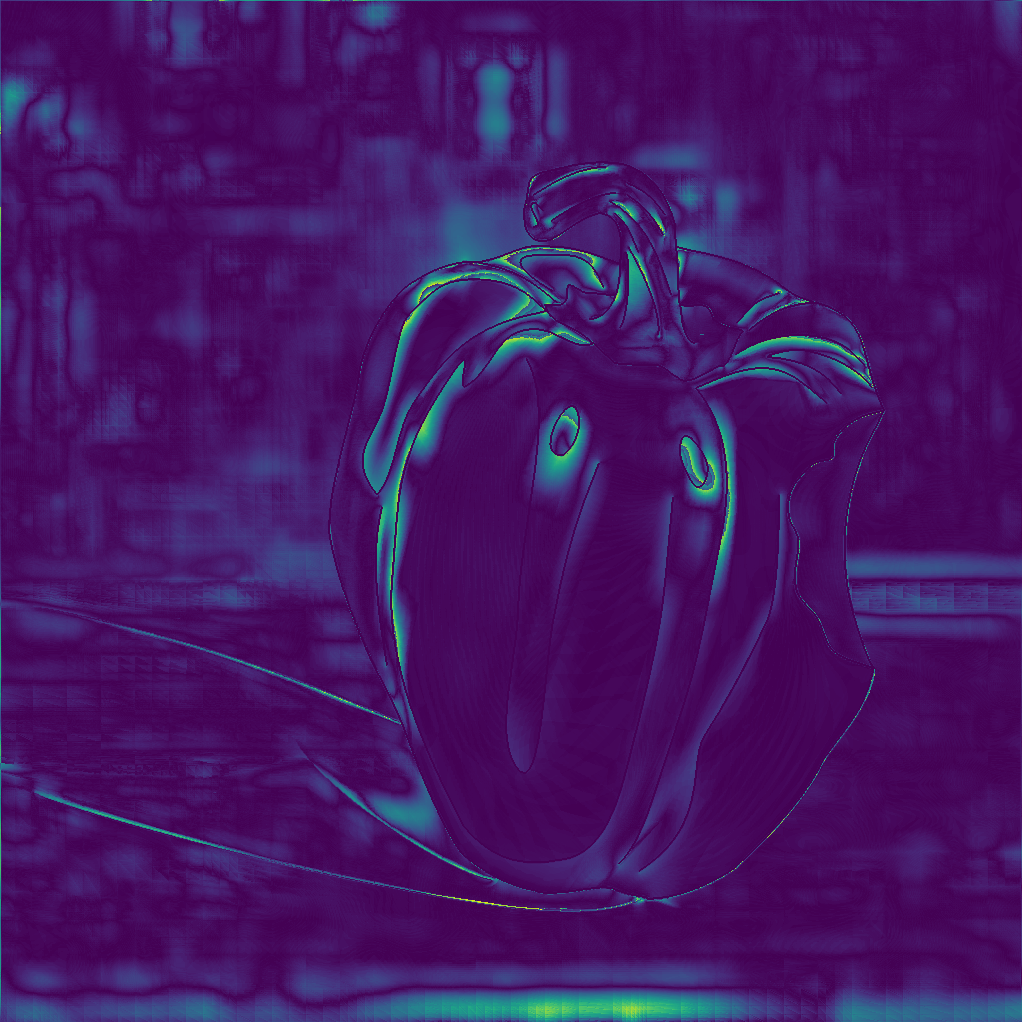}
  {0.026}
  \hfill%
  \compareentry
  {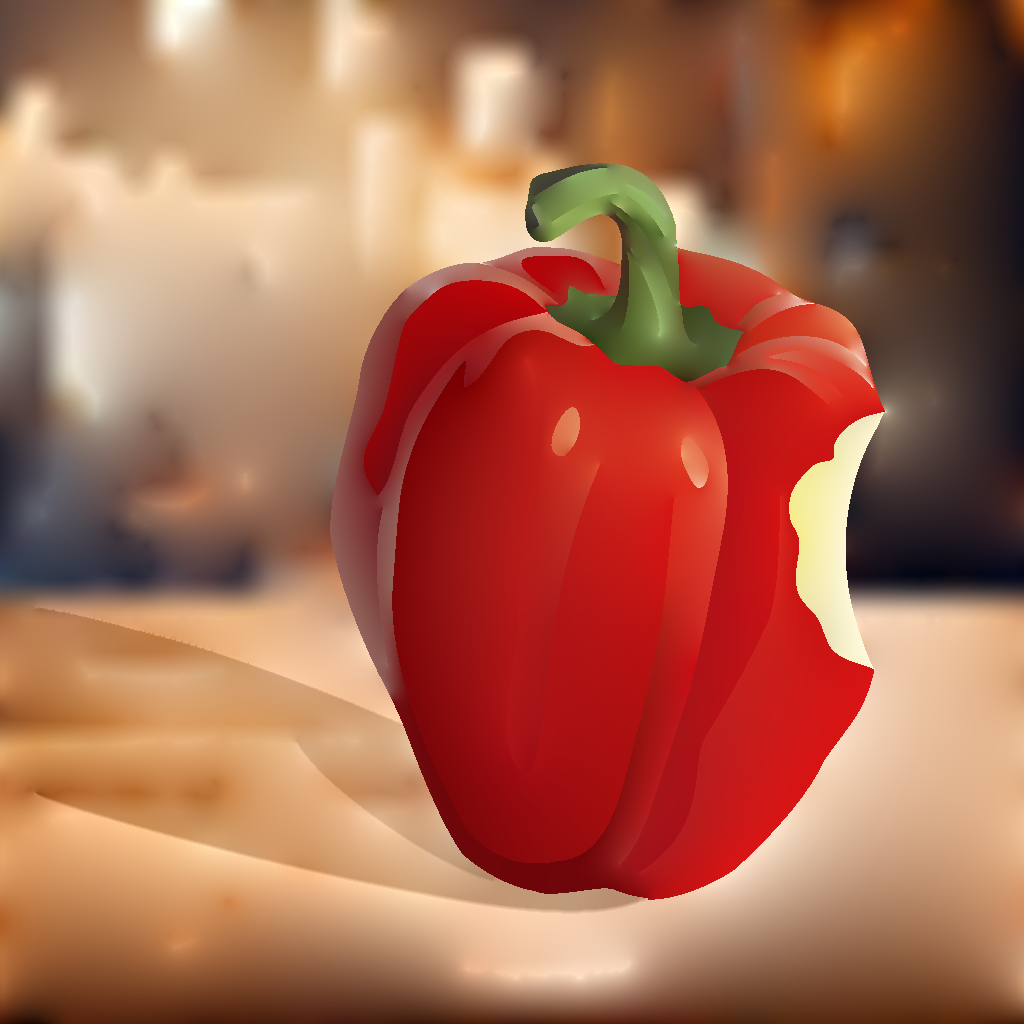}
  {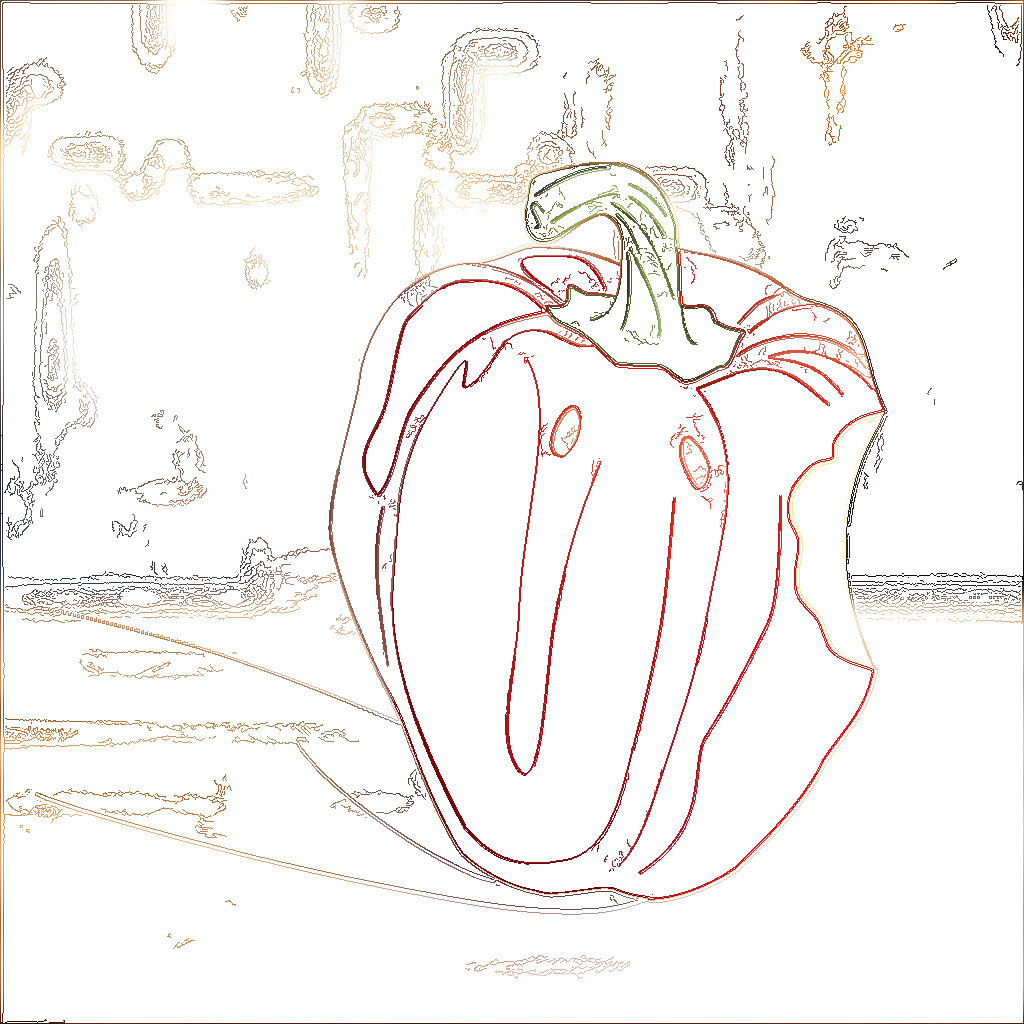}
  {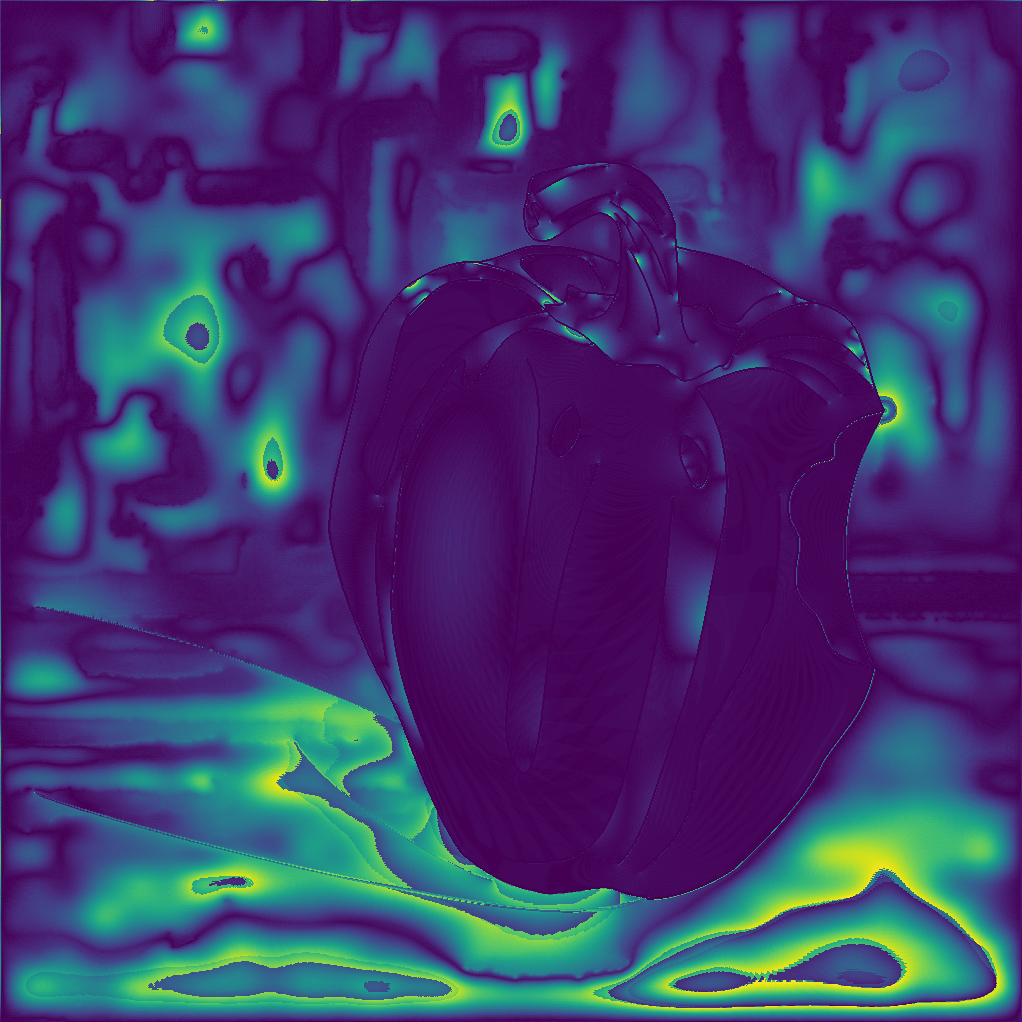}
  {0.089}
  \hfill%
  \compareentry
  {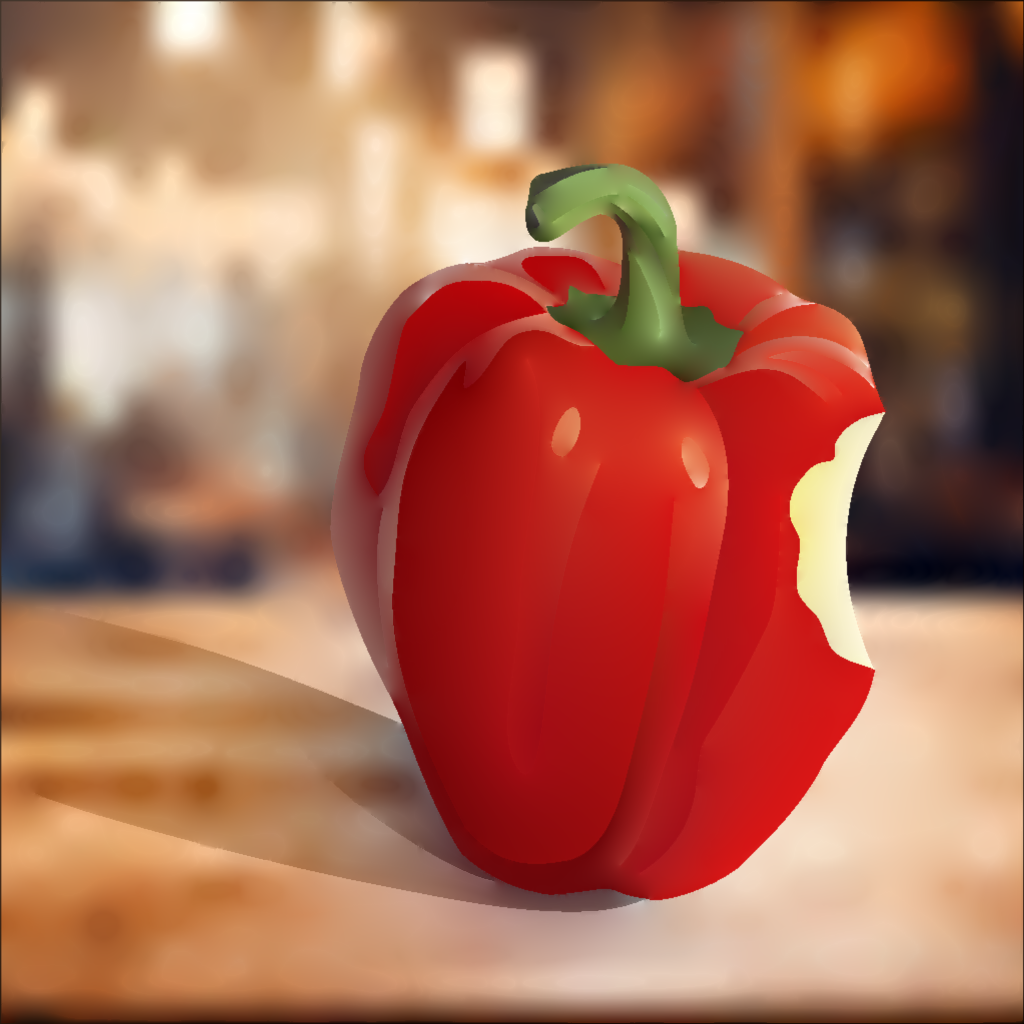}
  {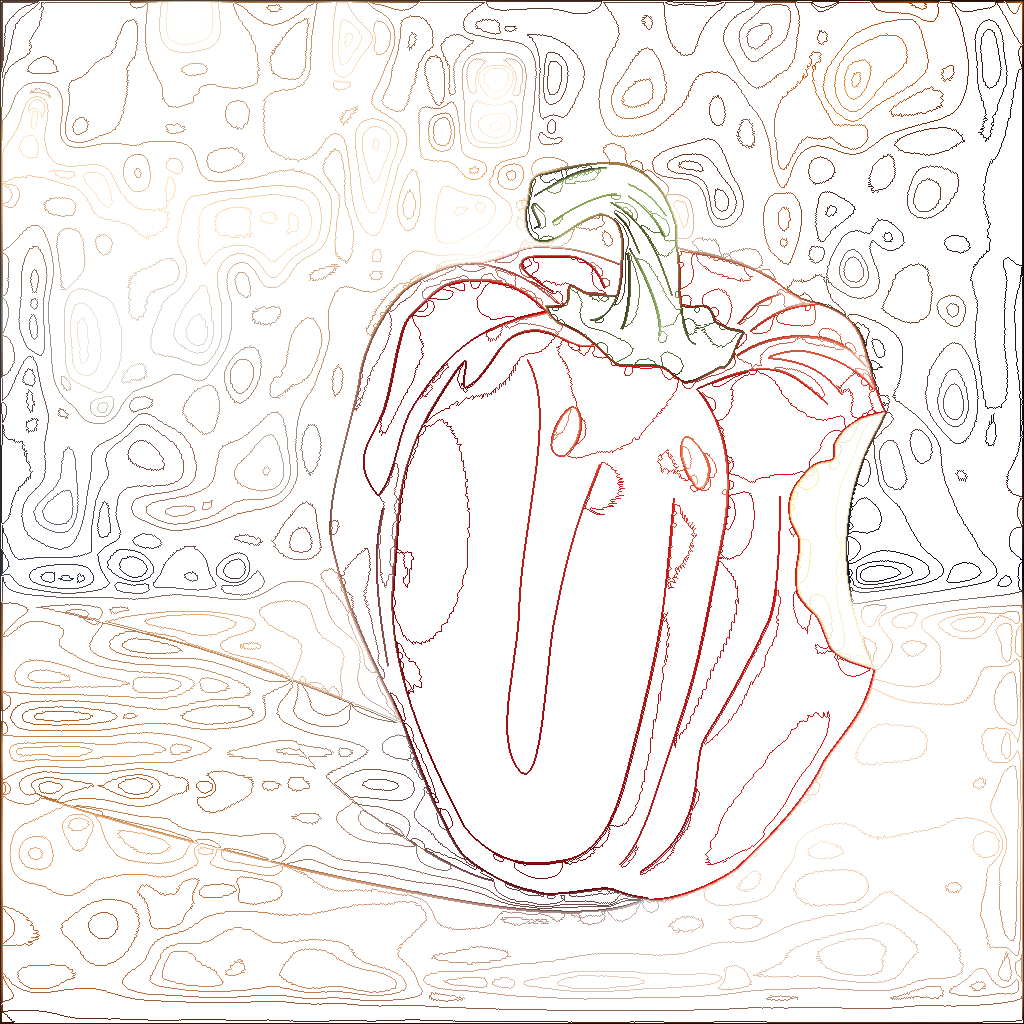}
  {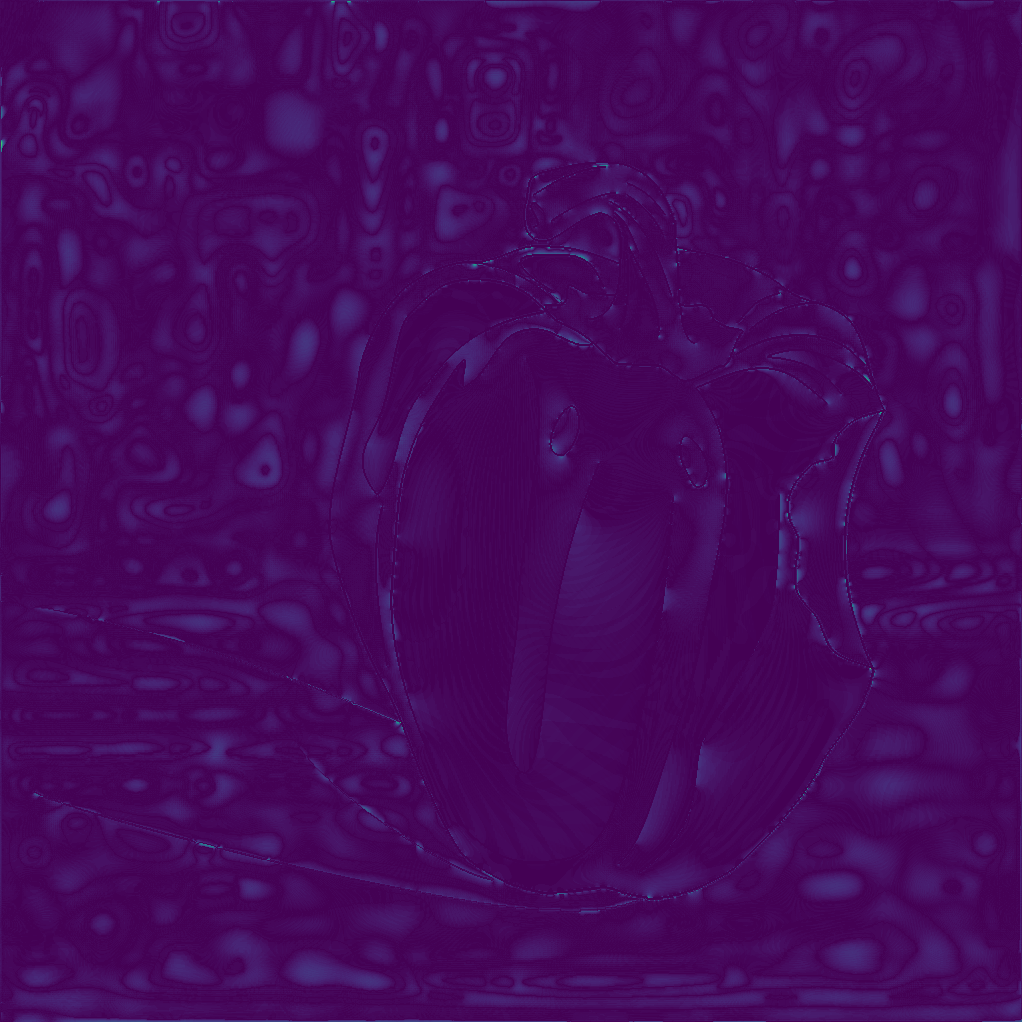}
  {0.008}
  \\%
  \vspace{0.4em}%
  \begin{minipage}{0.16\linewidth}%
      \includegraphics[width=\linewidth]{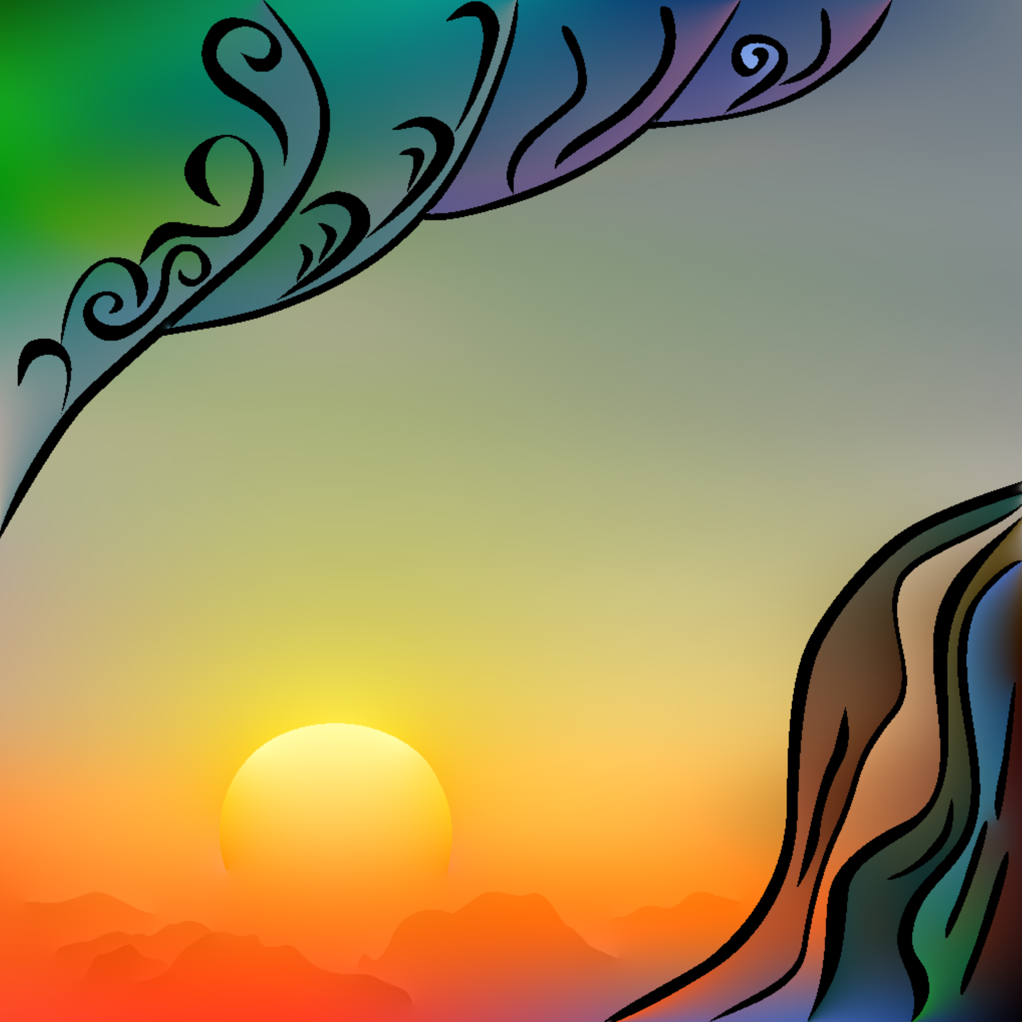}
  \end{minipage}\hfill%
  \compareentry
  {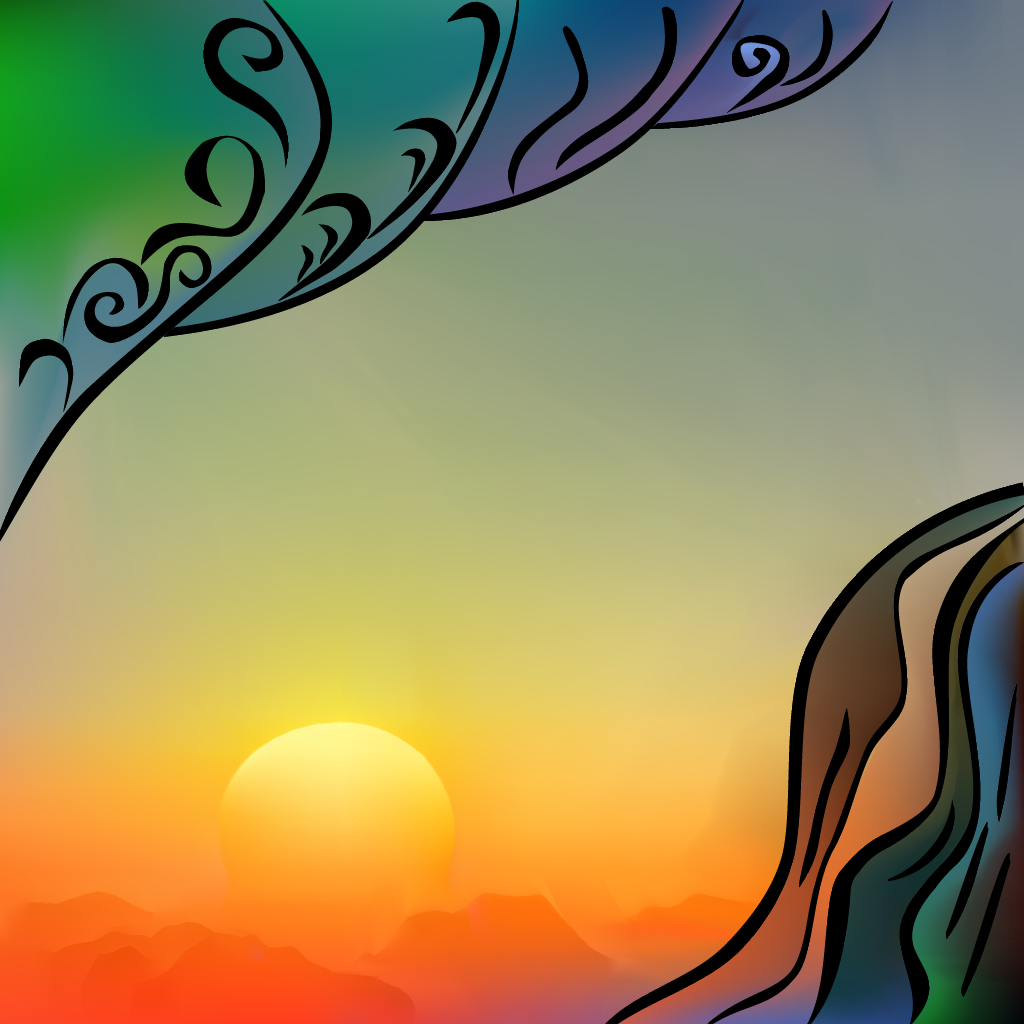}
  {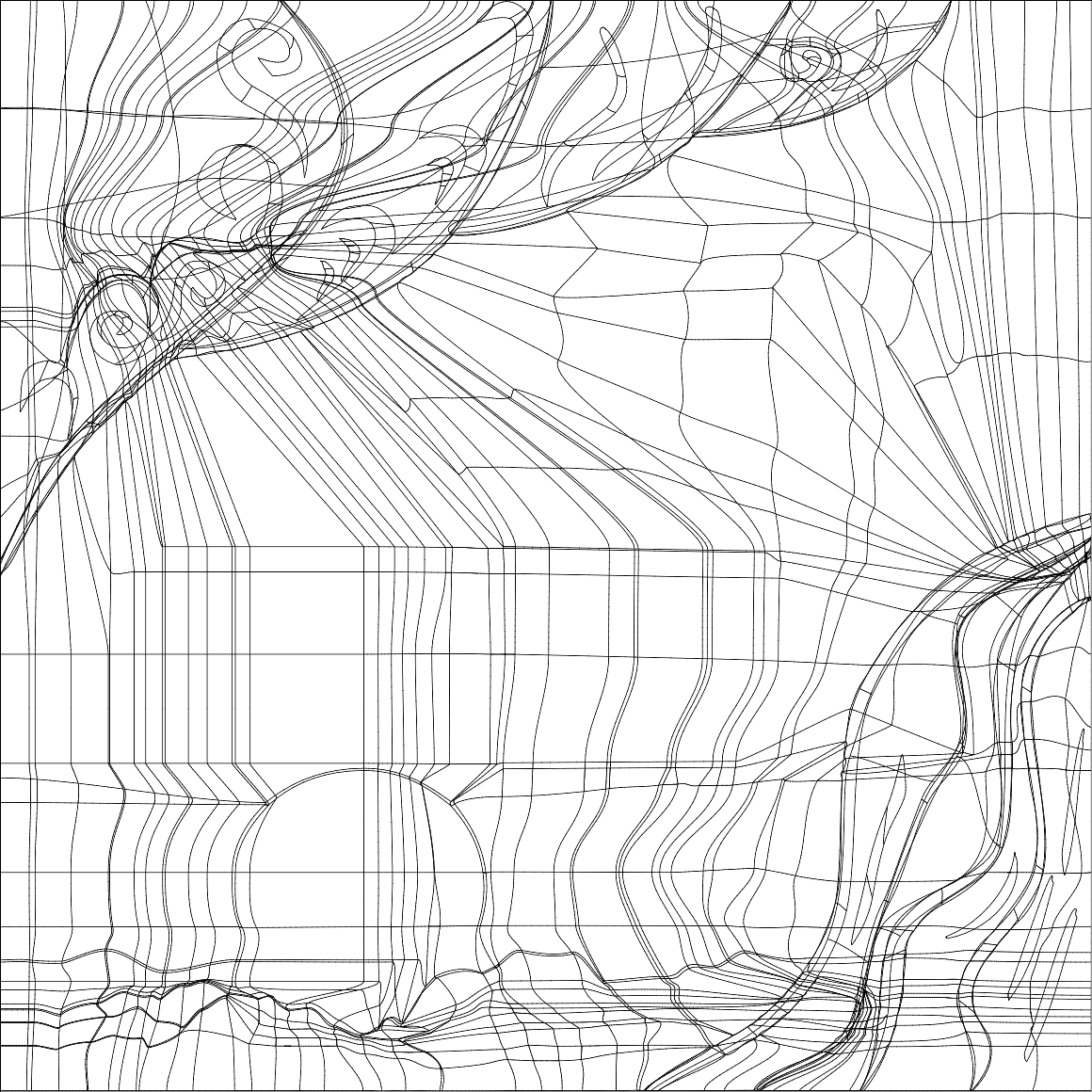}
  {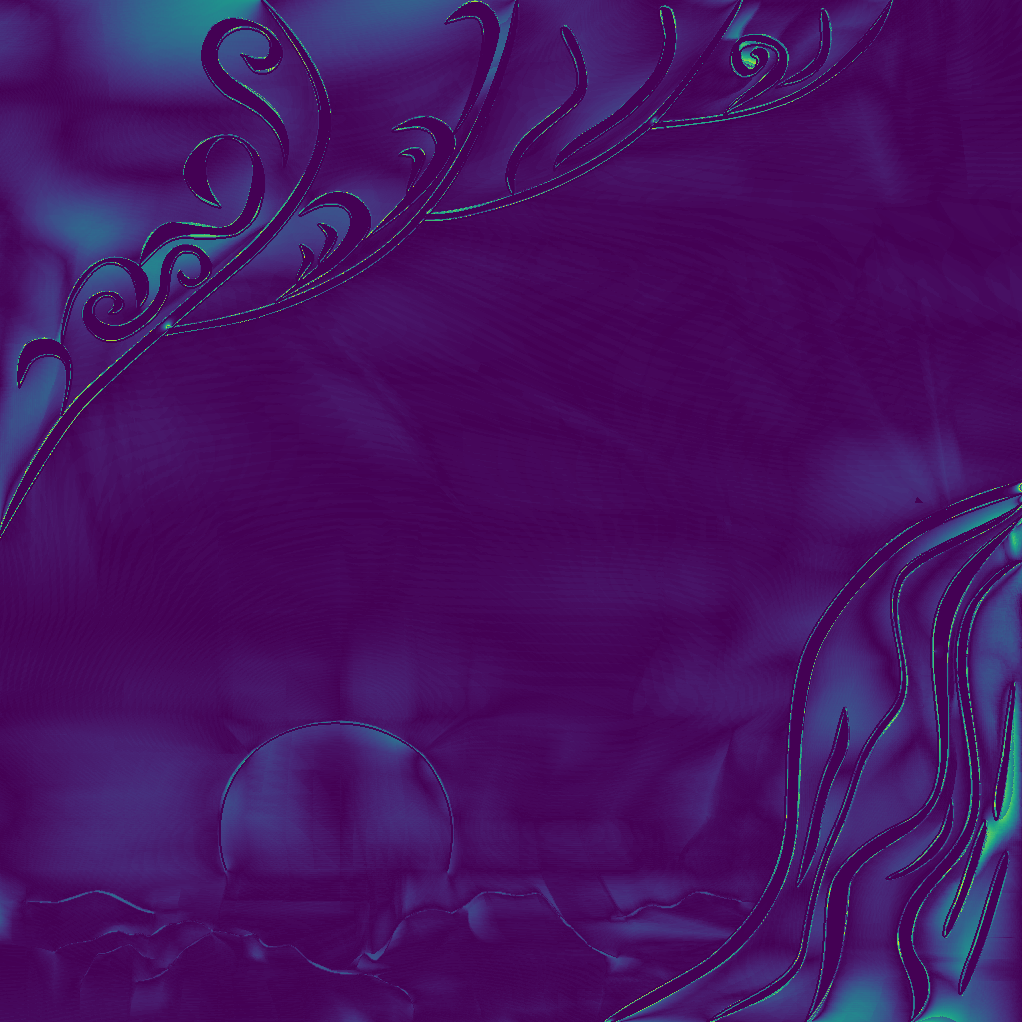}
  {0.032}
  \hfill%
  \compareentry
  {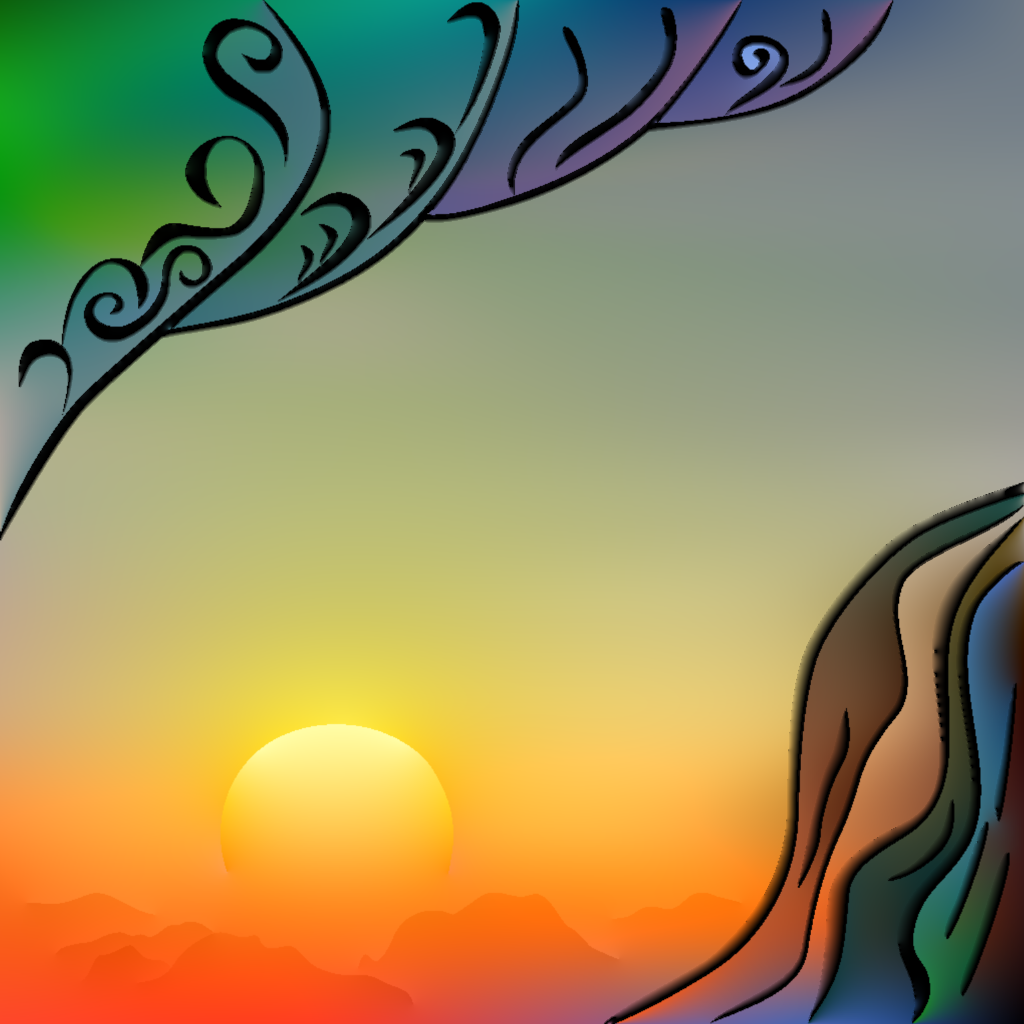}
  {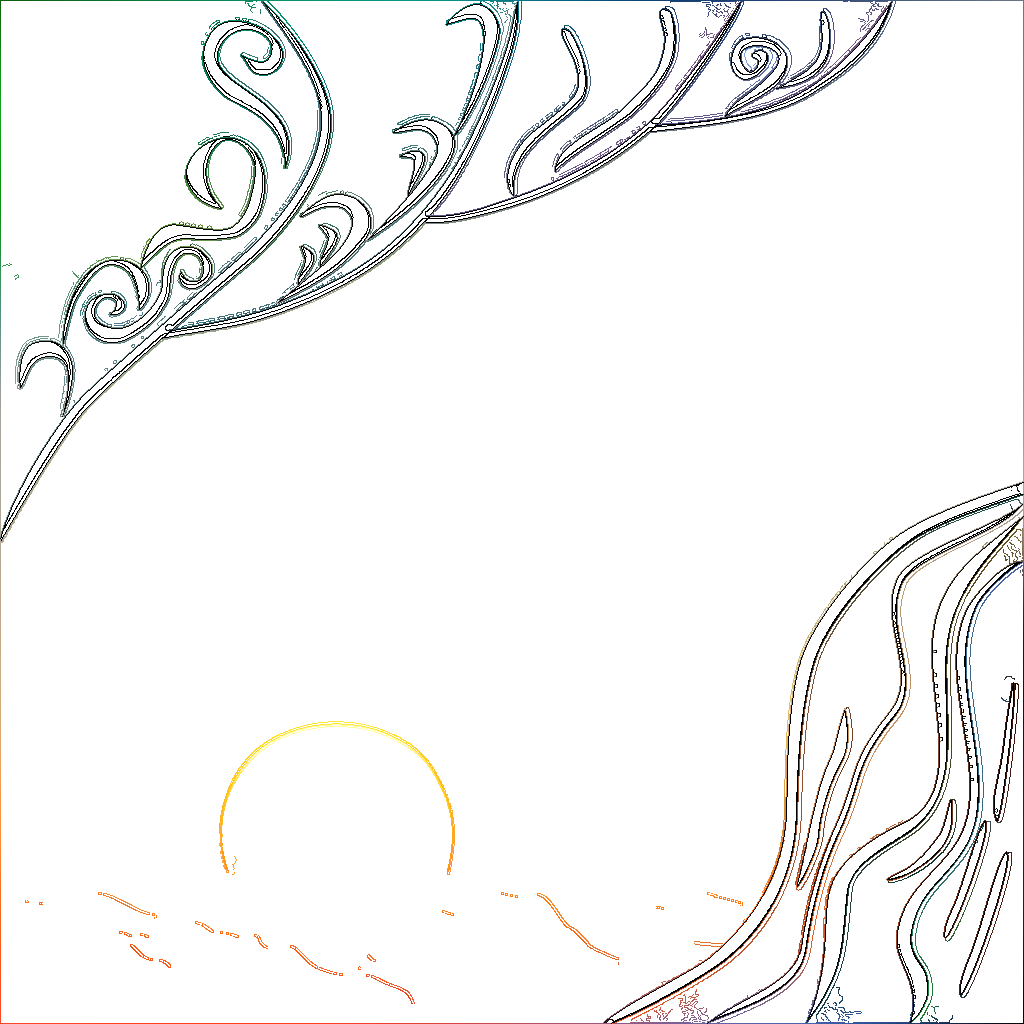}
  {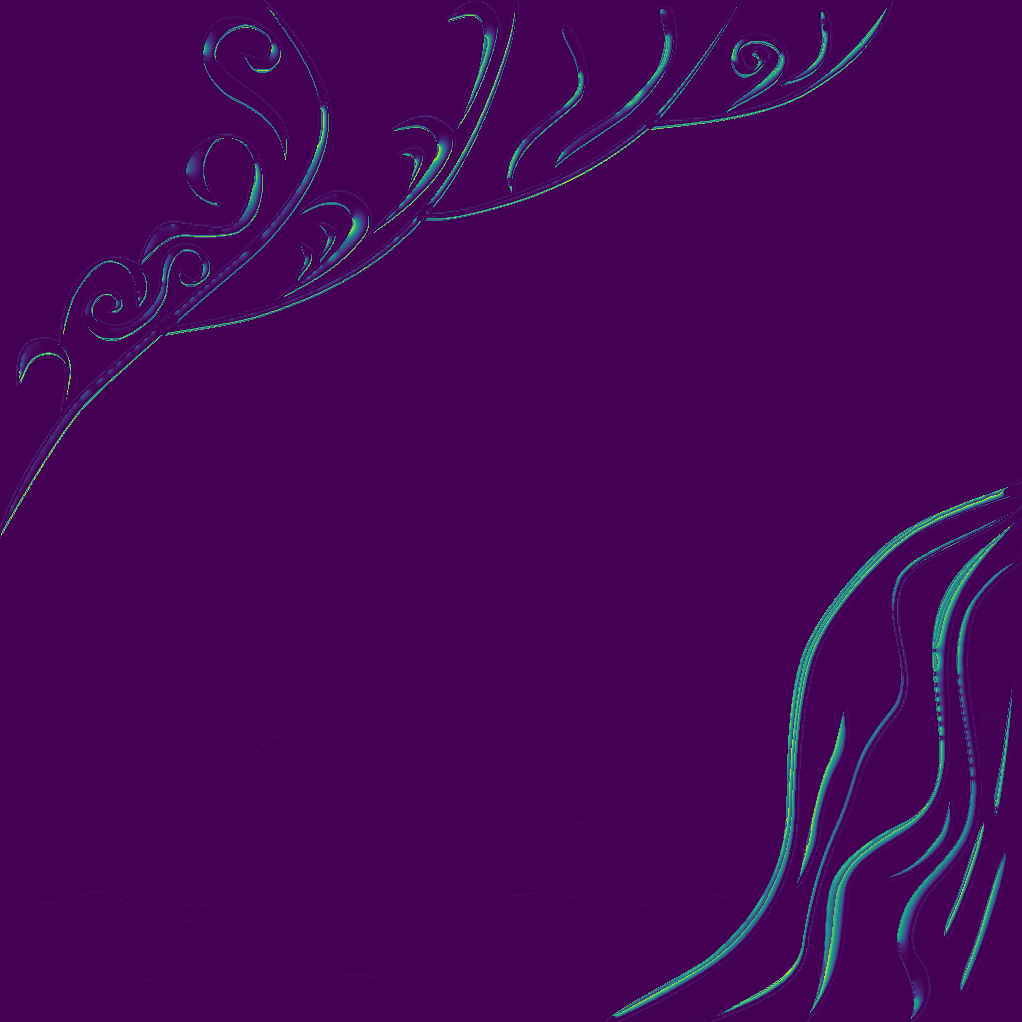}
  {0.033}
  \hfill%
  \compareentry
  {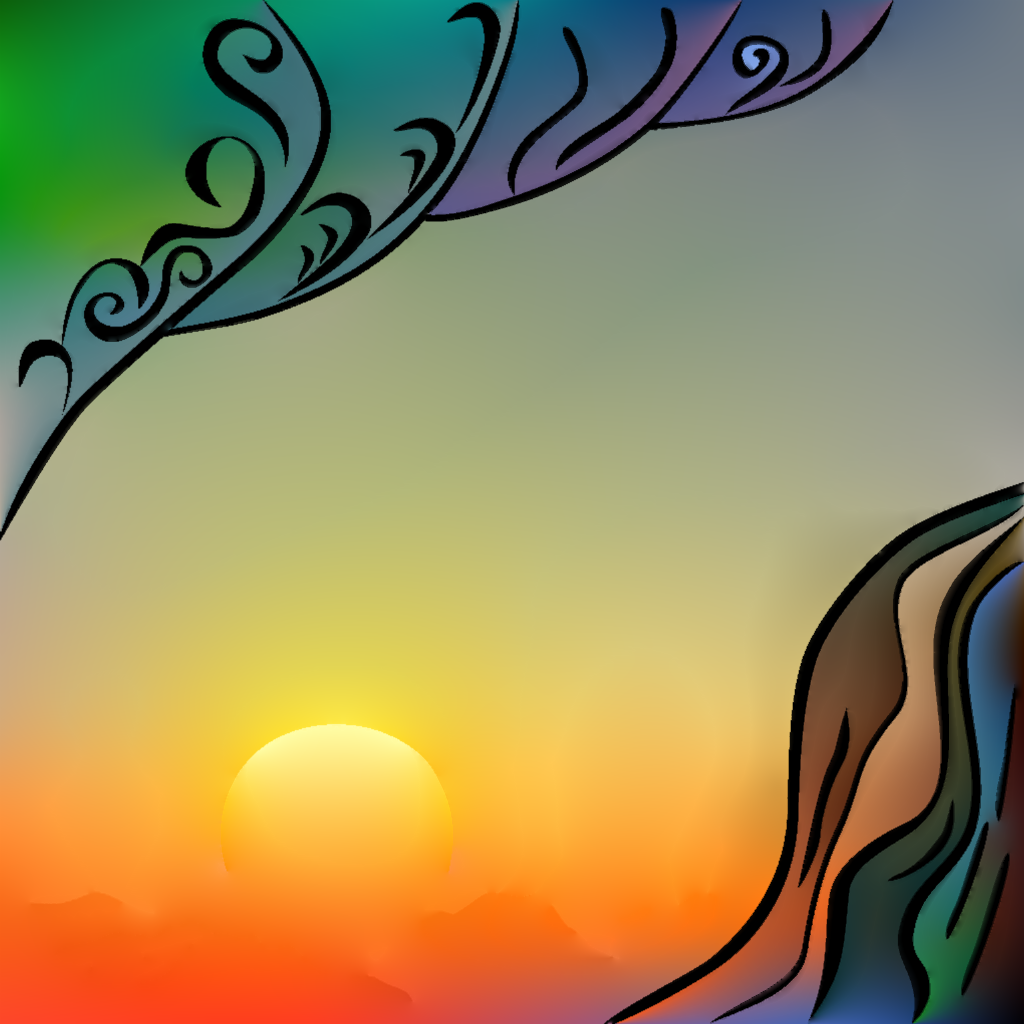}
  {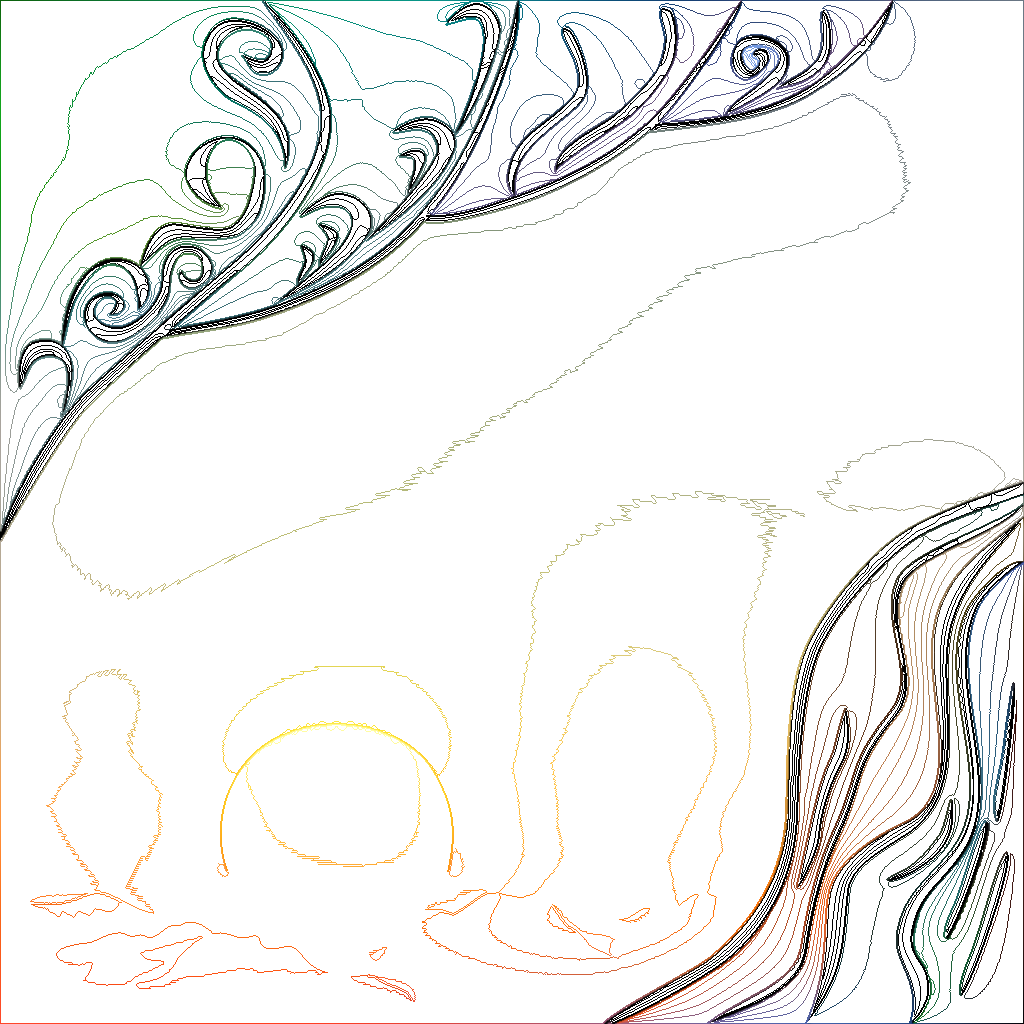}
  {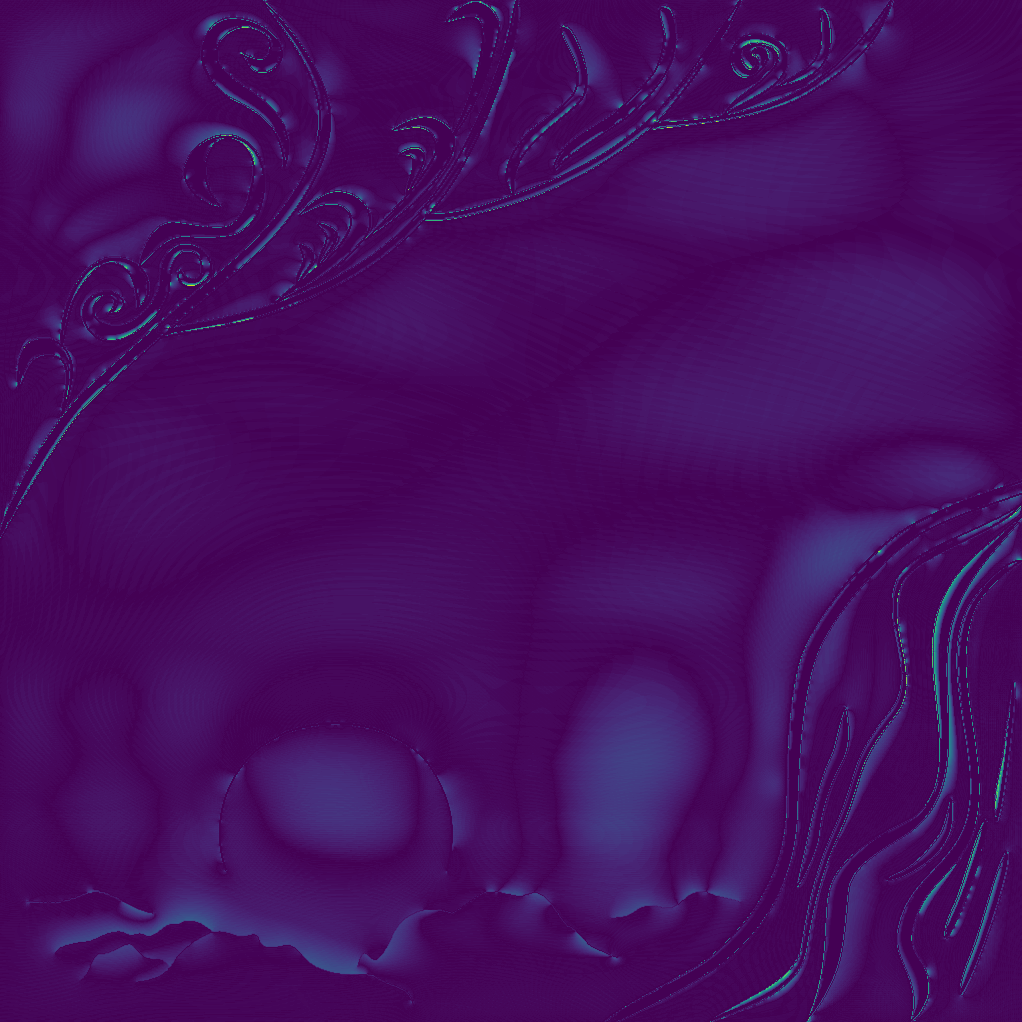}
  {0.013}
  \\%
  \vspace{0.4em}%
  \begin{minipage}{0.16\linewidth}%
      \includegraphics[width=\linewidth]{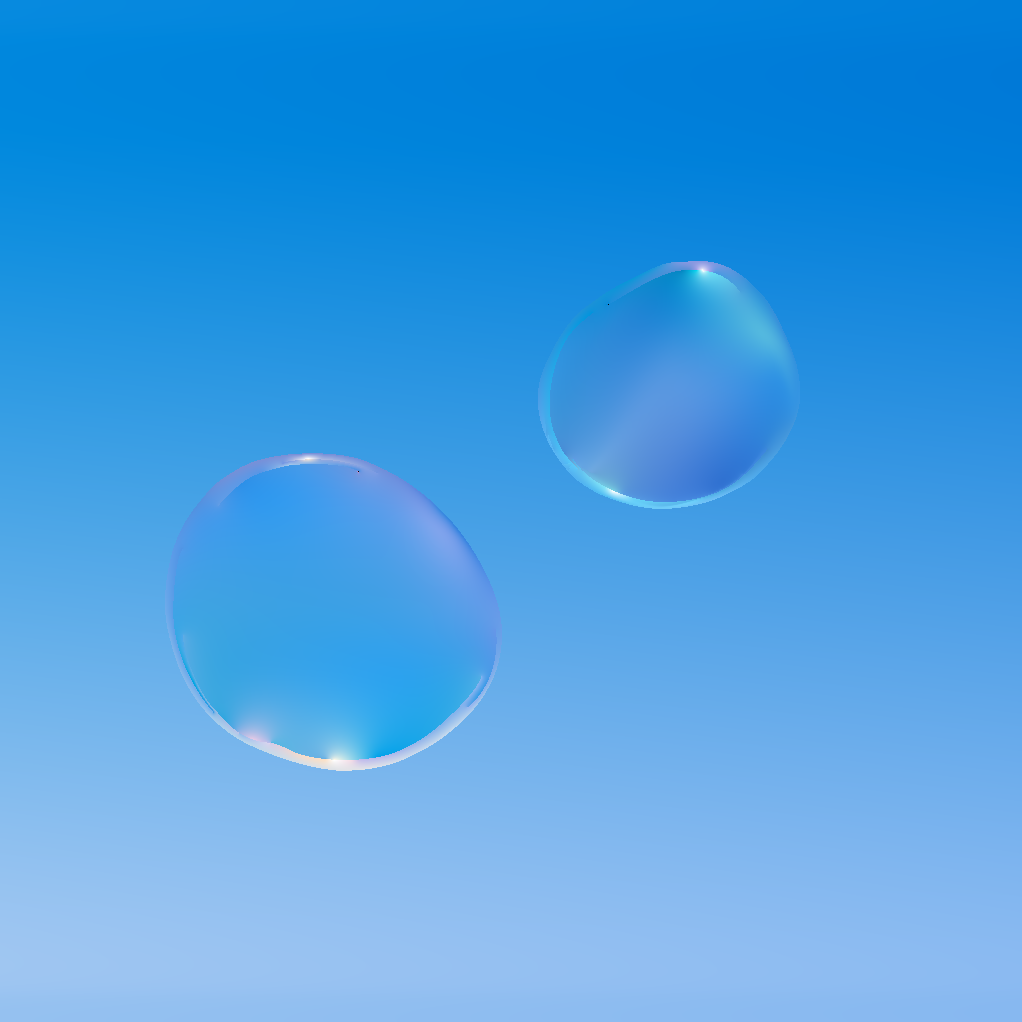}
  \end{minipage}\hfill%
  \compareentry
  {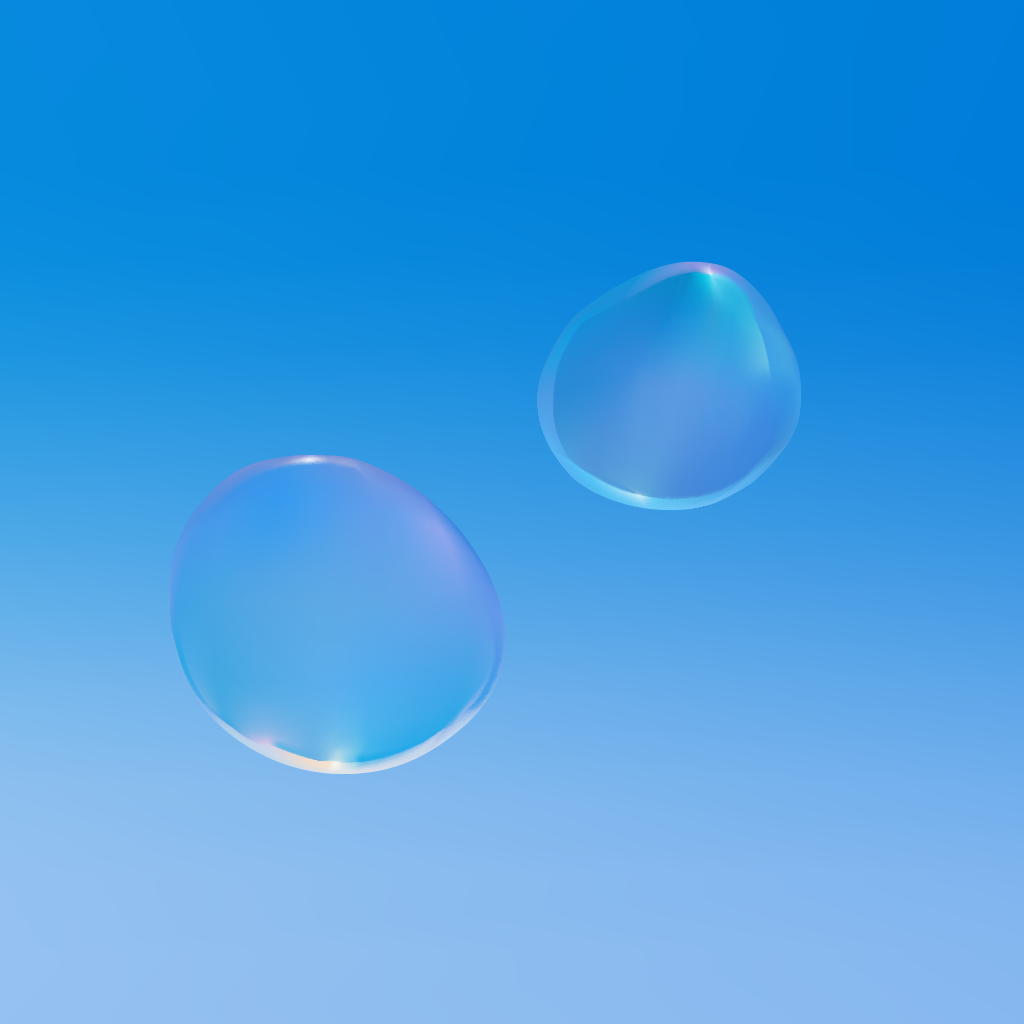}
  {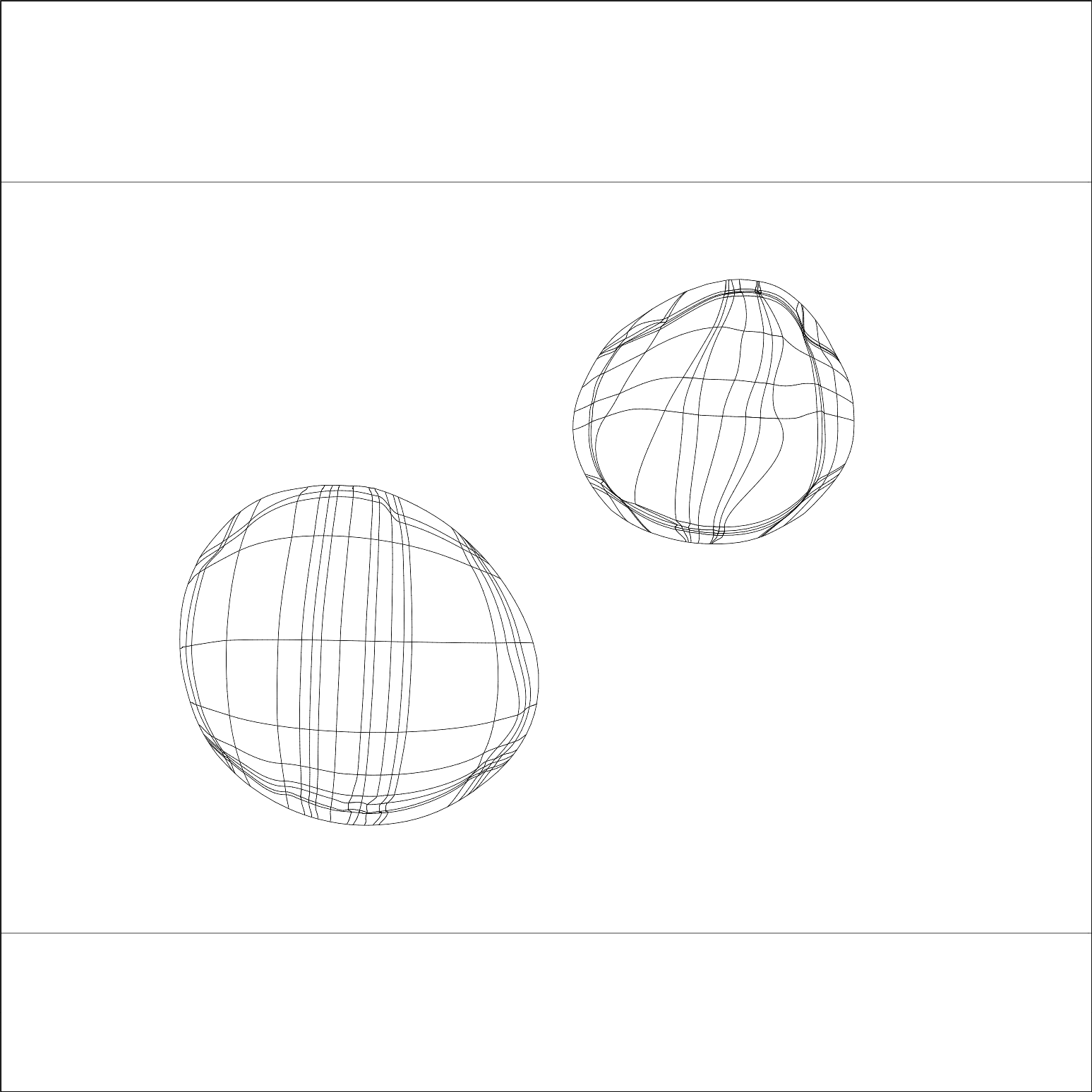}
  {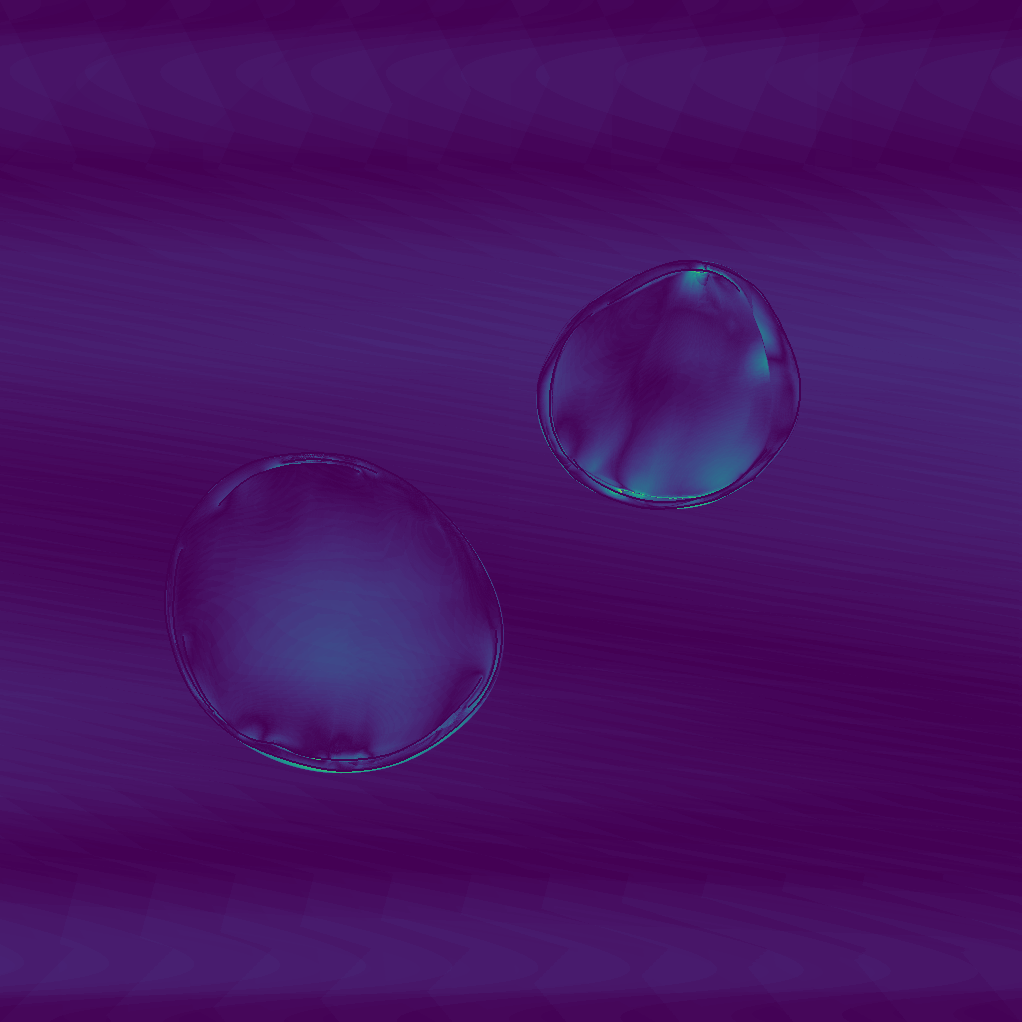}
  {0.022}
  \hfill%
  \compareentry
  {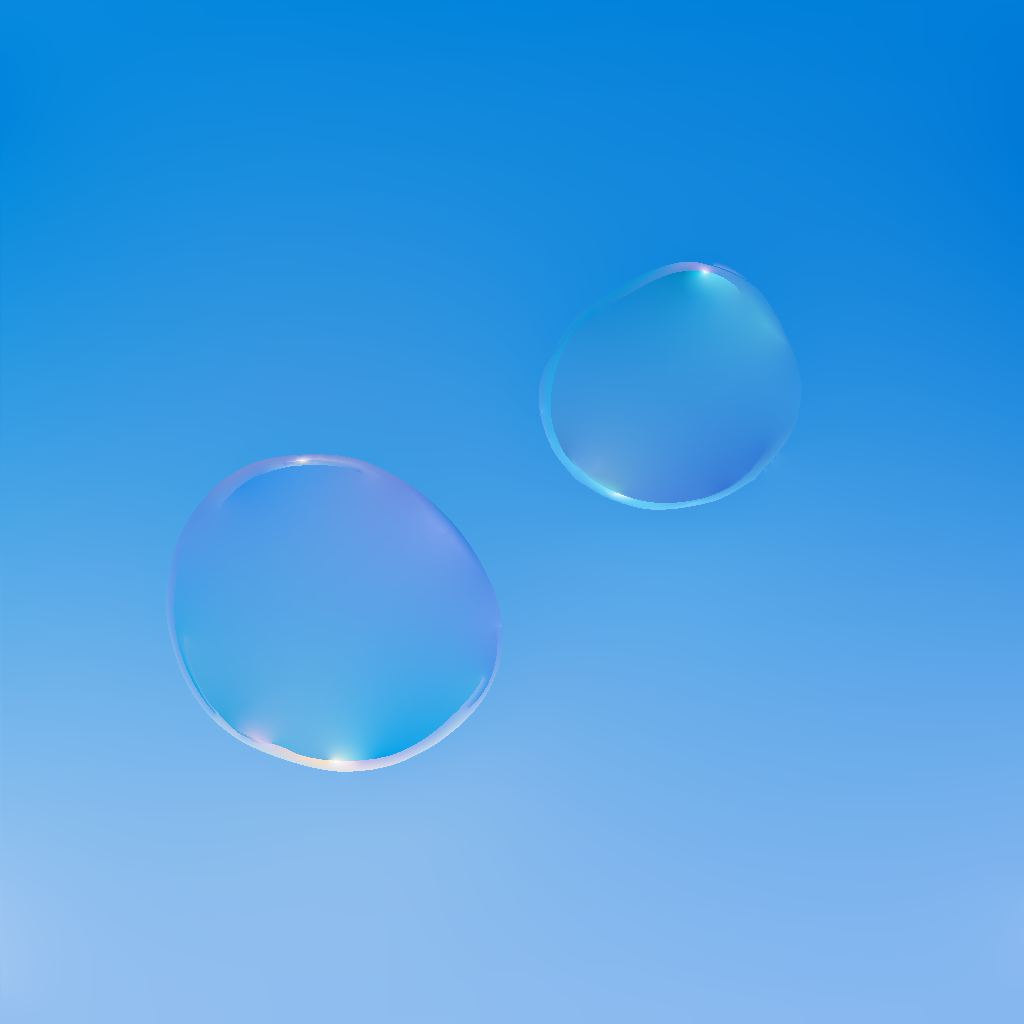}
  {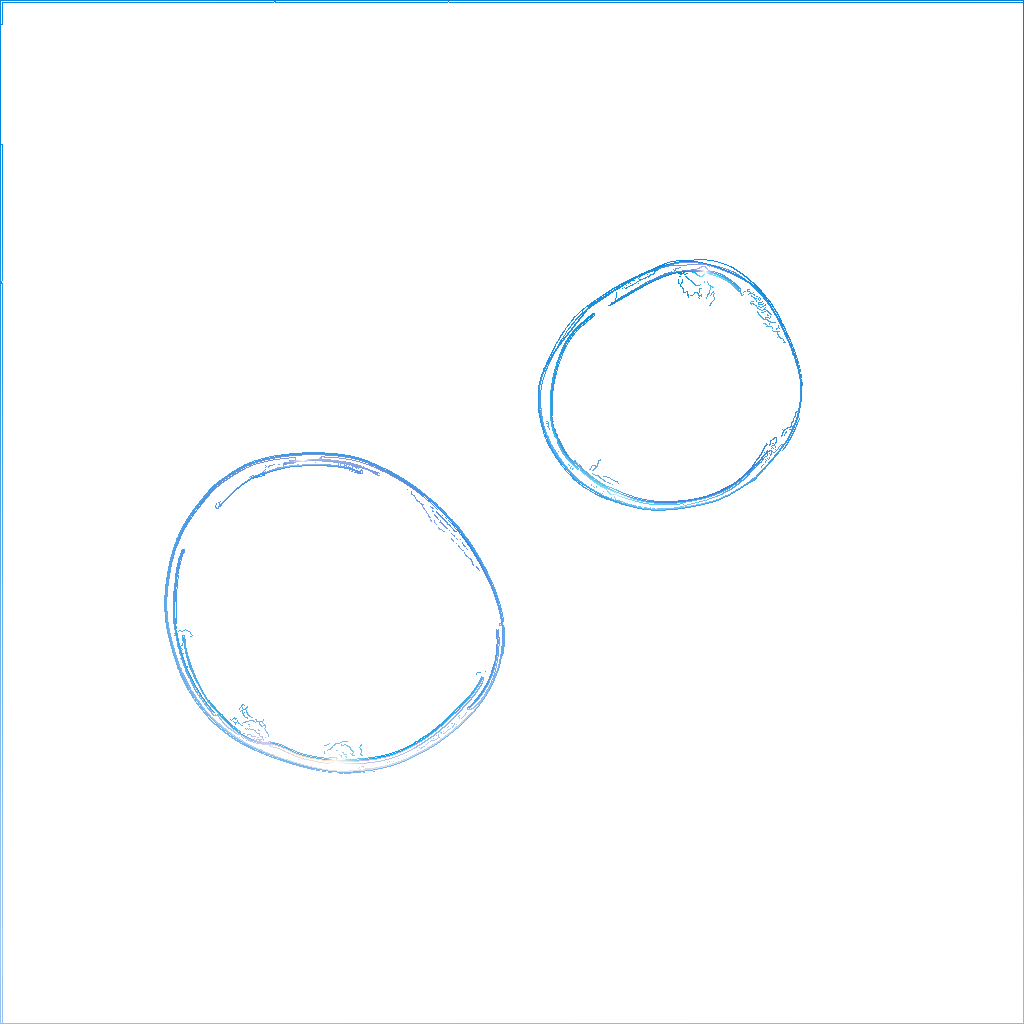}
  {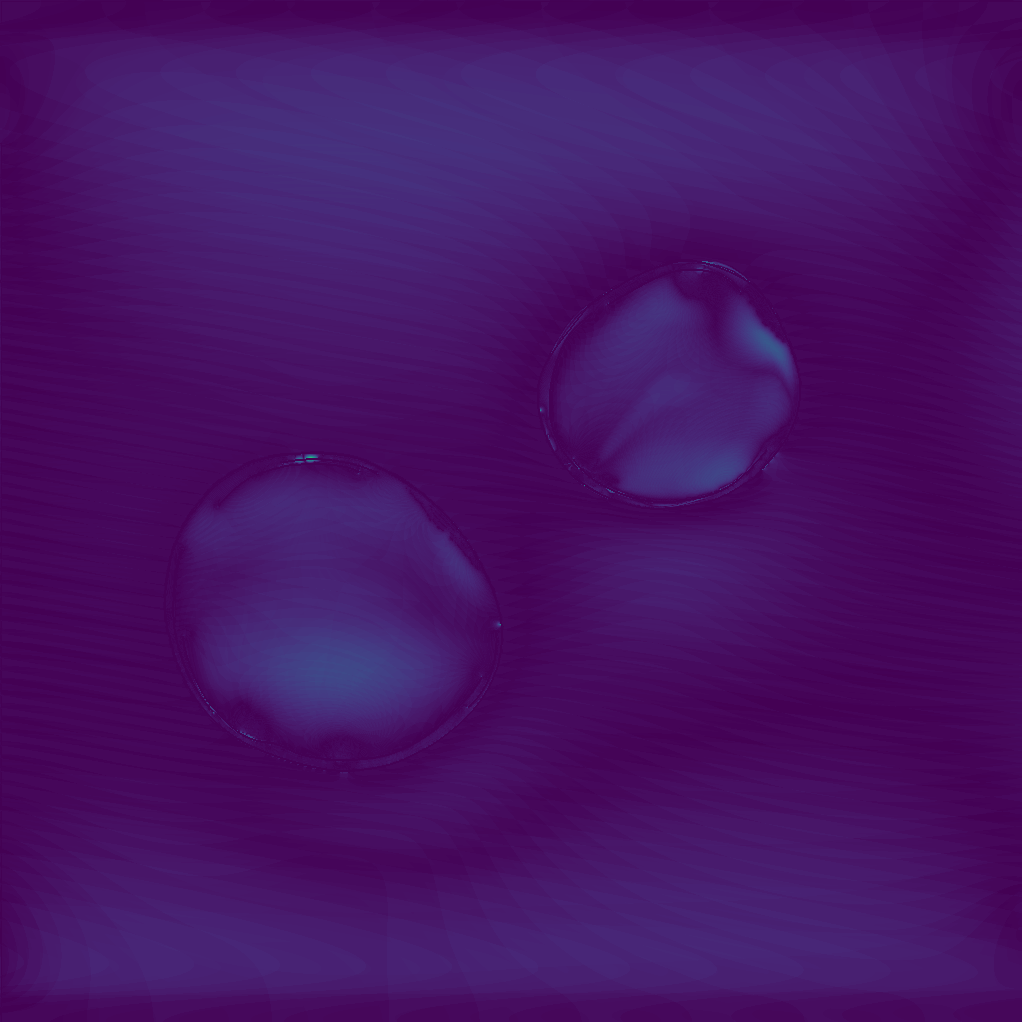}
  {0.024}
  \hfill%
  \compareentry
  {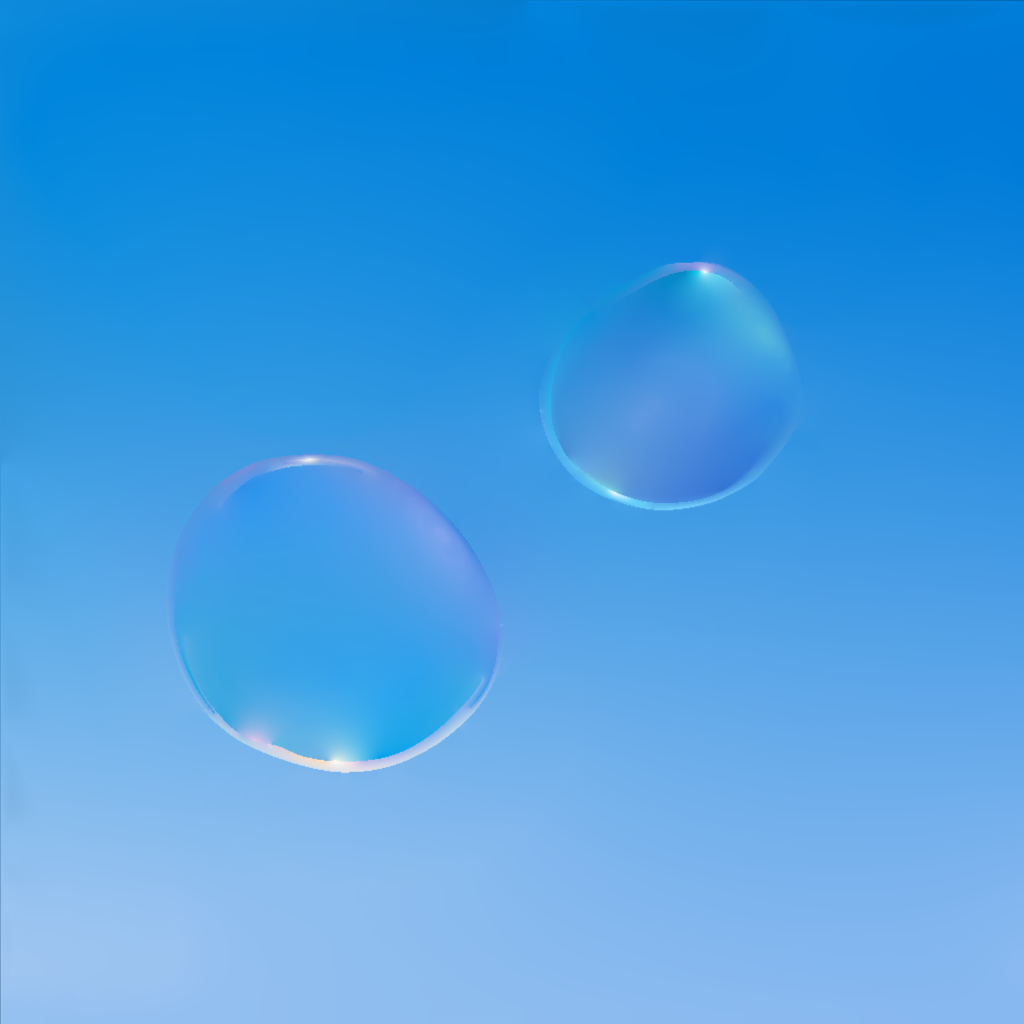}
  {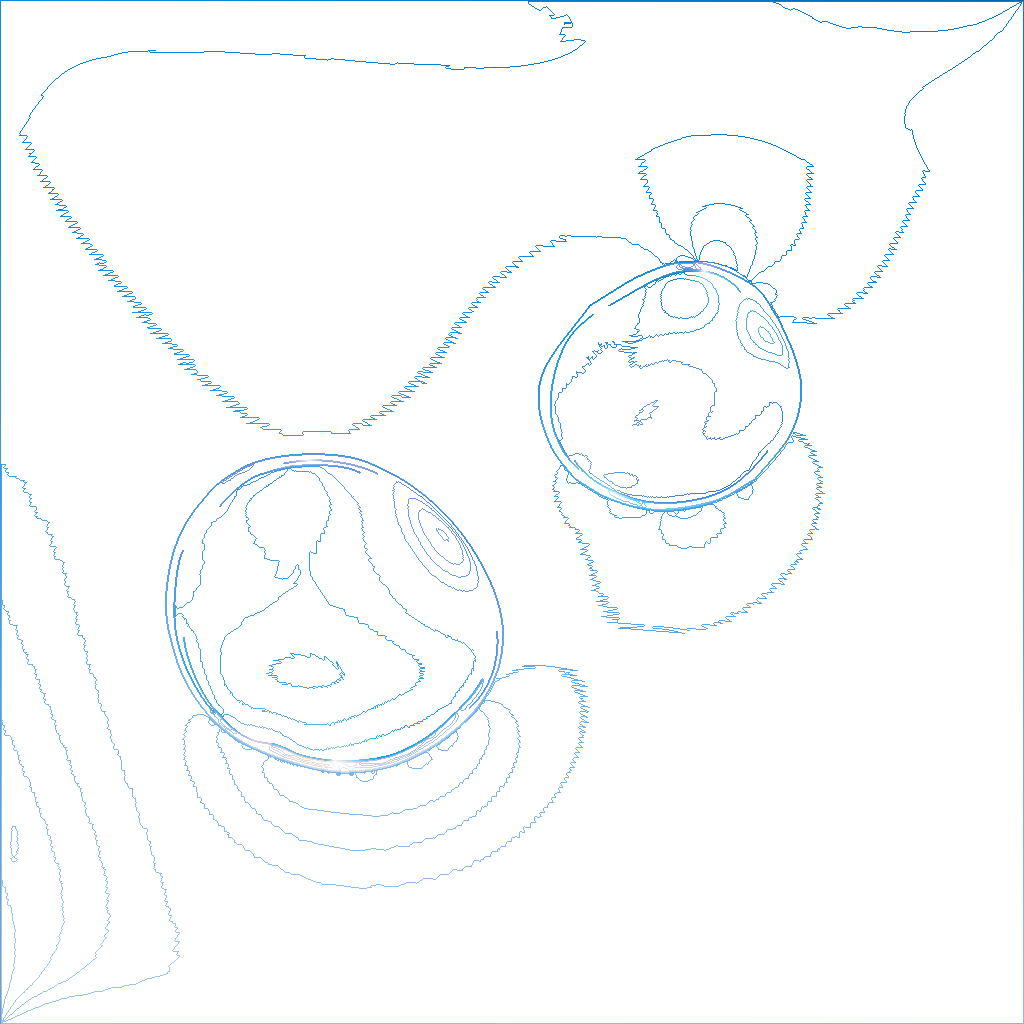}
  {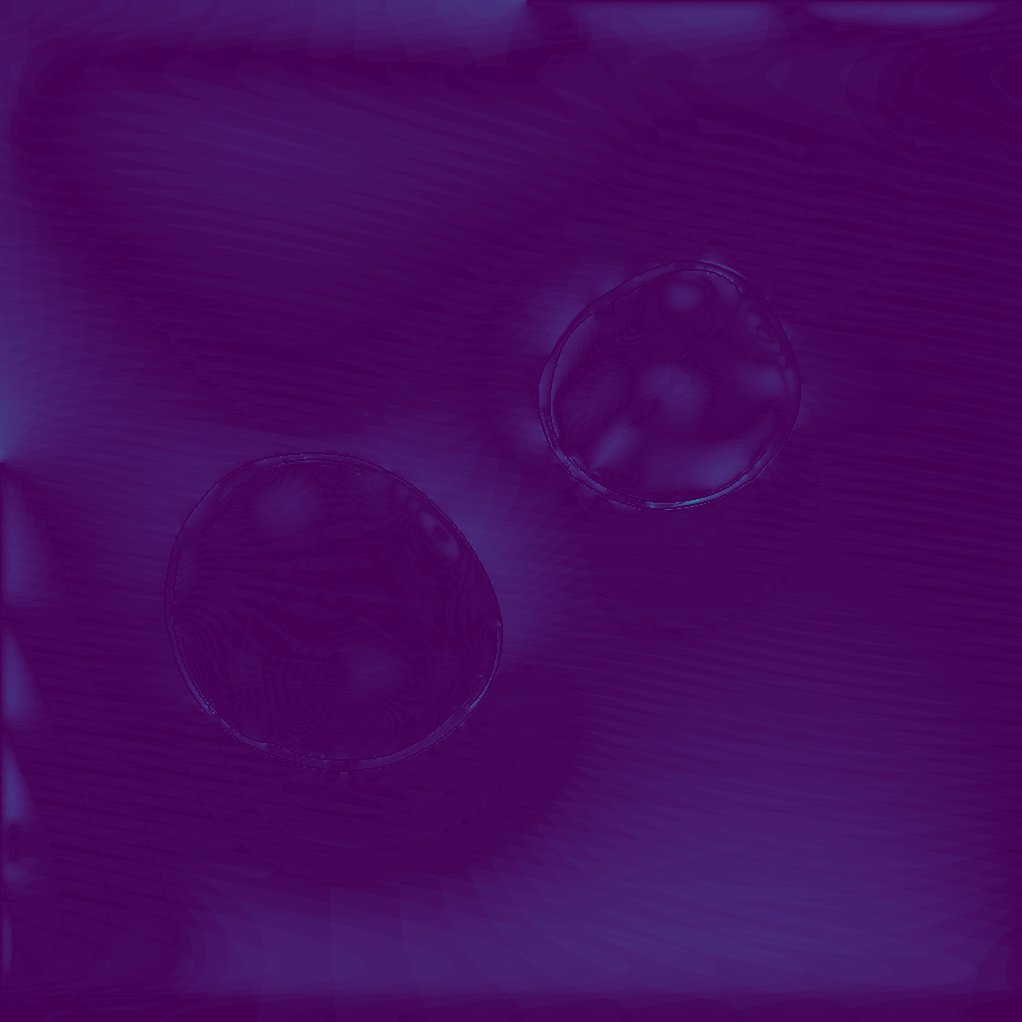}
  {0.011}
  \\%
  \vspace{0.4em}%
  \begin{minipage}{0.16\linewidth}%
      \includegraphics[width=\linewidth]{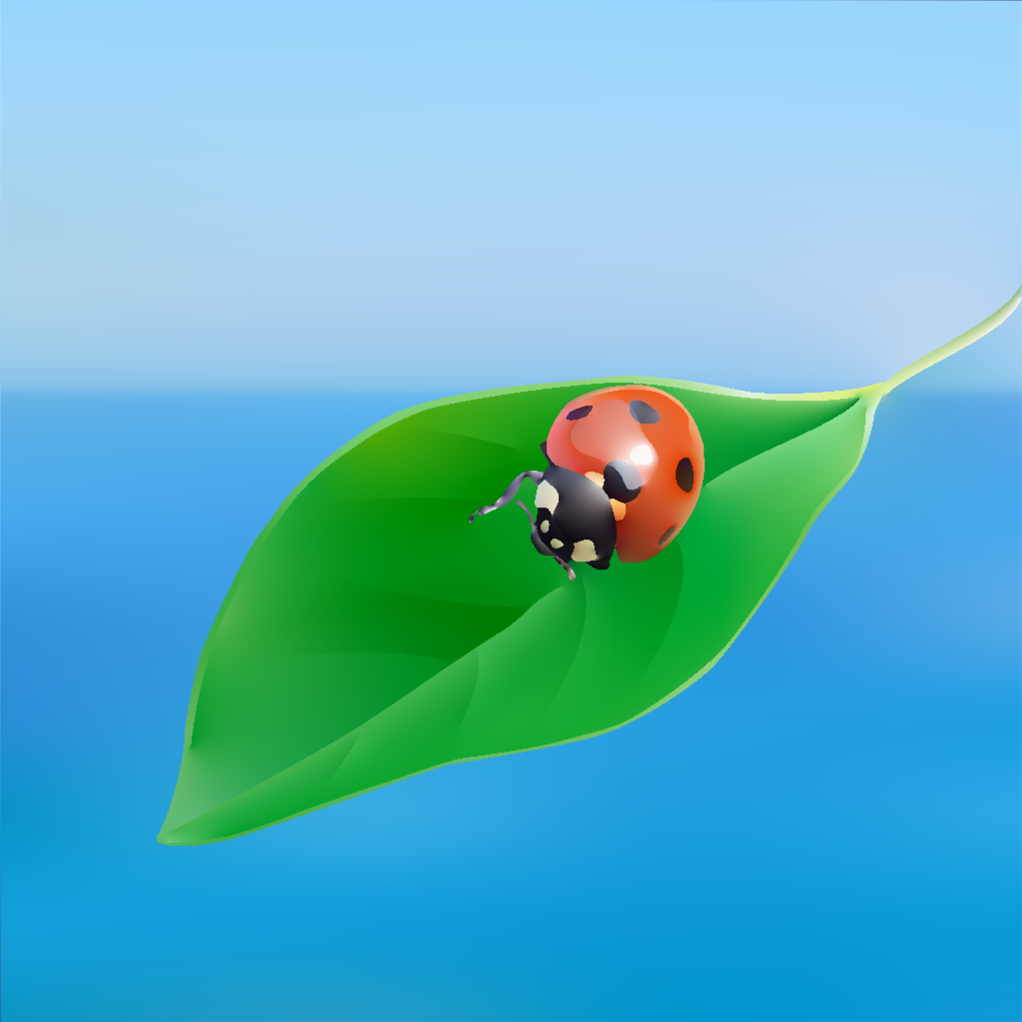}
  \end{minipage}\hfill%
  \compareentry
  {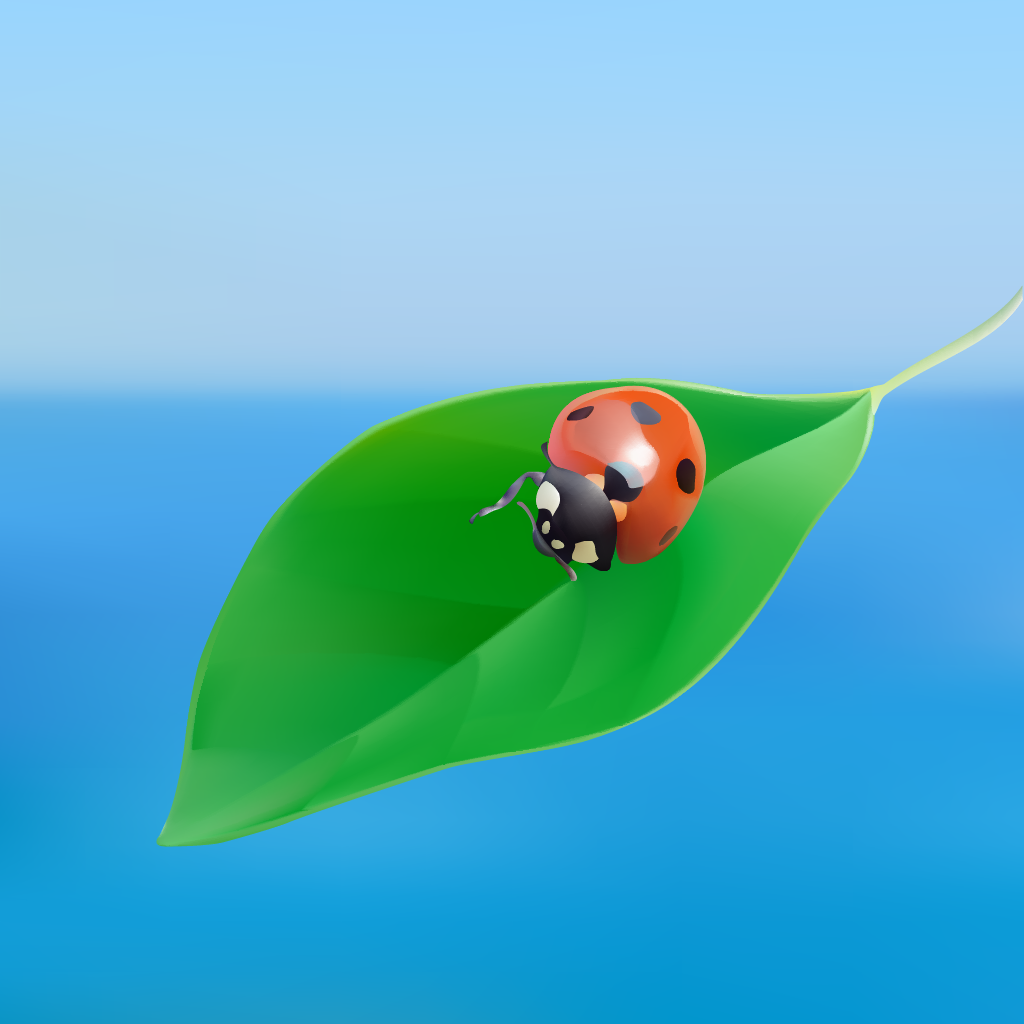}
  {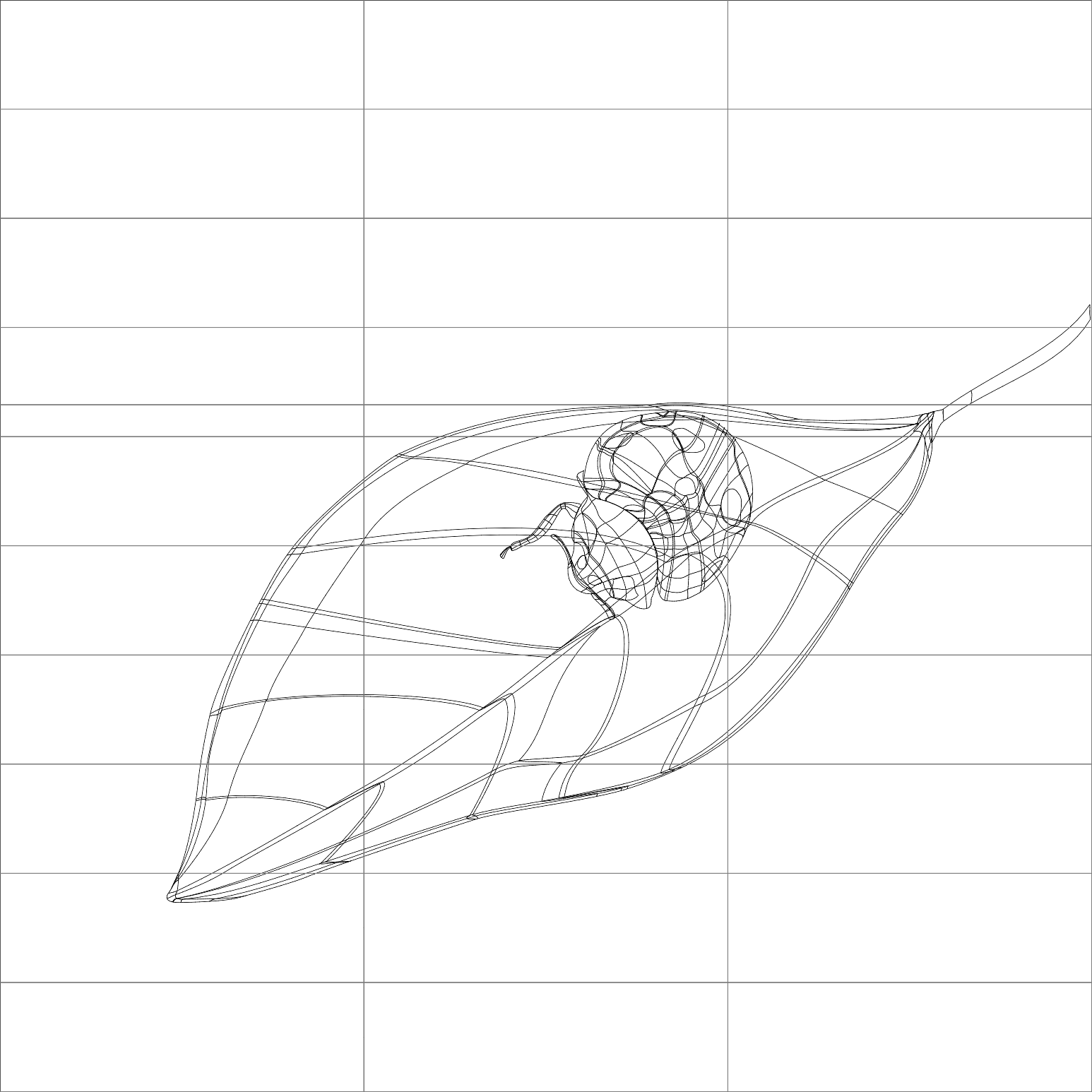}
  {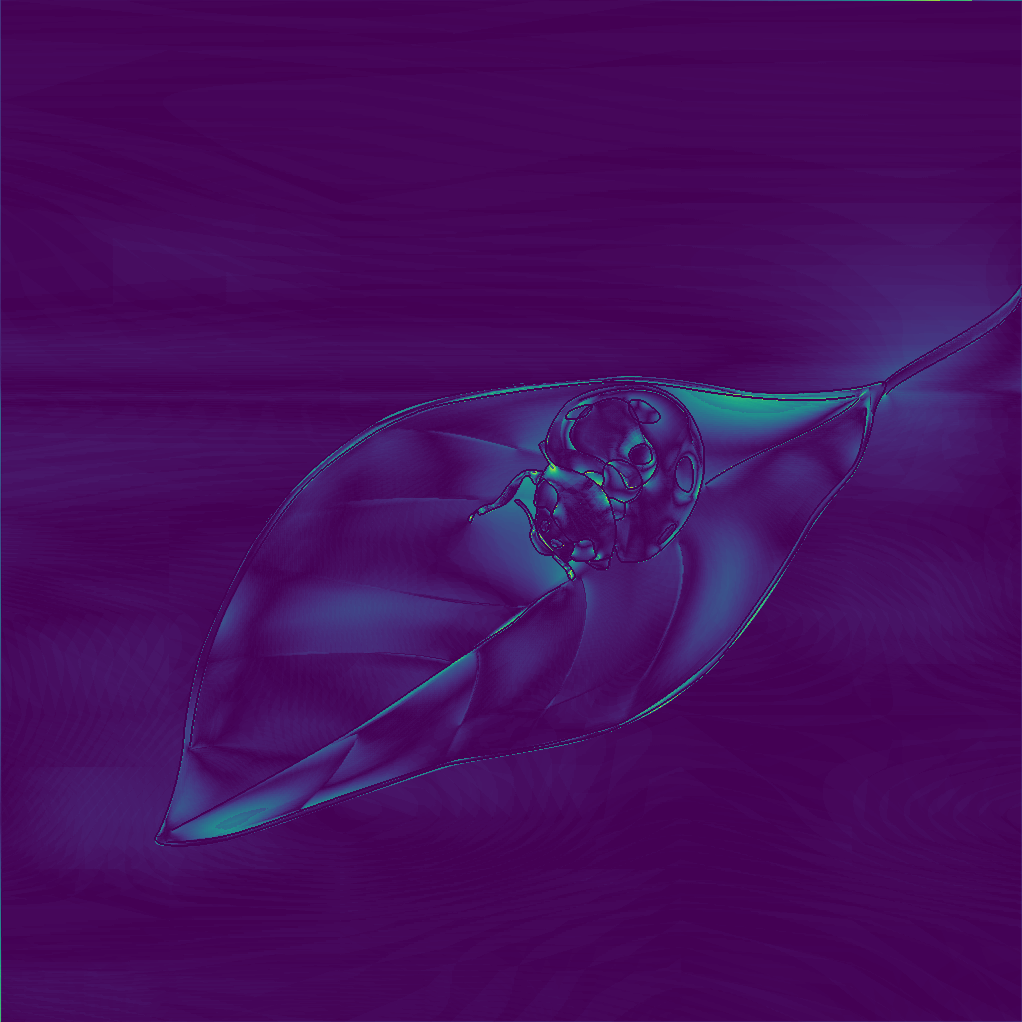}
  {0.021}
  \hfill%
  \compareentry
  {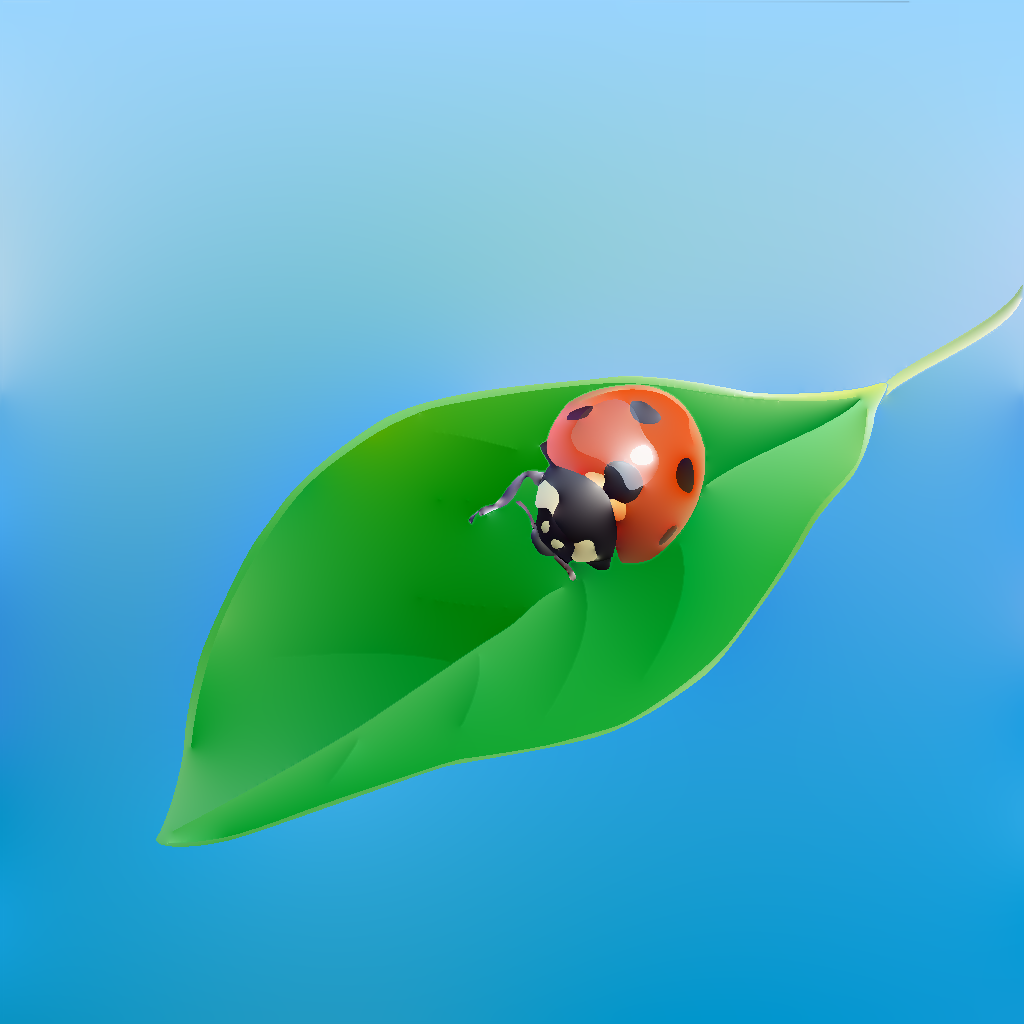}
  {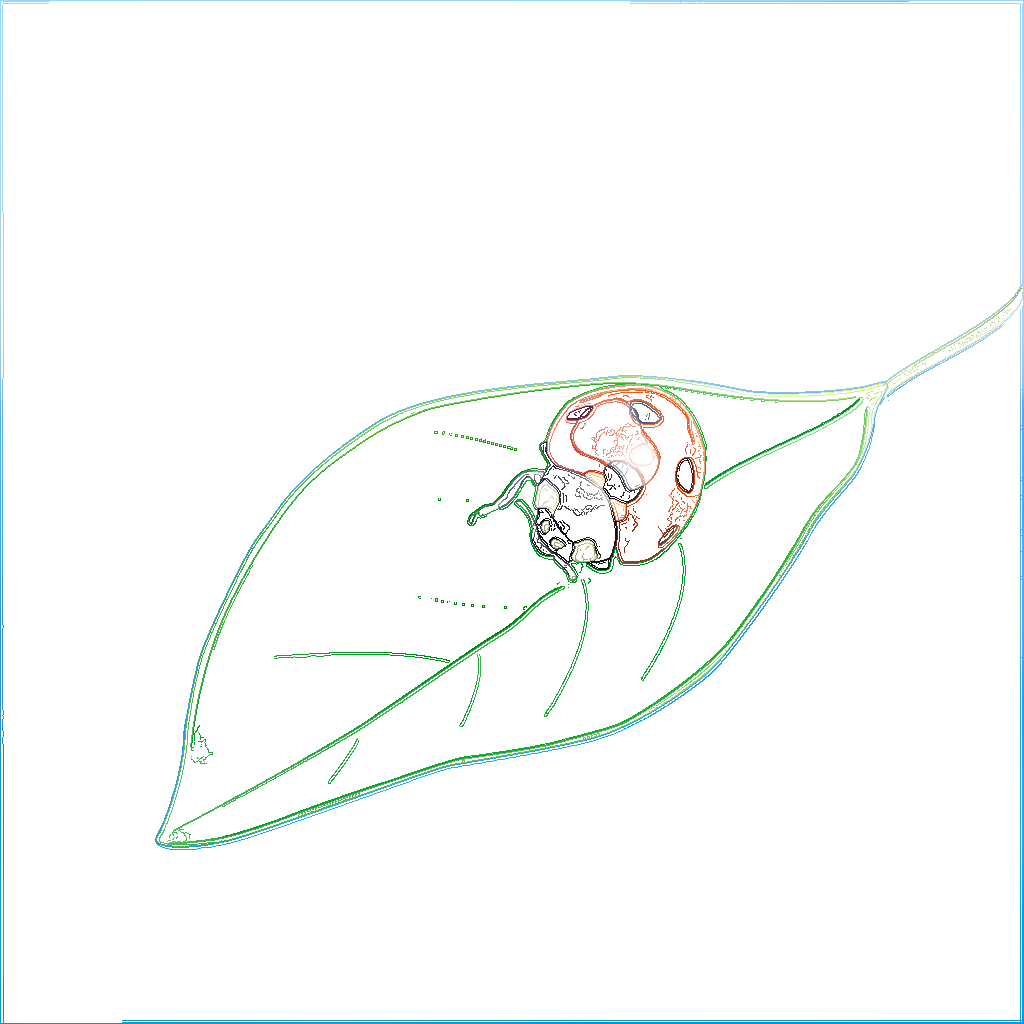}
  {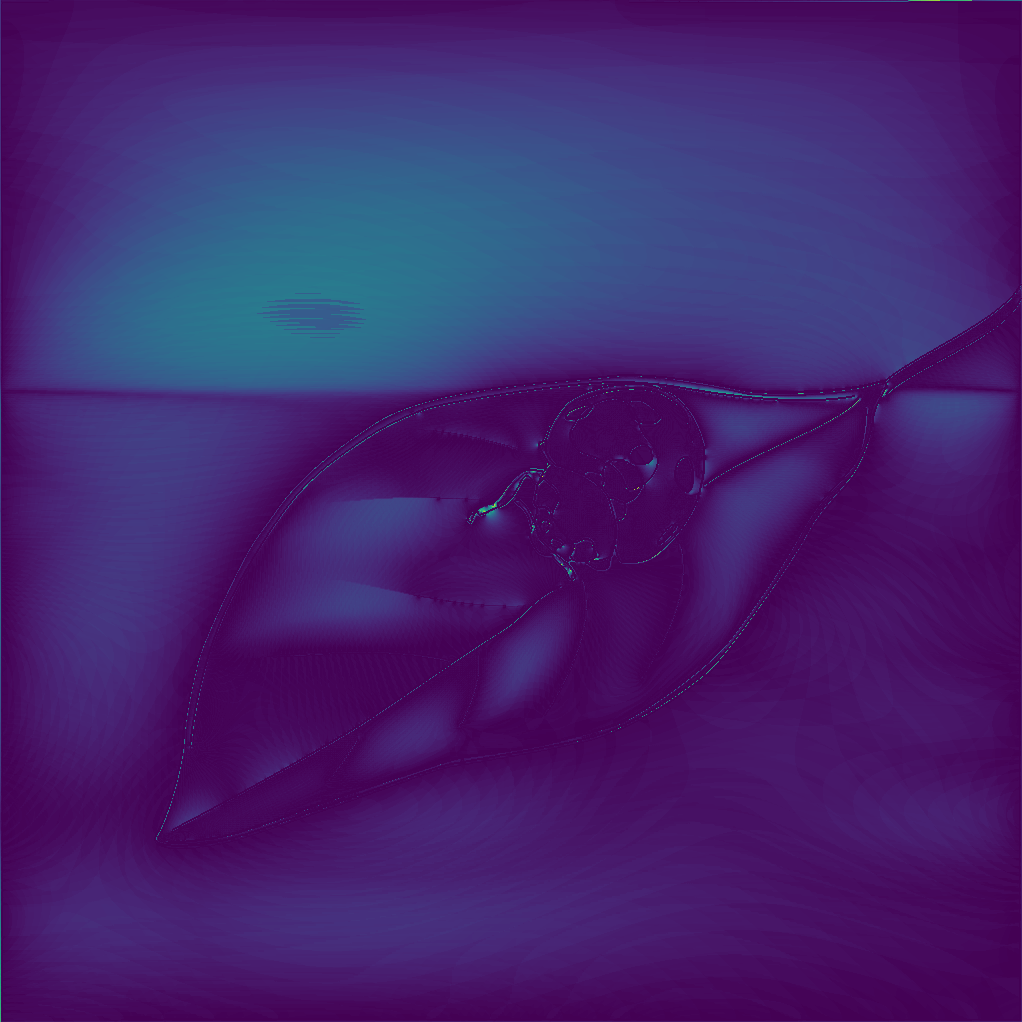}
  {0.043}
  \hfill%
  \compareentry
  {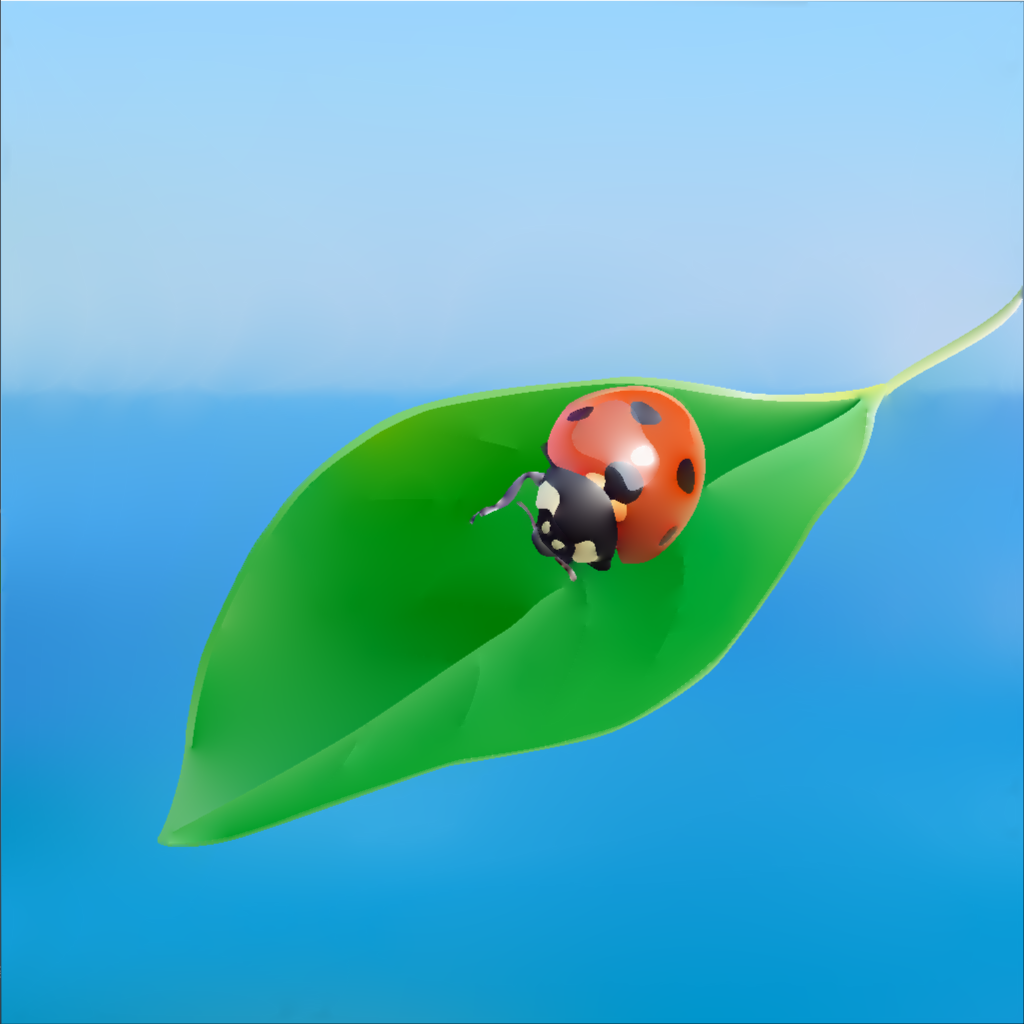}
  {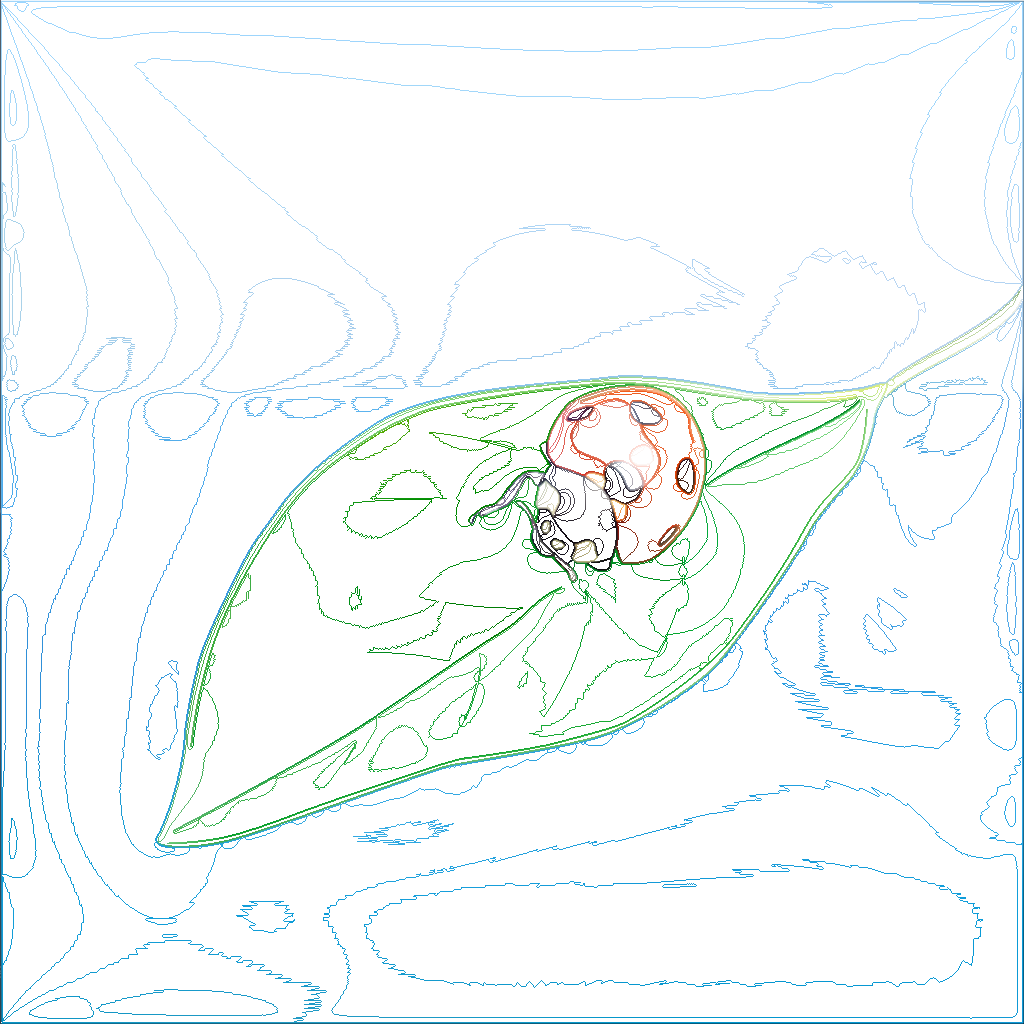}
  {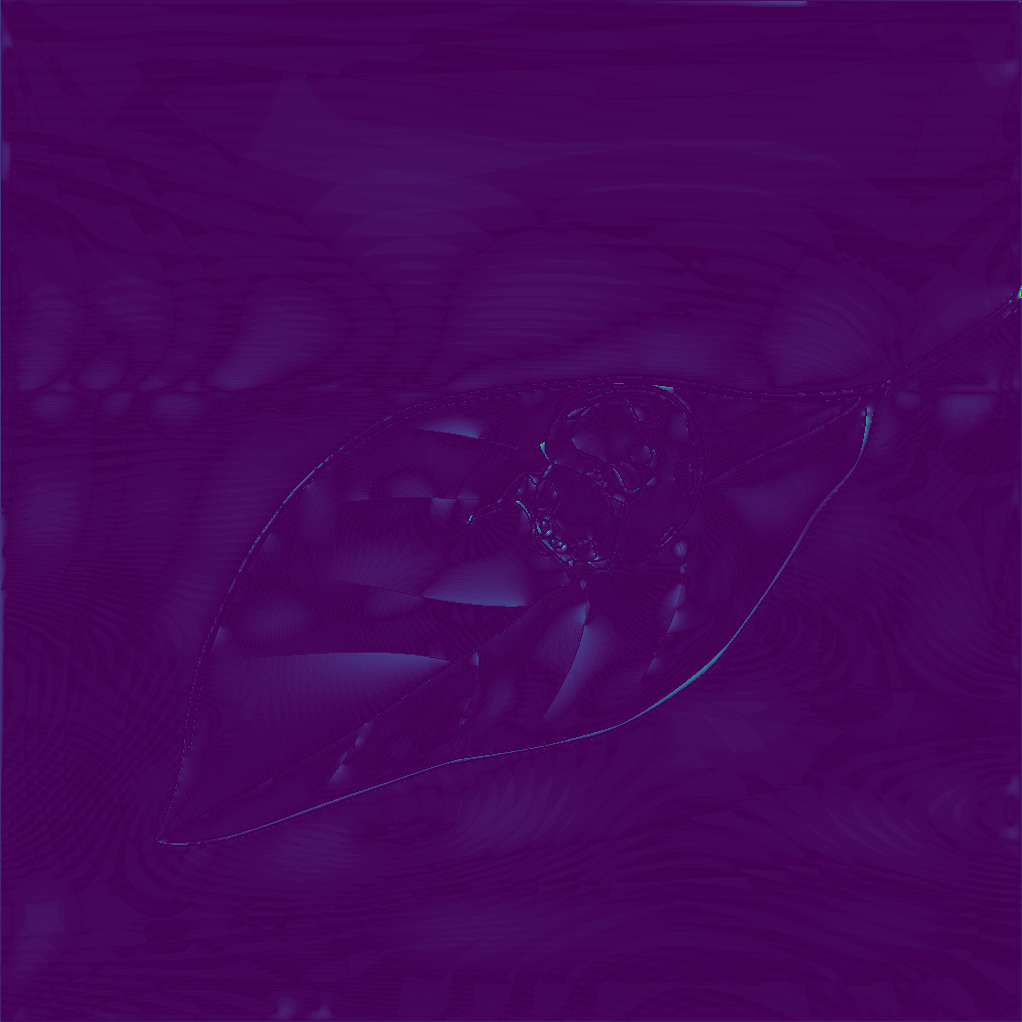}
  {0.007}
  \\%
  \vspace{0.4em}%
  \begin{minipage}{0.16\linewidth}%
      \includegraphics[width=\linewidth]{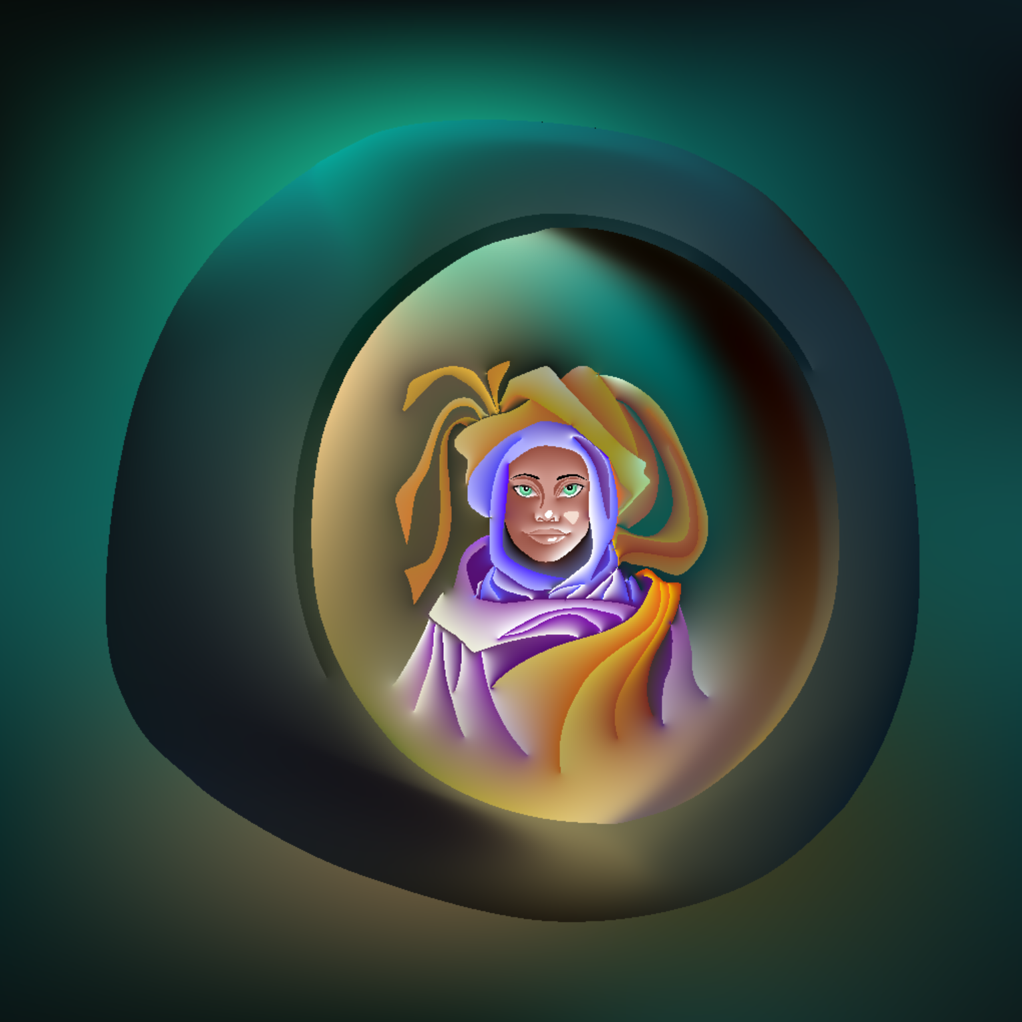}
  \end{minipage}\hfill%
  \compareentry
  {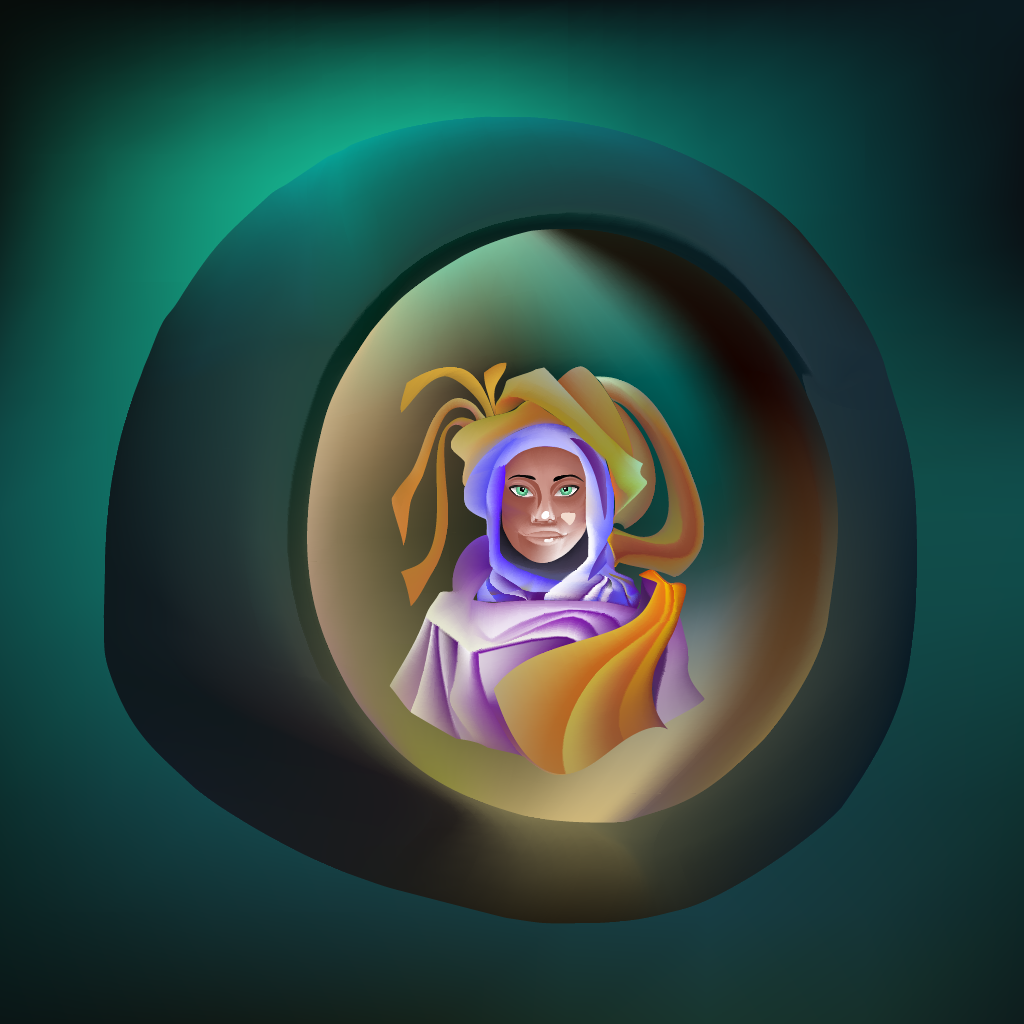}
  {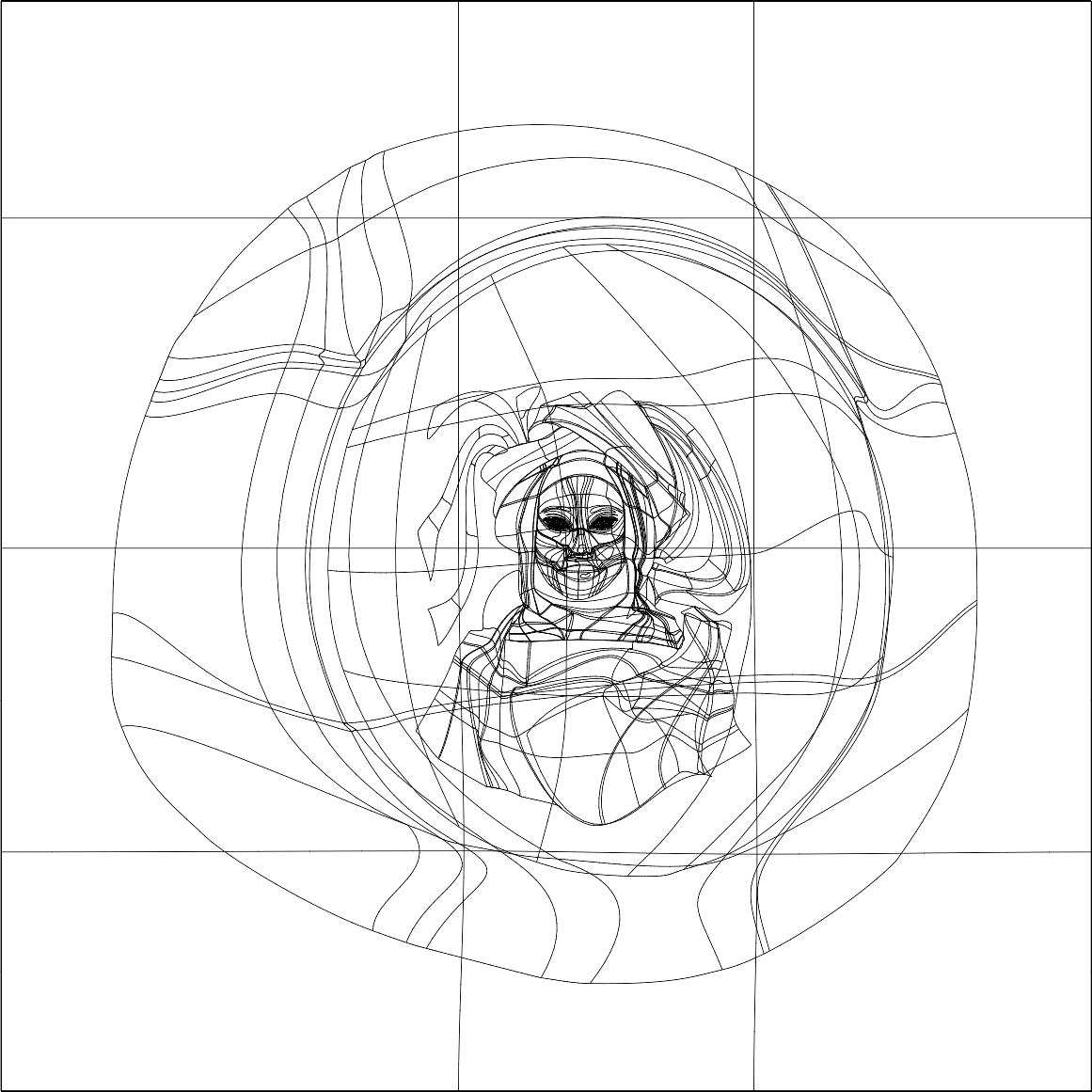}
  {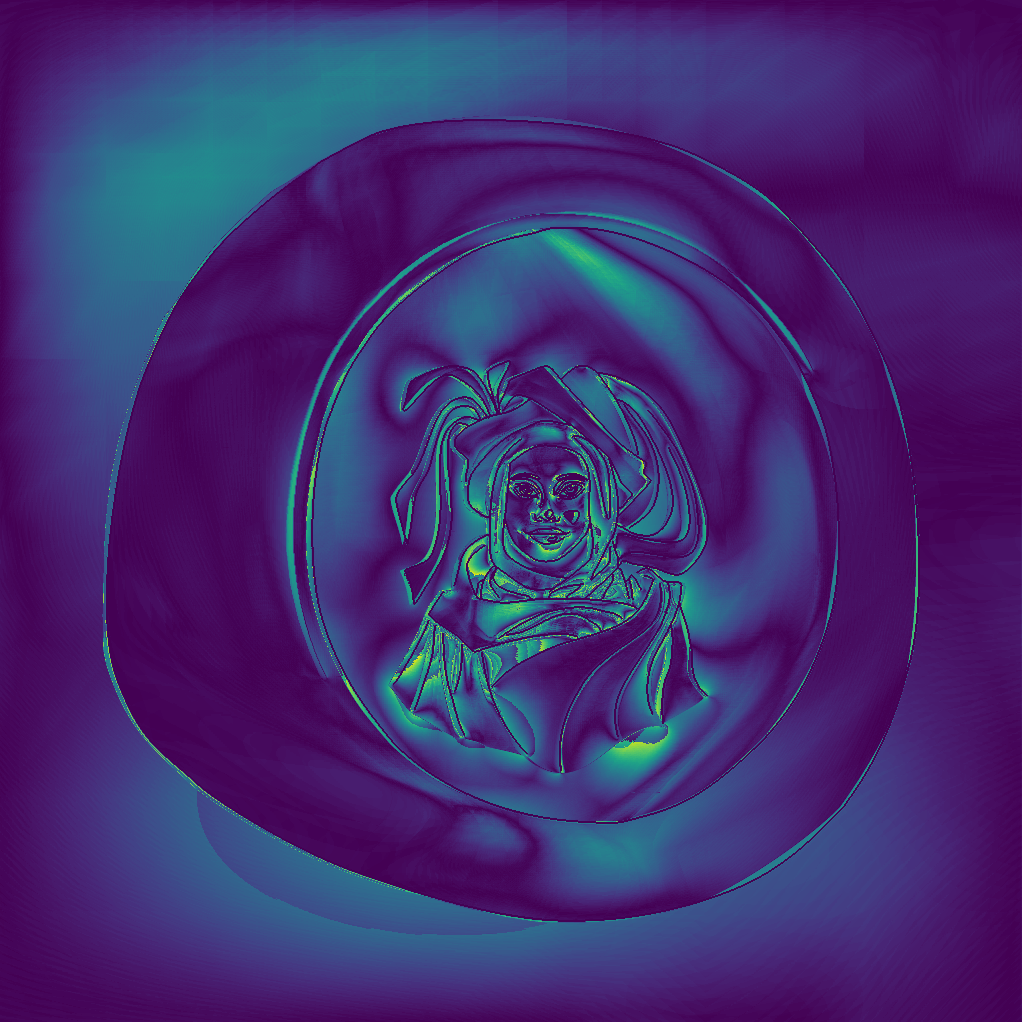}
  {0.064}
  \hfill%
  \compareentry
  {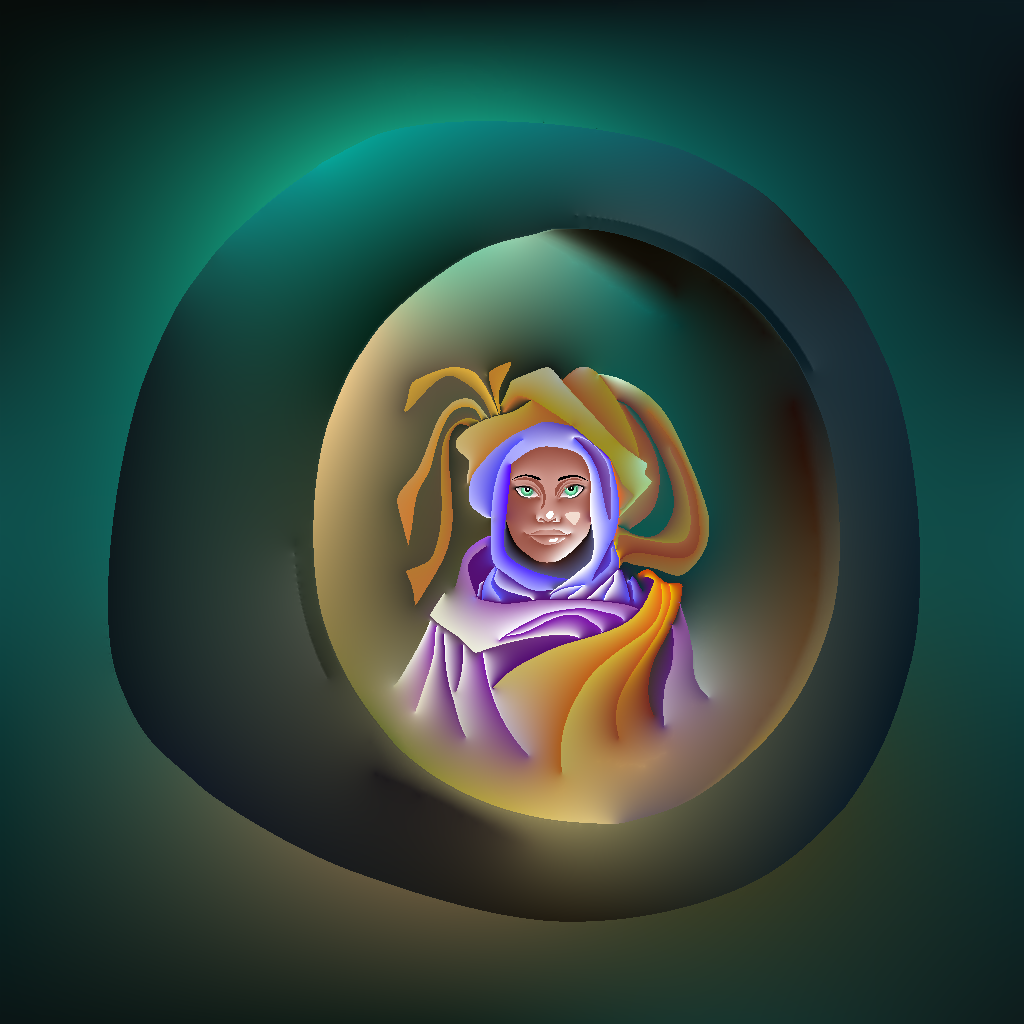}
  {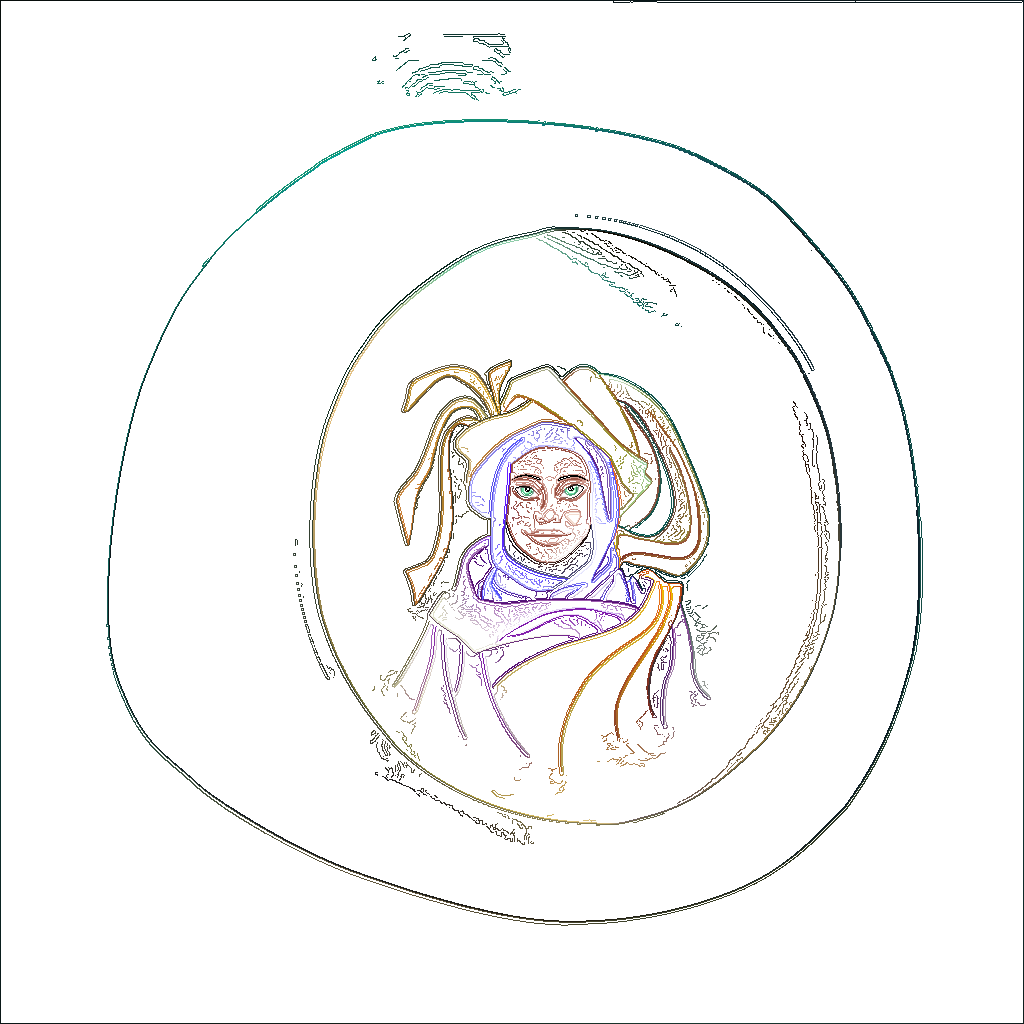}
  {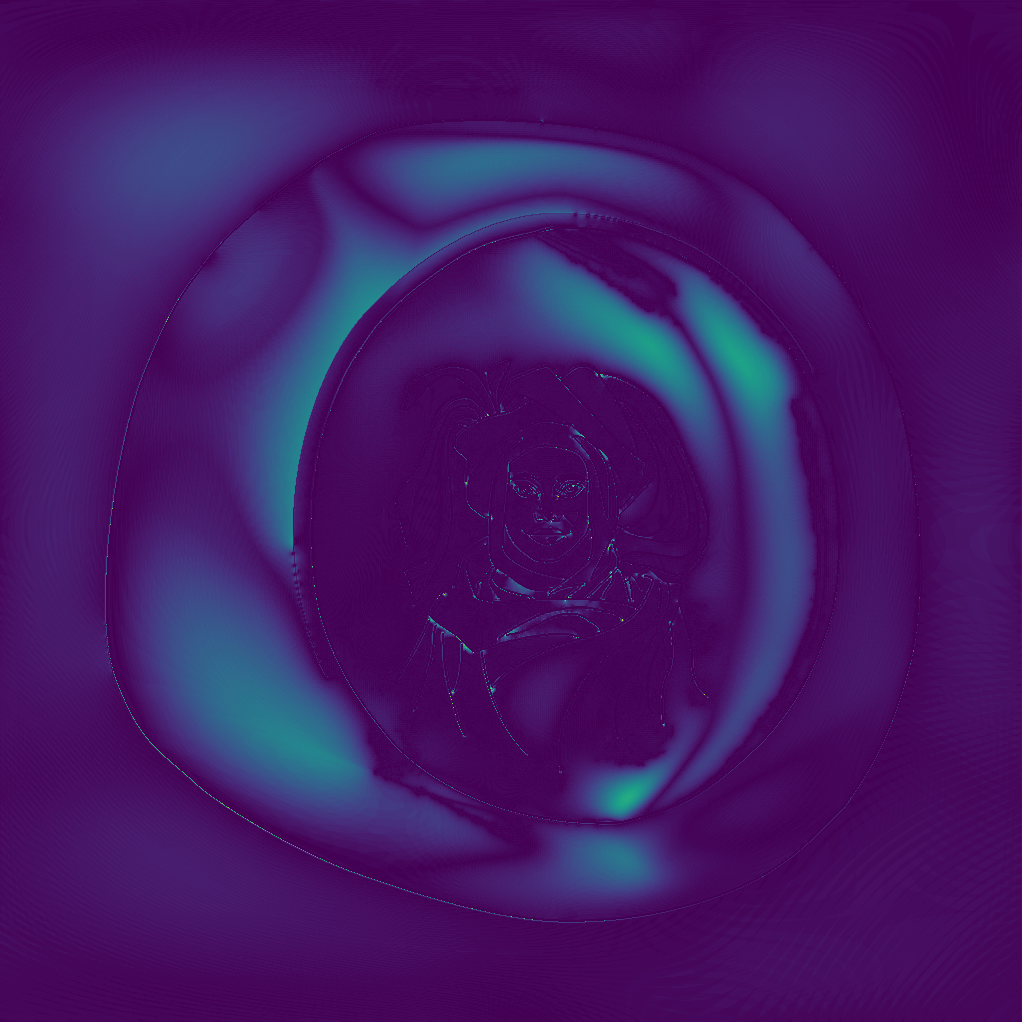}
  {0.022}
  \hfill%
  \compareentry
  {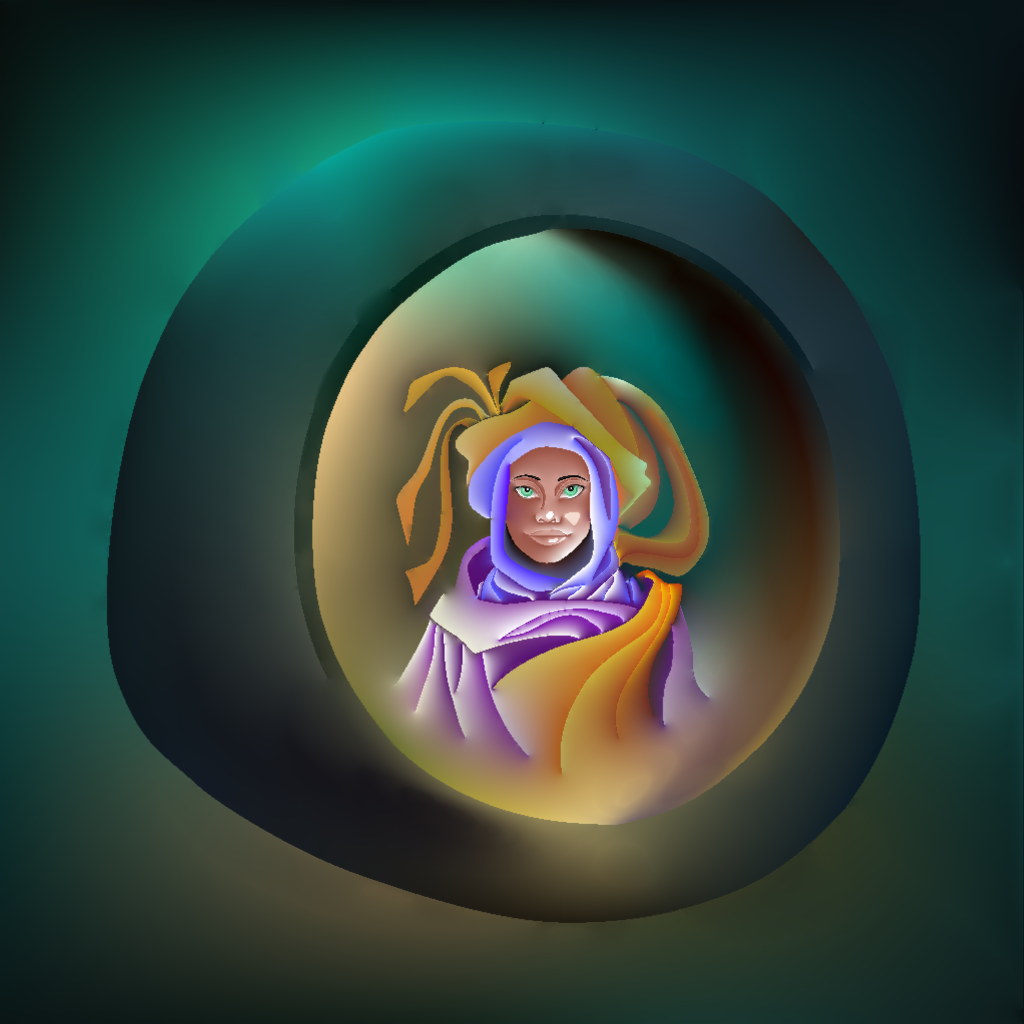}
  {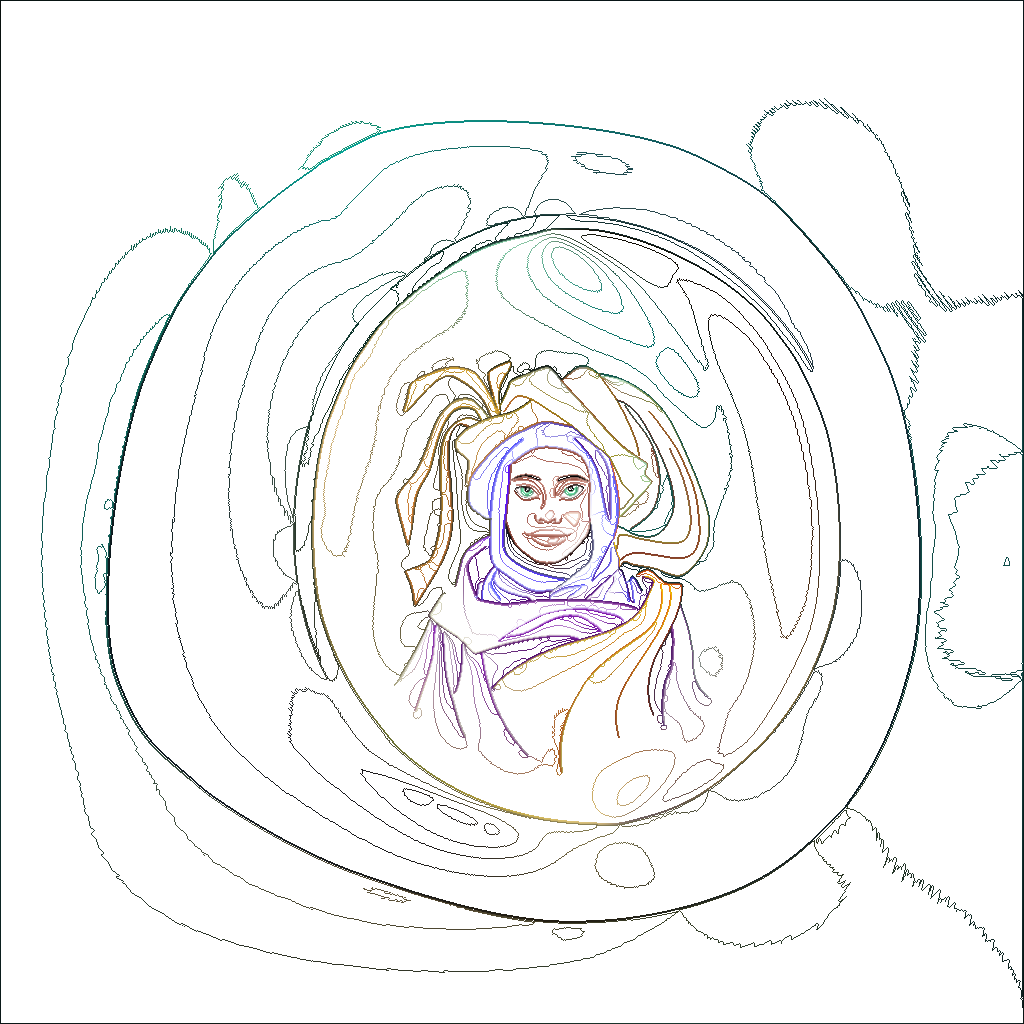}
  {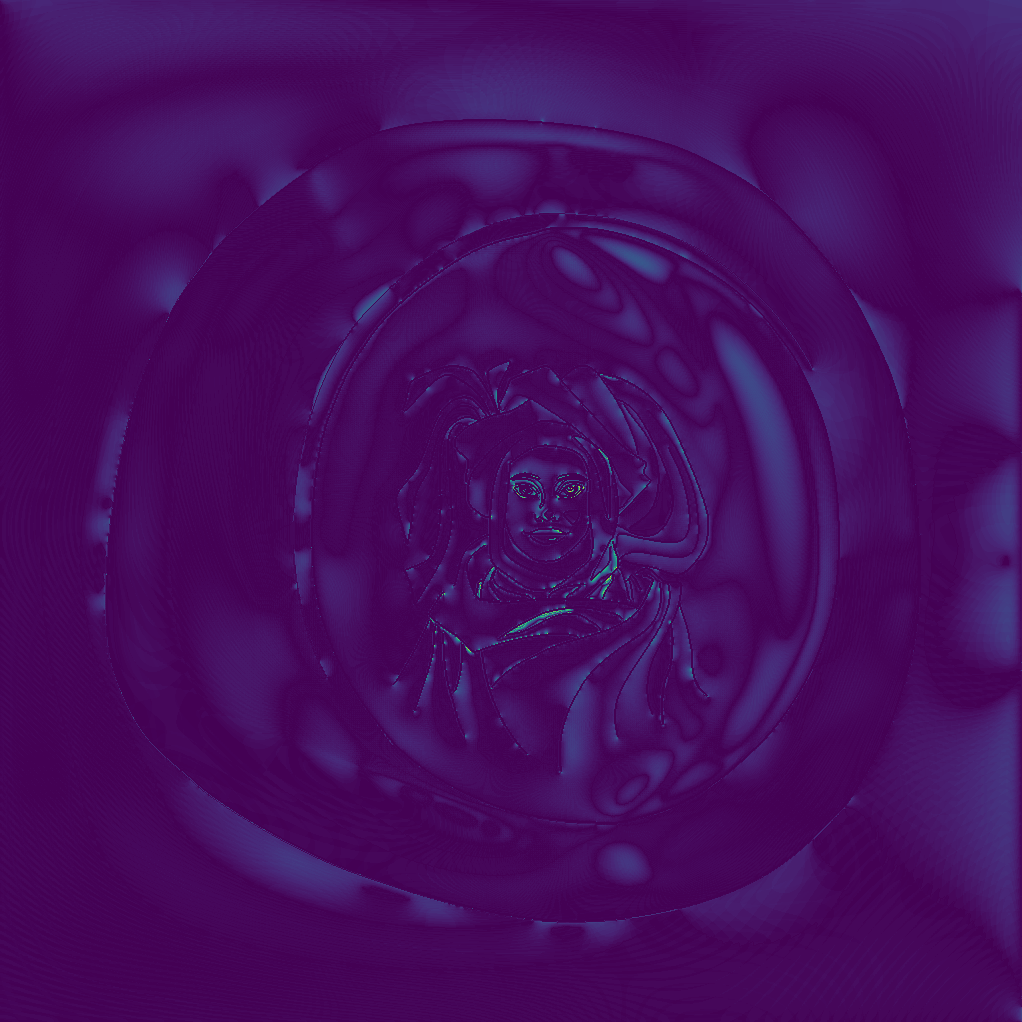}
  {0.011}
    \caption{Taking our result (a) as a reference, we test how well gradient meshes only (b) and diffusion curves only (c)-(d) approximate our result. Difference images show that gradient meshes often better capture inhomogeneous gradients in the background, while diffusion curves capture details in the foreground.}%
    \label{fig:comparison-eval}%
\end{figure*}

\begin{figure*}[t]%
    \centering%
    \includegraphics[width=0.17\linewidth]{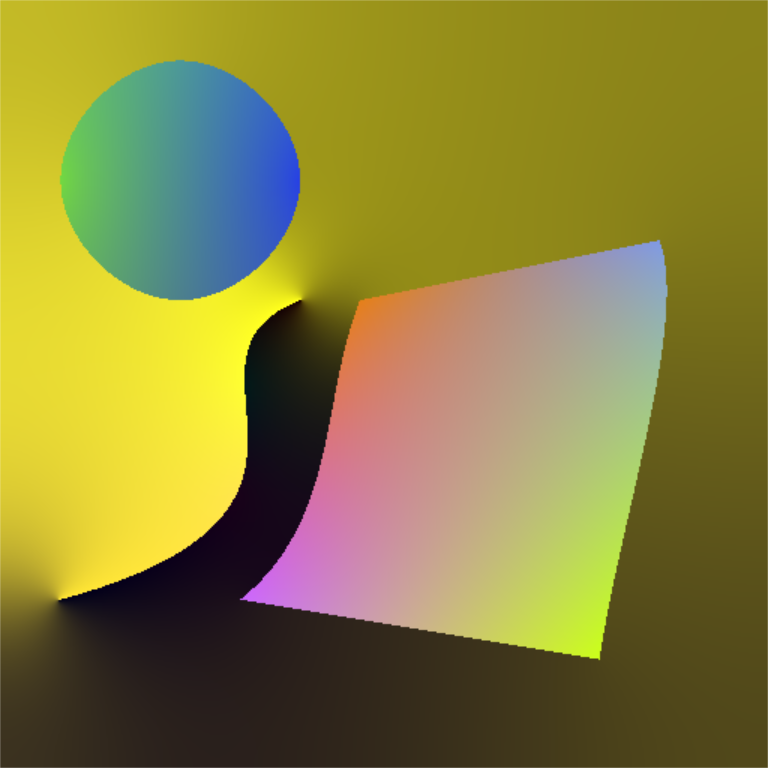}~~%
    \includegraphics[width=0.17\linewidth]{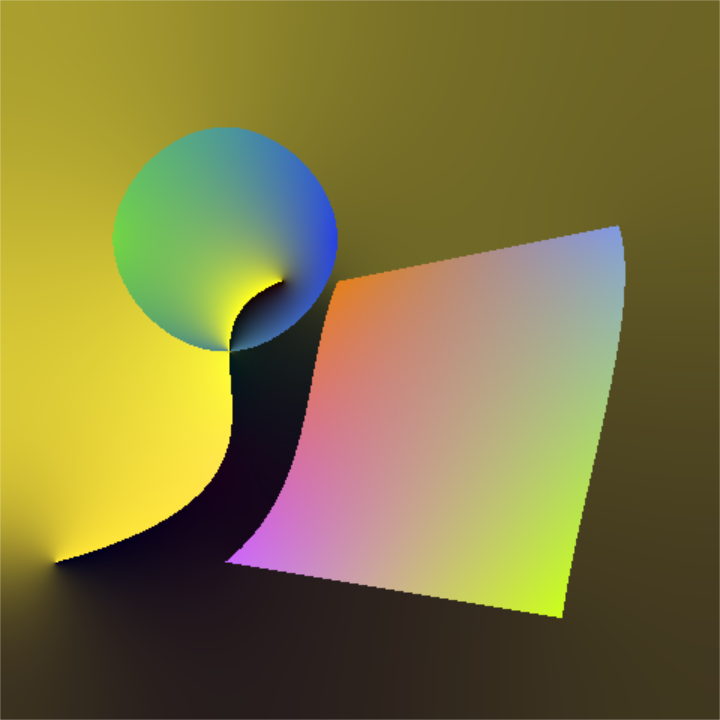}~~%
    \includegraphics[width=0.17\linewidth]{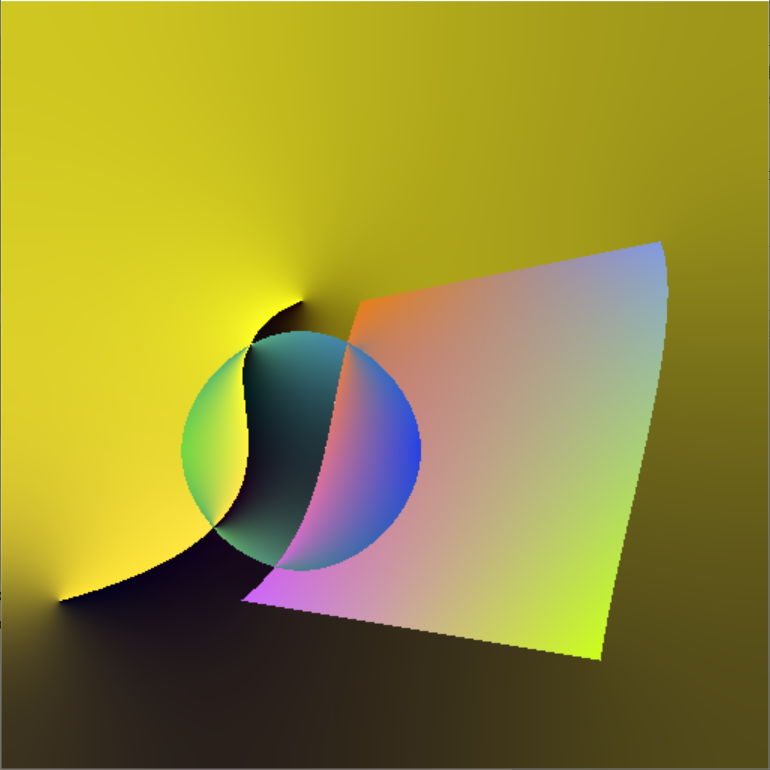}~~%
    \includegraphics[width=0.17\linewidth]{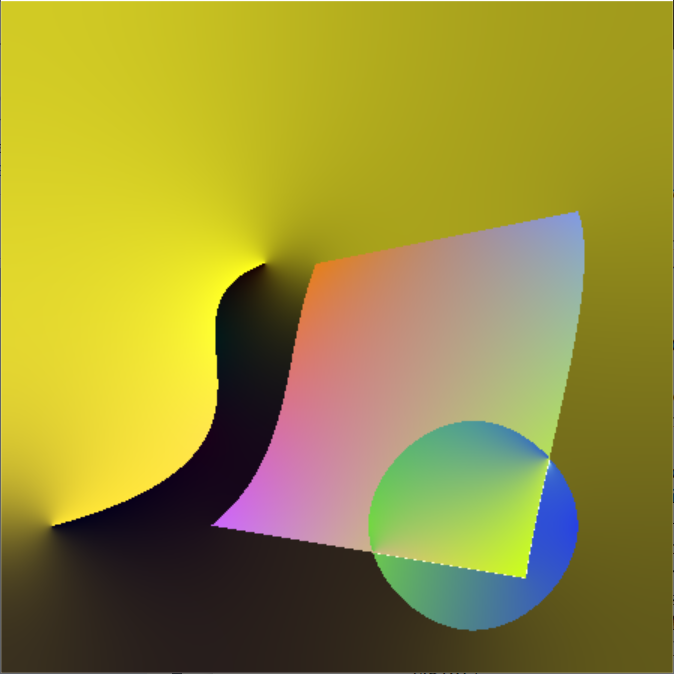}%
    \caption{During user interaction, new patches are formed and destroyed. To achieve a consistent scene editing experience, the user only edits input gradient meshes, diffusion curves, and Poisson curves. The edge graph and the unified patches are not exposed to the user and are created fully automatically.}
    \label{fig:movement}
\end{figure*}

\subsection{Comparison with Vectorizations}
So far, there exists no other method to compare with that supports the image synthesis for scenes that contain both gradient meshes and diffusion curves.
However, once an automatic vectorization algorithm is developed for our unified formulation, it should be possible to compare how well gradient meshes only, diffusion curves only, and our unified formulation are able to represent the same raster image.
For now, we can do a slightly different experiment. 
In Fig.~\ref{fig:comparison-eval}, we take the output from our unified renderer (a) and vectorize this image using gradient meshes (b) and diffusion curves (c)-(d).
For each image, the input primitives and a difference image to (a) are shown.
The gradient meshes are placed manually using Adobe Illustrator.
We can see that gradient meshes have a difficult time adjusting to local structures since they always subdivide entire rows or columns, see for example the swirls behind the crane.
Diffusion curves are useful for modeling sharp edges and details, which can be seen in the foreground objects across all test scenes.
In (c), the diffusion curves are vectorized using Canny edge detection~\citep{Orzan07:CannyLuminance} on the multi-channel color gradient of \cite{DiZenzo86}, as done by \cite{orzan2008diffusion}.
In (d), edges are found using isocontours in the residual field as done by \cite{zhao2017inverse}, although without their subsequent shape optimization.
For both (c) and (d), we use the vectorization methods to fill the areas of the gradient meshes, i.e., we take our diffusion curves as initial set to start from.
We can see in the difference images that many diffusion curves are needed to represent the bi-cubic color gradients of a gradient mesh, which is apparent in scenes with complex backgrounds, such as in the pepper scene. 
While (d) obtains a low reconstruction error, the high curve density is not necessarily user-friendly.
A gradient mesh could have represented the inhomogeneous color gradients more compactly.
With our unified approach, both primitives can be combined in the same scene.

\subsection{User Interaction}
When individual input primitives are moved in the scene by the user, intersections with other primitives will occur that lead to the creation or removal of patches.
To ensure a consistent editing experience, we made sure that all user input is specified on the extended gradient meshes, extended diffusion curves, and Poisson curves alone.
On the front end, the user is not accessing the edge graph or the patch representation, which are both computed fully automatically from the input primitives.
Fig.~\ref{fig:movement} shows the rendered output when taking a closed diffusion curve and moving it across a scene.
Intersections with other diffusion curves and gradient meshes are created, which causes new patches to form or disappear.
Using our data structures, the results remain consistent throughout the editing process.

\setlength{\tabcolsep}{2.8pt}

\begin{table*}[t]%
    \centering%
    \caption{Performance measurements for the various test scenes used in the paper. The columns list the corresponding figure, the number of input diffusion curves (\#DC), the number of input Poisson curves (\#PC), the number of input gradient meshes (\#GM) for which we list the grid resolution in brackets, the number of vertices (\#V) and edges (\#E) in the edge graph, the number of patches (\#P), the computation time to form the edge graph, the computation time to traverse the edge graph, the time needed to resolve the nesting of patches, and the number of Jacobi iterations for rendering. All timings are in milliseconds.}
    \begin{tabular}{c|c|c|c|c|c|c|c|c|c|c|c}%
        Scene & Figure & \#DCs & \#PCs & \#GMs & \#Vs  & \#Es & \#Ps& Form graph & Trav. graph & Form patches & It. \\\hline%
        Crane 
        & Fig.~\ref{fig:teaser} 
        & 68            
        & 5             
        & 3             
        & 160           
        & 124            
        & 8             
        & 16              
        & 5              
        & 13              
        & 1\,k 
        \\ \hline
        Pepper 
        & Fig.~\ref{fig:qualitative-eval} (first row)  
        & 36            
        & 3             
        & 1             
        & 184            
        & 165            
        & 8             
        & 11            
        & 9            
        & 12       
        & 10\,k 
        \\ \hline
        Sunset 
        & Fig.~\ref{fig:qualitative-eval} (second row) 
        & 38            
        & 6             
        & 1             
        & 88            
        & 108            
        & 37            
        & 16            
        & 2            
        & 7              
        & 15\,k 
        \\ \hline
        Bubble
        & Fig.~\ref{fig:qualitative-eval} (third row) 
        & 9            
        & 0             
        & 3      
        & 62            
        & 53            
        & 3            
        & 2            
        & 1              
        & 1              
        & 15\,k 
        \\ \hline%
        Ladybug 
        & Fig.~\ref{fig:qualitative-eval} (fourth row) 
        & 59            
        & 8             
        & 2      
        & 111            
        & 113            
        & 20            
        & 14            
        & 2              
        & 16              
        & 2\,k 
        \\ \hline%
        Portal
        & Fig.~\ref{fig:qualitative-eval} (fifth row) 
        & 87            
        & 1             
        & 3             
        & 211           
        & 167           
        & 17            
        & 20            
        & 3              
        & 14              
        & 15\,k 
    \end{tabular}%
    \label{tab:performance}
\end{table*}

\subsection{Performance Analysis}
We evaluated our method on an Intel Core i9-10980XE with 4.6\,Ghz and an Nvidia RTX 2080 TI GPU.
The performance measurements are reported in Table~\ref{tab:performance}.
To give context we report for each scene statistics about the complexity of the input data (number of diffusion curves, number of Poisson curves, number of gradient meshes), statistics on the complexity of the edge graph (number of vertices and edges), and the number of patches in our patch representation.
We report the time it takes to compute the edge graph, to traverse the edge graph during the patch construction, and to resolve the nesting relationships.

Across all test scenes, the edge graph construction was at around 12-17\,ms, the edge graph traversal at 4-10\,ms, and the patch creation at 13-45\,ms. In sum, we achieved across all scenes an interactive frame rate (29-67 milliseconds per frame).
While we incorporated bounding volume hierarchies to accelerate the intersection and containment tests, we think that further improvements could include partial updates of both the edge graph and the patch representation whenever an object moves.
Further, a hierarchical discretization could be used to accelerate the winding angle tests.
Despite us rebuilding the graph and patches from scratch, we obtained interactive results.

\begin{figure}[b]%
    \centering%
    \begin{minipage}{0.45\linewidth}
        \begin{tikzpicture}[scale=0.68]
            \tikzstyle{every node}=[font=\Large]
            \begin{axis}[
                xlabel={Total number of vertices in edge graph},
                ylabel={Time (ms)},
                yscale=0.66,
                xtick={0, 1000, 2000, 3000},
                ytick={0, 2000, 4000, 6000, 8000, 10000},
                grid style=dashed,
                scaled ticks=false,
                xticklabel style = {
                   /pgf/number format/fixed,
                   /pgf/number format/precision = 0,
                    font = \sffamily,
                },
            ]
            \addplot[
                color=lineOne, style={ultra thick},mark=*
                ]
                coordinates {(55,4.3807705)(199,59.7884)(461,337.514946)(860,1218.9363)(1290,2798.61361)(1854,5948.02925)(2544,11147.02166)};
            \end{axis}
        \end{tikzpicture}  
    \end{minipage}\quad%
    \begin{minipage}{0.45\linewidth}
        \begin{tikzpicture}[scale=0.68]
            \tikzstyle{every node}=[font=\Large]
            \begin{axis}[
                xlabel={Total number of vertices in polylines},
                ylabel={Time (ms)},
                yscale=0.66,
                xtick={0, 25000, 50000, 75000, 100000},
                grid style=dashed,
                scaled ticks=false,
                xticklabel style = {
                   /pgf/number format/fixed,
                   /pgf/number format/precision = 0,
                    font = \sffamily,
                },
            ]
            \addplot[
                color=lineThree, style={ultra thick},mark=*
                ]
                coordinates {(3222,29.259013)(4522,38.975281)(9812,90.89083)(13882,115.89891)(30828,266.07919)(42918,398.17891)(96746,1046.26542)};
            \end{axis}
        \end{tikzpicture}  
    \end{minipage}
    \caption{Scalability experiments (total time in milliseconds) for varying input complexity (left) and discretization (right) on an Intel Core i7-6700K.}%
    \label{fig:scalability}
\end{figure}%

To assess the scalability, we randomly generated a varying number of diffusion curves in the unit square. The x-coordinates of the control points were set to $(0,0.3,0.7,1.0)$. The y-coordinates are random numbers in $[0,1]$. This test leads to a large number of edge crossings. In Fig.~\ref{fig:scalability} (left), we plot the total construction time as a function of the number of vertices $|\cV|$ in the edge graph, which is a measure of the scene complexity. We observe that the runtime scales quadratically in the number of graph vertices, which can be expected since each edge is tested for intersections against each other edge.

Further, we examine the scalability of the algorithm for varying polyline discretizations in the crane scene. 
In Fig.~\ref{fig:scalability} (right), we varied the threshold $\tau\in\{0.1, 0.05, \dots, 0.0001\}$, which leads to an increasing number of polyline vertices.
We observe that the total construction time scales nearly linear in the number of polyline vertices.
This observation holds for the other scenes, as well, but with a different slope.

\subsection{Discussion}
The edge graph and our resulting patch construction directly depend on the quality of the input primitives.
If the input primitives contain tiny gaps or small crossings, then a corresponding patch representation would carry on those artifacts.
In recent years, more exact vector graphics rendering algorithms have been developed and utilized, such as Walk-on-Spheres~\citep{Sawhney20:WoS}, Walk-on-Boundaries~\citep{Sugimoto23:WoB}, or Walk-on-Stars~\citep{Sawhney23:WoSt}, or Walkin' Robin~\citep{Miller24:Robin}.
These approaches likewise suffer from inaccuracies in the scene description, since they are able to resolve the color transport through narrow gaps.
In Section~\ref{sec:edge-graph-implementation-detail}, we described our scene preprocessing, which closes gaps to prevent color leaking. To alleviate the need for such preprocessing, more research towards user interfaces that naturally steer the user towards more accurate scene descriptions would be helpful for both scene modeling and rendering.

\section{Conclusions}
In this paper, we unified the mathematical modeling of mesh-based vector graphics and curve-based vector graphics, which allows for the first time to include both types of primitives in the same scene.
For this, we rephrased the interpolation inside gradient meshes as solution to a Poisson problem by searching for a color field that matches the Laplacian of the gradient mesh.
We developed a four-stage pipeline that takes gradient meshes, diffusion curves, and Poisson curves as input.
First, we extended gradient meshes and diffusion curves by enabling the specification of Neumann conditions in addition to the usual Dirichlet conditions.
Second, our approach handles arbitrary intersections of diffusion curves and gradient meshes, since all intersections are resolved through the construction of an edge graph.
Third, from the edge graph, non-overlapping patches with well-defined boundary conditions and a target Laplacian are derived.
Fourth, a standard Poisson problem is solved in the interior of each patch to synthesize the output image.
The final rasterized images can be computed with any off-the-shelf Poisson solver.

In the past, research on mesh-based and curve-based smooth vector graphics followed independent research threads, which concentrated on the editing, rasterization, or vectorization of the individual primitives.
We hope that the unified treatment will spur further research that promotes synergies of the two approaches, for example regarding novel image synthesis or vectorization methods.
The next step in the generalization will include the incorporation of bi-harmonic diffusion curves~\cite{finch2011freeform}, for which harmonic solutions can be searched in the Laplacian domain.
Similar to common content creation tools, it could be useful to introduce multiple layers of smooth vector graphics.
Further, it would be interesting to develop a vectorization algorithm for the unified patch description. It would be imaginable to divide the input domain into regions. Per region, the algorithm could decide whether it is better to fit a gradient mesh or a set of diffusion curves.
We leave the development of a vectorization algorithm to future work.

\appendix

\section{Laplacian of Gradient Mesh}
\label{sec:gradient-mesh-fill}
Gradient meshes consist of one or multiple Ferguson patches, each containing bi-cubic color patches, whose Laplacian is not necessarily homogeneous.
In fact, the ability to add more variation to the color gradient is what makes gradient meshes artistically expressive.
In the following, we let $\vu=(u,v)\in[0,1]^2$ be the UV coordinate that parameterizes a single Ferguson patch with colors $\vc(\vu)$ and 2D spatial coordinates $\xx(\vu)$, following Eq.~\eqref{eq:gradient-mesh}.
Since a Ferguson patch is parameterized in UV coordinates $\vu$ and since the Poisson equation in Eq.~\eqref{eq:poisson-problem} is in spatial coordinates $\xx$, coordinate transformations are needed, including the mapping from $\xx$ to $\vu$ and its coordinate partials, as well as the gradient and the Laplacian of color with respect to $\xx$, as explained in the following.

\paragraph{Pre-Image}
The pre-image from image space coordinates $\xx$ back to $\vu$ is later for simplicity referred to as $\vu(\xx)$.
The pre-image exists if the coordinates $\vx(\vu)$ do not contain folds, i.e., if $\vx(\vu)$ is a homeomorphism.
Inside that function, we find the $\vu_0$ coordinate that maps to image space coordinate $\xx_0$ by solving the root finding problem $\xx(\vu_0)-\xx_0 = \vNull$, which is done using B\'ezier clipping~\citep{Sederberg90:BezierClipping}.
Thus, given an image space coordinate $\xx$, we can evaluate the color of a Ferguson patch at the corresponding UV coordinate via $\vc(\vu(\xx))$.

\paragraph{Coordinate Jacobian}
The partial derivatives of the inverse coordinate transformation are for $\xx(\vu) = (x(u,v), y(u,v))^\Transp$:
\begin{align}
&
\begingroup
\setlength\arraycolsep{1.5pt}
\begin{pmatrix}
\frac{\partial}{\partial x} u &
\frac{\partial}{\partial x} v &
0 & 0 & 0 \\
\frac{\partial}{\partial y} u &
\frac{\partial}{\partial y} v &
0 & 0 & 0 \\
\frac{\partial^2}{\partial x^2} u &
\frac{\partial^2}{\partial x^2} v &
\frac{\partial}{\partial x} u \frac{\partial}{\partial x} u &
\frac{\partial}{\partial x} v \frac{\partial}{\partial x} v &
2\frac{\partial}{\partial x} u \frac{\partial}{\partial x} v \\
\frac{\partial^2}{\partial y^2} u &
\frac{\partial^2}{\partial y^2} v &
\frac{\partial}{\partial y} u \frac{\partial}{\partial y} u &
\frac{\partial}{\partial y} v \frac{\partial}{\partial y} v &
2\frac{\partial}{\partial y} u \frac{\partial}{\partial y} v \\
\frac{\partial^2}{\partial x \partial y} u &
\frac{\partial^2}{\partial x \partial y} v &
\frac{\partial}{\partial x} u \frac{\partial}{\partial y} u &
\frac{\partial}{\partial x} v \frac{\partial}{\partial y} v &
\frac{\partial}{\partial x} u \frac{\partial}{\partial y} v + \frac{\partial}{\partial y} u \frac{\partial}{\partial x} v
\end{pmatrix} 
\endgroup
\end{align}
\begin{align}
=& 
\begingroup
\setlength\arraycolsep{1.5pt}
\begin{pmatrix}
\frac{\partial}{\partial u} x &
\frac{\partial}{\partial u} y &
0 &
0 &
0 \\
\frac{\partial}{\partial v} x &
\frac{\partial}{\partial v} y &
0 &
0 &
0 \\
\frac{\partial^2}{\partial u^2} x &
\frac{\partial^2}{\partial u^2} y &
\frac{\partial}{\partial u} x \frac{\partial}{\partial u} x &
\frac{\partial}{\partial u} y \frac{\partial}{\partial u} y &
2\frac{\partial}{\partial u} x \frac{\partial}{\partial u} y \\
\frac{\partial^2}{\partial v^2} x &
\frac{\partial^2}{\partial v^2} y &
\frac{\partial}{\partial v} x \frac{\partial}{\partial v} x &
\frac{\partial}{\partial v} y \frac{\partial}{\partial v} y &
2\frac{\partial}{\partial v} x \frac{\partial}{\partial v} y \\
\frac{\partial^2}{\partial u \partial v} x &
\frac{\partial^2}{\partial u \partial v} y &
\frac{\partial}{\partial u} x \frac{\partial}{\partial v} x &
\frac{\partial}{\partial u} y \frac{\partial}{\partial v} y &
\frac{\partial}{\partial u} x \frac{\partial}{\partial v} y + \frac{\partial}{\partial v} x \frac{\partial}{\partial u} y
\end{pmatrix}^{-1}
\endgroup
\end{align}

\paragraph{Gradient}
The spatial color gradient is by chain rule:
\begin{align}
    \frac{\partial \vc(\vu(\xx))}{\partial \xx} = \frac{\partial c(\vu(\xx))}{\partial \vu} \frac{\partial \vu(\xx)}{\partial \xx}.
\end{align}

\paragraph{Laplacian}
The Poisson problem in Eq.~\eqref{eq:poisson-problem} requires the Laplacian of color with respect to image space $\Delta_\xx \vc(\vu(\xx))$, which is likewise computed via chain rule:
\begin{align}
    \Delta_\xx \vc(\vu(\xx)) &= 
    \frac{\partial^2 \vc(\vu(\xx))}{\partial x^2} + \frac{\partial^2 \vc(\vu(\xx))}{\partial y^2}  \\
    &= 
    \frac{\partial \vu(\xx)}{\partial x}^\Transp \, \frac{\partial^2 \vc(\vu(\xx))}{\partial \vu^2}  \, \frac{\partial \vu(\xx)}{\partial x} 
    + 
    \frac{\partial \vc(\vu(\xx))}{\partial \vu}^\Transp \frac{\partial^2 \vu}{\partial x^2} \nonumber \\
    &+
    \frac{\partial \vu(\xx)}{\partial y}^\Transp \, \frac{\partial^2 \vc(\vu(\xx))}{\partial \vu^2}  \, \frac{\partial \vu(\xx)}{\partial y} 
    + 
    \frac{\partial \vc(\vu(\xx))}{\partial \vu}^\Transp \frac{\partial^2 \vu}{\partial y^2}. \nonumber
\end{align}
Using the above ingredients, we can now add the Laplacian of each Ferguson patch of a gradient mesh $\Delta_\xx\vc(\vu(\xx))$ to the patch Laplacian $\vf(\xx)$, see Eq.~\eqref{eq:patch-Laplacian}.
The symbolic derivatives as described above are an alternative to the numerical computation via finite differences from a rasterized gradient mesh image.
Unlike symbolic derivatives, the error of the numerical computation depends on the image resolution, which is likewise the case when the PDE solver estimates the derivatives with finite differences.
The symbolic derivatives could also be useful for differentiable rasterizers that might be employed in the future for automatic vectorization.


\end{document}